Abstract

# Circuit Quantum Electrodynamics

Lev Samuel Bishop

2010

Circuit Quantum Electrodynamics (cQED), the study of the interaction between superconducting circuits behaving as artificial atoms and 1-dimensional transmission-line resonators, has shown much promise for quantum information processing tasks. For the purposes of quantum computing it is usual to approximate the artificial atoms as 2-level qubits, and much effort has been expended on attempts to isolate these qubits from the environment and to invent ever more sophisticated control and measurement schemes. Rather than focussing on these technological aspects of the field, this thesis investigates the opportunities for using these carefully engineered systems for answering questions of fundamental physics. The low dissipation and small mode volume of the circuits allows easy access to the strong-coupling regime of quantum optics, where one can investigate the interaction of light and matter at the level of single atoms and photons. A signature of strong coupling is the splitting of the cavity transmission peak into a pair of resolvable peaks when a single resonant atom is placed inside the cavity—an effect known as vacuum Rabi splitting. The cQED architecture is ideally suited for going beyond this linear response effect. This thesis shows that increasing the drive power results in two unique nonlinear features in the transmitted heterodyne signal: the supersplitting of each vacuum Rabi peak into a doublet, and the appearance of additional peaks with the characteristic $\sqrt{n}$ spacing of the Jaynes–Cummings ladder. These constitute direct evidence for the coupling between the quantized microwave field and the anharmonic spectrum of a superconducting qubit acting as an artificial atom. This thesis also addresses the idea of Bell tests, which are experiments that aim to disprove certain types of classical theories, presenting a proposed method for preparing maximally entangled 3-qubit states via a 'preparation by measurement' scheme using an optimized filter on the time-dependent signal obtained via homodyne monitoring of the transmitted microwave field.

# Circuit Quantum Electrodynamics

A Dissertation
Presented to the Faculty of the Graduate School
of
Yale University
in Candidacy for the Degree of
Doctor of Philosophy

by
Lev Samuel Bishop

Dissertation Director: Professor Steven M. Girvin

May 2010

# Contents

# List of Figures

*for Cleo*

# Acknowledgements

This thesis would never have been possible without the ideas, friendship, support and assistance of numerous people. Foremost among them is of course Steve Girvin, who is infinitely more patient than the stereotypical infinitely-patient doctoral advisor.* He is a constant source of ideas and inspiration, and I feel enormously privileged to have him as my advisor.

I owe much to Jens Koch, who between numerous cups of coffee showed me how to write a scientific article. (I was slow to learn, so he taught me twice.)

My experimental colleague Jerry Chow went off on his own tangent to produce the beautiful data presented in chapter 4. He then put up with my demands for ever more precise numbers to feed into my simulations, which he was able to provide without breaking off from his $l \times m \times n$ multitasking in $l$ windows on $m$ virtual desktops on $n$ monitors.

I have bounced many a crazy-sounding idea off Andrew Houck in order to gauge *exactly how crazy* the idea may be. Dave Schuster has coefficient of restitution larger than unity: my crazy ideas bounce off him and come back at me *much more crazy*. I can guarantee to overcome any mental blocks by talking to him for an hour.

Eran Ginossar, Andreas Nunnenkamp and Lars Tornberg each contributed aspects of the work presented in chapter 5. Our weekly (later, daily) group meetings and pairwise problem-solving sessions during that time are a fond memory of mine, despite my spending much of the time in a state of confusion.

Jay Gambetta advocated the quantum trajectories approach that proved very fruitful

* Akin to TeX `fill` versus `fil`.

in chapter 5, and I learned much of what I know about circuit QED by talking to him and reading his papers.

I thank all of my friends, especially my friends on the 4th floor of Becton—the community that provided the experimental motivation for this thesis. It is a special environment where theorists and experimentalists can collaborate so closely.

Most importantly, I must thank my family, for their love and encouragement. It is impossible to record how grateful I am to my parents, nor can I imagine that I could have completed this thesis without the love of my wife and best friend June.

# Publication list

This thesis is based in part on the following published articles:

1. L. S. Bishop, J. M. Chow, J. Koch, A. A. Houck, M. H. Devoret, E. Thuneberg, S. M. Girvin, and R. J. Schoelkopf, "Nonlinear response of the vacuum Rabi resonance," *Nature Physics* **5**, 105–109 (2009).
2. J. M. Chow, J. M. Gambetta, L. Tornberg, J. Koch, L. S. Bishop, A. A. Houck, B. R. Johnson, L. Frunzio, S. M. Girvin, and R. J. Schoelkopf, "Randomized benchmarking and process tomography for gate errors in a solid-state qubit," *Physical Review Letters* **102**, 090502 (2009).
3. L. S. Bishop, L. Tornberg, D. Price, E. Ginossar, A. Nunnenkamp, A. A. Houck, J. M. Gambetta, J. Koch, G. Johansson, S. M. Girvin, and R. J. Schoelkopf, "Proposal for generating and detecting multi-qubit GHZ states in circuit QED," *New Journal of Physics* **11**, 073040 (2009).
4. L. DiCarlo, J. M. Chow, J. M. Gambetta, L. S. Bishop, B. R. Johnson, D. I. Schuster, J. Majer, A. Blais, L. Frunzio, S. M. Girvin, and R. J. Schoelkopf, "Demonstration of two-qubit algorithms with a superconducting quantum processor," *Nature* **460**, 240–244 (2009).

# Nomenclature

*Abbreviations:*

CPB Cooper Pair Box, see section 2.4.

CPW Coplanar waveguide.

cQED Circuit quantum electrodynamics, see chapter 2.

HEMT High electron mobility transistor: a low-noise-temperature semiconductor amplifier (still a factor of ~ 20 noisier than the quantum limit), see section 4.2.2.

LHV Local hidden variable. Describes a *classical* theory which tries to reproduce some quantum results by making use of additional *unobserved* degrees of freedom.

RWA Rotating wave approximation.

SME Stochastic master equation.

∼

*Symbols:*

$\{\cdot,\cdot\}$ Anticommutator: $\{x, y\} = xy + yx$, see (3.12).

$[\cdot,\cdot]$ Commutator $[x, y] = xy - yx$, see (2.8).

$|n, j\rangle$ The *bare* state with $n$ cavity excitations and $j$ transmon excitations, see (2.55).

$|\cdot\rangle_j$ Denotes the state of the $j$th qubit, see chapter 5.

$|n, \pm\rangle$ The Jaynes–Cummings eigenstates, see (2.51).

$|\Uparrow\rangle$ The state with all qubits in their excited state: $|\Uparrow\rangle = |\uparrow\uparrow\cdots\uparrow\rangle$, see section 5.6.

∼

*Latin Letters:*

$A$ The heterodyne amplitude $A = \sqrt{I^2 + Q^2}$, see (4.8).

$A_1$ The heterodyne signal resulting from a coherent state with a mean cavity occupancy of one photon $A_1^2 = 4V_0^2$, see (4.8).

$a^\dagger, a$ Creation, annihilation operators for a resonator, see equations (2.13) and (2.24).

$b^\dagger, b$ Creation, annihilation operators for the harmonic part of the transmon Hamiltonian, see (2.35).

$b_{\text{in}}, b_{\text{out}}$ The incoming, outgoing combination of bath modes that interacts with the system at time $t$, see (4.6).

$c^\dagger, c$ Ladder operators for the transmon: $c = \sum_j \frac{n_{j,j+1}}{n_{0,1}} |j\rangle \langle j+1|$, see (2.43).

$\mathcal{D}[\cdot]\cdot$ Dissipator: $\mathcal{D}[A]\rho = A\rho A^\dagger - \{A^\dagger A, \rho\}/2$, see (3.12).

$d$ Dipole moment $d = eL$, see section 4.1.

$E_\text{C}$ Single-electron charging energy $E_\text{C} = e^2/2C$, see (2.26).

$E_\text{J}$ Josephson energy, see (2.25).

$E_\text{J}^{\text{max}}$ Maximum value for the effective $E_\text{J}$ under flux tuning, see (2.46).

$E_m$ Energy of $m$th eigenstate of the transmon Hamiltonian, see (2.29).

$|\text{GHZ}\rangle$ The GHZ state $|\text{GHZ}\rangle = \big(|\uparrow\uparrow \cdots \uparrow\rangle + |\downarrow\downarrow \cdots \downarrow\rangle\big)/\sqrt{2}$, see (5.2).

$g$ Coupling strength: $g = g_{01}$, see (2.49). For $\omega_{01} = \omega_\text{r}$ this is the vacuum Rabi frequency, see chapter 4.

$g_{ij}$ Coupling strength: $g_{ij} = \beta\langle\, i \,|\, n \,|\, j \,\rangle$, see (2.48).

$H_\text{TL}$ Effective two-level Hamiltonian describing supersplitting, see (4.10).

$I, Q$ The quadratures of the electromagnetic field leaving the output port $I = V_0\langle a + a^\dagger\rangle$, $Q = V_0\langle \mathrm{i}a^\dagger - \mathrm{i}a\rangle$, see (4.8).

$i_b$ Current through branch $b$, see section 2.1.

$J(t)$ The measurement trace $J(t) = \sqrt{\Gamma_\text{ci}} \sum_j \langle \delta_j \sigma_j^z \rangle + \zeta(t)$, see (5.11).

$M$ The Bell–Mermin operator. For the case of 3 qubits, $M = \sigma_1^x\sigma_2^x\sigma_3^x - \sigma_1^x\sigma_2^y\sigma_3^y - \sigma_1^y\sigma_2^x\sigma_3^y - \sigma_1^y\sigma_2^y\sigma_3^x$. LHV theories satisfy $-2 \leq M \leq 2$, see equations (5.6) and (5.7).

$\mathcal{M}[\cdot]\cdot$ The measurement superoperator $\mathcal{M}[c]\rho_J = (c - \langle c\rangle)\rho_J/2 + \rho_J(c - \langle c\rangle)/2$, see (5.12).

$N$ Total number of qubits, see chapter 5.

$n$ — Number operator $n = -q/2e$. Counts Cooper pairs, see (2.26).

$n_g$ — Offset charge, see (2.27).

$n_{ij}$ — Matrix elements of number operator $n_{ij} = \langle i \,|\, n \,|\, j \rangle$, see (2.41).

$P_W$ — Projector keeping up to $W$ excitations of the transmon–cavity Hilbert space: $P_W = \sum_{0 \leqslant n+j \leqslant W} |n, j\rangle \langle n, j|$, see section 4.6.1.

$Q_b$ — Charge for branch $b$: $Q_b(t) = \int_{-\infty}^{t} i_b(t')\, \mathrm{d}t'$, see (2.1).

$q^{\mathrm{L}}$, $q^{\mathrm{R}}$ — Denotes the two transmons present in the sample described in section 4.2.

$q_n$ — Charge of node $n$, see (2.6).

$\vec{r}$ — Bloch vector: $\vec{r} = \{x, y, z\}$, $\rho = (\mathbb{1} + x\sigma_x + y\sigma_y + z\sigma_z)/2$, see (2.75).

$s$ — The time-integrated signal $s \propto \int_0^t \mathrm{d}t' \langle b + b^\dagger \rangle$, see section 5.3.1 and (5.14).

$T_1$, $T_2$ — Bloch equation coherence times, see (3.39).

$\mathsf{T}$ — Spanning tree for a circuit, see (2.4).

$\mathrm{tr}_R[\cdot]$ — Partial trace over $\mathcal{R}$, see (3.6).

$V_0$ — a voltage related to the gain of the experimental amplification chain, see (4.8).

$v_b$ — Voltage across branch $b$, see section 2.1.

$W$ — Number of excitations to keep in the truncation of the transmon–cavity Hilbert space, see section 4.6.1.

$X$ — The operator that is implemented by the dispersive readout $X = \sum_j \delta_j \sigma_j^z$, see (5.3).

$Z$ — Characteristic impedance of a resonator, see (2.16b).

*Greek Letters:*

$\alpha$ — Fine structure constant $\alpha \simeq 1/137$, see (4.3).

$\alpha_m$, $\alpha_m^{\mathrm{r}}$ — Absolute, relative anharmonicity of the $m$th transmon level, see equations (2.37) and (2.38).

$\beta$ — The transmon–cavity coupling constant, which can be calculated from the capacitance network, see (2.47).

$\Gamma_{\mathrm{ci}}$ — The effective measurement rate, reduced by an efficiency factor from the maximum rate, see (5.11).

$\gamma$ — The total transmon relaxation rate: $\gamma = \gamma_- + \gamma_+$, see section 4.5.

| | |
|---|---|
| $\gamma_-, \gamma_+$ | Qubit relaxation rate, excitation rate, see (3.43). |
| $\gamma_\varphi$ | Qubit dephasing rate, see (3.43). For transmon $\gamma_\varphi = \gamma_\varphi^\Phi + \gamma_\varphi^\mathrm{C}$, see section 4.5. |
| $\gamma_\varphi^\mathrm{C}$ | Transmon dephasing rate via charge noise, leading to a term in the master equation $\frac{\gamma_\varphi^\mathrm{C}}{2}\mathcal{D}\Big[\sum_j \frac{2\epsilon_j}{\epsilon_1-\epsilon_0}\lvert j\rangle\langle j\rvert\Big]\rho$, see (3.50). |
| $\gamma_\varphi^\Phi$ | Transmon dephasing rate via flux noise, leading to a term in the master equation $\frac{\gamma_\varphi^\Phi}{2}\mathcal{D}\Big[\sum_j 2j\lvert j\rangle\langle j\rvert\Big]\rho$, see (3.51). |
| $\Delta$ | Detuning between the drive and a vacuum Rabi peak: $\Delta = \omega_{01} \mp g - \omega_\mathrm{d}$, see (4.10). |
| $\Delta_j$ | Drive-atom detuning: $\Delta_j = \omega_j - j\omega_\mathrm{d}$, see (2.62). |
| $\Delta_\mathrm{r}$ | Drive-cavity detuning: $\Delta_\mathrm{r} = \omega_\mathrm{r} - \omega_\mathrm{d}$, see (2.62). |
| $\delta$ | Qubit-cavity detuning: $\delta = \omega_\mathrm{q} - \omega_\mathrm{r}$, see (2.52). |
| $\delta_j$ | Fractional contribution for the $j$th qubit to the mean dispersive shift $\delta_j = \chi_j/\bar{\chi}$, see section 5.3.1. |
| $\delta n_{ij}$ | Matrix element dispersion. The peak-to-peak variation in $n_{ij}$ as $n_\mathrm{g}$ is varied, see figure 2.6. |
| $\epsilon_m$ | Charge dispersion for the $m$th transmon level, see (2.32). |
| $\zeta(t)$ | Gaussian white noise with zero mean and $\langle\zeta(t)\zeta(t')\rangle = \delta(t-t')$, see (5.11). |
| $\kappa$ | The total photon relaxation rate: $\kappa = \kappa_- + \kappa_+$, see section 4.5. |
| $\kappa_-, \kappa_+$ | Photon decay, excitation rates, see (3.35). |
| $\xi$ | External drive strength, see (2.58). |
| $\rho$ | Density matrix of the system $\mathcal{S}$, see (3.16). |
| $\rho_\mathrm{s}$ | Steady-state density matrix: $\dot{\rho}_\mathrm{s} = 0$, see section 4.4. |
| $\sigma$ | Density matrix of the universe (both $\mathcal{S}$ and $\mathcal{R}$), see (3.16). |
| $\sigma_\bullet$ | Pauli matrices: $\sigma_\pm = \frac{1}{2}(\sigma_x \pm \mathrm{i}\sigma_y)$, see (2.49). |
| $\sigma_n$ | The measurement noise in each of the $I$ and $Q$ channels, see (4.19). |
| $\tilde{\sigma}_\bullet$ | Pauli operators for the reduced two-level system, see (4.9). |
| $\tilde{\Phi}$ | Externally-applied magnetic flux, see section 2.1. |
| $\Phi_0$ | Superconducting flux quantum $\Phi_0 = h/2e = 2.068 \times 10^{-15}$ Wb, see (2.25). |
| $\Phi_b$ | Flux for branch $b$: $\Phi_b(t) = \int_{-\infty}^{t} v_b(t')\,\mathrm{d}t'$, see (2.1). |

$\phi_n$ Flux of node $n$, see (2.4).

$\varphi$ Gauge invariant phase $\varphi = (2e/\hbar)\phi$, see (2.26).

$\bar{\chi}$ The mean dispersive shift, $\bar{\chi} = \sum_j^N \chi_j/N$, see section 5.3.1.

$\chi_{ij}$ Dispersive coupling, see (2.71).

$\Omega/2\pi$ Rabi frequency: $\Omega = 2\xi g/\Delta_\mathrm{r}$, see (2.67). Additionally, for the two-level model of the supersplitting, the effective drive strength $\Omega = \sqrt{2}\xi$, see (4.10).

$\omega_\mathrm{d}/2\pi$ Drive frequency, see (2.56).

$\omega_i/2\pi$ Frequency of $i$th transmon level $\hbar\omega_i = E_i$, see (2.48).

$\omega_\mathrm{IF}/2\pi$ Intermediate frequency $\omega_\mathrm{IF} = \omega_\mathrm{d} - \omega_\mathrm{LO}$, see figure 4.3 and (4.7).

$\omega_{ij}/2\pi$ Transition frequency between transmon levels $i$ and $j$, $\omega_{ij} = \omega_j - \omega_i$, see (2.49).

$\omega_\mathrm{LO}/2\pi$ Local oscillator frequency, see figure 4.3 and section 4.4.

$\omega_\mathrm{q}/2\pi$ Qubit transition frequency $\omega_\mathrm{q} = \omega_{01}$, see (2.49).

$\omega_R$ Density matrix for the reservoir $\mathcal{R}$, see equations (3.5) and (3.16).

$\omega_\mathrm{r}/2\pi$ Cavity frequency, see (2.24).

*Superscripts:*

L, R Denotes quantities associated with the two transmons $q^\mathrm{L}$ and $q^\mathrm{R}$, see chapter 4.

*Subscripts:*

$b$ Branch label, see section 2.1.

$j$ Denotes quantities relating to the $j$th qubit, see chapter 5.

$m$ Transmon eigenstate, see (2.29).

$n$ Node label, see (2.4).

# CHAPTER 1

# Introduction

Beginning with a suggestion from Feynmann in 1982 [1], and inspired by an argument in 1985 by Deutsch [2], scientists and engineers in a variety of disciplines have been excited by the idea of quantum information processing, in which a computation is carried out by controlling a complex collection of quantum objects. This idea seeks to combine two of the greatest advances in science and technology of the twentieth century: quantum mechanics and the digital computer. The discovery of the celebrated Shor algorithm [3] for discrete logarithms and integer factorization led to the realization that a quantum computer has the possibility to provide huge advances in computational power.

Unfortunately, the practical challenges to making a quantum information device are daunting. To build a quantum computer, the classical bits that store information in an ordinary computer must first be replaced with quantum bits (qubits). These qubits can be composed of any quantum system with two distinct states ('0' and '1'). To exceed truly the capabilities of conventional computers, the quantum engineer must acquire extremely precise control over the quantum domain, prevent any unknown evolution that affects the quantum states (decoherence), and amass many thousands of qubits. These qubits must then be 'wired up' in complex and prescribed arrangements, so that they can interact and communicate their quantum information back and forth during the computation [4]. These challenges are so daunting that many people have wondered whether building a quantum computer is possible at all. Although it would be disappointing to learn that quantum computing is for some fundamental reason impossible, this would be a very important result in itself, since it would indicate that our understanding of quantum mechanics is incomplete. Scott Aaronson has put this very nicely by describing *Shor's trilemma* [5], which states that the existence of the Shor factoring algorithm implies that either

1. The Extended Church–Turing Thesis—the foundation of theoretical computer science for decades—is wrong;
2. Textbook quantum mechanics is wrong; or
3. There exists a fast classical factoring algorithm.

A researcher expressing any one of these opinions is liable to attract the label *crackpot*, but nevertheless at least one of them must be true!

A number of systems have been proposed for implementing such a quantum computer, the most obvious ones being 'natural' quantum systems, such as single atoms, ions or spins, for which the quantum description is well established, and which are routinely manipulated in many laboratories. A more intriguing possibility is to use solid state systems, such as superconducting circuits and quantum dots. These have a technological appeal because they can be designed and fabricated using techniques borrowed from conventional electronics. Being many orders of magnitude larger than the natural quantum systems, for example a typical superconducting qubit comprises some $10^9$ atoms and can easily be seen with the naked eye, the quantum description of these systems is much less familiar.

It is certainly an interesting task to try to build a quantum computer, but this thesis does not take up that challenge, except briefly in the final chapter. Rather, my aim is to show that we can take the technology that has been developed in pursuit of this goal and we can apply it to studying fundamental physics.

## 1.1 Outline of thesis

In order to formulate a quantum description of a physical system it is usual to start from the classical Hamiltonian. Since it is somewhat unusual for electrical circuits to be analyzed in these terms, the first task of this thesis, in chapter 2 is to introduce the general scheme for forming a classical Hamiltonian, which can then be quantized in the canonical fashion. A straightforward example of this formalism is to apply it to the LC oscillator, which seems almost trivially simple, but forms the basis for all the circuits of this thesis: we can analyze the transmission line resonator by representing it as a sum of LC oscillators; similarly the transmon may be viewed as a slightly nonlinear LC oscillator. Adding the nonlinearity to the transmon has a number of non-obvious effects and I devote some space to investigating the anharmonicity, which is the effect that allows the transmon to behave as a qubit or artificial atom, as well as the charge dispersion and the matrix elements. One advantage that artificial atoms have over real atoms is that we can engineer them to have adjustable parameters, and I show how to make the transmon frequency depend on an externally-applied magnetic field.

With these fundamental building blocks we can start to build more complex circuits—the simplest involves one transmon and one resonator, and in the appropriate limit can be described by the well-known Jaynes–Cummings Hamiltonian, which is probably the simplest non-trivial quantum Hamiltonian. Despite the fact that the Jaynes–Cummings Hamiltonian can be solved analytically, it displays a rich set of phenomena, investigated in the remainder of the thesis. We probe and control the circuits by sending microwave frequency signals, so we need to understand how to incorporate this driving into our models. Fortunately this can be done quite simply by moving to a frame that is rotating at the drive frequency. Although the drive is applied to a port connected to the resonator, we are frequently using the drive as a way to control the transmon and hence it is helpful to make a displacement transformation to a frame where the drive term acts directly on the transmon. Finally, although the Jaynes–Cummings Hamiltonian is already quite simple, we can simplify the description even further in the so-called *dispersive limit*, where the qubit and the resonator are far detuned in frequency. In this limit, we can perform 1-qubit gates and we can use the fact that the resonator frequency becomes dependent on the qubit state, in order to measure that state of the qubit.

So far, we have discussed superconducting circuits in isolation, but of course there is always an unavoidable coupling to the environment. Chapter 3 presents the standard ways of formulating an 'open-system' description, as they apply to superconducting circuits. We can strongly constrain the dynamics by requiring the quite-reasonable condition of complete positivity. However, this is not sufficient to allow us to formulate a unique description. We can further simplify matters by making a Markovian approximation, which effectively means that any information leaking out of the system into the environment is instantly forgotten, and which leads directly to the Lindblad and Kossakowski formulations of master equations. These are general forms for allowed master equations, and give no guidance on how to derive a master equation for a specific situation. The weak coupling formalism, due to Davies, provides a means to proceed from a microscopic description of a system weakly coupled to a reservoir, to a Lindblad–Kossakowski master equation with dissipation terms that can be related to positive and negative frequency components of the reservoir correlation function, related by a Boltzmann factor in the event that the reservoir is a heat bath in thermal equilibrium. We can apply this formalism in a straightforward way to the simple cases of a damped harmonic oscillator, or a 2-level system, in the latter case reproducing the standard Bloch equation description. However, the weak-coupling formalism requires a strict separation of frequency scales, and thus does not directly apply to the more complicated situation of the Jaynes–Cummings Hamiltonian. A further problem relates to the fact that the microscopic relaxation processes affecting the transmon are currently not well understood, so we need to make some educated guesses in order to formulate a master equation for this system.

With a master equation for the transmon–cavity system, we can attempt to describe experimental results, such as the observed 'supersplitting' and multiphoton transitions, when the vacuum Rabi splitting is driven so hard that in the absence of the anharmonicity of the Jaynes–Cummings Hamiltonian there would be more than 1000 photons in the cavity, but due to the 'photon blockade' effect there are in fact only around 5 excitations. In order to analyze this situation, we should first understand how the strong coupling limit is reached, being essentially set by the fine structure constant, due to the quasi-one-dimensional nature of cQED. I introduce input-output theory in order to describe the experimentally relevant heterodyne amplitude, and explain how to solve the master equation numerically, thereby explaining the supersplitting and the multiphoton spectrum in exquisite detail. A simple two-level model can also be used to gain insight into the supersplitting, demonstrating that it is primarily a strong-driving effect. The precise agreement between the experiment and theory allows us to draw some conclusions about the system parameters, including the dephasing rates, and allows a stringent upper bound to be placed on the effective system temperature.

Chapter 5 turns to the situation of multiple qubits, describing an interesting way to prepare highly entangled states via 'preparation by measurement', an elegant probabilistic method of state preparation that seems especially well-suited to cQED. Since such states are intimately related to Bell tests (experiments that attempt to disprove certain types of classical theories), I present a short overview of Bell tests in theory and practice, hopefully dismissing some persistent myths. I introduce the concept of quantum trajectories, a powerful theoretical tool for simulating quantum systems and describing the measurement process, and I present an optimized filter for the time-domain experimental homodyne signal, that performs much better for the preparation-by-measurement of entangled states than a simple boxcar filter. The main result of this section is that preparation of a 3-qubit Greenberger–Horne–Zeilinger state by this method is entirely feasible with the experimental parameters available today. I use the same techniques to show that improvements to the experiments, specifically in the noise of the amplifiers, are necessary before a convincing Bell test can be performed.

In chapter 6 I speculate on directions for further research, inspired by the results of this thesis, including some ideas that are relevant to quantum computing.

# CHAPTER 2

# Circuit QED

Circuit Quantum Electrodynamics (cQED) [4] borrows techniques from the field of atomic cavity Quantum Electrodynamics (QED), which studies the interaction of light and matter at the quantum level, in the context of placing one or more atoms inside a high-finesse optical cavity [6, 7]. Amazingly, some circuits, though containing billions of atoms, behave very much like a single atom. This metaphor allows many familiar phenomena from atomic optics to be observed in a rather different context. Although circuits can behave much like artificial atoms, their properties can be quite extreme, allowing cQED to explore regimes of cavity QED that are difficult to reach with ordinary atoms.

The purpose of this chapter is to provide some background on the general topic of cQED at the level that will be needed for the rest of the thesis. The thesis of D. Schuster [8] is an excellent introduction to this topic, and some further aspects are covered in the thesis of L. Tornberg [9].

## 2.1 Circuit quantization

This section briefly reviews the general scheme for circuit quantization, discussed with more pedagogical detail by Devoret [10] and in a more systematic way by Burkard *et al.* [11]. The idea is to be able to start from a lumped-element circuit diagram for a non-dissipative circuit and systematically proceed first to the classical Hamiltonian and thence to the quantum Hamiltonian.

The lumped-element approximation is appropriate when all length scales are much smaller than the electromagnetic wavelength for frequencies of interest. (An explicit example of this is in section 2.3, where we will consider transmission line resonators.) Within the lumped-element approximation we will describe a circuit as a network, where nodes are joined by two-terminal circuit components such as capacitors and inductors. (Circuits may incorporate components with three or more terminals, but for this thesis these are unnecessary.) Each two-terminal component $b$ has a voltage $v_b(t)$ across it and a current $i_b(t)$ through it. The ordinary description of circuits makes use of these voltages and currents. However, for purposes of deriving a Hamiltonian description of the circuit, it is more convenient to work in terms of charges $Q_b(t)$ and fluxes $\Phi_b(t)$ defined as the time integrals of the voltages and currents:

$$\Phi_b(t) = \int_{-\infty}^{t} v_b(t')\,\mathrm{d}t', \tag{2.1a}$$

$$Q_b(t) = \int_{-\infty}^{t} i_b(t')\,\mathrm{d}t', \tag{2.1b}$$

where it is assumed that initially the circuit is at rest, $v_b(-\infty) = i_b(-\infty) = 0$, and that any external bias is switched on adiabatically from $t = -\infty$.

We work with two categories of components: they may be either of capacitive type (possibly nonlinear)

$$v_b = f(Q_b) \tag{2.2}$$

or of inductive type (again, possibly nonlinear)

$$i_b = g(\Phi_b). \tag{2.3}$$

'Real' components can be represented as a combination of such inductors and capacitors. For example, a physical tunnel junction can be modeled as a nonlinear inductor (Josephson element) in parallel with a linear capacitor. External voltage and current sources may be included as an infinite limit of very large capacitors and inductors.

With the above background out of the way, here is a recipe for translating a circuit diagram into a classical Hamiltonian:

1. Represent the circuit as a network of two-terminal capacitors and inductors;

2. Optionally use the usual rules for series and parallel combinations of linear components to simplify the circuit;

3. Choose any one node of the circuit as *ground*. Describe the remaining nodes as *active*;

4. Choose a spanning tree $\mathsf{T}$ of the network (*i.e.*, a loop-free graph that includes all nodes);

5. Introduce a node flux for each active node $n$ as the time-integral of the voltage on the (unique) path on $\mathsf{T}$ from that node to ground:

$$\phi_n(t) = \sum_b S_{nb} \int_{-\infty}^{t} v_b(t')\,\mathrm{d}t', \tag{2.4}$$

where $S_{nb}$ is 0 if the path on $\mathsf{T}$ from ground to $n$ does not pass through $b$ or otherwise it is ±1 depending on the orientation of the path;

6. Find the energy of the capacitive elements $T$ in terms of the branch voltages, and the energy of the inductive elements $V$ in terms of the branch fluxes;

7. Write $T$ and $V$ in terms of the node fluxes (and their time derivatives). For a branch $b$ linking nodes $n$ and $n'$, the branch voltage is the time derivative of the branch flux $v_b = \dot{\Phi}_b$. The branch flux is $\Phi_b = \phi_n - \phi_{n'} + \tilde{\Phi}_{l(b)}$, with $\tilde{\Phi}_{l(b)} = 0$ for $b \in \mathsf{T}$ and otherwise $\tilde{\Phi}_{l(b)}$ is the externally-applied magnetic flux through the loop $l(b)$ that is produced by adding $b$ to $\mathsf{T}$. (Ref. [11] describes how to incorporate mutual inductances);

8. Form the Lagrangian

$$L(\phi_1, \dot{\phi}_1, \ldots, \phi_N, \dot{\phi}_N) = T - V; \tag{2.5}$$

9. Define node charges as the conjugate momenta of the node fluxes, in the usual way:

$$q_n = \frac{\partial L}{\partial \dot{\phi}_n}; \tag{2.6}$$

10. Perform the Legendre transform to obtain the Hamiltonian

$$H(\phi_1, q_1, \ldots, \phi_N, q_N) = \sum_{i=1}^{N} \dot{\phi}_i q_i - L. \tag{2.7}$$

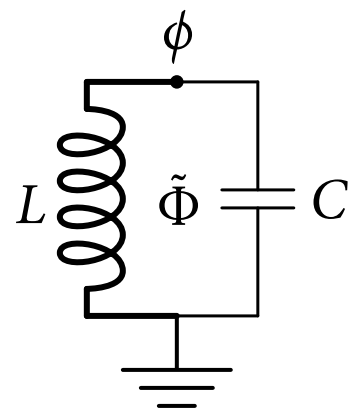

**Figure 2.1: The LC oscillator.** The ground node is chosen as the bottom node of the diagram. There is one active node, with node flux $\phi$. The spanning tree, denoted by heavier lines, is chosen passing through the inductive element. There is an externally applied flux $\tilde{\Phi}$ through the loop formed by the inductor $L$ and the capacitive element $C$.

To proceed from the classical Hamiltonian to the quantum version, replace the classical variables by corresponding quantum operators obeying the proper commutation relations

$$\left[\phi_n, q_n\right] = \mathrm{i}\hbar. \tag{2.8}$$

In section 2.5 we will deal with superconducting circuits that contain *islands*, namely pieces of superconductor with only capacitors and Josephson junctions connecting them to the rest of the circuit, and no d.c. connections. In such cases, it is meaningful to speak of the number of Cooper pairs that have tunneled to the island, and correspondingly it can be shown that the potential energy term in the Hamiltonian is a purely periodic function of the flux. Denoting the angular frequency of this periodicity as $\kappa_n$, we should write the commutation relation (2.8) in the form

$$\left[\exp(\mathrm{i}\kappa_n\phi_n), q_n\right] = -\hbar\kappa_n \exp(\mathrm{i}\kappa_n\phi_n). \tag{2.9}$$

The next sections apply the formalism to some simple circuits that are used in the rest of this thesis. In section 2.2 we warm up on the lumped-element LC oscillator. Trivial though the LC oscillator may seem, it is the basis of all the circuits considered in this thesis: section 2.3 examines the transmission line cavity resonator, showing that it can be treated as a set of infinitely many LC oscillators; section 2.5 represents the transmon as a slightly nonlinear LC oscillator.

## 2.2 Quantum LC oscillator

As a trivial example of the formalism of the previous section, consider the LC oscillator of figure 2.1. This circuit has one active node (thus we suppress the node index $n$ in this section).

Choosing the spanning tree to be the inductive branch, and assuming the externally applied magnetic flux $\tilde{\Phi}$ is constant, the Lagrangian is

$$L(\phi, \dot{\phi}) = \frac{C\dot{\phi}^2}{2} - \frac{\phi^2}{2L}, \tag{2.10}$$

and the Hamiltonian is simply the textbook harmonic oscillator

$$H = \frac{q^2}{2C} + \frac{\phi^2}{2L}, \tag{2.11}$$

which we can quantize in the usual way as

$$H = \hbar\omega\left(a^\dagger a + \frac{1}{2}\right), \tag{2.12}$$

by introducing creation and annihilation operators obeying

$$[a, a^\dagger] = 1, \tag{2.13}$$

$$\phi = \sqrt{\frac{\hbar Z}{2}}\left(a + a^\dagger\right), \quad \text{and} \tag{2.14}$$

$$q = -\mathrm{i}\sqrt{\frac{\hbar}{2Z}}\left(a - a^\dagger\right). \tag{2.15}$$

and where the resonant frequency $\omega$ and characteristic impedance $Z$ are, as expected, given by

$$\omega = \sqrt{\frac{1}{LC}}, \quad \text{and} \tag{2.16a}$$

$$Z = \sqrt{\frac{L}{C}}. \tag{2.16b}$$

## 2.3 Transmission line resonator

A transmission line of length $d$, with capacitance per unit length $c$ and inductance per unit length $l$, may be treated as the continuum limit of a chain of LC oscillators [12]. Such a circuit is shown in figure 2.2. The ground node is marked, and as the spanning tree we choose the capacitive branches. Assuming there are no externally-applied magnetic fluxes,* the

* The effects of *static* externally-applied fluxes can be removed by a canonical transformation.

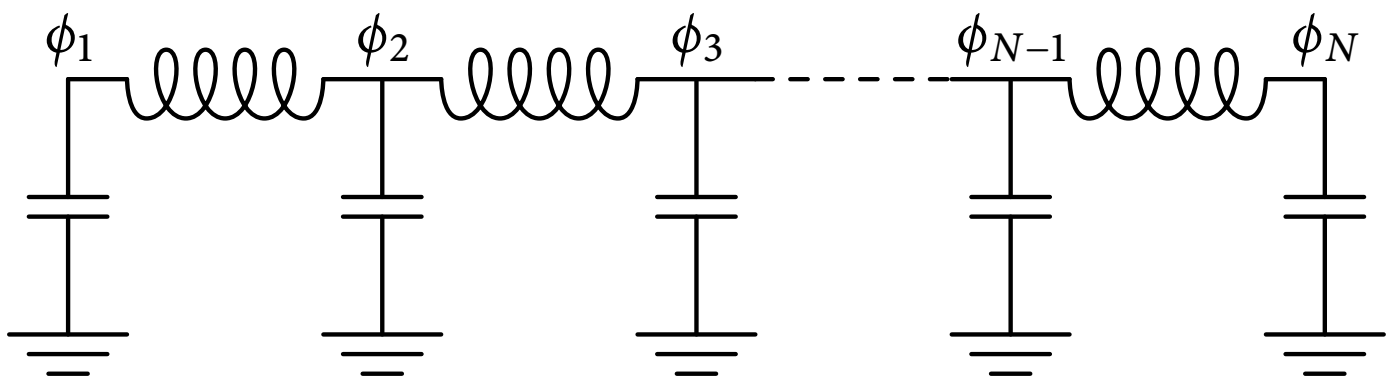

**Figure 2.2: The transmission line.** The figure shows a transmission line with open-circuit boundary conditions, represented as the continuum limit of a chain of LC oscillators.

Lagrangian is

$$L(\phi_1, \dot{\phi}_1, \dots, \phi_N, \dot{\phi}_N) = \sum_{i=1}^{N} \frac{\Delta C \dot{\phi}_i^2}{2} - \sum_{i=1}^{N-1} \frac{(\phi_{i+1} - \phi_i)^2}{2\Delta L}, \tag{2.17}$$

with $\Delta C = cd/N$, $\Delta L = ld/N$. In the continuum limit, $N \to \infty$, this becomes the integral

$$L[\phi(x,t), \dot{\phi}(x,t)] = \int_0^d \frac{c\dot{\phi}(x,t)^2}{2} - \frac{1}{2l}\left(\frac{\partial \phi(x,t)}{\partial x}\right)^2 \mathrm{d}x. \tag{2.18}$$

The Euler–Lagrange equation for $\phi(x,t)$ is thus

$$\frac{\partial^2 \phi}{\partial t^2} - v^2 \frac{\partial^2 \phi}{\partial x^2} = 0, \tag{2.19}$$

where $v = 1/\sqrt{lc}$ is the wave velocity. This has solutions

$$\phi(x,t) = \sum_{n=1}^{\infty} A_n \cos(k_n x + \alpha_n) \cos(k_n v t + \beta_n), \tag{2.20}$$

where $A_n$, $k_n$, $\alpha_n$ and $\beta_n$ depend on the boundary conditions. For the case of open-circuit boundary conditions at $x = 0$ and $x = d$, as shown in the figure, we have

$$\left.\frac{\partial \phi}{\partial x}\right|_{x=0} = \left.\frac{\partial \phi}{\partial x}\right|_{x=d} = 0, \tag{2.21}$$

which gives $\alpha_n = 0$, $k_n = n\pi/d$. ($A_n$ and $\beta_n$ will be determined by the initial conditions.) Substituting (2.20) into (2.18) and integrating out the $x$ dependence yields

$$L(\Phi_1, \dot{\Phi}_1, \dots) = \sum_{n=1}^{\infty} \frac{C_n \dot{\Phi}_n^2}{2} - \frac{\Phi_n^2}{2L_n}, \tag{2.22}$$

where $\Phi_n(t) = A_n \cos(k_n v t + \beta_n)$ keeps the time dependence of the solution. Thus this is an effective Lagrangian for a circuit consisting of uncoupled LC oscillators with effective capaci-tances $C_n = cd/2$ and effective inductances $L_n = 2dl/n^2\pi^2$, and hence resonant frequencies $\omega_n = n\nu\pi/d$. The quantum Hamiltonian for a transmission line cavity is therefore

$$H = \hbar \sum_n \omega_n \left( a_n^\dagger a_n + \frac{1}{2} \right). \tag{2.23}$$

Generally we are only interested in the behavior of a circuit in the vicinity of a particular frequency. In such cases we can pull out only one mode (often the fundamental, $n = 1$) and ignore the dynamics of the other modes. In such cases the cavity Hamiltonian is simply

$$H = \hbar\omega_\mathrm{r} \left( a^\dagger a + \frac{1}{2} \right), \tag{2.24}$$

where $\omega_\mathrm{r}$ is the frequency of the relevant cavity mode, with creation and annihilation operators $a^\dagger$ and $a$. In the cQED literature, the cavity Hamiltonian is usually written as (2.24) without any further explanation.

## 2.4 Qubits and artificial atoms: the need for anharmonicity

Harmonic oscillators, whether resulting from discrete capacitors and inductors or as modes of cavities, are one of the building blocks of cQED. However, harmonic oscillators are not sufficient for all the tasks we would like to perform. The limitation results because although the quantized harmonic oscillator has discrete energy levels, these levels have uniformly increasing energy. This means that it is not possible to address a specific pair of levels and selectively drive a transition between only those levels. To make the physics more interesting, we introduce some anharmonicity into the system, and this will invariably involve adding a Josephson element to the circuit, because it is the only known dissipation-free nonlinear circuit element [13]. There are various schemes for incorporating a junction, leading to what are known as *phase qubits* [14], *flux qubits* [15, 16] and *charge qubits* [17, 18]. An overview of these different topologies is given by Clarke and Wilhelm [19]. These circuits have 'qubit' in the names, which highlights the potential for observing *two-level* physics in these systems. This terminology emphasizes one of the intended uses, namely quantum information processing and quantum computing, where qubits (short for *quantum bits*) fill the rôle played in classical computation by ordinary bits. However, none of these systems is strictly two-level. It is more accurate to say that they have sufficient anharmonicity that they exhibit effective two-level physics within a restricted frequency range. In the cQED viewpoint, these circuits may be described as *artificial atoms*, explicitly allowing for the possibility that more than two levels are relevant.

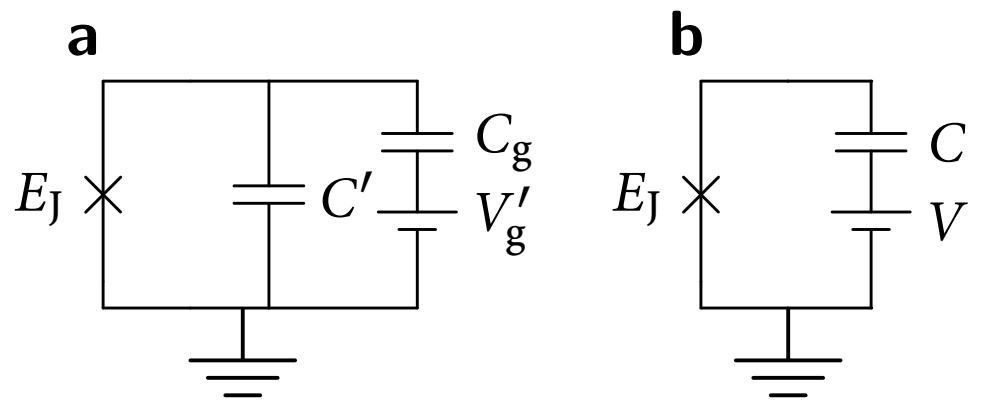

**Figure 2.3: The charge qubit.** **a**, This is the same circuit as in figure 2.1, except the linear inductor $L$ has been replaced by a nonlinear Josephson element $E_J$, and in addition there is a capacitively-coupled gate electrode. **b**, An equivalent circuit, with effective capacitance $C = C_g + C'$ and voltage $V = \frac{C_g}{C_g + C'} V_g$.

The charge qubit gains its anharmonicity by taking the LC oscillator of section 2.2 and replacing the linear inductor by a nonlinear inductor, in the form of a Josephson junction. In one extreme of the charge qubit, known as the Cooper Pair Box (CPB) regime, the anharmonicity is very large, dominating over all other energy scales. In the other extreme, known as the transmon regime, the anharmonicity is a small perturbation on the harmonic behavior. It is this latter situation that is important for this thesis, and is the focus of the remainder of this section.

## 2.5 Charge qubit Hamiltonian

The Josephson element behaves as a nonlinear inductor, with an energy that, due to the discreteness of the Cooper pair charge, is periodic in the flux $-E_J \cos\big((2e/\hbar)\phi\big)$, where the Josephson energy $E_J$ is a property of the junction and depends on the superconducting gap and the barrier transparency. The scale of the nonlinearity is set by the superconducting flux quantum $\Phi_0 = h/2e$. Simply replacing the inductive term in (2.11) gives

$$H = \frac{q^2}{2C} - E_J \cos\left(\frac{2e}{\hbar}\phi\right). \tag{2.25}$$

Introducing the dimensionless *gauge invariant phase* $\varphi = (2e/\hbar)\phi$, which directly corresponds to the phase difference across the junction of the superconducting condensate; the number operator, $n = -q/2e$, which counts how many Cooper pairs have crossed the junction;

and the charging energy $E_C = e^2/2C$, we can write this in a simpler form as

$$H = 4E_C n^2 - E_J \cos\varphi. \tag{2.26}$$

However, this is not quite right because there may be some offset: this could arise due to the intentional presence of a capacitively-coupled *gate electrode* with a d.c. bias voltage, as shown in figure 2.3, or it could be due to other stray couplings. For the linear LC oscillator such a static offset could be removed via a canonical transformation of coordinates. For the nonlinear oscillator this offset must be included explicitly:

$$H = 4E_C(n - n_g)^2 - E_J \cos\varphi, \tag{2.27}$$

where $n_g = Q_r/2e + C_g V_g/2e$ is the effective offset charge, measured in units of the Cooper pair charge, and $Q_r$ represents offset charge due to environmental sources other than the gate electrode. In this circuit, the number operator $n$ has discrete eigenvalues, corresponding to an integer number of Cooper pairs tunneling across the junction, and the phase operator $\varphi$ is a compact variable such that the wavefunction satisfies $\psi(\varphi + 2\pi) = \psi(\varphi)$. Thus, as described in section 2.1, the commutation relation between the conjugate variables $n$ and $\varphi$ is

$$\left[e^{i\varphi}, n\right] = -e^{i\varphi}. \tag{2.28}$$

The Hamiltonian (2.27) may be solved analytically, in terms of special functions: the eigenenergies $E_m$ can be written as

$$E_m(n_g) = E_C a_{2[n_g + k(m,n_g)]}(-E_J/2E_C), \tag{2.29}$$

where $a_\nu(q)$ denotes Mathieu's characteristic value, and $k(m, n_g)$ is a integer-valued function that orders the eigenvalues [20, 21]. For doing numerical calculations, the $a_\nu(q)$ are not easily evaluated. Instead it is preferable to solve (2.27) numerically, diagonalizing in a truncated charge basis

$$H = 4E_C \sum_{j=-N}^{N} (j - n_g)^2 |j\rangle\langle j| - E_J \sum_{j=-N}^{N-1} \big(|j+1\rangle\langle j| + |j\rangle\langle j+1|\big). \tag{2.30}$$

In this form it is clear that the Josephson term in the Hamiltonian describes the tunneling of Cooper pairs. The number of charge basis states that needs to be retained, $2N + 1$, depends on the ratio $E_J/E_C$ and on the number of eigenstates that are relevant to a given situation.

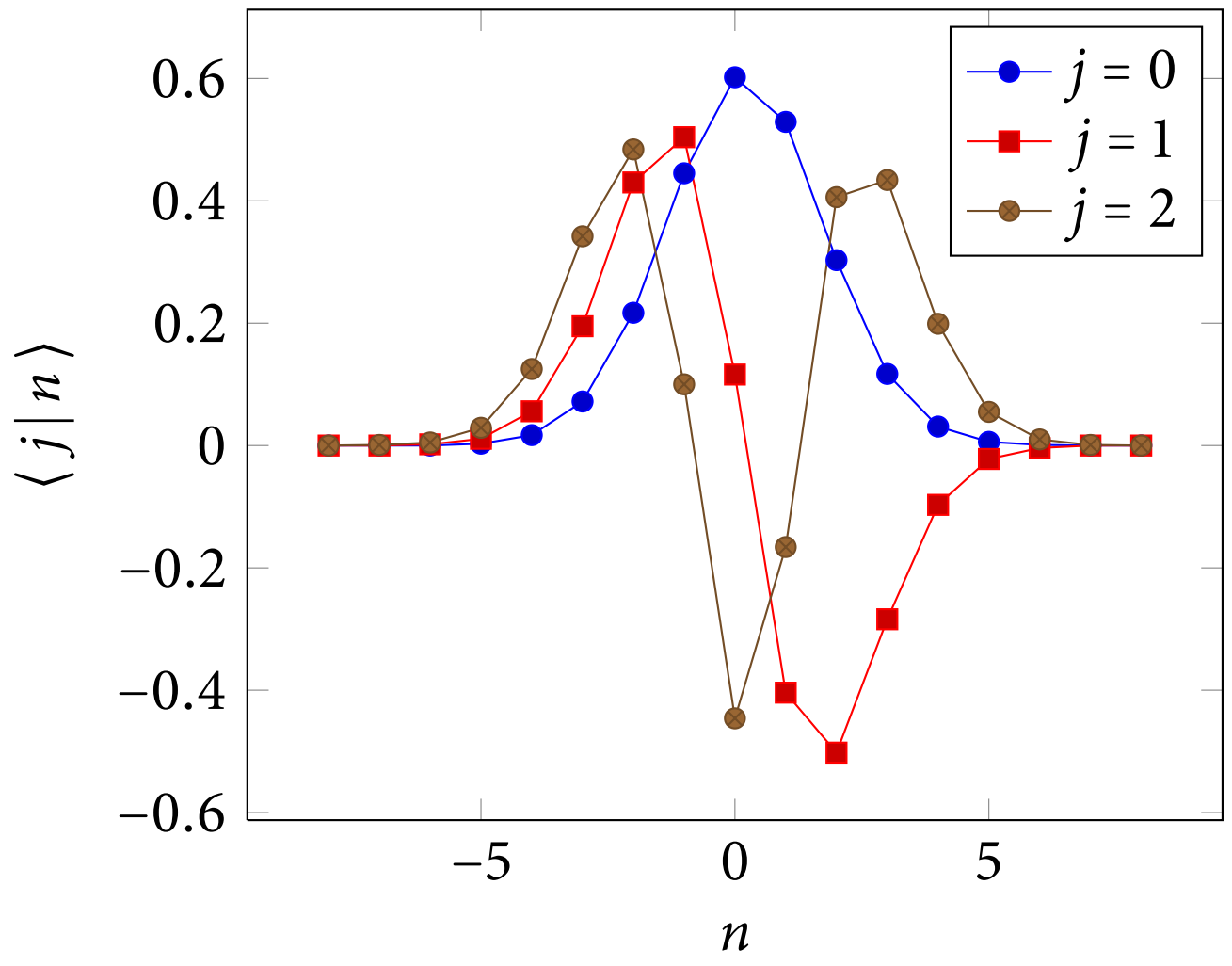

**Figure 2.4: Wavefunctions of the transmon.** The lowest 3 eigenfunctions of the transmon, in the charge basis, for $E_J/E_C = 50$ and $n_g = 0.2$.

Figure 2.4 shows the first 3 eigenvectors in the charge basis for $E_J/E_C = 50$, showing that the charge states with $n \simeq -5, \ldots, 5$ participate strongly. (For the calculations in this thesis, for describing the lowest 3 or 4 levels to sufficient accuracy, approximately 40 charge-basis states were retained. Appendix A contains Mathematica code.) The rest of this section examines the properties of these eigenfunctions of (2.27).

### 2.5.1 Charge dispersion

The original charge qubits [17, 18] operated in the regime $E_J/E_C \simeq 1$. As shown in figure 2.5a, in this regime the energy levels are approximately quadratic with $n_g$ except in the vicinity of level crossings, where a gap of size approximately $E_J$ opens. This is undesirable, because it is experimentally difficult to control $n_g$ to the extremely precise level that is needed to avoid unwanted drifts in the transition frequencies between levels, even when operating at the so-called *sweet spots* (first used in experiments with *quantronium* qubits [22]) where $\partial E_j/\partial n_g = 0$. To avoid this problem, the transmon was introduced [20], which uses a much larger capacitor so as to achieve $E_J/E_C \gg 1$. As can be seen in figure 2.5b–d, when the $E_J/E_C$ ratio increases, the levels flatten greatly and the $n_g$ dependence disappears [21]. We can make this statement more precise by introducing the *charge dispersion*, $\epsilon_m$. For the $m$th energy

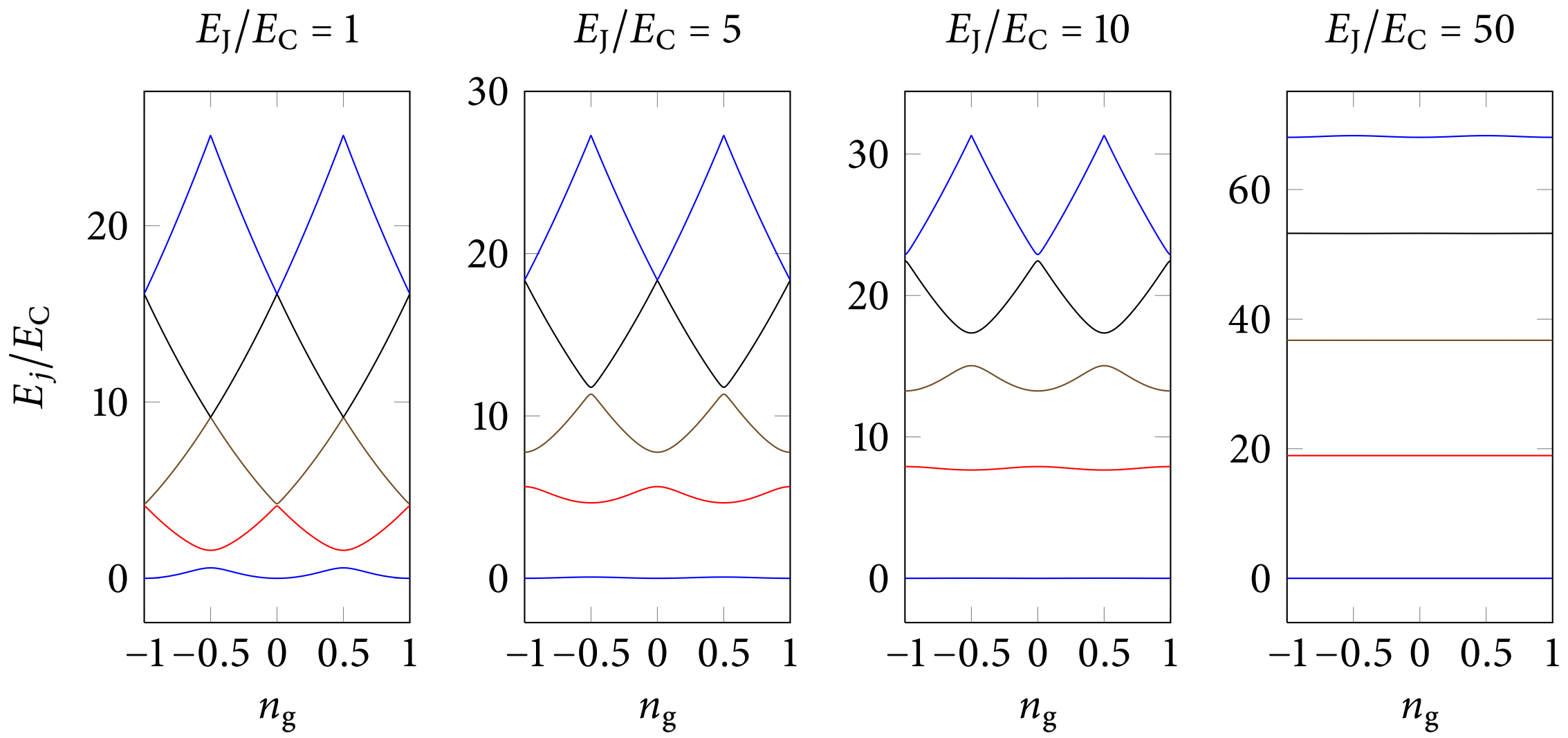

**Figure 2.5: Charge dispersion.** The energies of the lowest 5 levels of the transmon Hamiltonian (2.30), in units of the charging energy $E_C$. For low $E_J/E_C$ ratio, the energies are parabolic functions of the offset charge $n_g$, with avoided crossings; as the ratio is increased the levels become exponentially flatter.

level this is defined as the peak-to-peak energy range as $n_g$ is varied

$$\epsilon_m = E_m(n_g = 1/2) - E_m(n_g = 0). \tag{2.31}$$

In the limit of small charge dispersion, the dispersion relation $E_m(n_g)$ is well approximated as a cosine:

$$E_m(n_g) \simeq E_m(n_g = 1/4) - \frac{\epsilon_m}{2} \cos(2\pi n_g). \tag{2.32}$$

The asymptotics of the Mathieu solution (2.29) give the result that

$$\epsilon_m \simeq (-1)^m E_C \frac{2^{4m+5}}{m!} \sqrt{\frac{2}{\pi}} \left(\frac{E_J}{2E_C}\right)^{\frac{m}{2}+\frac{3}{4}} e^{-\sqrt{8E_J/E_C}}, \tag{2.33}$$

valid for $E_J/E_C \gg 1$. The important thing to note about this expression is that the charge dispersion decreases *exponentially* with $\sqrt{E_J/E_C}$, as was first noted by Averin *et al.* in 1985 [23].

### 2.5.2 Anharmonicity

Since the transmon has only a weak anharmonicity, it is reasonable to treat it as a perturbation of a harmonic oscillator. We expand the cosine in (2.27) to 4th order to obtain

$$H = 4E_C n^2 - E_J + \frac{E_J \varphi^2}{2} - \frac{E_J \varphi^4}{24}, \tag{2.34}$$

where the $n_g$ dependence has been removed since, as described in the previous section, it is exponentially small for the transmon.* Introducing creation and annihilation operators $b^\dagger$ and $b$ for the harmonic oscillator described by the quadratic part of (2.34), we can rewrite this in the form of a Duffing oscillator

$$H = \sqrt{8E_C E_J}\left(b^\dagger b + 1/2\right) - E_J - \frac{E_C}{12}\left(b^\dagger + b\right)^4. \tag{2.35}$$

Performing perturbation theory in the quartic term gives the first-order approximation to the energies

$$E_m \simeq -E_J + \sqrt{8E_J E_C}\left(m + \frac{1}{2}\right) - \frac{E_C}{12}\left(6m^2 + 6m + 3\right). \tag{2.36}$$

Define the absolute anharmonicity $\alpha_m$ of a level as the difference of the transition energy from the next level lower $E_{m-1,m}$ and the transition energy to the next higher level $E_{m,m+1}$, where $E_{mn} = E_n - E_m$ is the transition energy between levels $m$ and $n$. Using (2.35) gives

$$\alpha_m = E_{m+1,m} - E_{m,m-1} \simeq -E_C. \tag{2.37}$$

This absolute anharmonicity should be compared to the transition energy $E_{01} \simeq \sqrt{8E_J E_C}$ of the transmon, giving a relative anharmonicity

$$\alpha_m^r = \alpha_m / E_{01} \simeq -(8E_J/E_C)^{-1/2}. \tag{2.38}$$

This justifies the statement that 'the anharmonicity is weak' when $E_J/E_C \gg 1$. However, the anharmonicity decreases only *algebraically* with $E_J/E_C$, as compared to the exponential dependence of the charge dispersion. A typical example: a transmon with energy ratio $E_J/E_C = 60$ and transition frequency $E_{01}/h = 5\,\text{GHz}$ has anharmonicity of 271 MHz and charge dispersion of 1.8 kHz.

* The charge dispersion for *any* perturbative expansion of (2.27) is identically zero. This results from the fact that a perturbative expansion cannot preserve the periodicity of the Hamiltonian with respect to $\varphi$.

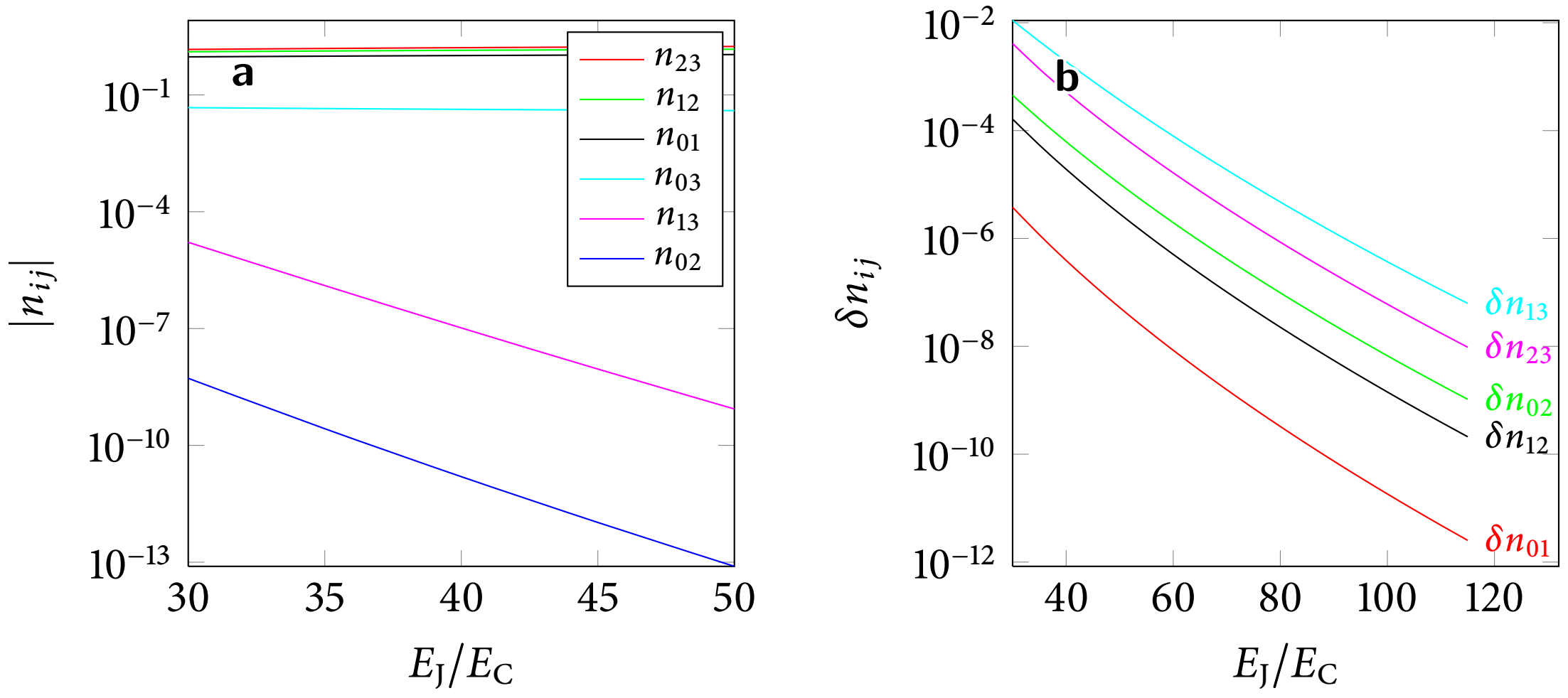

**Figure 2.6: Charge dependence of the matrix elements.** **a**, The mean value of $n_{ij}$ over $n_g$, showing the selection rule that exponentially suppresses matrix elements $n_{i,i+k}$ for $k$ even. The matrix elements for $n_{i,i+1}$ are all of order unity and are therefore difficult to distinguish at the scale of this graph. **b**, The peak-to-peak range in $n_{ij}$ as $n_g$ is varied, showing that for $E_J/E_C \gg 1$ the dispersion is exponentially suppressed.

### 2.5.3 Matrix elements and selection rules

We shall couple the transmon to other circuit components in later sections. If the coupling is via a transmon operator $A$ then we need to calculate matrix elements of the form

$$A_{ij} = \langle\, i \,|\, A \,|\, j \,\rangle. \tag{2.39}$$

These elements may be found numerically using the eigenvectors $|i\rangle$ obtained from diagonalizing (2.30). The most common case is a 'dipole-like' coupling via a linear electric field. For example, this situation applies when the transmon is placed in a cavity resonator (section 2.6). In this case, the relevant operator is the charge operator $n$, and it is instructive to look at the perturbative result for matrix elements. The number operator is given asymptotically for large $E_J/E_C$ by

$$n = -\mathrm{i}\left(\frac{E_J}{8E_C}\right)^{1/4} \frac{b - b^\dagger}{\sqrt{2}}. \tag{2.40}$$

The perturbation theory result for the eigenstates yields

$$n_{j+1,j} = \langle j+1 | n | j \rangle \simeq \sqrt{\frac{j+1}{2}} \left( \frac{E_J}{8E_C} \right)^{1/4}, \tag{2.41a}$$

$$n_{j+k,j} = \langle j+k | n | j \rangle \simeq 0, \quad |k| > 1. \tag{2.41b}$$

Thus the matrix elements are approximately those of a harmonic oscillator. With the full numeric solution it is possible to see that there is a selection rule: the matrix elements with even $k$ fall off exponentially, whereas the matrix elements with odd $k$ fall off only algebraically. This can be understood because all the terms in the expansion of the $\cos(\varphi)$ of (2.27) are even and do not mix odd and even states. As an example, for $E_J/E_C = 60$, $n_g = 0.25$ we have

$$\frac{n_{2,0}}{n_{1,0}} \simeq 1.7 \times 10^{-6}, \tag{2.42a}$$

$$\frac{n_{3,0}}{n_{1,0}} \simeq 0.03. \tag{2.42b}$$

Figure 2.6 shows how the matrix elements depend on $n_g$. In the same way that we defined the charge dispersion, we can define a dispersion of the matrix elements, $\delta n_{ij}$, as the peak-to-peak variation in $n_{ij}$ caused by varying $n_g$. Similar to the charge dispersion, $\delta n_{ij}$ is also exponentially suppressed for large $E_J/E_C$.

For $E_J/E_C \simeq 1$, the selection rule still holds at special values of $n_g$: in the case that $n_g = 0.5$ (or $n_g = 0$) parity is preserved and the matrix elements with even $k$ are identically zero. By adjusting $n_g$ very slightly away from this special value, it is possible to break the symmetry in a controlled way. Similar effects have been observed in flux qubits: applied flux $\tilde{\Phi}_x = 1.5\Phi_0$ is a symmetry point, where the transition between the ground and first-excited states via two-photon processes ($\omega_d = E_{01}/2$) is forbidden, but changing the applied flux very slightly to $\tilde{\Phi}_x = 1.4995\Phi_0$ breaks the symmetry and the transition becomes allowed [24].

It is useful for future sections to define ladder operators $c^\dagger$ and $c$ satisfying

$$c = \sum_j \frac{n_{j,j+1}}{n_{0,1}} |j\rangle \langle j+1|, \tag{2.43a}$$

$$c^\dagger = \sum_j \frac{n_{j+1,j}}{n_{0,1}} |j+1\rangle \langle j|. \tag{2.43b}$$

In the harmonic limit, these become the usual bosonic creation and annihilation operators.

### 2.5.4 Flux tuning: the split transmon

It is useful to be able to vary the properties of the transmon. In particular, being able to bring the transition frequency in resonance with a transmission line cavity allows observing the vacuum Rabi splitting, the topic of chapter 4, and precise adjustment of the frequencies is necessary to bring the dispersive shifts into the correct ratios for the preparation-by-measurement of chapter 5. As was shown in section 2.5.2, unlike for the CPB, the gate voltage $n_g$ is not useful for tuning the transmon frequency. The remaining transmon parameters are $E_C$ and $E_J$. Varying $E_C$ is conceivable, for example using a mechanical linkage to move one capacitor electrode, but it is generally more convenient to alter $E_J$. This is done by replacing the Josephson junction by a parallel-connected pair of Josephson junctions. The Hamiltonian of this loop is

$$H = -E_{J1}\cos(\varphi) - E_{J2}\cos(\varphi + 2\pi\tilde{\Phi}/\Phi_0), \tag{2.44}$$

where $E_{J1}$, $E_{J2}$ are the Josephson energies of the two junctions, $\tilde{\Phi}$ is the externally applied flux threading the loop formed by the junctions, and as before $\Phi_0 = h/2e$ is the flux quantum. Using trigonometric identities this is easily rewritten as

$$H = -E_{J\Sigma}\cos\left(\frac{\pi\tilde{\Phi}}{\Phi_0}\right)\sqrt{1 + d^2\tan^2\left(\frac{\pi\tilde{\Phi}}{\Phi_0}\right)}\cos(\varphi - \varphi_0), \tag{2.45}$$

with $E_{J\Sigma} = E_{J1} + E_{J2}$ being the sum of the Josephson energies, $d = \frac{E_{J2}-E_{J2}}{E_{J1}+E_{J2}}$ being the junction asymmetry. The phase offset $\varphi_0$ is given by $\tan(\varphi_0 + \pi\tilde{\Phi}/\Phi_0) = d\tan(\pi\tilde{\Phi}/\Phi_0)$. Thus this pair of junctions behaves like a single junction with an effective $E_J$ that in the limit of small asymmetry behaves as

$$E_J(\tilde{\Phi}) \simeq E_J^{\max}\cos(\pi\tilde{\Phi}/\Phi_0). \tag{2.46}$$

For typical experimental scenarios $d \simeq 0.1$.

## 2.6 Coupling a transmon to a resonator

Consider the situation depicted in figure 2.7 where a transmon is located at the center of a coplanar waveguide (CPW) transmission line resonator with open-circuit boundary conditions. Thus the transmon is at a voltage antinode for the $l = 2$ mode of the resonator and can couple to this mode. It is fairly obvious that the correct Hamiltonian for this circuit is a sum of the transmon Hamiltonian (2.27), the transmission line resonator Hamiltonian (2.24) and a dipole coupling term which is a product of the voltage in the cavity, proportional to $a + a^\dagger$, and the charge of the transmon, proportional to $n$:

$$H = 4E_\mathrm{C}(n - n_\mathrm{g})^2 - E_\mathrm{J}\cos(\varphi) + \hbar\omega_\mathrm{r} a^\dagger a + \beta n(a^\dagger + a). \tag{2.47}$$

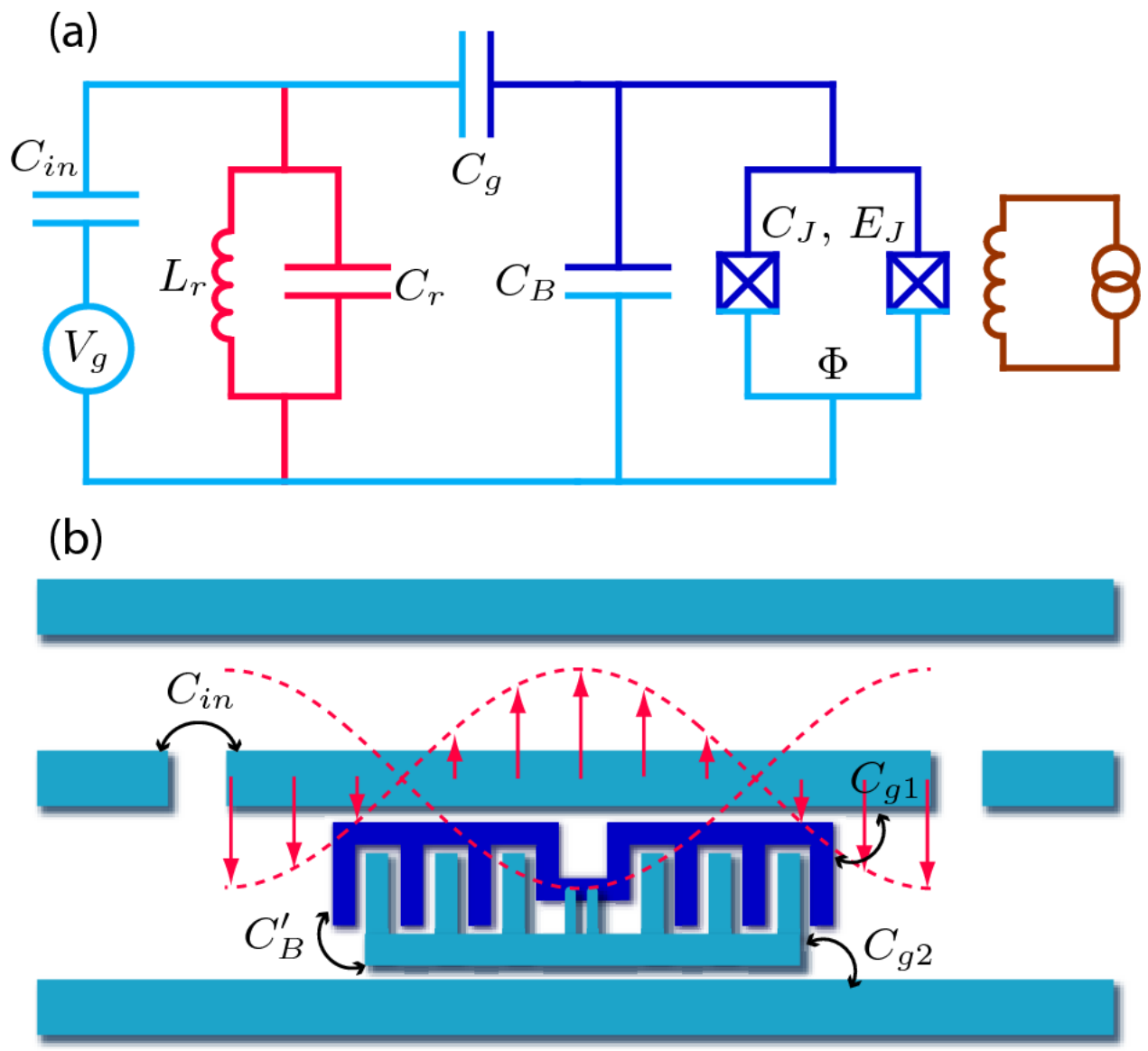

**Figure 2.7: Coupling a transmon to a resonator. a**, Effective circuit diagram showing the transmon (dark blue: $C_\mathrm{J}$, $E_\mathrm{J}$), resonator (red: $L_\mathrm{r}$, $C_\mathrm{r}$), flux-biasing circuit (brown), voltage-biasing circuit (cyan, not usually present); **b**, Simplified schematic of the device design (not to scale) showing large interdigitated capacitors that produce the transmonic $E_\mathrm{J}/EC \gg 1$. In this version of the design, the transmon sits at the center of the transmission line, coupling to the second harmonic, $l = 2$, of the cavity. (Figure used with permission from [20]. See Copyright Permissions.)

However, it is necessary to go through the detailed calculation in order to take account of the full capacitance network (as indicated in figure 2.8) and to obtain the effective parameters $E_\mathrm{J}$, $E_\mathrm{C}$, $\omega_\mathrm{r}$ and $\beta$ in terms of the bare parameters of the problem. This rather tedious calculation is outlined in [20, appendix A]. It is worth mentioning that because of this step, it is not really possible to speak of $E_\mathrm{C}$ as a 'transmon parameter' nor $\omega_\mathrm{r}$ as a 'resonator parameter'. This is not like cavity QED with real atoms, where one can measure the resonator frequency when it is empty of atoms, or do spectroscopy experiments on atoms outside the cavity. In atomic cavity QED, a shift of the atom frequency when it is put into the cavity would be described as a *Lamb shift*, and the parameters appearing in the Hamiltonian (2.47) already include these frequency renormalizations. (On the other hand, the *change* in the shift in the transmon frequency, as the effective $E_J$ of the transmon is tuned, has indeed been observed [25].)

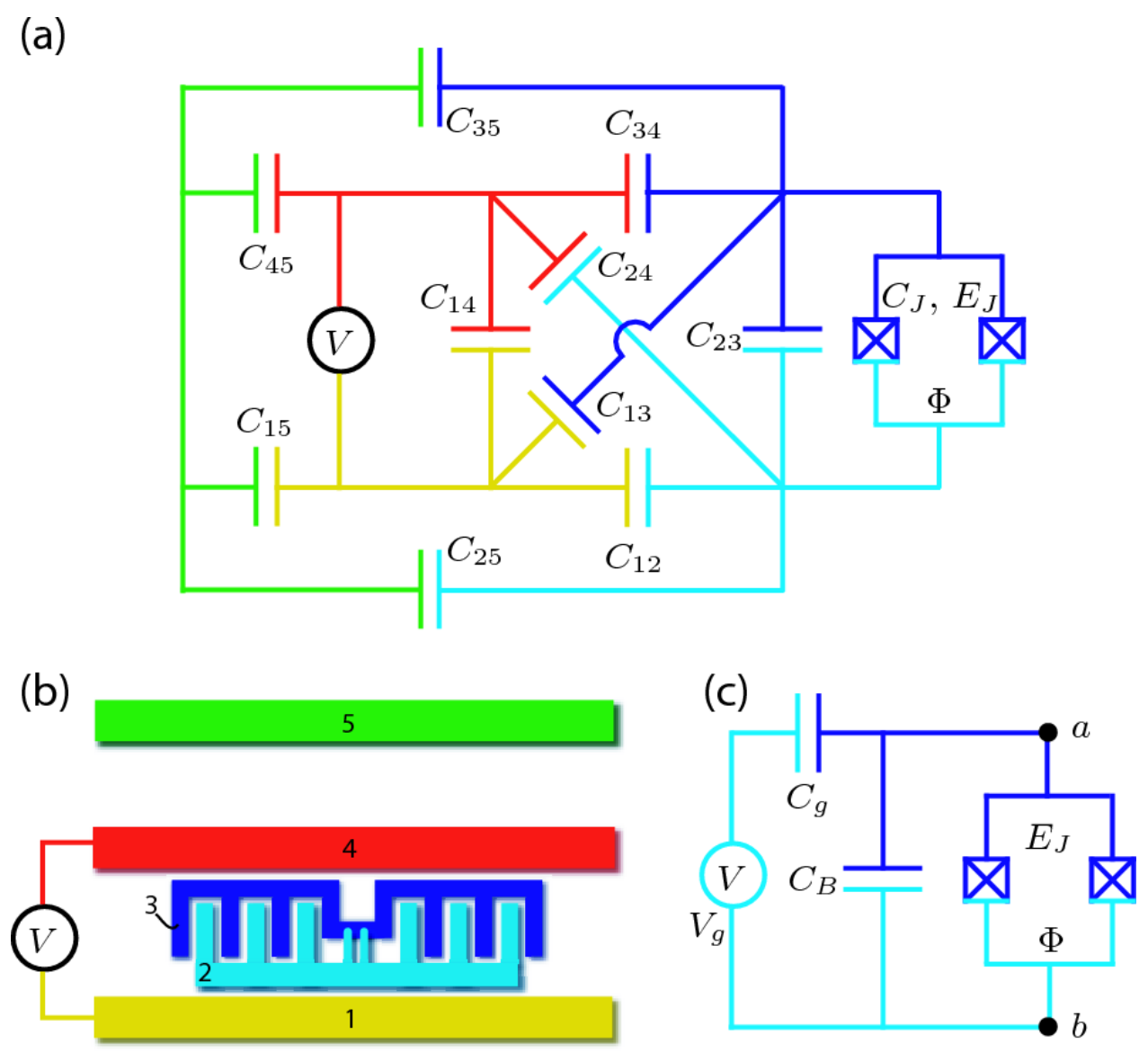

**Figure 2.8: The capacitance network for the transmon in a coplanar waveguide resonator. a**, The complete circuit diagram, showing all the capacitances, designed and parasitic, between the 5 metallic areas of the transmon-cavity circuit, shown in **b** (not to scale). **c**, The simplified equivalent circuit which can be found by using the electrical engineer's rules for series and parallel capacitors. (Figure used with permission from [20]. See Copyright Permissions.)

By introducing transmon frequencies $\omega_i = E_i/\hbar$ and coupling strengths $\hbar g_{ij} = \beta\langle\, i\,|\,n\,|\,j\,\rangle$, denoting the transmon eigenkets as $|j\rangle$, and from now on setting $\hbar = 1$, we can rewrite (2.47) in the form

$$H = \omega_r a^\dagger a + \sum_j \omega_j \,|j\rangle\langle j| + \sum_{i,j} g_{i,j}\,|i\rangle\langle j|\,(a + a^\dagger). \tag{2.48}$$

In the case that the anharmonicity is sufficiently large that the transmon can be treated as a (two-level) qubit, (2.48) takes the form

$$H = \omega_r a^\dagger a + \omega_q \sigma_z/2 + g\sigma_x(a + a^\dagger), \tag{2.49}$$

where the qubit is represented as a spin-½ system, making the identifications* $|0\rangle \to |\downarrow\rangle$ and

* This is the quantum optics convention. In NMR it is usual to choose the opposite convention, which occasionally leads to confusion.

$|1\rangle \to |\uparrow\rangle$. The qubit frequency is $\omega_\mathrm{q} = \omega_{01}$ (where $\omega_{ij} = \omega_j - \omega_i$ is the transition frequency between levels $i$ and $j$) and the coupling strength is $g = g_{01}$. The $\sigma_i$ are the Pauli matrices. Equation (2.49) is known as the *Rabi Hamiltonian*,* and therefore it is reasonable to call (2.48) a *generalized Rabi Hamiltonian.*

## 2.7 Jaynes–Cummings Physics

In the case that $\omega_\mathrm{r} \simeq \omega_\mathrm{q}$ and $\omega_r \gg g$ it is reasonable to make a rotating wave approximation (RWA) and drop from (2.49) the so-called *counter-rotating* terms, $a^\dagger\sigma_+$ and $a\sigma_-$, where $\sigma_\pm = \frac{1}{2}(\sigma_x \pm \mathrm{i}\sigma_y)$. The resulting expression is the well-known Jaynes–Cummings Hamiltonian [26]

$$H = \omega_\mathrm{r} a^\dagger a + \omega_\mathrm{q}\sigma_z/2 + g(a\sigma_+ + a^\dagger\sigma_-). \tag{2.50}$$

Although the Jaynes–Cummings Hamiltonian is probably the simplest non-trivial Hamiltonian imaginable, it contains a lot of interesting physics and the next few sections discuss it in further detail.

The coupling term $g(a\sigma_+ + a^\dagger\sigma_-)$ only connects the states $|n-1,\uparrow\rangle$ and $|n,\downarrow\rangle$, which leads to the Hamiltonian being block-diagonal, with $2\times 2$ blocks. This allows for an exact analytic solution, giving eigenstates

$$|0\rangle = |0,\downarrow\rangle\,, \tag{2.51a}$$
$$|n,+\rangle = \cos(\theta_n)\,|n-1,\uparrow\rangle + \sin(\theta_n)\,|n,\downarrow\rangle\,, \tag{2.51b}$$
$$|n,-\rangle = -\sin(\theta_n)\,|n-1,\uparrow\rangle + \cos(\theta_n)\,|n,\downarrow\rangle\,, \tag{2.51c}$$

and eigenenergies

$$E_0 = -\frac{\delta}{2}, \tag{2.52a}$$
$$E_{n,\pm} = n\omega_\mathrm{r} \pm \frac{1}{2}\sqrt{4g^2 n + \delta^2}, \tag{2.52b}$$

for $n = 1, 2, \ldots$ and where $\delta = \omega_\mathrm{q} - \omega_\mathrm{r}$ is the qubit-cavity detuning and $\theta_n$ satisfies

$$\tan(2\theta_n) = \frac{2g\sqrt{n}}{\delta}. \tag{2.53}$$

---

* Some authors call (2.49) the Jaynes–Cummings Hamiltonian, but I reserve that name for (2.50).

These solutions of the full Hamiltonian are called *dressed states* to emphasize that although $|n,\pm\rangle$ connect smoothly to the *bare states* $|n-1,\uparrow\rangle$, $|n,\downarrow\rangle$ in the limit $g/\delta \to 0$, in general they contain a combination of both bare states.

When the cavity and qubit are in resonance $\delta = 0$, $\omega_\mathrm{r} = \omega_\mathrm{q} = \omega$, the above expressions simplify even further to

$$|0\rangle = |0,\downarrow\rangle, \qquad E_0 = 0, \tag{2.54a}$$
$$|n,\pm\rangle = \big(|n-1,\uparrow\rangle \pm |n,\downarrow\rangle\big)/\sqrt{2}, \qquad E_{n,\pm} = n\omega \pm g\sqrt{n}. \tag{2.54b}$$

We can also apply the RWA to the generalized Rabi Hamiltonian (2.48) to obtain a *generalized Jaynes–Cummings Hamiltonian*

$$H = \omega_\mathrm{r} a^\dagger a + \sum_j \omega_j |j\rangle\langle j| + \sum_j g_{j,j+1}\big(|j+1\rangle\langle j|\, a + \text{h.c.}\big) \tag{2.55a}$$
$$= \omega_\mathrm{r} a^\dagger a + \sum_j \omega_j |j\rangle\langle j| + \sum_j g\big(ac^\dagger + a^\dagger c\big). \tag{2.55b}$$

Equation (2.55) has the same block-diagonal property as the ordinary Jaynes–Cummings Hamiltonian, with the coupling term only connecting the states within each $m$-excitation subspace $\{|n,j\rangle \mid n+j=m\}$, using $|n,j\rangle$ to denote the bare state with $n$ cavity excitations and $j$ transmon excitations. The $m\times m$ size of the blocks generally precludes writing the dressed states in any simple form, however.

## 2.8 Driving

### 2.8.1 Introducing the drive

One way to introduce a classical drive into the system is to imagine there being a second cavity, with frequency $\omega_\mathrm{d}$ and creation and annihilation operators $d^\dagger$ and $d$. If the two cavities are allowed to interact via a capacitor this will produce a Hamiltonian

$$H = H_\mathrm{sys} + H_\mathrm{d} + \omega_\mathrm{d} d^\dagger d, \tag{2.56}$$

where $H_\mathrm{sys}$ represents the undriven system Hamiltonian, $H_\mathrm{d}$ describes the interaction between the two cavities

$$H_\mathrm{d} = \epsilon\big(a + a^\dagger\big)\big(d + d^\dagger\big) \tag{2.57}$$

and $\epsilon$ describes the coupling. Imagining the drive cavity to be initialized in a very highly-excited coherent state $|\beta\rangle$ and $\epsilon$ to be very small, we can treat this cavity as remaining in the same state $|\beta\rangle$ for all time. This gives

$$H_\mathrm{d} = \left(a + a^\dagger\right)\left(\xi \mathrm{e}^{-\mathrm{i}\omega_\mathrm{d} t} + \xi^* \mathrm{e}^{\mathrm{i}\omega_\mathrm{d} t}\right), \tag{2.58}$$

where $\xi = \epsilon\beta$ defines the strength of the driving. In the case that the driving is not too strong, such that $\xi \ll \omega_{l,m}$, for all transition frequencies $\omega_{l,m}$ between states that are connected by the action of $a + a^\dagger$—*i.e.*, all states such that $\langle\, l \,|\, a + a^\dagger \,|\, m \,\rangle \neq 0$—it is possible to make an RWA:

$$H_\mathrm{d} = a\xi^* \mathrm{e}^{\mathrm{i}\omega_\mathrm{d} t} + a^\dagger \xi \mathrm{e}^{-\mathrm{i}\omega_\mathrm{d} t}. \tag{2.59}$$

### 2.8.2 The rotating frame of the drive

It is inconvenient to have a time-dependent Hamiltonian, so we make the transformation given by the time-dependent operator

$$U(t) = \exp\left[\mathrm{i}\omega_\mathrm{d} t\left(a^\dagger a + \sum_j |j\rangle\, j \,\langle j|\right)\right], \tag{2.60}$$

applied to the driven generalized Jaynes–Cummings Hamiltonian (2.55)

$$\tilde{H} = U(H + H_\mathrm{d})U^\dagger - \mathrm{i}U\dot{U}^\dagger \tag{2.61a}$$

$$= (\omega_\mathrm{r} - \omega_\mathrm{d})a^\dagger a + \sum_j (\omega_j - j\omega_\mathrm{d})\,|j\rangle\langle j| + g\left(a^\dagger c + a c^\dagger\right) + \left(a\xi^* + a^\dagger \xi\right). \tag{2.61b}$$

Finally, introducing the frequency differences $\Delta_\mathrm{r} = \omega_\mathrm{r} - \omega_\mathrm{d}$ and $\Delta_j = \omega_j - j\omega_\mathrm{d}$, and allowing for the possibility that the drive strength is a slow function of time $\xi(t)$ we can write the driven generalized Jaynes–Cummings Hamiltonian in the rotating frame, suppressing the tilde on the $\tilde{H}$:

$$H = \Delta_\mathrm{r} a^\dagger a + \sum_j \Delta_j\,|j\rangle\langle j| + g\left(a^\dagger c + a c^\dagger\right) + \left(a\xi(t)^* + a^\dagger \xi(t)\right). \tag{2.62}$$

## 2.9 Displacement transformation

When we are using the drive term to cause transitions of the transmon, it is useful to rewrite the Hamiltonian in a frame where the drive terms act directly on the transmon. We can do this using the Glauber displacement operator

$$D(\alpha) = \exp\left[\alpha(t)a^\dagger - \alpha^*(t)a\right]. \tag{2.63}$$

This gives a displaced Hamiltonian

$$\tilde{H} = D^\dagger H D - \mathrm{i}D^\dagger \dot{D} \tag{2.64a}$$

$$\begin{aligned} &= \Delta_\mathrm{r} a^\dagger a + \sum_j \Delta_j |j\rangle\langle j| + g\left(a^\dagger c + a c^\dagger\right) + g\left(\alpha^* c + \alpha c^\dagger\right) \\ &\quad + \left(a\xi(t)^* + a^\dagger \xi(t)\right) + \Delta_\mathrm{r}\left(\alpha a^\dagger + \alpha^* a\right) - \mathrm{i}\left(\dot{\alpha} a^\dagger + \dot{\alpha}^* a\right). \end{aligned} \tag{2.64b}$$

Choose $\alpha(t)$ as a solution of the differential equation

$$-\mathrm{i}\dot{\alpha}(t) + \Delta_\mathrm{r}\alpha(t) + \xi(t) = 0. \tag{2.65}$$

The terms in the second line of (2.64b) cancel for this choice of $\alpha$, and the Hamiltonian becomes, suppressing the tilde on the $\tilde{H}$,

$$H = \Delta_\mathrm{r} a^\dagger a + \sum_j \Delta_j |j\rangle\langle j| + g\left(a^\dagger c + a c^\dagger\right) + \frac{1}{2}\left(\Omega^*(t)c + \Omega(t)c^\dagger\right), \tag{2.66}$$

where we have introduced the *Rabi frequency* $\Omega(t) = 2g\alpha(t)$. For constant, time-independent drive strength, the Rabi frequency is

$$\Omega = \frac{2\xi g}{\Delta_\mathrm{r}}. \tag{2.67}$$

## 2.10 Dispersive limit

When the cavity and transmon are sufficiently detuned compared to their coupling strength, $g_{j,j+1}/(\omega_{j+1,j} - \omega_\mathrm{r}) \ll 1$, we can make a unitary transformation on the Hamiltonian

$$\tilde{H} = U H U^\dagger, \tag{2.68}$$

with

$$U = \exp\left[\sum_j \lambda_j |j\rangle\langle j+1| a^\dagger - \text{h.c.}\right], \tag{2.69}$$

and expand to second order in the small parameter $\lambda_j = g_{j,j+1}/(\omega_{j,j+1} - \omega_\mathrm{r})$, dropping two-photon terms:

$$\begin{aligned}\tilde{H} = \Delta_j \sum_j |j\rangle\langle j| + \Delta_\mathrm{r} a^\dagger a + \sum_j \chi_{j,j+1} |j+1\rangle\langle j+1| - \chi_{01} a^\dagger a |0\rangle\langle 0| \\ + \sum_{j=1} \big(\chi_{j-1,j} - \chi_{j,j+1}\big) a^\dagger a |j\rangle\langle j| \\ + \frac{1}{2}\big(\Omega^*(t) c + \Omega(t) c^\dagger\big),\end{aligned} \tag{2.70}$$

with dispersive couplings $\chi_{ij}$ given by

$$\chi_{ij} = \frac{g_{ij}^2}{\omega_{ij} - \omega_\mathrm{r}}. \tag{2.71}$$

If we now treat the transmon as a qubit, truncating to the lowest two levels, we obtain the driven dispersive Hamiltonian in the rotating frame

$$H' = \Delta_\mathrm{q}' \sigma_z/2 + \big(\Delta_\mathrm{r}' + \chi\sigma_z\big) a^\dagger a + \big(\Omega^*(t)\sigma_- + \Omega(t)\sigma_+\big). \tag{2.72}$$

We see that the qubit frequency acquires a Lamb shift, $\Delta_\mathrm{q}' = \Delta_\mathrm{q} + \chi_{01}$, and the cavity frequency is shifted, $\Delta_\mathrm{r}' = \Delta_\mathrm{r} - \chi_{12}/2$. The qubit-cavity interaction can be interpreted either as a shift of the qubit frequency, dependent on the number of photons in the cavity (dynamical Stark shift), or as a shift of the cavity frequency dependent on the qubit state, with the dispersive shift $\chi$ given by

$$\chi = \chi_{01} - \chi_{12}/2. \tag{2.73}$$

It is important to note that it matters that we made the dispersive transformation *before* making the two-level approximation. If we had performed these operations in the opposite order we would have found a dispersive Hamiltonian with the same form as (2.72), but instead of (2.73) we would have found

$$\chi = \frac{g^2}{\delta}, \tag{2.74}$$

where $\delta = \omega_\mathrm{q} - \omega_\mathrm{r}$ is the qubit-cavity detuning.

### 2.10.1 One-qubit gates

The *Bloch vector* $\vec{r} = \{x, y, z\}$ is a compact way to represent an arbitrary density matrix $\rho$ of a qubit, via

$$\rho = (\mathbb{1} + x\sigma_x + y\sigma_y + z\sigma_z)/2. \tag{2.75}$$

Examining (2.72) it is clear that by choosing the phase of the drive, we can directly perform rotations of $\vec{r}$ about any axis in the $x$-$y$ plane. By chaining several such rotations, it is thus possible to perform arbitrary rotations about any axis. In practice, with careful calibration and shaping of the pulses, these rotations can be performed with excellent fidelity [27].

### 2.10.2 Readout

The state-dependent shift $\chi$ of the cavity frequency allows for readout of the qubit state. The amplitude and phase of the transmitted and reflected waves from the cavity are dependent on the cavity frequency, thus by driving the cavity close to its bare frequency, and measuring the reflected or transmitted wave we can determine the qubit state. We return to examine this point in more detail in chapter 5.

# CHAPTER 3

# Master equation

The previous chapter was concerned with the behavior of the transmon coupled to a cavity resonator, at the level of the Hamiltonian. This chapter concerns the coupling of the transmon-cavity system to uncontrolled environmental degrees of freedom. Whereas the Hamiltonian dynamics preserves the purity of the wavefunction, the interaction with the environment can cause the system density matrix to become mixed, via such processes as *relaxation* and *dephasing*. These phenomena can be described in the framework of quantum Markovian master equations.

This chapter begins with a definition of *quantum operations*. Then we show how by imposing the Markov property, we can derive the standard Kossakowski and Lindblad forms for the master equation, which describe the most general type of master equation usually considered. The following section then discusses how to derive a master equation for a particular case of a known microscopic coupling between a system and its environment, by the argument of *weak coupling*. Next we make use of this framework to derive standard master equations, for the damped harmonic oscillator and for the qubit (the Bloch equations). The discussion until this point is largely borrowed from standard texts in the fields of decoherence and open systems [28–30]. For simplicity of exposition, the system is implicitly assumed to have a discrete, finite Hilbert space, although many of the results generalize to bounded operators over countably infinite Hilbert spaces (and practitioners routinely ignore these restrictions and apply the results to general operators over general Hilbert spaces). The interested reader should consult the literature. The final section of this chapter discusses the appropriate way to apply these methods to derive the master equation for the transmon-cavity system.

## 3.1 Quantum operations

### 3.1.1 Positive maps

We wish to discuss the temporal evolution of the state of a quantum mechanical system. The most general description of the state is via the density matrix $\rho$ which has by definition the properties:

1. Hermitian: $\rho = \rho^\dagger$;
2. positive semidefinite: $\lambda \geq 0$, for all eigenvalues $\lambda$ of $\rho$;
3. unit trace: $\operatorname{tr}\rho = 1$.

(By way of analogy, a classical probability distribution is required to be real, everywhere positive, and to integrate to unity.) We write the transformation describing the state change of the system as

$$\rho \mapsto \Lambda\rho \tag{3.1}$$

using the map $\Lambda$. The density matrix may be formed as a convex probabilistic mixture of other density matrices $\rho = p\rho_1 + (1-p)\rho_2$, for $p \in [0,1]$ and we thus require that

$$\Lambda(p\rho_1 + (1-p)\rho_2) = p\Lambda\rho_1 + (1-p)\Lambda\rho_2. \tag{3.2}$$

Thus, $\Lambda$ must be a linear map that preserves the density matrix properties. Such a map is called a *positive, trace-preserving* map.

### 3.1.2 Complete positivity

It is possible to derive a much more stringent requirement on quantum dynamical maps by imagining that there is an $n$-level ancilla system with trivial Hamiltonian $H = 0$ placed far away from the open system $\mathcal{S}$. Because these systems do not interact, the joint dynamical map $\Lambda_n$ should be of the form $\Lambda \otimes \mathbb{1}_n$. Obviously, $\Lambda_n$ must be a positive map for all $n = 1, 2, \ldots$. This condition on $\Lambda$ is called *complete positivity* and is significantly stronger than plain positivity. It requires that the evolution of the universe must remain positive under $\Lambda$, even if $\mathcal{S}$ is entangled with some other system. A particular result [31] is that any completely positive trace preserving (CPTP) map, also known as a *quantum channel* or *quantum operation*, on an $N$-dimensional system, admits an *operator sum representation*

$$\Lambda\rho = \sum_{\alpha=1}^{N^2} W_\alpha \rho W_\alpha^\dagger, \tag{3.3}$$

with a completeness relation

$$\sum_{\alpha=1}^{N^2} W_\alpha^\dagger W_\alpha = \mathbb{1}_N. \tag{3.4}$$

The $W_\alpha$ are called *Kraus operators* and their choice is not unique.

The canonical example of a map which is positive but not completely positive is the transposition map $\rho \mapsto \rho^{\mathrm{T}}$. The non-complete positivity of the transposition map should be obvious, once we recall that partial transposition is the well-known Peres–Horodecki test for entanglement [32, 33]: if the density matrix of a bipartite system ceases to be positive semidefinite under transposition of one of the components, then this is a sufficient condition to prove that the system is entangled.

### 3.1.3 Reduced dynamics

Instead of the argument of the previous subsection (based on the imaginary ancilla system), an alternative derivation of (3.3) comes from considering the interaction of the open system $\mathcal{S}$ with the external world, reservoir $\mathcal{R}$. Assume that the initial state of the system-reservoir combination is given by a product: $\rho \otimes \omega_R$, and that the combined system undergoes reversible evolution denoted by the unitary operator $U$. The transformation describing the state change of $\mathcal{S}$ is given by

$$\rho \mapsto \mathrm{tr}_R(U\rho \otimes \omega_R U^\dagger), \tag{3.5}$$

where $\mathrm{tr}_R$ denotes the *partial trace* over $\mathcal{R}$, defined via

$$\langle \phi \,|\, \mathrm{tr}_R\, \gamma \,|\, \psi \rangle = \sum_\nu \langle \phi \otimes f_\nu \,|\, \gamma \,|\, \psi \otimes f_\nu \rangle \tag{3.6}$$

for all states $|\phi\rangle, |\psi\rangle$ of $\mathcal{S}$, all operators $\gamma$ in $\mathcal{S} \oplus \mathcal{R}$, and any arbitrary orthonormal basis $\{f_\nu\}$ in $\mathcal{R}$. By decomposing $\omega_R = \sum_\nu \lambda_\nu |f_\nu\rangle \langle f_\nu|$ it is possible [29] to construct explicitly the operator sum representation (3.3), finding the $W_\alpha$ in terms of $U$ and $\omega_R$.

### 3.1.4 Aside: Not completely positive maps?

Not everyone is convinced by the above arguments justifying complete positivity. Shaji and Sudarshan [34] point out that the argument of section 3.1.2 based on the possible presence of an ancilla requires that on the one hand we assume the ancilla to be completely isolated from $\mathcal{S}$ but on the other hand it be entangled with $\mathcal{S}$. Similarly, they note that the argument of section 3.1.3 based on reduced dynamics, requires on the one hand we assume the initial state of the combined system is a product $\rho \otimes \omega_R$ and on the other hand the unitary evolution $U$ be such that it causes entanglement between the systems. They thus conclude that the standard arguments are unconvincing and they recommend that positive maps are as good candidates as completely positive maps for describing open quantum evolution. Although this is an interesting point, the remainder of this thesis follows the standard approach of requiring complete positivity. (As we shall see, the class of master equations that maintains complete positivity, even after making the Markov restriction, is still much more general than we need—our problem is to find reasonable assumptions that allow us to restrict the class of maps sufficiently to allow comparisons with experiment.)

## 3.2 Markovian dynamics

The previous sections discussed the properties of a single dynamical map $\Lambda$. In order to describe the time evolution of an open system we need a one-parameter family of such maps $\{\Lambda_t | t \geq 0\}$. Generally $\Lambda_t$ satisfies a complicated integro-differential equation. However, by making the Markov approximation the evolution becomes quite simple. Specifically, we wish to describe the *quantum dynamical semigroup*, defined as the family $\{\Lambda_t | t \geq 0\}$ satisfying

1. $\Lambda_t$ is a dynamical map (*i.e.*, completely positive and trace preserving);
2. $\Lambda_t \Lambda_s = \Lambda_{t+s}$: the semigroup condition or Markov property. (It is a semigroup rather than a group simply because each element may not have a unique inverse—the purpose is to describe *irreversible* dynamics);
3. $\mathrm{tr}[(\Lambda_t \rho) A]$ is a continuous function of $t$ for any density matrix $\rho$ and observable $A$.

These conditions imply the existence of linear map $L$ called a *generator of the semigroup*, such that

$$\dot{\rho}_t = L\rho_t, \tag{3.7}$$

where $\rho_t = \Lambda_t \rho$. Equation (3.7) is called a *quantum Markovian master equation*. The generator $L$ can be written in *Kossakowski normal form* [35] as

$$L\rho = -\mathrm{i}\left[H, \rho\right] + \frac{1}{2}\sum_{i,k=1}^{N^2-1} A_{ik}\Big(\left[F_i, \rho F_k^\dagger\right] + \left[F_i\rho, F_k^\dagger\right]\Big). \tag{3.8}$$

Here $H = H^\dagger$ is the Hamiltonian describing the dynamics of the open system, including reversible effects due to the environment (for example renormalizations of transition frequencies). To make the decomposition into Hamiltonian and non-Hamiltonian parts unique, we impose

$$\mathrm{tr}(H) = 0. \tag{3.9}$$

The *orthonormal matrix set* $\{F_i | i = 1, 2, \ldots, M = N^2 - 1\}$ comprises $M$ matrices of dimension $(N \times N)$ and has the properties

$$\mathrm{tr}(F_i) = 0, \qquad \mathrm{tr}(F_i F_k^\dagger) = \delta_{ik}. \tag{3.10}$$

The quantities describing the irreversible dynamics: the lifetimes, longitudinal and transverse relaxation times, and so on, are contained in the Hermitian $(M \times M)$-matrix $A$, which is constrained to be positive. The requirement that $A$ be positive puts non-trivial constraints among the decoherence parameters, such as the well-known $T_2 \le 2T_1$ constraint for two-level systems (as we shall derive in section 3.5). The choice of $A$ and $H$ is unique, given (3.9) and a particular choice of the orthonormal set $\{F_i\}$.

It is frequently helpful to use an alternative representation of the generator $L$, the *Lindblad normal form*,* which can be found by diagonalizing the matrix $A$

$$L\rho = -\mathrm{i}\left[H, \rho\right] + \sum_{i=1}^{M} \mathcal{D}[V_i]\rho, \tag{3.11}$$

* Lindblad [36] proved that (3.11) is the most general form of the generator, even for infinite Hilbert spaces, with $H$ and $V_i$ bounded operators.

where the *dissipator* $\mathcal{D}$ is defined via

$$\mathcal{D}[A]\rho = A\rho A^\dagger - \{A^\dagger A, \rho\}/2, \tag{3.12}$$

$\{\cdot,\cdot\}$ denoting the anticommutator, and the $V_i$ are called *Lindblad operators* (they are also known as *jump operators*, which terminology will become clear when we consider quantum trajectories in section 5.2). Note that by construction $\mathrm{tr}(\mathcal{D}[A]\rho) = 0$ so that $L$ preserves the trace of $\rho$.

The choice of the set $\{V_i\}$ is not unique. In particular, the generator is invariant under unitary mixing

$$V_i \to U_{ij} V_j. \tag{3.13}$$

## 3.3 Weak coupling

The previous sections gave the general form of the Markovian master equation. The Kossakowski normal form (3.8) makes it clear that for an $N$-dimensional system, in addition to the unitary dynamics, we need in general to give $(N^2-1)^2$ parameters in order to specify fully the irreversible dynamics. However, the discussion so far gives no guidance on how to choose these parameters, beyond the requirement that the matrix $A$ should be positive. This section presents the rigorous microscopic derivation of these parameters, originally due to Davies [37].

Consider the system $\mathcal{S}$ in contact with the reservoir $\mathcal{R}$ and assume the coupling is 'weak'* so that a perturbative treatment of the interaction is possible. That is, write the Hamiltonian as

$$H_{\mathrm{tot}} = H_{\mathrm{S}} + H_{\mathrm{R}} + \lambda V. \tag{3.14}$$

It is convenient to work in the interaction picture, denoted by tilde, in which the von Neumann equation for the state $\sigma$ of the total system reads

$$\dot{\tilde{\sigma}}(t) = -\mathrm{i}\lambda\left[\tilde{V}(t), \tilde{\sigma}(t)\right] \tag{3.15a}$$

$$= -\mathrm{i}\lambda\left[\tilde{V}(t), \tilde{\sigma}(0)\right] - \lambda^2 \int_0^t \mathrm{d}s \left[\tilde{V}(t), \left[\tilde{V}(s), \tilde{\sigma}(s)\right]\right]. \tag{3.15b}$$

* The specific condition is given below, before (3.25).

In (3.15b), which is still exact, the von Neumann equation in its integral version was inserted into the differential equation version.

The first assumption is the *Born approximation*, which states $\lambda V$ is sufficiently small that we should treat the total state as factorized,

$$\sigma(t) \simeq \rho(t) \otimes \omega_{\mathrm{R}}. \tag{3.16}$$

Here $\rho(t)$ is the state of $\mathcal{S}$ and $\omega_{\mathrm{R}}$ is the state of $\mathcal{R}$, taken to be constant because the reservoir is supposed to be unaffected by the system. The idea of this approximation is not that there are no excitations in the reservoir due to the system, but rather that any such excitations decay extremely rapidly and that we are describing the dynamics on a coarse-grained timescale.

The interaction term $V$ can always be expanded in the form

$$V = \sum_{\alpha} A_{\alpha} \otimes B_{\alpha}, \tag{3.17}$$

where $A_{\alpha} = A_{\alpha}^{\dagger}$ act only on $\mathcal{S}$ and $B_{\alpha} = B_{\alpha}^{\dagger}$ act only on $\mathcal{R}$. Additionally we require that $\mathrm{tr}_{\mathrm{R}}(\omega_{\mathrm{R}} B_{\alpha}) = 0$.

Taking the partial trace over $\mathcal{R}$ we can obtain the integro-differential equation for the system density matrix

$$\begin{aligned}\dot{\tilde{\rho}}(t) &= -\lambda^2 \,\mathrm{tr}_{\mathrm{R}} \int_0^t \mathrm{d}s \left[\tilde{V}(t), \left[\tilde{V}(s), \tilde{\rho}(s) \otimes \omega_{\mathrm{R}}\right]\right] \\ &= -\lambda^2 \sum_{\alpha,\beta} \left\{ \int_0^t \mathrm{d}s \left\langle \tilde{B}_{\alpha}(t) \tilde{B}_{\beta}(s) \right\rangle \left[\tilde{A}_{\alpha}(t) \tilde{A}_{\beta}(s) \tilde{\rho}(s) - \tilde{A}_{\beta}(s) \tilde{\rho}(s) \tilde{A}_{\alpha}(t)\right] + \mathrm{h.c.} \right\}.\end{aligned} \tag{3.18}$$

The next assumption is that the reservoir correlation functions $\left\langle \tilde{B}_{\alpha}(t) \tilde{B}_{\beta}(s) \right\rangle$ decay on a timescale $t - s \sim \tau_{\mathrm{R}}$ which is much shorter than the timescale on which $\tilde{\rho}(t)$ evolves significantly. In this case we substitute $s = t - \tau$ and let the upper limit of the integral go to infinity. We also replace $\tilde{\rho}(s) \to \tilde{\rho}(t)$. This is the *Markov approximation* and the resulting equation reads

$$\dot{\tilde{\rho}}(t) = -\lambda^2 \,\mathrm{tr}_{\mathrm{R}} \int_0^{\infty} \mathrm{d}\tau \left[\tilde{V}(t), \left[\tilde{V}(t-\tau), \tilde{\rho}(t) \otimes \omega_{\mathrm{R}}\right]\right] \tag{3.19a}$$

$$= -\lambda^2 \sum_{\alpha,\beta} \left\{ \int_0^{\infty} \mathrm{d}\tau \left\langle \tilde{B}_{\alpha}(t) \tilde{B}_{\beta}(t-\tau) \right\rangle \left[\tilde{A}_{\alpha}(t) \tilde{A}_{\beta}(t-\tau) \tilde{\rho}(t) - \tilde{A}_{\beta}(t-\tau) \tilde{\rho}(t) \tilde{A}_{\alpha}(t)\right] + \mathrm{h.c.} \right\}. \tag{3.19b}$$

The Markov approximation effectively means that the reservoir 'has no memory', or that once information from the system enters the reservoir it keeps traveling towards infinity, never to return. Indeed a prototypical example of a Markovian reservoir would be a infinitely long transmission line,* such that once a photon leaves the system it never reflects off anything and returns, in which case $\tau_\mathrm{R}$ is of the order of the inverse photon frequency $\tau_\mathrm{R} \sim \omega^{-1}$. For cQED with typical frequencies $\omega/2\pi \simeq 5\,\mathrm{GHz}$ and decoherence rates $\gamma/2\pi \simeq 1\,\mathrm{MHz}$ we see that the Markov approximation is well justified in this case. Conversely a typical example of a non-Markovian environment could be the same transmission line with a kink in it such that a small portion of the outgoing field is reflected back after a delay that is comparable in magnitude to the system decoherence time. (Such an experiment has been performed by Turlot *et al.* [38]).

Equation (3.19) is indeed a Markovian master equation, but unfortunately it does not (except in a few trivial cases) generate a quantum dynamical semigroup. In order to produce a completely positive master equation it is necessary to assume that $H_\mathrm{S}$ has a discrete spectrum

$$H_\mathrm{S} = \sum_E E\,|E\rangle\langle E|\,. \tag{3.20}$$

We can expand $\tilde{A}_\alpha(t)$ in terms of *eigenoperators of the Hamiltonian*, $A^\alpha_\omega$, satisfying

$$\tilde{A}_\alpha(t) = \sum_\omega A^\alpha_\omega \mathrm{e}^{-\mathrm{i}\omega t}, \tag{3.21}$$

where $\{\omega\}$ is the set of energy differences $\{E - E'\}$ and

$$A^\alpha_\omega = \sum_{E-E'=\omega} |E\rangle\langle E|A_\alpha|E'\rangle\langle E'|\,, \tag{3.22a}$$

$$A^\alpha_{-\omega} = \left(A^\alpha_\omega\right)^\dagger. \tag{3.22b}$$

Inserting into (3.19) yields

$$\dot{\tilde{\rho}}(t) = -\lambda^2 \sum_{\alpha,\beta}\sum_{\omega,\omega'} \mathrm{e}^{\mathrm{i}(\omega-\omega')t}\Gamma_{\alpha\beta}(\omega')\left[A^\alpha_{-\omega}A^\beta_{\omega'}\tilde{\rho}(t) - A^\beta_{\omega'}\tilde{\rho}(t)A^\alpha_{-\omega}\right] + \text{h.c.}\,, \tag{3.23}$$

* Such a transmission line is also the proper way to include a resistive circuit element, by setting the appropriate characteristic impedance (2.16b).

with

$$\Gamma_{\alpha\beta}(\omega) = \int_0^\infty e^{i\omega t}\big\langle \tilde{B}_\alpha(t)\tilde{B}_\beta(0)\big\rangle\, dt, \tag{3.24}$$

where we assumed that $\omega_R$ is a stationary state of the reservoir Hamiltonian $[H_R, \omega_R] = 0$ in order to be able to write $\Gamma_{\alpha\beta}(\omega)$ as time-independent, although this assumption is not strictly necessary.*

Finally, we assume that only those terms with $\omega = \omega'$ survive, due to all other terms being rapidly rotating. This is equivalent to the assumption that $\lambda$ is small enough that the decoherence is slow compared to $\tau_S$, the slowest timescale of the system dynamics, given by the largest $|\omega - \omega'|^{-1}$, $\omega \neq \omega'$. We can rewrite $\Gamma$ as

$$\Gamma_{\alpha\beta}(\omega) = \frac{1}{2}\gamma_{\alpha\beta}(\omega) + iS_{\alpha\beta}(\omega), \tag{3.25}$$

with $\gamma_{\alpha\beta}(\omega)$ given by the Fourier transform of the reservoir correlation function,

$$\gamma_{\alpha\beta}(\omega) = \Gamma_{\alpha\beta}(\omega) + \Gamma^*_{\beta\alpha}(\omega) = \int_{-\infty}^{\infty} e^{i\omega t}\big\langle \tilde{B}_k(t)\tilde{B}_l(0)\big\rangle\, dt, \tag{3.26}$$

and the Hermitian matrix $S_{\alpha\beta}$ defined by

$$S_{\alpha\beta}(\omega) = \frac{1}{2i}\big[\Gamma_{\alpha\beta}(\omega) - \Gamma^*_{\beta\alpha}(\omega)\big]. \tag{3.27}$$

In terms of these definitions, the master equation can be rewritten

$$\dot{\tilde{\rho}}(t) = -i\big[H', \tilde{\rho}(t)\big] + \frac{1}{2}\sum_{\alpha,\beta}\sum_{\omega}\gamma_{\alpha\beta}(\omega)\Big\{\big[A^\alpha_\omega\tilde{\rho}, A^\beta_{-\omega}\big] + \big[A^\alpha_\omega, \tilde{\rho}A^\beta_{-\omega}\big]\Big\}. \tag{3.28}$$

The Hermitian operator

$$H' = \sum_{\alpha,\beta}\sum_{\omega} S_{\alpha\beta}(\omega)A^\alpha_{-\omega}A^\beta_\omega, \tag{3.29}$$

commuting with the system Hamiltonian, $[H_S, H'] = 0$, describes a renormalization of the system energies due to the coupling to the environment, the *Lamb shift*. We can write (3.28)

* An important example with $[H_R, \omega_R] \neq 0$ is when the reservoir is in a squeezed vacuum state, as discussed in [30, section 3.4.3].

in the Schrödinger picture as

$$\dot{\rho}(t) = -\mathrm{i}\big[H_S + H', \rho(t)\big] + \frac{1}{2}\sum_{\alpha,\beta}\sum_{\omega}\gamma_{\alpha\beta}(\omega)\Big\{\big[A^{\alpha}_{\omega}\rho, A^{\beta}_{-\omega}\big] + \big[A^{\alpha}_{\omega}, \rho A^{\beta}_{-\omega}\big]\Big\}. \tag{3.30}$$

It is possible to use Bochner's theorem to show that $\gamma_{\alpha\beta}(\omega)$ is a positive matrix, and thus (3.30) is in the standard form (3.8), and it describes completely positive Markovian evolution.

We see that the positive-frequency components of the reservoir correlation function are associated with *relaxation* processes with the reservoir absorbing energy from the system, while the negative-frequency components describe a transfer of energy from the reservoir to the system and the d.c. component describes *dephasing* processes, where no energy is transferred. This concept is expanded further, in the context of a *quantum noise* approach to measurement and amplification, in the pedagogical review of A. A. Clerk *et al.* [39].

It is not immediately obvious, but (3.30) is significantly less general than (3.8), due to the nontrivial constraints among $H'$, $\gamma_{\alpha\beta}$ and $A^{\alpha}_{\omega}$. There is an example of this in relation to two level systems in section 3.5.

### 3.3.1 Heat bath

The discussion so far has made no particular assumption about the state $\omega_R$ of the reservoir $\mathcal{R}$. If $\mathcal{R}$ is to be an equilibrium *heat bath* of inverse temperature $\beta = 1/kT$, then this implies

$$\gamma_{\alpha\beta}(-\omega) = \mathrm{e}^{-\beta\omega}\gamma_{\beta\alpha}(\omega). \tag{3.31}$$

This in turn implies that the most general form of the heat bath generator is

$$L\rho = -\mathrm{i}\,[H_{\text{eff}}, \rho] + \sum_{\omega>0}\big\{\mathcal{D}[V_\omega]\rho + \mathrm{e}^{-\beta\omega}\mathcal{D}[V^{\dagger}_{\omega}]\rho\big\}, \tag{3.32}$$

with $[H_{\text{eff}}, H_S] = 0$, and $\mathrm{e}^{\mathrm{i}H_S t}V_\omega \mathrm{e}^{-\mathrm{i}H_S t} = \mathrm{e}^{-\mathrm{i}\omega t}V_\omega$. Equation (3.32) is a type of detailed balance condition, in the following sense: If the spectrum of $H_S$ is non-degenerate then under (3.30) the diagonal elements of $\rho$ evolve *independently* of the off-diagonal elements, obeying a classical *Pauli master equation*

$$\dot{\rho}_{ii} = \sum_j\Big(W_{ij}\rho_{jj} - W_{ji}\rho_{ii}\Big), \tag{3.33a}$$

where the transition rates $W_{ij}$ are exactly those that could be obtained from Fermi's golden rule

$$W_{ij} = \sum_{\alpha,\beta} \gamma_{\alpha\beta}(E_i - E_j)\langle\, j \,|\, A_\alpha \,|\, i \,\rangle\langle\, i \,|\, A_\beta \,|\, j \,\rangle. \tag{3.33b}$$

Even if the reservoir is not in thermal equilibrium, it is possible to take (3.31) as defining an *effective temperature*, as is done in NMR with the *spin temperature*. Of course the effective temperature might in general be frequency dependent, in the case that the system has more than two levels.

## 3.4 Damped harmonic oscillator

A simple example of the formalism is for the harmonic oscillator, with the coupling to the environment linear in the position and momentum

$$H_S = \omega a^\dagger a, \tag{3.34a}$$

$$\lambda V = \gamma_1(a + a^\dagger)B_1 + i\gamma_2(a - a^\dagger)B_2, \tag{3.34b}$$

with coupling constants $\gamma_i$. Because there is only one frequency of the system, the $A_\omega^\alpha$ are simply $a$ and $a^\dagger$, and the master equation is*

$$\dot{\rho} = -i\omega[a^\dagger a, \rho] + \kappa_- \mathcal{D}[a]\rho + \kappa_+ \mathcal{D}[a^\dagger]\rho, \tag{3.35}$$

with constants $\kappa_\pm \geq 0$ that are determined by $\gamma_i$ and the reservoir spectral density at $\mp\omega$. Since experiments in cQED are generally performed with cavities having $\omega/2\pi \simeq 5\,\mathrm{GHz}$ in a dilution refrigerator at $T \simeq 20\,\mathrm{mK}$, corresponding to $\kappa_+/\kappa_- \simeq \exp(-\beta\omega) \simeq 10^{-5}$, it is usual to set $\kappa_+ = 0$ and drop the $\mathcal{D}[a^\dagger]$ term (and drop the subscript on $\kappa_-$).

## 3.5 Bloch equations

As can be seen from (3.8), the most general Markovian master equation for the two-level system has 3 parameters describing the Hamiltonian evolution, and 9 parameters describing the dissipation. One way [29] to write this is in terms of the Bloch vector (2.75), $\vec{r} = \{x, y, z\}$,

$$\dot{x} = -\gamma_3 x + (\alpha - \omega_0)y + (\beta - \omega_2)z - \sqrt{2}\lambda, \tag{3.36a}$$

$$\dot{y} = (\alpha + \omega_0)x - \gamma_2 y + (\delta - \omega_1)z + \sqrt{2}\mu, \tag{3.36b}$$

$$\dot{z} = (\beta + \omega_2)x + (\delta + \omega_1)y - \gamma_1 z - \sqrt{2}\nu. \tag{3.36c}$$

---

* We naïvely ignore the fact that $H_S$ is unbounded.

Here, $\omega_i$ are the Hamiltonian parameters, and $\alpha$, $\beta$, $\delta$, $\gamma_i$, $\lambda$, $\mu$, $\nu$ parameterize the dissipation, with constraints

$$0 \leq \gamma_i \leq \gamma_j + \gamma_k, \quad \{i, j, k\} \text{ a permutation of } \{1, 2, 3\}, \tag{3.37a}$$

$$4(\alpha^2 + \nu^2) \leq \gamma_1^2 - (\gamma_2 - \gamma_3)^2, \tag{3.37b}$$

$$4(\beta^2 + \mu^2) \leq \gamma_2^2 - (\gamma_1 - \gamma_3)^2, \tag{3.37c}$$

$$4(\delta^2 + \lambda^2) \leq \gamma_3^2 - (\gamma_1 - \gamma_2)^2, \tag{3.37d}$$

$$\begin{aligned} &16(\alpha\beta\delta + \alpha\lambda\mu + \delta\mu\nu) + 4\gamma_1(\alpha^2 + \nu^2) + 4\gamma_2(\beta^2 + \mu^2) + 4\gamma_3(\delta^2 + \lambda^2) \\ &\quad + \gamma_1^2(\gamma_2 + \gamma_3) + \gamma_2^2(\gamma_1 + \gamma_3) + \gamma_3^2(\gamma_1 + \gamma_2) \geq 16\beta\lambda\nu + 2\gamma_1\gamma_2\gamma_3 + \gamma_1^2 + \gamma_2^2 + \gamma_3^2 \\ &\quad + 4\gamma_1(\beta^2 + \delta^2 + \lambda^2 + \mu^2) + 4\gamma_2(\alpha^2 + \delta^2 + \lambda^2 + \nu^2) + 4\gamma_3(\alpha^2 + \beta^2 + \mu^2 + \nu^2). \end{aligned} \tag{3.37e}$$

Performing a change of basis to rotate the Hamiltonian part to be proportional to $\sigma_z$ is always possible so without loss of generality we may choose $\omega_1 = \omega_2 = 0$. Additionally, choosing as a very special case $\alpha = \beta = \delta = \lambda = \mu = 0$, $\gamma_2 = \gamma_3$, and renaming

$$T_1 \triangleq 1/\gamma_1, \qquad T_2 \triangleq 1/\gamma_2, \qquad \bar{z} \triangleq -\sqrt{2}\nu T_1, \tag{3.38}$$

we obtain the usual Bloch equations

$$\dot{x} = -\frac{x}{T_2} - \omega_0 y, \tag{3.39a}$$

$$\dot{y} = \omega_0 x - \frac{y}{T_2}, \tag{3.39b}$$

$$\dot{z} = -\frac{1}{T_1}(z - \bar{z}), \tag{3.39c}$$

and in this case, the inequalities (3.37) simply reproduce the well-known condition

$$T_2 \leq 2T_1. \tag{3.40}$$

Although the Bloch equations are a very special case of the general Markovian master equation for a two-level system, they in fact describe the most general master equation that can be derived within the weak coupling formalism. To see that this is true, observe that for a two-level system with bare Hamiltonian

$$H_{\mathrm{S}} = \omega_{\mathrm{q}}\sigma_z/2 \tag{3.41}$$

there are only three relevant frequencies, $\pm\omega_{\mathrm{q}}$ and 0, and that each frequency has only one non-trivial associated eigenoperator $A_\omega$ evolving at that frequency:

$$A_{\pm\omega_{\mathrm{q}}} = \sigma_{\pm}, \tag{3.42a}$$
$$A_0 = \sigma_z. \tag{3.42b}$$

This gives the master equation

$$\dot{\rho} = -\mathrm{i}\left[\omega_0\sigma_z/2, \rho\right] + \gamma_- \mathcal{D}[\sigma_-]\rho + \gamma_+ \mathcal{D}[\sigma_+]\rho + \frac{\gamma_\varphi}{2}\mathcal{D}[\sigma_z]\rho. \tag{3.43}$$

Here, $\gamma_\pm$ are related to the $\pm\omega_{\mathrm{q}}$ components of the reservoir correlation function and $\gamma_\varphi$ to the d.c. component. Equation (3.43) is identical to (3.39) after we make the identification

$$T_1 = \frac{1}{\gamma_- + \gamma_+}, \tag{3.44a}$$
$$T_2 = \frac{2}{\gamma_- + \gamma_+ + 2\gamma_\varphi}, \tag{3.44b}$$
$$\bar{z} = \frac{\gamma_+ - \gamma_-}{\gamma_+ + \gamma_-}. \tag{3.44c}$$

Equation (3.43) is generally taken as the master equation for a qubit in cQED. For the same reasons as described in the previous section, the $\mathcal{D}[\sigma_+]$ term is usually dropped, corresponding to $\bar{z} = -1$.

## 3.6 Master equation for the transmon-cavity system

The Bloch equations give an excellent description of the behavior of two-level systems in many experimental scenarios, ranging from NMR to quantum optics—there are very few experiments where phenomena have been observed that would require the additional generality of equations (3.36). It is also quite hard to imagine how one could reduce the unitary dynamics of a combined system-with-reservoir to the irreversible dynamics of an open system and arrive at a master equation of the form (3.36), in any kind of rigorous way. It would therefore be pleasant if we could also use a weak-coupling argument to derive a master equation for the combined transmon and cavity system: this would give a microscopic explanation for the damping, and as an additional advantage there would be far fewer parameters than for the fully-general case. Unfortunately, there are two problems with this approach. The first problem is that we do not know *a priori* which transmon operators $A_\alpha$ are the relevant ones that couple reservoir degrees of freedom, producing dissipation. Unfortunately, most research to date has focussed on the behavior of only the lowest 2 levels of transmons, due to the interest in using transmons as qubits, and there has been little in the way of systematic investigation of dissipation of the higher levels. The behavior of the lower 2 levels is well-described by Bloch equations with $\bar{z} = 0$, but the microscopic origin of the $T_1$ and $T_2$ is a topic of active research. The next subsection outlines some plausible microscopic sources of dissipation. The second problem with the weak-coupling argument is of a more fundamental character: the weak-coupling argument requires that there is a separation of frequency scales such that either the frequency differences $\omega - \omega'$ are very large compared to the dissipation, or otherwise that the frequencies are identical. Davies recognized this problem very early on [40] and derived a more sophisticated version of the weak-coupling argument, where the system Hamiltonian is separated into two commuting pieces

$$H_\mathrm{S} = H_\mathrm{S}^0 + H_\mathrm{S}^1, \tag{3.45a}$$

$$\left[H_\mathrm{S}^0, H_\mathrm{S}^1\right] = 0, \tag{3.45b}$$

where for $H_\mathrm{S}^0$ all the 'small' terms $\omega - \omega'$ vanish, and where $H_\mathrm{S}^1$ is small and is treated perturbatively. Although this is an improvement, in practise there are always some frequency differences in the Hamiltonian that are neither very large compared to the dissipation nor negligibly small. For example, for dispersive readout (section 2.10.2) the state-dependent shift $\chi$ of the cavity frequency is typically intentionally chosen to be approximately equal to the cavity linewidth. Since a rigorous application of the weak coupling argument is impossible in this situation, it is usual in cQED to take a rather simplistic approach to this problem and just assume that the master equation for the combined system can be formed by adding the terms from each of the components. For example, for a qubit coupled to a cavity, we combine (2.50), (3.35) and (3.43) to get

$$\dot{\rho} = -\mathrm{i}\left[H, \rho\right] + \kappa\mathcal{D}[a]\rho + \gamma_-\mathcal{D}[\sigma_-]\rho + \frac{\gamma_\varphi}{2}\mathcal{D}[\sigma_z]\rho, \tag{3.46a}$$

$$H = \omega_\mathrm{r} a^\dagger a + \omega_\mathrm{q}\sigma_z/2 + g\left(a\sigma_+ + a^\dagger\sigma_-\right), \tag{3.46b}$$

assuming zero temperature. To reiterate, this is explicitly a Markovian master equation, but it is not one that could be derived from weak coupling. It is worth noting that in cQED we have no direct access to the parameters of the uncoupled system, unlike in atomic cavity QED where it is possible to measure the transition frequencies of the real atoms when they are outside the cavity and to measure the cavity parameters when it is empty. In cQED the cavity and 'atom' are permanently attached and this is not possible. Thus $\omega_\mathrm{r}$, $\omega_\mathrm{q}$ and $g$ of (3.46b) should perhaps be considered more as renormalized parameters, already including the first-order effects of any Lamb-shift-like effects.* Similarly, there is no experimental access to the relaxation parameters of the uncoupled qubit and cavity, so $\kappa$, $\gamma_-$ and $\gamma_\varphi$ should be thought of as renormalized relaxation parameters not necessarily having the same values as we would measure if we could somehow uncouple the qubit from the cavity.

### 3.6.1 Possible microscopic mechanisms of decoherence for transmons

Koch *et al.* [20] discuss a number of different decoherence mechanisms for the transmon. The ones which are likely to be most important are:

**Dipole-like coupling and multimode Purcell effect.** An obvious choice for the coupling operator is the charge operator $n$. This applies in particular to the situation that the transmon couples to an electromagnetic mode, for example an unintended mode of the sample holder. Even more specifically, it applies to the *multimode Purcell effect*. The Purcell effect [41] describes the relaxation of the transmon *via the cavity*. It refers to the fact that the dressed-state solutions of the (generalized) Jaynes–Cummings Hamiltonian have both transmon and cavity character, and so the relaxation rates are modified compared to the bare qubit. As such, the Purcell effect is explicitly included in (3.46). However, recall that in proceeding from (2.23) to (2.24) we dropped all but one modes of the cavity. It is true that at the Hamiltonian level the ignored higher cavity levels have little influence, due to being far detuned from the frequencies of interest, but they can still be relevant to the dissipation. Properly including the multimode Purcell effect is, unfortunately, not as simple as taking (3.46) and inserting a sum over cavity modes $\mathcal{D}[a]\rho \to \sum_n \mathcal{D}[a_n]\rho$, etc. The problem is that the sum does not converge, analogous to other situations in QED where care is needed with high-frequency cutoffs. Houck *et al.* [42] showed that a semiclassical expression, based on a circuit model in which the cavity transforms the 50 Ω impedance of the environment as seen by the transmon, fits experimental data rather well, as seen in figure 3.1. For the results reported in chapter 4, the parameters are such that the multimode Purcell effect is expected to be the dominant cause of relaxation.

---

* Since the dissipation terms in (3.46) do not come from a weak-coupling master derivation, it is not obvious what Lamb shifts should be associated with them.

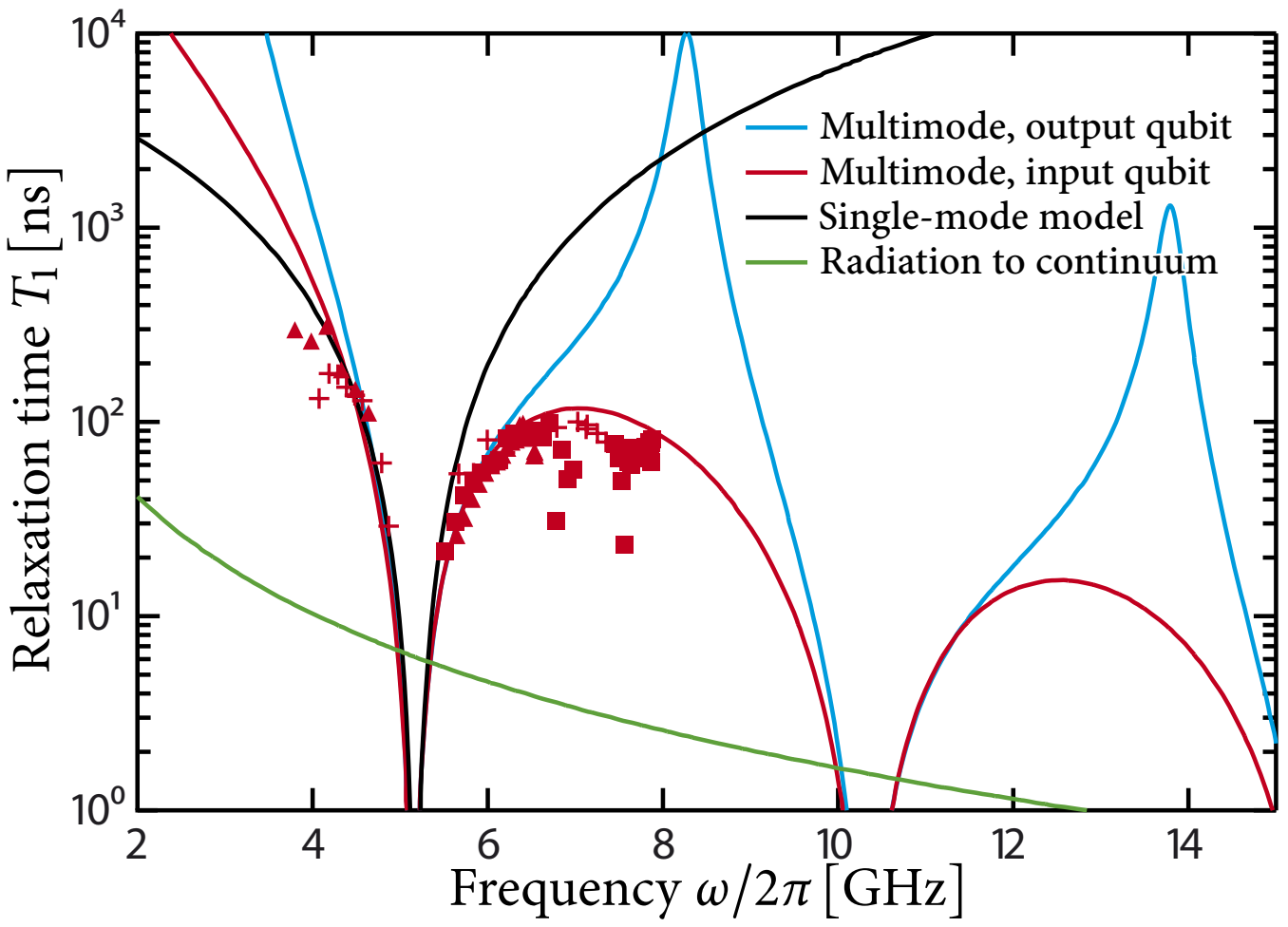

**Figure 3.1: Multimode Purcell effect.** Comparison of multimode and single-mode models of relaxation. Spontaneous emission lifetimes into a single-mode cavity are symmetric about the cavity frequency, while within the circuit model lifetimes below the cavity are substantially longer than above. The measured $T_1$ for three similar qubits deviates substantially from the single-mode prediction, but agrees well with the circuit model. Two different curves for the multimode model are shown, to illustrate that the multimode model predictions depend on the position of the qubit within the cavity. The expected decay time for radiation into a continuum is shown for comparison. (Figure used with permission from [42]. See Copyright Permissions.)

The component of the charge operator $n$ that oscillates at the transmon frequency is simply $c$ (defined in (2.43)) and the resulting dissipation term is

$$\gamma_- \mathcal{D}[c]\rho + \gamma_+ \mathcal{D}[c^\dagger]\rho = \gamma_- \mathcal{D}\Big[\sum_j \frac{n_{j,j+1}}{n_{01}} |j\rangle \langle j+1|\Big]\rho + \gamma_+ \mathcal{D}\Big[\sum_j \frac{n_{j+1,j}}{n_{01}} |j+1\rangle \langle j|\Big]\rho, \quad (3.47)$$

where the chosen normalization ensures that $\gamma_\pm$ have their usual two-level interpretation in the case that the sum is truncated at the lowest two levels. We ignore non-nearest-neighbor components of $n$ because for $E_J/E_C \gg 1$ these are small (section 2.5.3). Equation (3.47) assumes that the anharmonicity is small compared to the relaxation, which may or may not be true. In the opposite limit, it is more appropriate to split $n$ into components at each frequency $\omega_{j,j+1}$

$$\gamma_- \sum_j \mathcal{D}\Big[\frac{n_{j,j+1}}{n_{01}} |j\rangle\langle j+1|\Big]\rho + \gamma_+ \sum_j \mathcal{D}\Big[\frac{n_{j+1,j}}{n_{01}} |j+1\rangle\langle j|\Big]\rho, \tag{3.48}$$

although neither (3.47) nor (3.48) is rigorously justified, given that as described above the frequencies are also shifted by the coupling to the cavity, and there is no simple separation of frequency scales.

**Dephasing via charge noise.** For the Cooper pair box with $E_J/E_C \simeq 1$ the dephasing is known to be primarily caused by slow fluctuations of the offset charge $n_g$. The exponential suppression of this effect is precisely the motivation for using $E_J/E_C \gg 1$, so this mechanism is unlikely to be the dominant cause of dephasing for the lowest levels of the transmon. Due to their larger charge dispersion it is likely relevant for the higher levels, however. Dephasing due to fluctuations of any parameter, $\alpha$, can be incorporated as a coupling to the operator $\partial H/\partial\alpha$. For the case of charge noise, the d.c. component of $\partial H/\partial n_g$ is

$$\sum_j \frac{\mathrm{d}E_j(n_g)}{\mathrm{d}n_g} |j\rangle\langle j| \sim \sin(2\pi n_g) \sum_j \epsilon_j |j\rangle\langle j|, \tag{3.49}$$

where $\epsilon_j$ is the charge dispersion, and we have made use of (2.32). Absorbing* $\sin^2(2\pi n_g)$ into $\gamma_\varphi^C$ (squared because $\mathcal{D}[A]\rho$ is quadratic in $A$) and choosing the normalization such that $\gamma_\varphi^C$ has its two-level interpretation when truncating to the lowest two levels, the dissipation term is thus

$$\frac{\gamma_\varphi^C}{2}\mathcal{D}\Big[\sum_j \frac{2\epsilon_j}{\epsilon_1 - \epsilon_0} |j\rangle\langle j|\Big]\rho. \tag{3.50}$$

**Dephasing via flux noise.** For the transmon, the dephasing of the lowest levels can be caused by an effective fluctuation of the flux, through the squid loop that tunes the transmon. Away from the *flux sweet spots* $\tilde{\Phi} = j\Phi_0$, integer $j$, at which $\mathrm{d}E_J(\tilde{\Phi})/\mathrm{d}\tilde{\Phi} = 0$, this is known to be a significant source of dephasing for the lowest levels of the transmon. Ignoring the

* One might wonder that since there is typically no gate electrode for transmons, $n_g$ is uncontrolled and $\gamma_\varphi^C$ could vary in time. In this context it is worth noting that in order to improve the signal-to-noise ratio the experiments are typically repeated some $10^5$ to $10^6$ times with averaging, a process that takes some minutes. If the variations of $n_g$ are fast on this timescale, it is reasonable to treat the $\sin^2(2\pi n_g) \simeq \frac{1}{2}$.

anharmonicity, the dissipation is thus given by

$$\sin(\pi\tilde{\Phi}/\Phi_0)\mathcal{D}\Big[\sum_j j\,|j\rangle\langle j|\Big]\rho \sim \frac{\gamma_\varphi^\Phi}{2}\mathcal{D}\Big[\sum_j 2j\,|j\rangle\langle j|\Big]\rho, \tag{3.51}$$

where again the normalization allows the usual interpretation of $\gamma_\varphi^\Phi$ in the two-level truncation.

### 3.6.2 Putting the pieces together

Combining the driven generalized Jaynes–Cummings Hamiltonian (2.62) with the dissipation for the resonator (3.35) and for the transmon (3.47), (3.50) and (3.51) gives the master equation in the rotating frame

$$\begin{aligned}\dot{\rho} = &-\mathrm{i}\Big[\Delta_\mathrm{r} a^\dagger a + \sum_j \Delta_j |j\rangle\langle j| + g\big(a^\dagger c + a c^\dagger\big) + \big(a\xi^* + a^\dagger\xi\big), \rho\Big] \\ &+ \kappa_- \mathcal{D}[a]\rho + \kappa_+\mathcal{D}[a^\dagger]\rho + \gamma_-\mathcal{D}[c]\rho + \gamma_+\mathcal{D}[c^\dagger]\rho \\ &+ \frac{\gamma_\varphi^\mathrm{C}}{2}\mathcal{D}\Big[\sum_j \frac{2\epsilon_j}{\epsilon_1 - \epsilon_0}|j\rangle\langle j|\Big]\rho + \frac{\gamma_\varphi^\Phi}{2}\mathcal{D}\Big[\sum_j 2j\,|j\rangle\langle j|\Big]\rho.\end{aligned} \tag{3.52}$$

The parameters of (3.52) are not independent. Many of the parameters are set during fabrication of a sample, so their interdependence is difficult to see. However, when taking measurements on a given sample there are several control parameters or ‘knobs’ available: the flux $\tilde{\Phi}$ and the temperature, as well as the frequency $\omega_\mathrm{d}$ and amplitude $\xi$ of the drive. If the reservoirs can be considered as heat baths then

$$\kappa_+ = \mathrm{e}^{-\beta_\kappa \omega_\mathrm{r}}\kappa_-, \tag{3.53a}$$

$$\gamma_+ = \mathrm{e}^{-\beta_\gamma \omega_{01}}\gamma_-, \tag{3.53b}$$

where we have allowed for the possibility that the bath to which the cavity relaxes has an inverse temperature $\beta_\kappa$ different from the inverse temperature $\beta_\gamma$ of the bath to which the transmon relaxes. If the transmon relaxation is due to the multimode Purcell effect then there is just one bath and $\beta_\kappa = \beta_\gamma = \beta$. If $\omega_{01} \simeq \omega_\mathrm{r}$ then we can additionally assume that the Boltzmann factors are the same

$$r = \frac{\kappa_+}{\kappa_-} = \frac{\gamma_+}{\gamma_-}. \tag{3.54}$$

Varying the flux explicitly affects $\Delta_j$ and $c$, $c^\dagger$. It also affects $\gamma_-$ by changing the frequency at which the reservoir correlation functions should be evaluated. However, if we assume that the reservoir correlations are fairly 'white' over the frequency range of interest then it is reasonable to ignore this effect. In the case of the multimode Purcell effect, the higher cavity modes that are responsible for the dissipation are detuned by around $\omega_\mathrm{r}$, so if the transmon is being tuned over a frequency range that is small compared to $\omega_\mathrm{r}$, (as we have already assumed in writing the Jaynes–Cummings Hamiltonian) then this will hold. Dephasing due to flux noise is also a function of the applied flux $\gamma_\varphi^\Phi(\tilde{\Phi}) \sim \cos(\pi\tilde{\Phi}/\Phi_0)$. This dependence has been observed in experiment [43]. Nevertheless, when the tuning is only over a relatively small range in frequency, not too close to the flux sweet spot, it is reasonable to treat $\gamma_\varphi^\Phi$ as independent of $\tilde{\Phi}$.

Changing the drive frequency $\omega_\mathrm{d}$ only explicitly affects $\Delta_\mathrm{r}$ and $\Delta_j$ and there are no parameters other than $\xi$ with an explicit dependence on the drive amplitude.

In addition to these explicit effects, it is also possible to imagine that the dissipation constants might be functions of the control parameters via the reservoir state. For example, if there is some narrow resonance in the reservoir, then tuning the drive frequency $\omega_\mathrm{d}$ through this resonance might drive the reservoir far from equilibrium, significantly altering the reservoir correlation functions and thus potentially affecting the dissipation constants. Another example would be if the reservoir state were affected by the magnetic field that is used to tune the transmon. However, this type of effect has not been observed experimentally* and so we assume that apart from the explicit dependences described above, there is no dependence on the control parameters.

* A simple dependence of the reservoir state on magnetic field has not been observed. However, more complex hysteretic behaviors have been seen. In particular, after large magnetic field swings have been applied, occasionally the dissipation rates of the transmon can increase significantly, and this persists even if the magnetic field is returned to its original value. One way to undo this effect is to warm the sample above the superconducting transition temperature. From this evidence it seems likely that what is going on is creation of trapped vortices in the superconducting ground planes, and these are able to increase the dissipation somehow.

# CHAPTER 4

## Nonlinear response of the vacuum Rabi splitting

The Jaynes–Cummings Hamiltonian (2.50) provides a *fully quantum* description of the interaction between a single two-level system and a single mode of the electromagnetic field. It was originally developed in the context of the atomic cavity QED at optical frequencies, and previous chapters explained how it is also applicable for superconducting artificial atoms. It has also been used to model a number of other experimental scenarios. Amongst others, it has been applied to *Rydberg atoms* [6], which are highly-excited atomic states (principle quantum number $n \simeq 50$) that have very large dipole moments and transition frequencies in the 50 GHz range, and to semiconductor quantum dot systems [44, 45]. Perhaps less obviously, it can also be applied to the case of the coupling between the internal states and motional states of an ion in a trap [46, 47]—to the extent that the trapping potential is harmonic, the quantized phonons of the trap motion play the same rôle as the photons of the electromagnetic field. There are even proposals to implement the Jaynes–Cummings Hamiltonian using superconducting qubits coupled to nanomechanical oscillators [48], with very recent results reported in [49].

Recall that the Jaynes–Cummings Hamiltonian has the form

$$H = \omega_r a^\dagger a + \omega_q \sigma_z/2 + g(a\sigma_+ + a^\dagger \sigma_-). \tag{2.50}$$

The energy spectrum of this Hamiltonian is shown schematically in figure 4.1. One specific feature is the *vacuum Rabi splitting*, which is the effect that upon putting a *single* resonant ($\omega_q = \omega_r$) atom into a cavity, the cavity transmission peak splits into a pair of peaks, separated by twice the coupling strength* $g$, also known as the *vacuum Rabi frequency*. In order for this effect to be observable in experiment, the separation of the peaks must be larger than the linewidth of those peaks. For a superconducting qubit at low temperature the linewidth can be determined from the relaxation parameters of (3.46a): the photon leakage rate $\kappa$ and the qubit decoherence rates $\gamma$ and $\gamma_\varphi$. For the case of real atoms there is an additional source of decoherence, namely that the atoms have a finite lifetime, $\tau$ before they drift out of the cavity. Thus, in order to have clearly-resolvable vacuum Rabi splitting we need $g \gg \kappa, \gamma, \gamma_\varphi, \tau^{-1}$, which is known as the *strong coupling* regime of cavity QED. The strong-coupling regime has indeed been reached, the first time being in 1992 for single real atoms in optical cavities [50], and subsequently for Rydberg atoms in microwave cavities [51], artificial atoms in circuit QED [52] (although experiments as early as 1989 [53], using a transmission line with a movable absorber, are very closely related and could be interpreted as an early manifestation of strong coupling physics) and quantum-dot systems [44, 45].

In the case of atomic cavity QED, the vacuum Rabi frequency is a tiny fraction of the atomic frequency $g/\omega_q \ll 1$ and there is not much that can be done to alter this—experimental advances in reaching the strong-coupling regime mainly depend on reducing the sources of dissipation as far as possible, for example by using extremely reflective mirrors for the cavity and various schemes to reduce the motion of the atoms. Conversely, reaching the strong-coupling regime with transmons is significantly easier, due to the quasi-1d nature of the system allowing $g$ to be a much larger fraction of the atomic frequency. Section 4.1 discusses this point. Section 4.2 gives brief details of the experimental setup that was used for investigating the strong-coupling regime, using a sample containing two different transmons.

In cQED the atom cannot be physically removed from the cavity to switch off the vacuum Rabi splitting—instead the atom is tuned far away from the cavity in frequency space. The vacuum Rabi splitting is thus typically observed in the form of an avoided crossing, occurring as the qubit frequency is tuned through resonance with the cavity. Such an avoided crossing is certainly an expected behavior from the Jaynes–Cummings Hamiltonian but, as was pointed out by Zhu *et al.* [54] and Tian *et al.* [55], it is certainly not proof of the Jaynes–Cummings physics, nor is it a uniquely quantum behavior of the system, given that the resonance frequencies of coupled classical harmonic oscillators can display the same behavior. However,

* Recall that $g$ may be a function of $E_J$ and hence $\omega_q$. Thus, the vacuum Rabi frequency is $g(\omega_q = \omega_r)$.

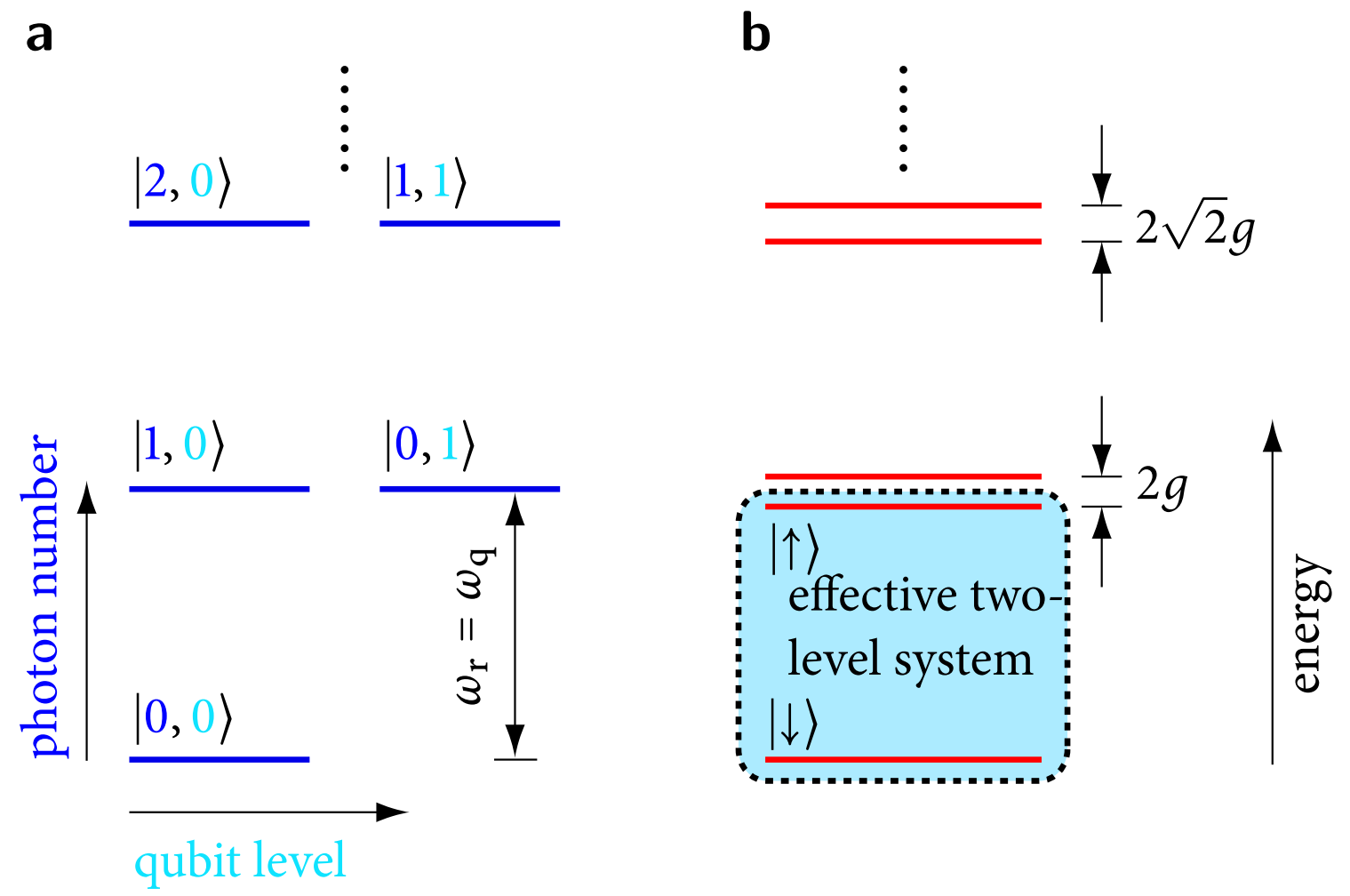

**Figure 4.1: Jaynes–Cummings level diagram of the resonator–qubit system. a**, Bare levels in the absence of coupling. The states are denoted $|n, j\rangle$ for photon number $n$ and occupation of qubit level $j$. **b**, Spectrum of the system including the effects of qubit–resonator coupling. The effective two-level system relevant to describing the lower vacuum Rabi peak comprises the ground state and the antisymmetric combination of qubit and photon excitations.

there is more to the Jaynes–Cummings Hamiltonian than merely an avoided crossing in a transmission spectrum, and the remainder of this chapter discusses two such characteristic aspects, both of which can conveniently be observed by driving the system very strongly, beyond linear response.

The first such characteristic aspect is that the vacuum Rabi splitting, unlike the avoided crossing in the interaction of a pair of harmonic oscillators, is caused by an avoided crossing between discrete quantum energy levels, one of which belongs to a two-level system. The combined qubit–cavity 'molecule' can thus display two-level physics of its own, as shown in figure 4.1b. In particular, under strong driving the transmission spectrum displays saturation effects. Saturation effects are well known from many quantum systems, giving rise to power-broadening in NMR and such phenomena as *resonance fluorescence* in atomic physics.* *Photon blockade* effects were observed with optical cavity QED [57], showing up in time-domain measurements as a photon anti-bunching.† Due to the phase-insensitive heterodyne detection scheme used in typical cQED experiments, the saturation appears in a somewhat

* Resonance fluorescence effects in the form of Autler–Townes splitting and the Mollow triplet have very recently also been observed with transmons [56].

† Anti-bunching is a typical behavior of strongly driven two-level systems, originally observed in the 1970s with atoms in free space [58–61].

unexpected way: as the drive power is increased, the peaks of the transmission spectrum, already split by the vacuum Rabi effect, split again. Section 4.5 discusses this phenomenon, which we have called *supersplitting*. This can be contrasted with the more usual power-broadening that would be observed in the same experiment, except using photon-counting detection.

The second characteristic aspect of the strongly-driven Jaynes–Cummings Hamiltonian is that there is a whole 'ladder' of energy states in (2.50), figure 4.1, and quantum mechanics gives rise to a distinct anharmonicity of these splittings*: the splitting of the $n$-excitation manifold is enhanced by a factor $\sqrt{n}$ as compared to the vacuum Rabi ($n = 1$) situation, (2.54b). This *quantum Rabi* anharmonicity has been observed in time-domain experiments using single Rydberg atoms in microwave [62] cavities, where peaks due to the $\sqrt{2}$, $\sqrt{3}$, $\sqrt{4}$ cases were visible in the Fourier-transformed experimental signal. In a cQED system using transmons, the position of the $n = 2$ levels was demonstrated in a two tone pump-probe measurement [63]. Using phase qubits, this $\sqrt{n}$ scaling has been observed in time-domain measurements for up to $n = 15$ [64, 65] (and refinements of these experiments are able to perform synthesis of arbitrary quantum states of the cavity [66]).

Transmission spectroscopy under strong driving, as in the experiments described in this chapter, allows multiphoton transitions to the higher levels of the Jaynes–Cummings ladder, as was predicted by Carmichael *et al.* [67]. With heroic effort, the 2-photon transition was observed in optical cavities [68]. These experiments are very difficult to perform with real atoms, because the very strong laser fields cause such effects as cavity birefringence, and have a tendency to create ponderomotive forces that push the atoms out of the cavity. Rather sophisticated postprocessing is therefore needed in order to select for only those atoms which have not been heated by the laser, causing them to oscillate too strongly in the cavity. By contrast to the situation with real atoms, with transmons there is no particular difficulty with increasing the drive power, and we were able to see the $\sqrt{n}$ splitting for $n = 1, \ldots, 5$. This is discussed in section 4.6.

## 4.1 Strong coupling: the fine structure limit

This section presents a simple calculation [4, 6, 69] showing that the coupling strength of an atom and a photon in cavity QED has an upper limit that can be related to fundamental

* To avoid any confusion: this is the anharmonicity of the Jaynes–Cummings ladder, which should not be mistaken for the anharmonicity of the bare transmon (section 2.5.2).

constants. A photon excites an atom by moving one of its electrons into a larger orbit (for an artificial atom, a Cooper pair is excited rather than an electron). The dipole moment $d = eL$, where $e$ is the electron charge and $L$ is a distance, is a measure of the size of the atom, and determines how strongly the atom interacts with a given electric field. The vacuum Rabi frequency is thus given by $g = dE_0$, where $E_0$ is the root-mean-square electric field at the location of the atom, due to vacuum fluctuations. The vacuum fluctuations exist in both electric and magnetic fields, and have an amplitude equal to that due to half a photon. A simple estimate of this electric field can be obtained from the density of energy, $\epsilon_0 E^2/2$, stored in the electric field, which accounts for half the energy (the other half being stored in the magnetic field):

$$\frac{\omega}{4} = \frac{\epsilon_0}{2} \int E^2 \, \mathrm{d}V = \frac{\epsilon_0}{2} E_0^2 V, \tag{4.1}$$

where $\epsilon_0$ is the permittivity of free space, $\omega$ is the transition frequency of the atom/cavity and $V$ is the volume of the cavity. Thus, the field strength increases as the volume of the cavity is decreased. A typical three-dimensional cavity used with real atoms will have a volume that is many cubic wavelengths. In circuit QED, we can use a one-dimensional transmission-line cavity, which must be half a wavelength long but can be much smaller in the transverse directions, giving a volume much less than a cubic wavelength and thus a greatly enhanced field strength. For concreteness, consider a coaxial transmission line of radius $r$, with volume, $V = \pi r^2 \lambda/2$, for which

$$E_0 = \frac{1}{r}\sqrt{\frac{\omega^2}{2\pi^2 \epsilon_0 c}}, \tag{4.2}$$

where the wavelength $\lambda = 2\pi c/\omega$ and $c$ is the speed of light. Multiplying this field strength by the dipole moment, we can express the vacuum Rabi frequency in dimensionless units:

$$\frac{g}{\omega} = \frac{L}{r}\sqrt{\frac{e^2}{2\pi^2 \epsilon_0 c}} = \frac{L}{r}\sqrt{\frac{2\alpha}{\pi}} \tag{4.3}$$

in which the dimensionless combination of the fundamental physical constants of electromagnetism, the fine structure constant $\alpha = e^2/4\pi\epsilon_0 c \simeq 1/137$, has appeared. The strongest coupling will result from a cavity whose transverse size is small enough that the atom completely fills the transverse dimension ($L/r \simeq 1$), and then the coupling can be several percent. In comparison, because the three-dimensional cavities in either optical or microwave experiments using real atoms have bigger sizes and the real atoms used have smaller dipole moments, the largest couplings possible so far have been much smaller, $g/\omega \sim 10^{-6}$. The large interactions achievable in the one-dimensional cavities of cQED make it significantly easier to attain the strong-coupling regime, although care is still needed to keep the dissipation low enough.

The above argument strictly only applies for a Cooper pair box, where a single excitation really means moving a single Cooper pair onto the island. For a transmon, several charge states are involved in each energy level, as shown in figure 2.4, and the coupling can be enhanced by an additional factor $(E_J/E_C)^{1/4}$, as shown in (2.40).

## 4.2 Experimental setup

The measurements described in the rest of this chapter have been performed in the Schoelkopf laboratory,* in a dilution refrigerator at 15 mK. The sample consists of two transmons, denoted $q^L$ and $q^R$, coupled to an on-chip CPW cavity. The CPW resonator has a half-wavelength resonant frequency of $\omega_r/2\pi = 6.92$ GHz and a photon decay rate of $\kappa_-/2\pi = 300$ kHz. Transmission measurements are performed using a heterodyne detection scheme: the transmitted RF voltage signal through the cavity is mixed down to a 1 MHz carrier signal, and then digitally mixed down to d.c. to obtain the transmitted voltage amplitude as a function of frequency. The vacuum Rabi coupling strengths for the two transmons are obtained as $g^L/\pi = 94.4$ MHz and $g^R/\pi = 347$ MHz. Time domain measurements of the transmons show that the $T_1$ is limited by the multimode Purcell effect (section 3.6.1) and completely homogenously broadened ($T_2 = 2T_1$) at their flux sweet spots [42]. The coherence times are $T_1^L = 1.4$ µs and $T_2^L = 2.8$ µs (measured at the flux sweet spot) and $T_1^R = 1.7$ µs and $T_2^R = 0.7$ µs (measured away from the flux sweet spot). The charging energies of the two transmons are measured to be $E_C^R/2\pi = 340$ MHz and $E_C^L/2\pi = 400$ MHz.

### 4.2.1 Details of the Sample

Fabrication of the sample followed the description given in [43]. The two transmons were fabricated with two separate lithography stages on a single-crystal sapphire substrate. The cavity was defined via optical lithography in a dry-etching process of 180 nm thick niobium. The transmons were patterned using electron-beam lithography and made in a double-angle deposition of evaporated aluminum consisting of a 100 nm and 20 nm thick layer [70]. The

* I am only analyzing the data here: I did not make the measurements myself....

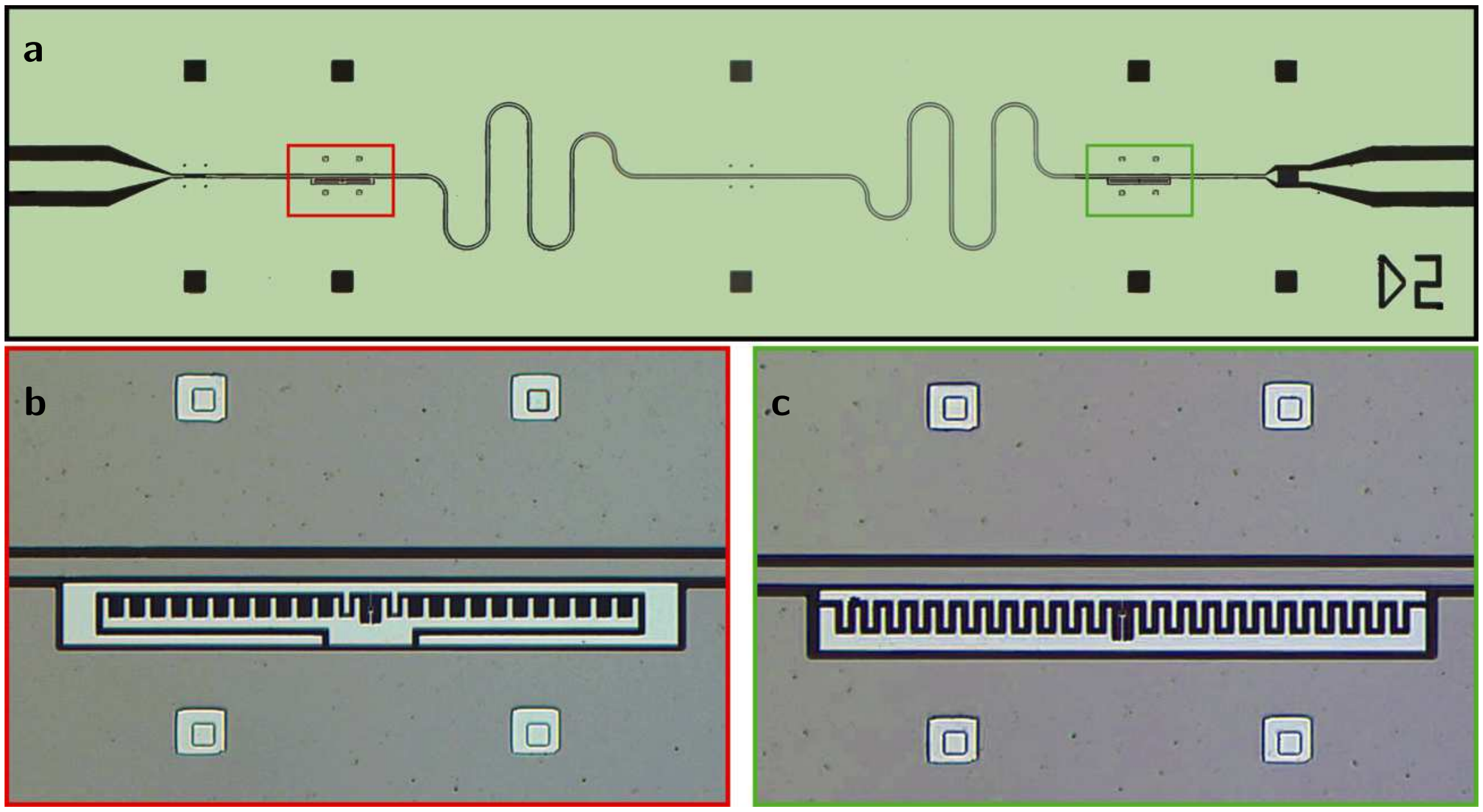

**Figure 4.2: Two-transmon circuit QED sample.** **a**, Optical micrograph of a chip with two different transmons coupled to a coplanar waveguide resonator. The cavity is operated as a half-wave resonator and the transmons are located at opposite ends of the cavity, where the electric field has an anti-node. The sample is 7 mm long. **b**, Optical micrograph of transmon with reduced cavity coupling $g^{\mathrm{L}}/\pi = 94.4\,\mathrm{MHz}$. Compared to **c**, which is the optical micrograph of a transmon with higher cavity coupling $g^{\mathrm{R}}/\pi = 347\,\mathrm{MHz}$, the transmon in **b** is designed to have a larger capacitance to the lower ground plane due to the arms which extend around the edge.

transmons use the split-junction design discussed in section 2.5.4, with junction areas of $\sim 0.20 \times 0.25\,\mu\mathrm{m}^2$, such that the effective Josephson energy may be tuned by an external magnetic field, $E_J^{\mathrm{L,R}} = E_{\mathrm{J\,max}}^{\mathrm{L,R}}|\cos(\pi\tilde{\Phi}^{\mathrm{L,R}}/\Phi_0)|$, and they can be tuned independently due to their different superconducting loop areas, $\tilde{\Phi}^{\mathrm{R}} = 0.625\tilde{\Phi}^{\mathrm{L}}$. The two transmons are designed to have significantly different cavity coupling parameters $g_0^{\mathrm{L,R}}$ by increasing the total capacitance of $q^{\mathrm{L}}$. This is effectively done with the 'cradle' design as shown in figure 4.2b, where the arms extending around the edge increase the capacitance to ground. A way to see that this design decreases the dipole moment of the transmon is to notice that extending the arms around increases the symmetry, making the 'upper' plate of the capacitor couple more equally to the center pin of the CPW and to the lower ground plane, compared to the more strongly-coupled transmon $q^{\mathrm{R}}$ shown in figure 4.2c.

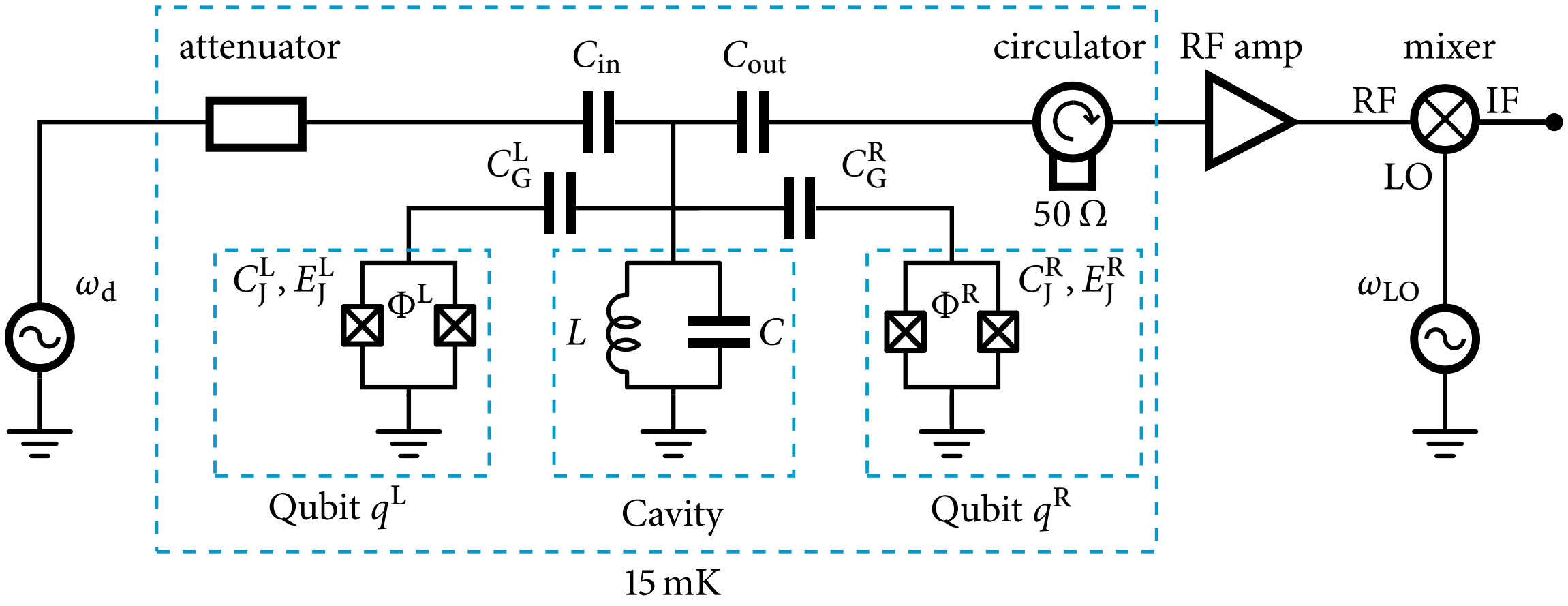

**Figure 4.3: Schematic for measurement setup.** Only a single RF drive tone is used. The HEMT is anchored at 4 K and has a noise temperature of ~ 5 K. Two circulators are used in series, each providing an isolation of ~ 20 dB over the frequency range 4 to 8 GHz.

### 4.2.2 Measurement Details

There are two electrical connections to the setup, an input line for the RF drive which is thermalized via attenuation to 15 mK before entering the sample, and an output line which is amplified via a low-noise-temperature high electron mobility transistor (HEMT). A chain of microwave circulators, thermally anchored at 15 mK,* precedes the HEMT to reduce the reflected noise entering the sample. Figure 4.3 shows a simplified circuit diagram of the measurement setup. This is a 'one-sided' cavity, in that the input capacitance is much smaller than the output capacitance $C_{\text{in}} \ll C_{\text{out}}$. Thus, any photons in the cavity will almost certainly leak out through the output side, towards the amplifier.

Transmission measurements are performed using a heterodyne detection scheme. An RF drive tone is applied to the input side of the cavity. The transmitted RF voltage signal from the cavity is amplified and mixed down to a 1 MHz IF signal, and the in-phase and quadrature components are extracted digitally. By detecting the transmitted voltage amplitude while sweeping both the RF drive frequency $\omega_d/2\pi$ and the external magnetic field, the transmission map shown in figure 4.4 can be obtained. From this map, the two transmons can be identified from the different avoided crossing splittings $2g^{L,R}$ as well as the different flux periodicities.

For observing the vacuum Rabi splitting, the magnetic field can be tuned to locations where only one of the transmons is in resonance with the cavity, and the other transmon can be ignored. Figure 4.4 provides a coarse location of the relevant splittings—the exact

* Section 4.6.3 discusses an experiment where the circulators were kept at ~ 100 mK.

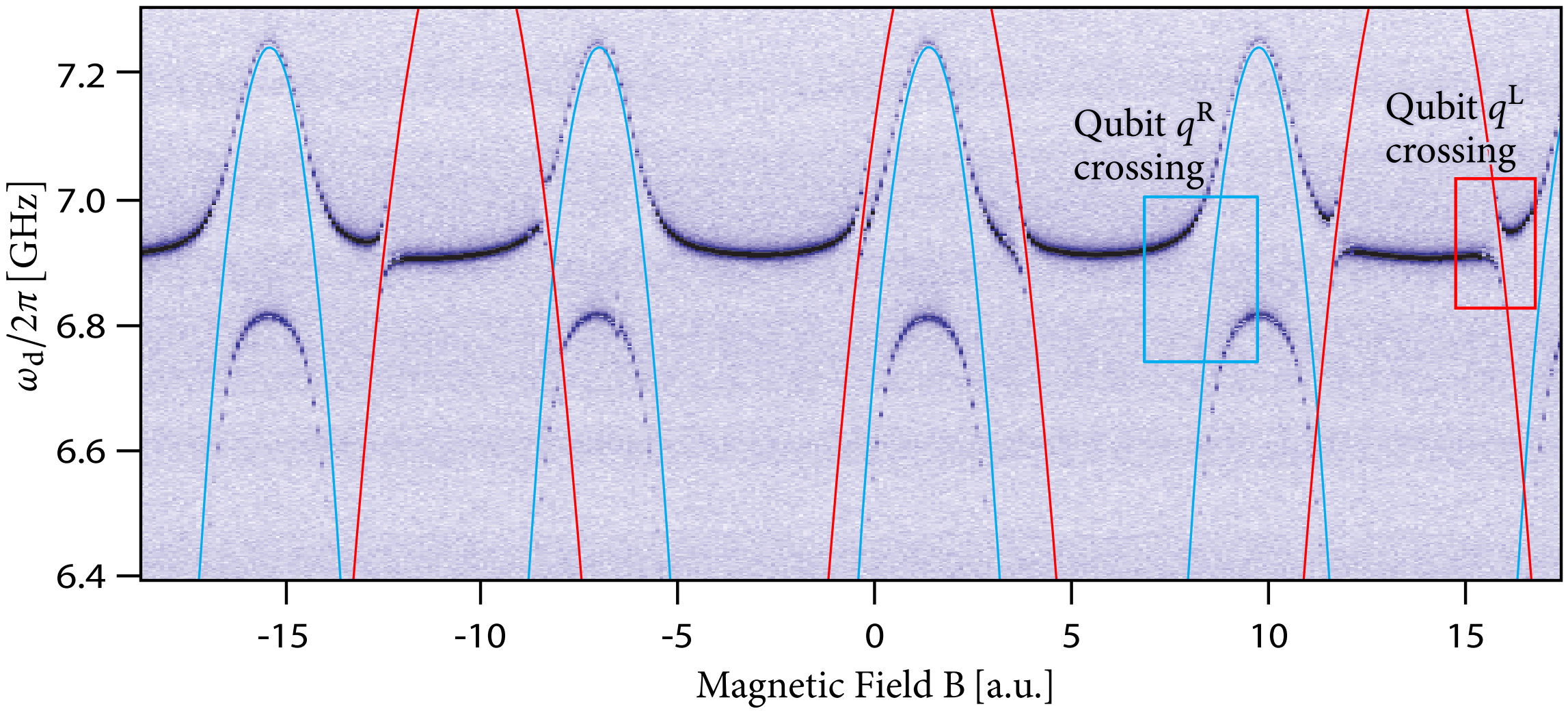

**Figure 4.4: Transmission versus magnetic field and drive frequency.** The experiment with the transmon $q^R$ is performed around the crossing at magnetic field $B = 8$. The experiment with transmon $q^L$ is performed around the crossing at $B = 15$.

resonance condition is then obtained by successive fine tuning of the magnetic field in small steps and checking the difference in frequency between two fitted Lorentzians through the vacuum Rabi splitting, until this difference is minimal.

## 4.3 Input-output theory

The open-systems approach of chapter 3 tells us how the leakage of photons from the cavity affects the dynamics of the system, via the master equation, but it does not tell us about measuring those outgoing photons. When discussing the measurement of the outgoing field from the cavity, the appropriate language is that of input-output theory. This is discussed in detail in [39, appendix D], so I give only a brief summary here. For a 'one-port' device, operated in reflection, the first thing is to introduce a boundary on the transmission line connected to that port, so that there is a discrete set of energy levels. We assume the boundary is far away such that there are many modes of the resonator, described by operators $b_\alpha$, $b_\alpha^\dagger$, with frequencies $\omega_\alpha$. We will later take the boundary to infinity such that there is in fact a continuum of such frequencies. The next step is to solve the Hamiltonian system for the composite system comprising the device coupled to the external transmission line, with a

coupling (in the RWA)

$$H_{\text{int}} = -\mathrm{i}\sum_{\alpha}\left(f_\alpha a^\dagger b_\alpha + f_\alpha^* a b_\alpha^\dagger\right). \tag{4.4}$$

The constants $f_\alpha$ describing the strength of the coupling depend on the coupling capacitance separating the system from the transmission line. After taking the boundary to infinity, the important result for the present discussion can be written in the form

$$b_{\text{out}}(t) = b_{\text{in}}(t) + \sqrt{\kappa}a(t), \tag{4.5}$$

where the notation has a specific meaning: $a(t)$ is simply the Heisenberg operator at time $t$; $b_{\text{in}}$ and $b_{\text{out}}$ are however specific time-dependent combinations of the Heisenberg operators $b_\alpha(t)$:

$$b_{\text{in}}(t) = \frac{1}{\sqrt{2\pi\eta}}\sum_{\alpha}\mathrm{e}^{-\mathrm{i}\omega_\alpha(t-t_0)}b_\alpha(t_0), \quad \text{and} \tag{4.6a}$$

$$b_{\text{out}}(t) = \frac{1}{\sqrt{2\pi\eta}}\sum_{\alpha}\mathrm{e}^{-\mathrm{i}\omega_\alpha(t-t_1)}b_\alpha(t_1), \tag{4.6b}$$

where $t_0$ represents a time in the distant past, well before the incident wave packet launched at the system has reached it, and $t_1$ represents a time far into the future, well after the packet has interacted with the cavity and reflected off towards infinity. Thus the modes $b_{\text{in}}(t)$ and $b_{\text{out}}(t)$ represent the particular combination of bath modes which is coupled to the system at time $t$, and for $\kappa = 0$ (corresponding to a fully reflecting mirror) $b_{\text{out}}(t) = b_{\text{in}}(t)$ showing the outgoing wave is simply the reflected incoming wave. The density of states $\eta$ is assumed to be constant over the range of frequencies relevant to the system (a Markov approximation) and $\kappa = 2\pi f_\alpha^2\eta$, where $f_\alpha$ is also assumed constant.

For the experiments described here, we have a two-sided cavity and apply no driving from the 'output' side. The driving from the 'input' side is included as in section 2.8 via a term in the Hamiltonian and we use input-output theory to describe the output port. Strictly, the incoming signal on the 'output' port, $b_{\text{in}}$, is not zero and we should take into account the reflected vacuum noise, but since the HEMT amplifier adds so much classical noise of its own, the quantum noise is not important and we take $b_{\text{out}} = \sqrt{\kappa}a$.

## 4.4 Heterodyne detection

The amplified outgoing wave is sent through a mixer, which can be thought of as multiplying the voltage with the signal from a local oscillator of frequency $\omega_{\mathrm{LO}}/2\pi$. Assuming a steady-state has been reached for the system, $\dot{\rho}_{\mathrm{s}} = 0$, this means that in the non-rotating frame the voltage oscillates at the drive frequency $\omega_{\mathrm{d}}/2\pi$. So, the mixer output is given by:

$$V_{\mathrm{m}} = \alpha\left\langle b_{\mathrm{out}} + b_{\mathrm{out}}^{\dagger}\right\rangle \cos\omega_{\mathrm{LO}} t \tag{4.7a}$$
$$= \alpha\sqrt{\kappa}\left\langle a\mathrm{e}^{-\mathrm{i}\omega_{\mathrm{d}} t} + a^{\dagger}\mathrm{e}^{\mathrm{i}\omega_{\mathrm{d}} t}\right\rangle \cos\omega_{\mathrm{LO}} t \tag{4.7b}$$
$$= \frac{\alpha\sqrt{\kappa}}{2}\left\langle\left(\mathrm{e}^{-\mathrm{i}(\omega_{\mathrm{d}}+\omega_{\mathrm{LO}})t} + \mathrm{e}^{-\mathrm{i}\omega_{\mathrm{IF}} t}\right)a + \left(\mathrm{e}^{+\mathrm{i}(\omega_{\mathrm{d}}+\omega_{\mathrm{LO}})t} + \mathrm{e}^{+\mathrm{i}\omega_{\mathrm{IF}} t}\right)a^{\dagger}\right\rangle, \tag{4.7c}$$

where the intermediate frequency is given by $\omega_{\mathrm{IF}} = \omega_{\mathrm{d}} - \omega_{\mathrm{LO}}$ and $\alpha$ describes the amplifier and mixer gain. After low-pass filtering to remove the fast oscillating terms, in principle it is easy to extract the quadratures $I = V_0\langle a + a^{\dagger}\rangle$ and $Q = V_0\langle \mathrm{i}a^{\dagger} - \mathrm{i}a\rangle$, where $V_0$ is a voltage related to the gain of the experimental amplification chain. However, the phase relation between the LO and the RF drive is not stable with respect to sweeping the drive frequency and only the heterodyne amplitude is useful, not the phase. We therefore use the steady-state transmission amplitude, expressed as

$$A = \sqrt{I^2 + Q^2} = 2V_0|\langle a\rangle| = 2V_0|\mathrm{tr}(a\rho_{\mathrm{s}})|. \tag{4.8}$$

Our calculations thus only require the steady-state solution $\rho_{\mathrm{s}}$ of the master equation. The intensity $A^2$ is conveniently expressed in units of $A_1^2 = 4V_0^2$, the intensity resulting from a coherent state with a mean cavity occupancy of one photon, $\langle a^{\dagger}a\rangle = 1$.

It is worth emphasizing that (4.8) does not imply that we are able to measure $\langle a\rangle$ in a *single* projective measurement, which is after all impossible since $a$ is non-Hermitian, unlike $I$ and $Q$. Nevertheless, it is true that the numerical value of $A$ is given by $\mathrm{tr}(a\rho_{\mathrm{s}})$.

## 4.5 Two-level behavior: Supersplitting

The experiment is performed using transmons, with their finite anharmonicity (section 2.5.2), rather than true two-level systems, so the Jaynes–Cummings Hamiltonian is modified by the presence of the higher transmon levels. Thus, the appropriate Hamiltonian describing the

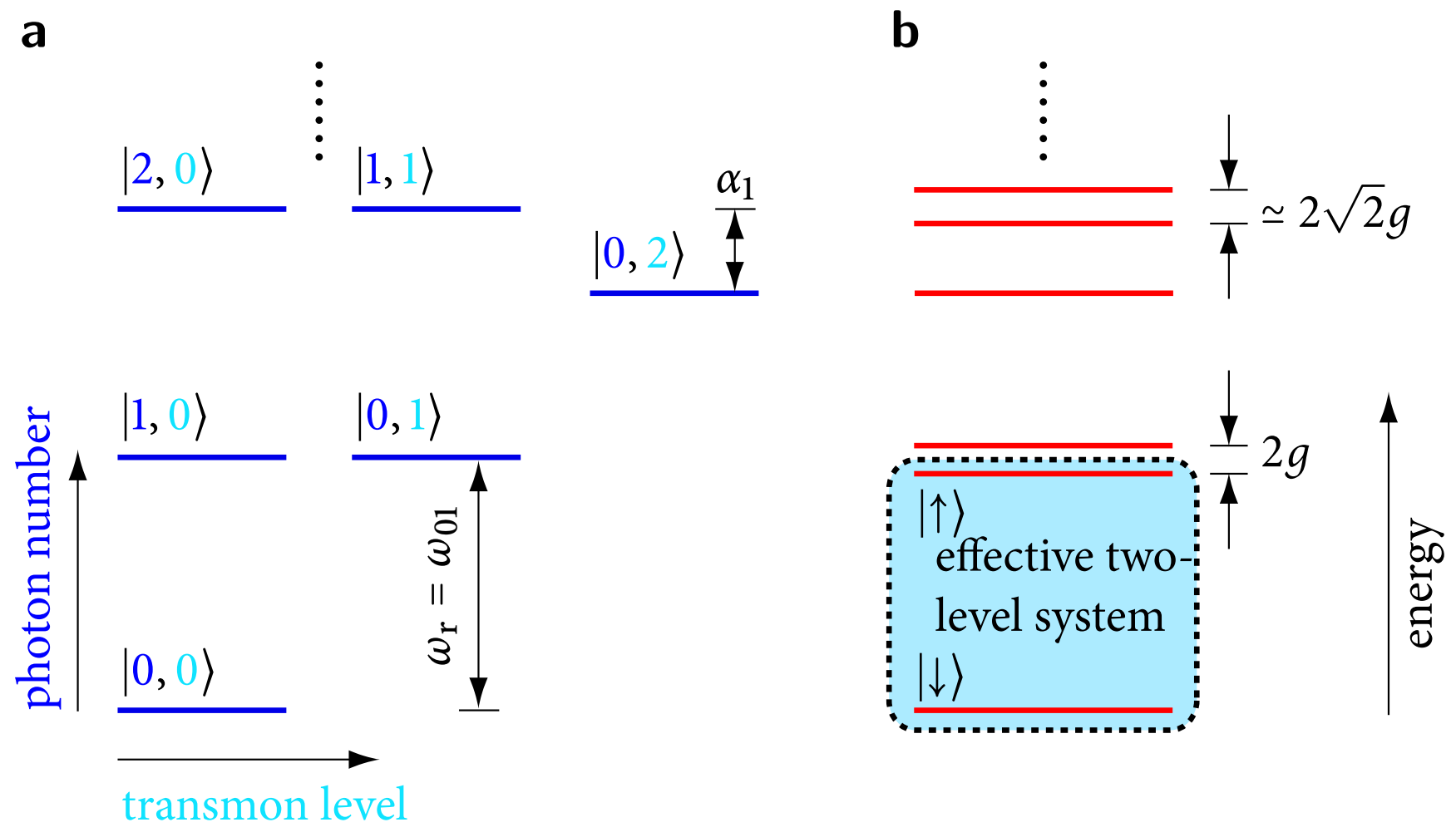

**Figure 4.5: Extended Jaynes–Cummings level diagram of the resonator–transmon system.** **a**, Bare levels in the absence of coupling. The states are denoted $|n, j\rangle$ for photon number $n$ and occupation of transmon level $j$. **b**, Spectrum of the system including the effects of transmon–resonator coupling. The effective two-level system relevant to describing the lower vacuum Rabi peak comprises the ground state and the antisymmetric combination of transmon and photon excitations. Compare to the case of the ordinary Jaynes–Cummings Hamiltonian, using two-level qubits, shown in figure 4.1.

system is the generalized Jaynes–Cummings Hamiltonian (2.62), reproduced here:

$$H = \Delta_{\mathrm{r}} a^\dagger a + \sum_j \Delta_j |j\rangle\langle j| + g\left(a^\dagger c + a c^\dagger\right) + \left(a\xi(t)^* + a^\dagger \xi(t)\right). \tag{2.62}$$

The resulting energy level diagram for vanishing drive, $\xi = 0$, is schematically shown in figure 4.5. In the linear-response regime, the vacuum Rabi peaks have a characteristic Lorentzian line shape. Their separation and width are given by $2g$ and $(\gamma + \kappa + 2\gamma_\varphi)/2$, respectively, where $\gamma = \gamma_- + \gamma_+$, $\kappa = \kappa_- + \kappa_+$ and $\gamma_\varphi = \gamma_\varphi^\Phi + \gamma_\varphi^{\mathrm{C}}$. Using the transmon with the larger coupling, $q^{\mathrm{R}}$, the splitting is observed to exceed 260 linewidths, shown in figure 4.6a. Thus the experiment is very far into the strong-coupling regime. When increasing the drive power beyond linear-response, the shape of the transmission curve changes drastically as shown in figure 4.6b–f. Each vacuum Rabi peak develops a central dip and eventually 'supersplits' into a doublet of peaks. By solving for the steady-state of the transmon master equation (3.52) and using the expression for the heterodyne amplitude (4.8) we find excellent agreement with the experiment. Details of the numerical solution of the master equation are given in section 4.6.1.

The supersplitting is fundamentally just a saturation effect of a strongly-driven two-level system. In the case of the left vacuum Rabi peak, represented in figure 4.5b, this comprises the Jaynes–Cummings ground state and the antisymmetric superposition of transmon and photon excitation. (The right vacuum Rabi peak may be modeled by instead taking the symmetric superposition.) We can label these states as

$$|\downarrow\rangle = |0,0\rangle\,, \quad \text{and} \tag{4.9a}$$
$$|\uparrow\rangle = (|1,0\rangle - |0,1\rangle)/\sqrt{2}. \tag{4.9b}$$

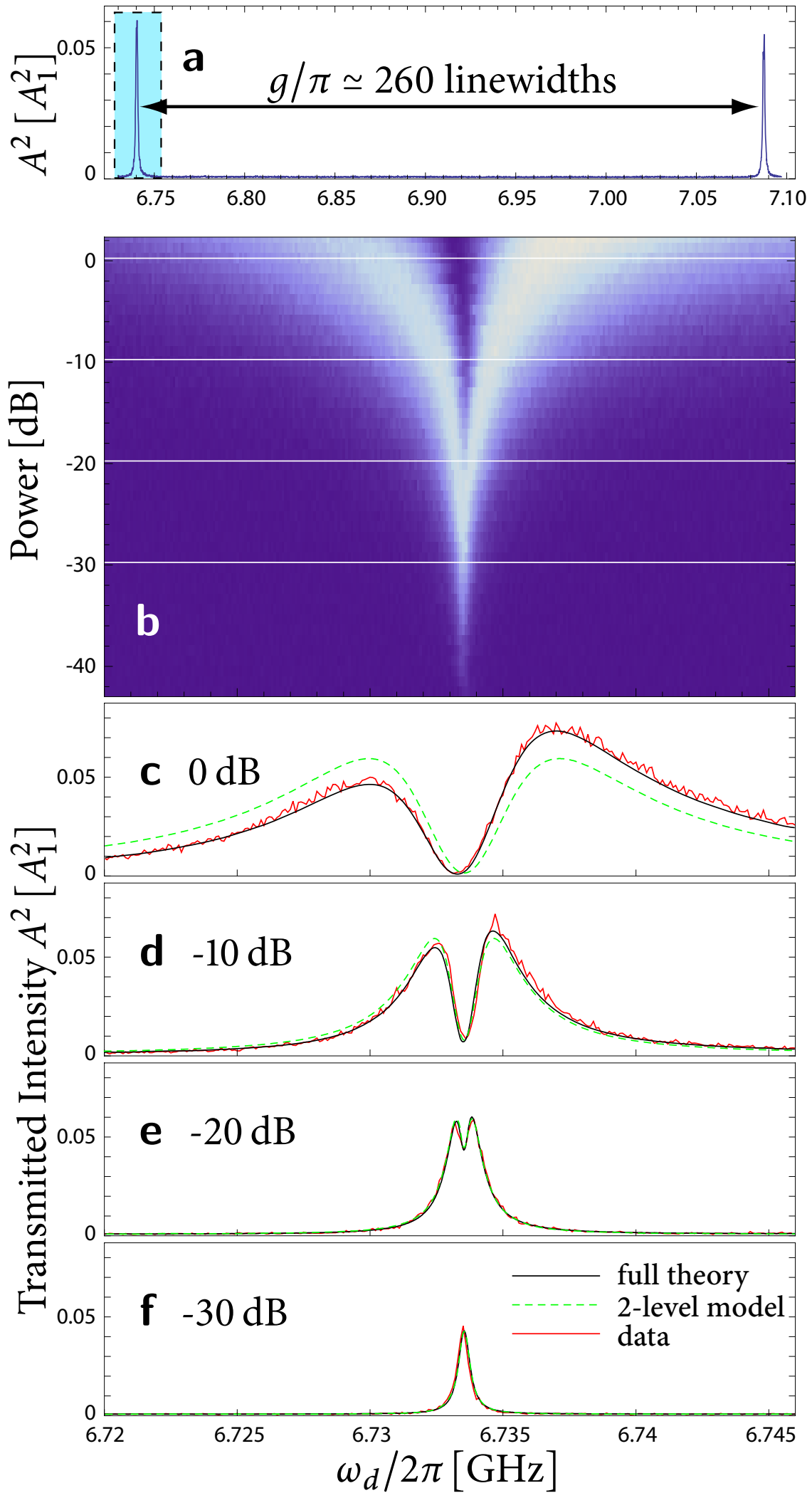

**Figure 4.6: Supersplitting of the vacuum Rabi resonance when probing heterodyne transmission beyond linear response.** The experimental data are obtained with a circuit QED system in the strong-coupling regime, where the vacuum Rabi splitting is observed to exceed 260 linewidths, see **a**. All plots show the heterodyne intensity $A^2$ in units of $A_1^2$. **b**, Measured intensity (color scale) for the left vacuum Rabi peak, as a function of drive frequency and power. The plot reveals the supersplitting of a single Lorentzian into a doublet of peaks. **c–f**, Cuts for constant power at the values indicated in **b**. In linear response, **f**, the vacuum Rabi peak is Lorentzian; as the power increases a central dip develops, **e**, leading to supersplitting of the peak, **d**, and eventually becoming asymmetric at the largest powers, **c**. The experimental data (red line) is in excellent agreement with theory (black line). The results from the 2-level approximation are shown for comparison (green dashed line).

The anharmonicity is so large that we can work within the effective two-level subspace, where it is easy to calculate the matrix elements [71] and show that the photon operators are mapped to Pauli operators $a \to \tilde{\sigma}_-/\sqrt{2}$, $a^\dagger \to \tilde{\sigma}_+/\sqrt{2}$, where the tilde denotes that these operators apply to the reduced two-level system, rather than a bare qubit. Thus, the effective Hamiltonian in the rotating frame is

$$H_{\mathrm{TL}} = \frac{\Delta}{2}\tilde{\sigma}_z + \frac{\Omega}{2}\tilde{\sigma}_x, \tag{4.10}$$

a scenario that Carmichael and coworkers [55, 71] have referred to as 'dressing of dressed states', dressed once because of the interaction between atom and cavity, dressed a second time by the driving field. The Hamiltonian $H_{\mathrm{TL}}$ refers to the frame rotating at the drive frequency; $\Delta = \omega_{01} - g - \omega_{\mathrm{d}}$ is the detuning between drive and vacuum Rabi peak; and $\Omega = \sqrt{2}\xi$ is the effective drive strength. With the notable exception of the work of I. Schuster *et al.* [68], previous investigations were primarily concerned with effects on photon correlations and fluorescence, as observed in photon-counting measurements [55, 57]. According to the operator mapping, photon counting can be related to the measurement of $\langle\tilde{\sigma}_z\rangle$, whereas detection of the heterodyne amplitude, $A$, corresponds to $|\langle\tilde{\sigma}_-\rangle|$. As a result, heterodyne detection fundamentally differs from photon counting and we will see that the vacuum Rabi supersplitting is a characteristic of heterodyne detection only.

After restricting the master equation (3.52) to the two-level subspace, the system evolution can be expressed in terms of Bloch equations (3.39)

$$\dot{x} = -x/T_2' - \Delta y, \tag{4.11a}$$
$$\dot{y} = \Delta x - y/T_2' - \Omega z,$$
$$\dot{z} = \Omega y - (z+1)/T_1'. \tag{4.11b}$$

Here, $T_1'$ and $T_2'$ are the effective relaxation and dephasing times, which are related to $\gamma$, $\gamma_\varphi$,

and $\kappa$ via $T_1'^{-1} = (\gamma + \kappa)/2$ and $T_2'^{-1} = (\gamma + 2\gamma_\varphi + \kappa)/4$ (see also (3.44)).* The steady-state solution of the Bloch equations for $x$ and $y$ gives the heterodyne amplitude

$$A = \frac{V_0 T_2' \Omega \sqrt{(\Delta^2 T_2'^2 + 1)/2}}{\Delta^2 T_2'^2 + T_1' T_2' \Omega^2 + 1}. \tag{4.12}$$

This expression describes the crossover from linear response at small driving strength, $\Omega \ll (T_1' T_2')^{-1/2}$, producing a Lorentzian of width $2T_2'^{-1}$, to the doublet structure observed for strong driving. As the drive power is increased, the response saturates and the peak broadens, until at $\Omega = (T_1' T_2')^{-1/2}$ the peak undergoes supersplitting with peak-to-peak separation $2T_2'^{-1}\sqrt{T_1' T_2' \Omega^2 - 1}$. The fact that we use heterodyne detection is indeed crucial for the supersplitting: It is easy to verify within the two-level approximation that photon counting always results in a Lorentzian. For photon counting, probing beyond the linear-response regime merely results in additional power-broadening; specifically, the width of the Lorentzian is given by $2T_2'^{-1}\sqrt{T_1' T_2' \Omega^2 + 1}$, as shown in figure 4.7. Figure 4.8 shows the $I$ and $Q$ quadratures of the response. While $A^2$ develops supersplitting, $I$ and $Q$ are observed mainly to change their relative magnitudes. Figure 4.9 compares the nonlinear responses due to heterodyne or photon-counting detection against the hypothetical linear response of a bare cavity.

That there is a difference between photon counting and heterodyne detection is a characteristic of a *single* atom. For a many-atom system, for which the relevant description is in terms of Maxwell–Bloch equations, both types of measurement would typically give the same result, and this many-atom nonlinear response would be rather different from the single-atom case, developing first as a frequency-pulling and eventually yielding hysteresis [72].

In figure 4.6c–e, the analytical expression (4.12) is plotted for comparison with the full numerical results and the experimental data. We find good agreement for low to moderate drive power, confirming that the supersplitting can be attributed to driving the vacuum Rabi transition into saturation while measuring the transmission with the heterodyne technique. For higher drive power a left-right asymmetry appears in the true transmission spectrum, which is not reproduced by (4.12), and which is partly due to the influence of levels beyond the two-level approximation. Before moving onto numerical calculations with higher levels, discussed in section 4.6, however, the next section simplifies things even further, in order to gain additional understanding of the supersplitting.

---

* At least the expression for $T_1'$ is intuitively obvious: that $T_1'^{-1}$ is given by the mean of $\kappa$ and $\gamma$ makes sense if we picture the (anti-)symmetric state as being 'half photon, half qubit'.

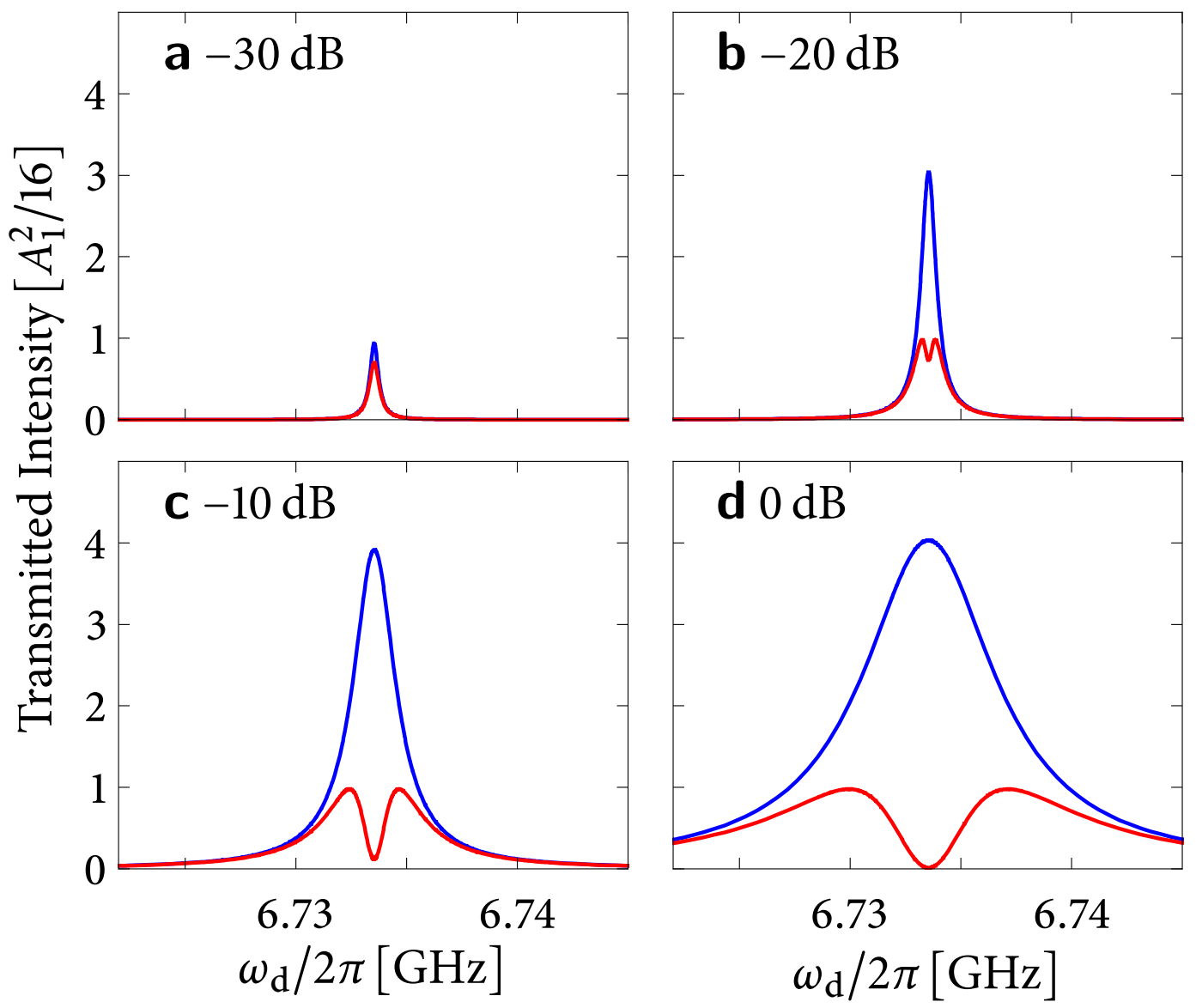

**Figure 4.7: Comparison of heterodyne detection and photon counting.** The Bloch equation solution for the squared heterodyne amplitude, $A^2$, (red) is compared with the intensity measured by photon counting, $A_1^2\langle a^\dagger a\rangle$, (blue). The same parameter values are used as in figures 4.8 and 4.11. Both types of measurement agree for low drive power, but for higher powers the heterodyne signal supersplits whereas the photon counting signal only power-broadens.

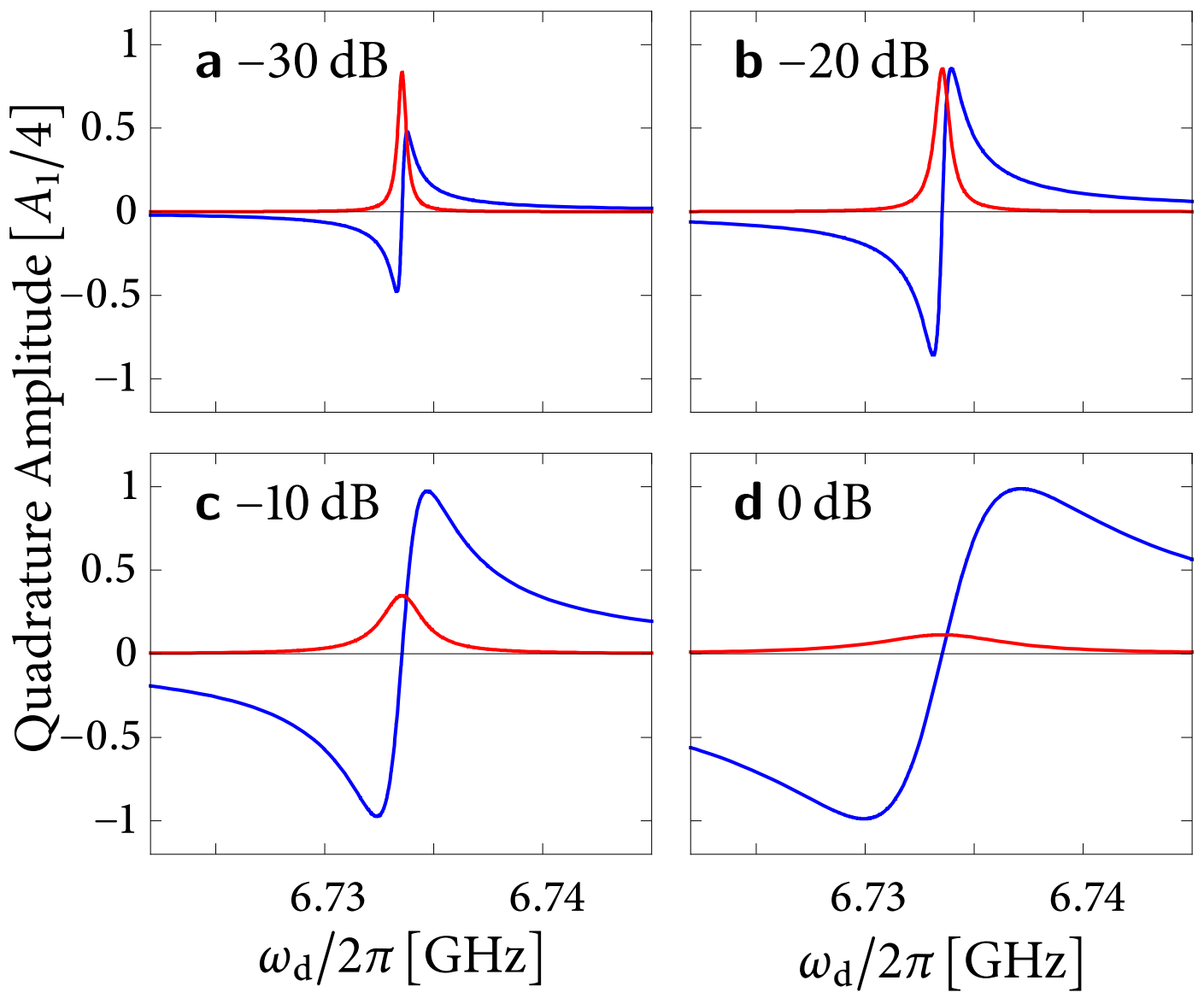

**Figure 4.8: Quadratures of the vacuum Rabi signal.** The $I$ (red) and $Q$ (blue) quadratures of the Bloch equation solution, for the same parameter values as used in figures 4.7 and 4.11. The $Q$ quadrature grows relative to the $I$ quadrature as the drive power increases.

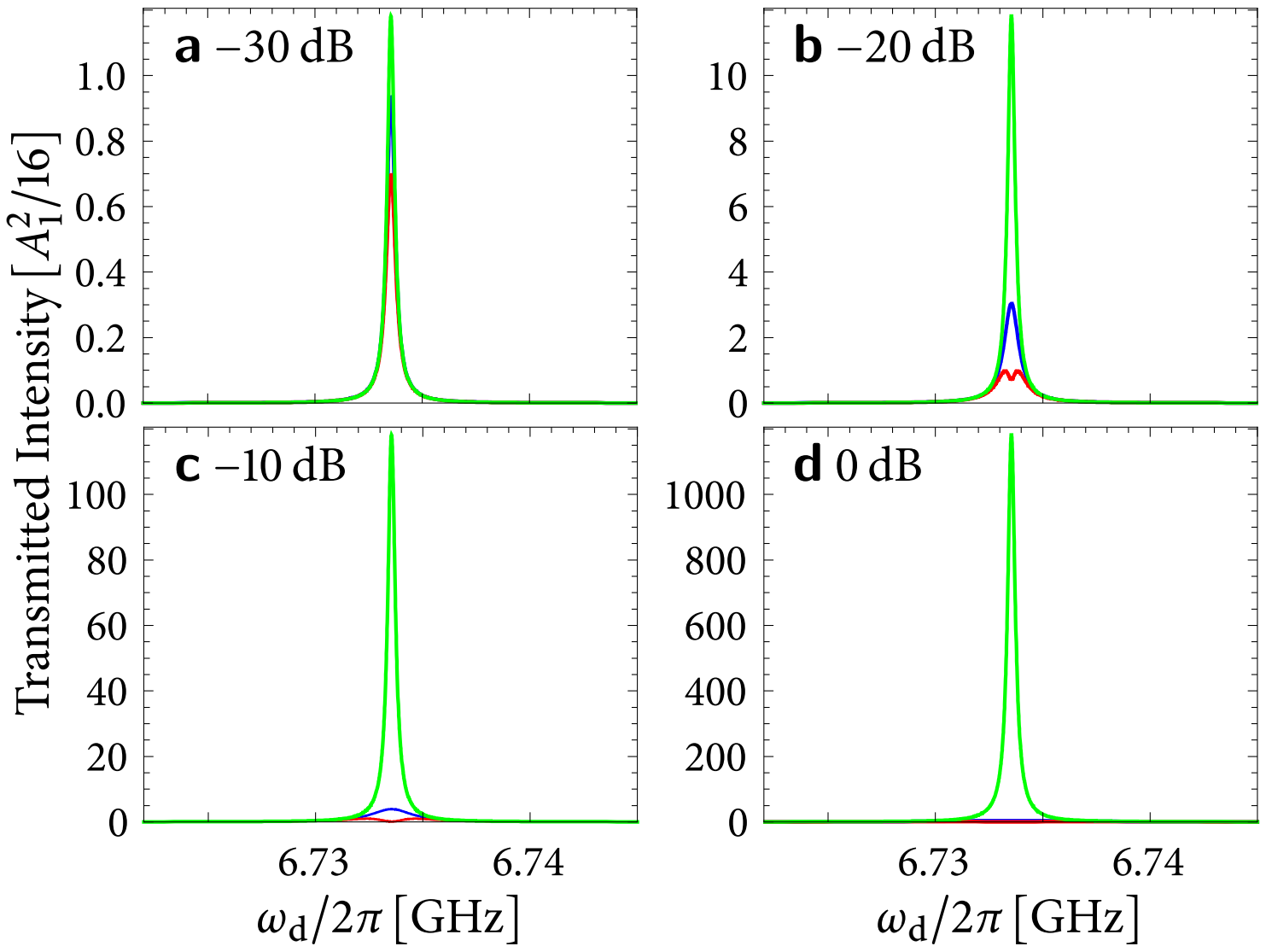

**Figure 4.9: Heterodyne vs photon counting vs linear response.** These are the same curves as in figure 4.7 but with the addition of the linear-response intensity (green curve). Note the change of scale on the $y$-axis! This emphasizes that the supersplitting and power broadening are saturation effects, where despite increasing the drive power by a factor of 1000, the transmitted power hardly changes.

### 4.5.1 Simple model of supersplitting

The supersplitting of vacuum Rabi peaks can be explained qualitatively within a simple intuitive model, based on the two-level approximation already introduced. This 'reduced Bloch model' disregards pure dephasing and is an approximation for large powers (to be defined more precisely below).

Assume that the relaxation channel is monitored so that the system always remains in a pure state. In this case the dynamics of the two-level system can be visualized on the surface of the Bloch sphere (figure 4.10), where the state of the system is represented by a unit arrow. The unitary evolution under the Hamiltonian of (4.10) corresponds to a rotation of the state arrow about a tilted axis in the $y = 0$ plane. Here, the tilt angle is determined by the detuning $\Delta$ and the drive strength $\Omega$, and the rotation frequency is given by $\sqrt{\Delta^2 + \Omega^2}/2\pi$.

At zero temperature, relaxation processes can be pictured as a resetting of the system state to the ground state (south pole of the Bloch sphere) at random times and with an average rate $1/T_1'$. We focus on the case where the rotation frequency is large compared to this rate, that is

$$\sqrt{\Delta^2 + \Omega^2} \gg 1/T_1', \tag{4.13}$$

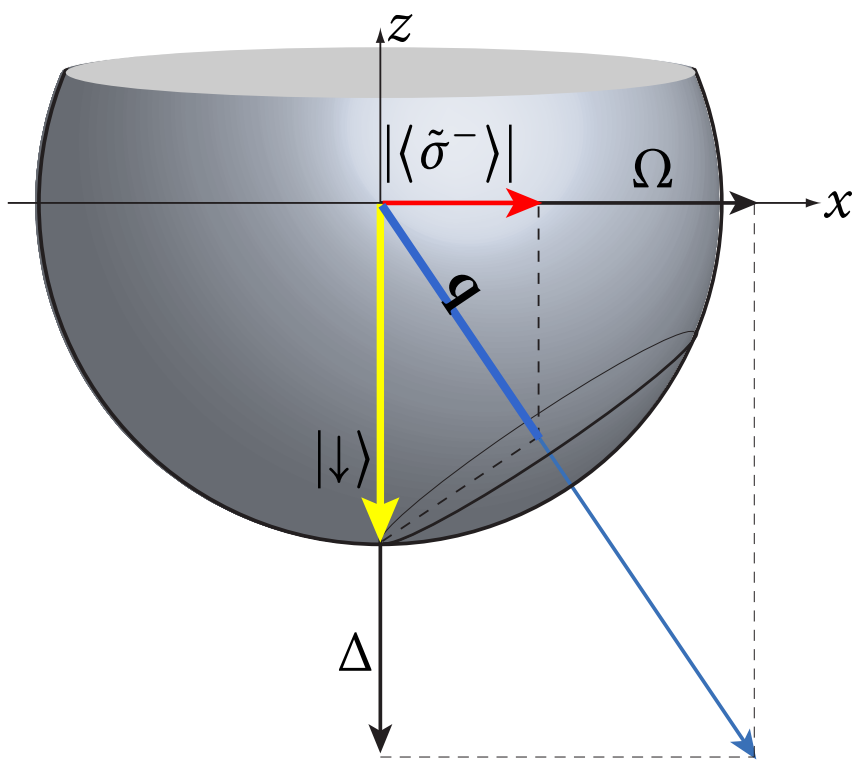

**Figure 4.10: Bloch sphere picture for the qubit–photon 2-level system.** Starting from the ground state $|\downarrow\rangle$ represented by the thick arrow pointing to the south pole of the sphere, the evolution under the Hamiltonian is nutation around the tilted axis whose $x$ and $y$ components are determined by the drive strength $\Omega$ and detuning $\Delta = \omega_{01} - g - \omega_{\mathrm{d}}$. The measured heterodyne amplitude is proportional to $|\langle\tilde{\sigma}^-\rangle|$. This quantity can be approximated by the time-averaged projection of this motion onto the $x$-axis.

satisfied for sufficiently large drive strength and/or detuning. Under these conditions, the expectation value $|\langle\tilde{\sigma}^-\rangle| = |\langle\tilde{\sigma}_x\rangle - \mathrm{i}\langle\tilde{\sigma}_y\rangle|/2$ can be approximated by the time-averaged projections of the rotation onto the $x$- and $y$-axes, see figure 4.10, which results in

$$\langle\tilde{\sigma}_x\rangle = \frac{\Delta\Omega}{\Delta^2 + \Omega^2}, \quad \langle\tilde{\sigma}_y\rangle = 0. \tag{4.14}$$

Accordingly, the approximation of the reduced Bloch model results in a heterodyne amplitude of

$$A \propto \frac{|\Delta|\,\Omega}{\Delta^2 + \Omega^2}. \tag{4.15}$$

As shown in figure 4.11, this reproduces the shape of the experimental data very well for high enough drive powers. In particular, it reproduces the surprising dip at zero detuning. Within the reduced Bloch model this corresponds to the case of a rotation about the $x$-axis, such that the projection onto the $x$-axis always vanishes. From this it is also clear that the predicted squared amplitude does not reduce to the linear-response Lorentzian shape in the limit $\Omega \to 0$, which is to be expected given the assumption (4.13) made in deriving the result (4.15).

We note that (4.15) does not completely agree with the corresponding asymptotic limit of the solution of the Bloch equations, (4.12). (This can be corrected if the above argument is modified and made rigorous as an unraveling of the master equation, in the language of quantum trajectories, section 5.2.) The above analysis may be generalized to the full driven Hamiltonian of (2.62), as is discussed below in section 4.6.3.

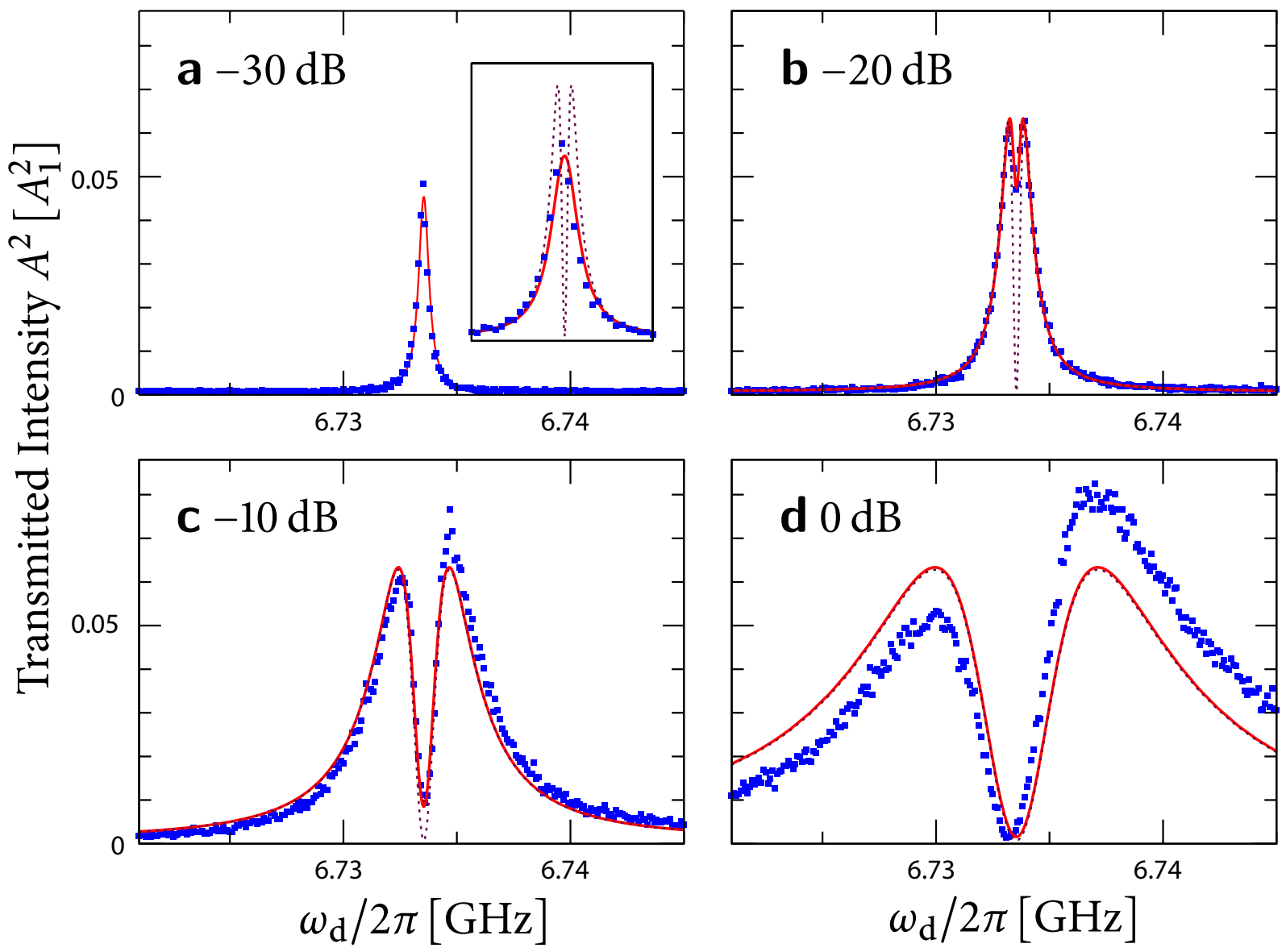

**Figure 4.11: Supersplitting of the vacuum Rabi peak in experiment and theory.** Comparison between experimental data (filled squares); the reduced Bloch model, (4.15) (dotted lines); and the Bloch equation solution, (4.12), (solid lines) for the line cuts shown in Fig. 2c. **a–d** show the squared transmitted amplitude $A^2$ as a function of drive frequency $\omega_d/2\pi$, for 4 decades of drive power. For clarity, in **a** the reduced model result is only shown in the magnified inset. The Bloch equation calculation agrees with the data except at the highest power. As expected, the reduced model fails at lower powers but follows the Bloch equation result for moderate and high power.

## 4.6 Multi-photon transitions: Climbing the Jaynes–Cummings ladder

Higher levels of the extended Jaynes–Cummings Hamiltonian become increasingly important as the drive power is raised. Figure 4.12 shows the emergence of additional peaks in the transmission spectrum. Each of the peaks can be uniquely identified with a multiphoton transition from the ground state to an excited Jaynes–Cummings state. For simplicity, we consider the situation where the anharmonicity $\alpha$ and the coupling strength $g$ are sufficiently different that mixing between higher transmon levels and the regular Jaynes–Cummings states $|n,\pm\rangle = (|n,0\rangle \pm |n-1,1\rangle)/\sqrt{2}$ is minimal for the low-excitation subspaces. Accordingly, the experiments are carried out using the lower $g$ transmon in the sample, $q^{\mathrm{L}}$, with a smaller coupling of $g/\pi = 94.4\,\mathrm{MHz}$. In this case, the $n$-photon transitions to the $n$-excitation subspace occur at frequencies $E_{n\pm}/2\pi n = (\omega_r \pm n^{-1/2} g)/2\pi$, and thus reveal the anharmonicity of the Jaynes–Cummings ladder. The features associated with unsaturated $n$-photon transitions have width set by the characteristic decay rates from the $|n,\pm\rangle$ states, $[(2n-1)\kappa_- + \gamma_-]/4\pi \simeq 1\,\mathrm{MHz}$. As the drive increases, each $n$-photon transition begins to saturate in turn and develop additional structure analogous to the supersplitting of the vacuum Rabi peaks. The detailed comparison between experimental data and numerical simulation in figure 4.12c–e shows superb agreement down to the narrowest features observed. Figure 4.13 shows the quadratures of the simulated spectrum.

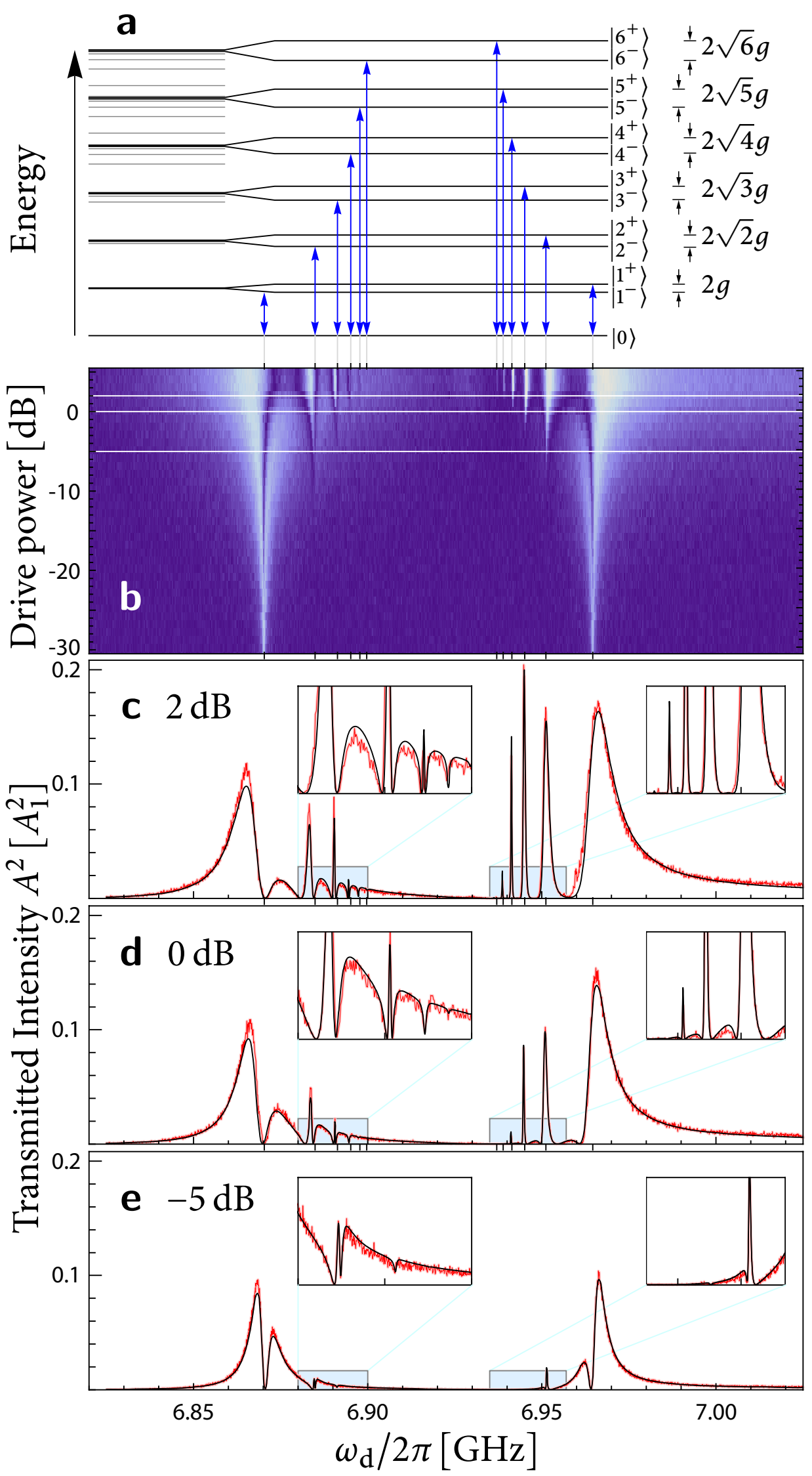

**Figure 4.12: Emergence of $\sqrt{n}$ peaks under strong driving of the vacuum Rabi transition.** **a**, The extended Jaynes–Cummings energy spectrum. All levels are shown to scale in the left part of the diagram: black lines represent levels $|n,\pm\rangle \simeq (|n,0\rangle \pm |n-1,1\rangle)/\sqrt{2}$ with only small contributions from higher ($j > 1$) transmon states; grey lines represent levels with large contributions from higher transmon states. In the right part of the diagram, the $\sqrt{n}$ scaling of the splitting between the $|n,\pm\rangle$ states is exaggerated for clarity, and the transitions observed in plots **b–e** are indicated at the $x$-coordinate $E_{n\pm}/2\pi n$ of their $n$-photon transition frequency from the ground state. **b**, Measured intensity ($A^2$, heterodyne amplitude squared) in color scale as a function of drive frequency and power. The multiphoton transitions shown in **a** are observed at their calculated positions. **c–e**, Examples of cuts for constant power, at the values indicated in **b** (results from the master equation (3.52) in black; experimental results in red), demonstrating excellent agreement between theory and experiment, which is reinforced in the enlarged insets. Good agreement is found over the full range in drive power from −45 dB to +3 dB, for a single set of parameters.

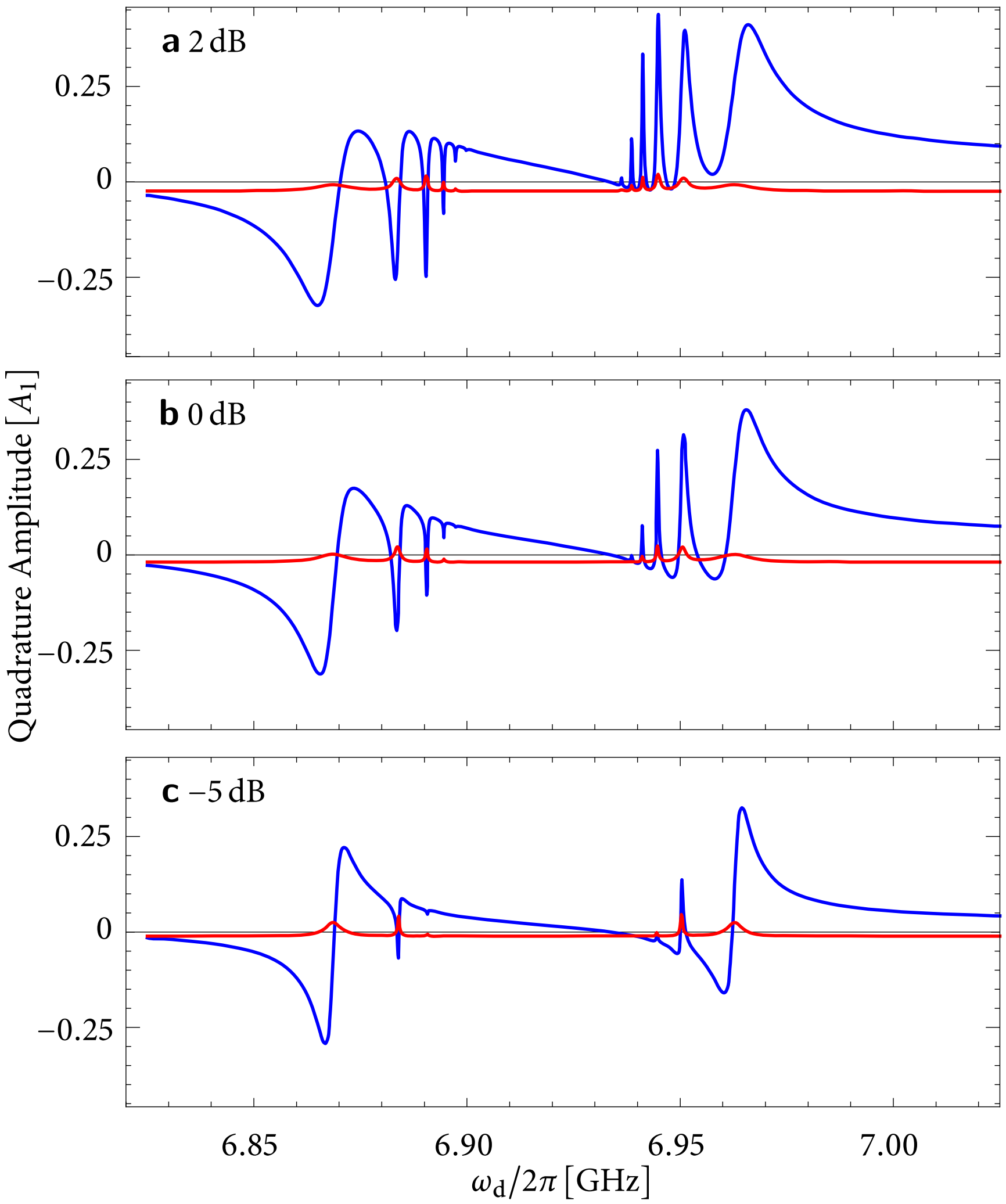

**Figure 4.13: Quadratures of $\sqrt{n}$ peaks under strong driving of the vacuum Rabi transition.** The $I$ (red) and $Q$ (blue) quadratures of the master equation (3.52) solution, using the same parameter values as in figure 4.12. We also include the small leakage of the drive past the cavity described in section 4.6.2, *i.e.* we show real and imaginary components of $2V_0\,\mathrm{tr}(a\rho_s) + b\xi$. Features due to multiphoton transitions appear predominantly in the $Q$ quadrature. The zero crossings of the $Q$ quadrature result in some of the narrowest structures visible in figure 4.12c–e.

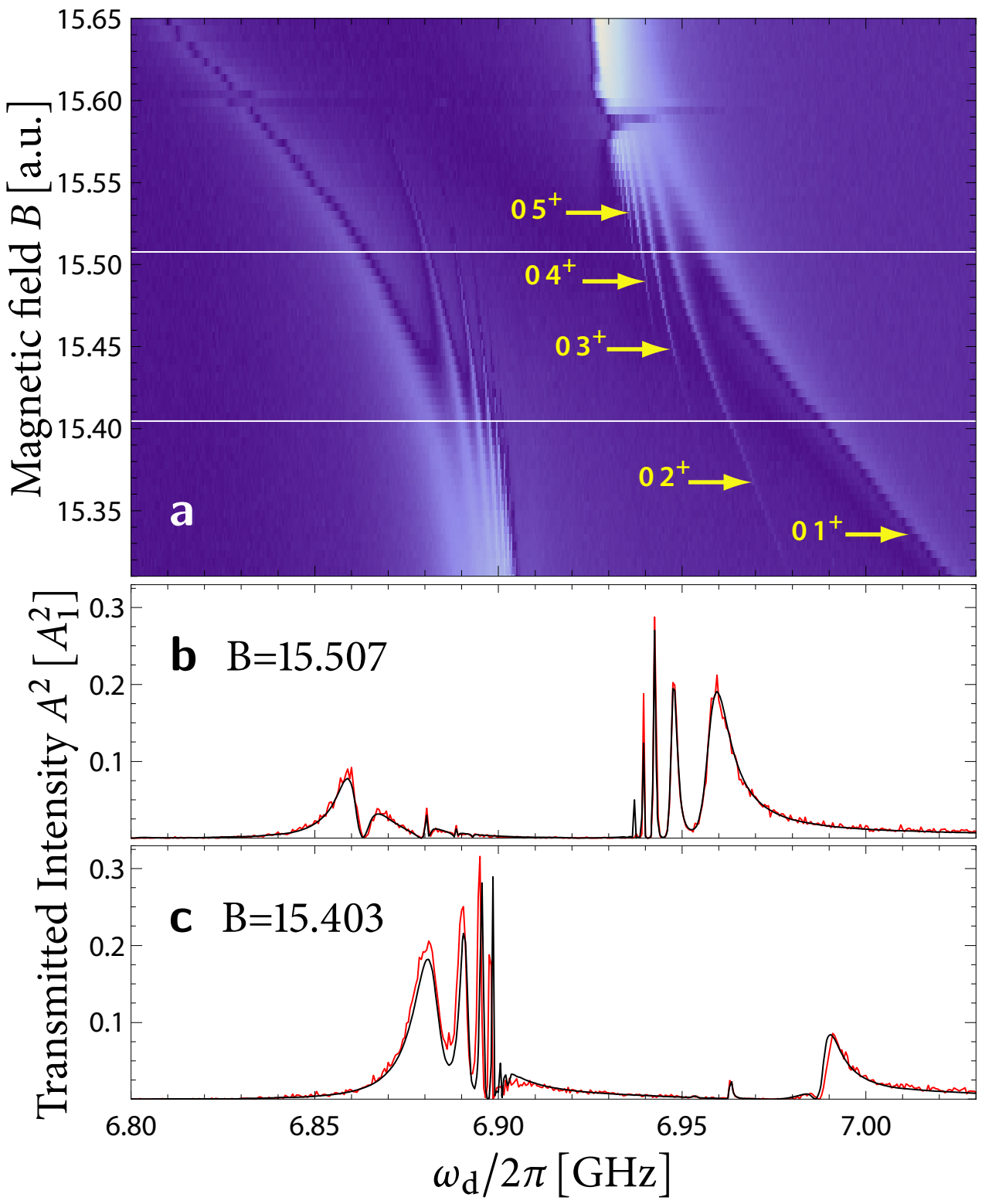

**Figure 4.14: Qubit–cavity avoided crossing at high drive power.** Transmission measurement when tuning the transmon frequency through resonance for a drive power of +1 dB. **a**, Measured intensity as a function of drive frequency and magnetic field. As the field is increased, the transmon frequency is tuned through resonance with the cavity, and anticrossing behavior is observed. The multiphoton transitions shown in figure 4.12a are visible. The anomaly at $B \simeq 15.59$ is most likely due to the crossing of a higher level of the second transmon present in the same cavity. **b–c**, Example cuts at constant magnetic field, at the values indicated in **a** (master equation results, calculated using the same parameters as for figure 4.12, are shown in black; measured results in red).

The possibility of multiphoton transitions at sufficiently large drive powers also affects the shape of the vacuum Rabi splitting when tuning the transmon frequency $\omega_{01}$ through resonance with the cavity, shown in figure 4.14. Instead of the simple avoided crossing commonly observed at low drive powers [52], figure 4.4, the presence of multiphoton transitions leads to a fan-like structure where individual branches can again be identified one-to-one with the possible transitions in the Jaynes–Cummings ladder. In the experimental data of figure 4.14a, processes up to the 5-photon transition are clearly visible. Detailed agreement with the theory verifies that the more general situation of non-zero detuning between transmon and resonator is correctly described by our model.

A different kind of multiphoton spectroscopy was reported by Deppe *et al.* [24]. In their experiment with flux qubits, instead of observing $n$-photon transitions $|0\rangle \leftrightarrow |n, \pm\rangle$, they observed 2-photon $|0\rangle \leftrightarrow |1, \pm\rangle$ transitions. They did this both for the vacuum Rabi case $\omega_q \simeq \omega_r$ and also for the case that the qubit is far-detuned from the cavity. These transitions would be very difficult to see with transmons, due to the parity selection rules of (2.41), whereas with the flux qubits, these selection rules only hold for certain special values of the applied flux. In addition to the small matrix elements, there is another reason why this type of 2-photon spectroscopy is strongly suppressed. In perturbation theory, an $n$-photon transition $|0\rangle \leftrightarrow |n, \pm\rangle$ can proceed via virtual intermediate states $|1, \pm\rangle, \ldots, |n-1, \pm\rangle$ and due to the limited anharmonicity of the Jaynes–Cummings Hamiltonian each such transition is only of order $g$ off-resonance; by contrast there is no close-by intermediate state with energy $\omega_q/2$ for the 2-photon $|0\rangle \leftrightarrow |1, \pm\rangle$ transition. This explains why compared to our 'strong driving' $\xi \simeq 5\,\text{MHz}$, in Ref. [24] the driving was much stronger, of order 1 GHz, which is already of such a strength that one might wonder about the validity of the RWA in deriving (2.59), when it comes to making detailed predictions.

### 4.6.1 Solving the master equation

For the steady-state solution of (3.52), the Hilbert space is truncated to a subspace with maximum number of excitations* $W$, using the projector $P_W = \sum_{0 \leqslant n+j \leqslant W} |n, j\rangle \langle n, j|$. In the

* In hindsight, truncating to a constant number of excitations was not necessarily the best (most efficient) choice. It means we end up keeping a lot of states describing the very highly-excited transmon states which are not playing any rôle (and which we are surely not describing accurately given that for the $E_J/E_C \simeq 52$ used here, the charge-dispersion of even the 4th excited transmon level is already quite significant, $\epsilon_4 \simeq 70\,\text{MHz}$, and we are not controlling $n_g$).

simulations, we keep up to $W = 7$ excitations, corresponding to keeping a Hilbert space of dimension $N = 1 + 2 + \cdots + 8 = 36$.

The question of the existence and uniqueness of the steady-state solution of (3.52) is an interesting one. It is certainly possible to invent situations where the solution of a master equation depends on the initial conditions (for example if there is no relaxation, only dephasing), or where the solutions of the master equation are oscillatory for all time. Situations where the steady-state solution exists and is unique are called 'uniquely' or 'genuinely' relaxing. In the case of master equations that result from a weak-coupling argument and describe a system coupled to a heat bath, as in (3.32), this is the question of whether the system *returns to equilibrium*. In that case it is equivalent to the condition [29]:

$$\begin{aligned} &\text{if} \quad \left[V_\omega, X\right] = \left[V_\omega^\dagger, X\right] = 0 \\ &\text{for all } \omega \geq 0 \text{ then} \quad X = c\mathbb{1}, \end{aligned} \tag{4.16}$$

for $c \in \mathbb{R}$ and $V_\omega$ as defined in section 3.3.1. As was emphasized in section 3.6.2, (3.52) is *not* a weak coupling master equation, and we cannot make use of (4.16) to decide if there is a unique solution. For $\kappa > 0$, $\gamma > 0$, however, we can intuit that there are no sets of states that do not couple to each other, (for example there is no symmetry that produces a *decoherence free subspace*). It can be checked numerically that for $\kappa > 0$, $\gamma > 0$, (3.52) is indeed uniquely relaxing.

Equation (3.52) is a linear equation in $\rho$ and thus we can write the condition $\dot{\rho} = 0$ as a linear-algebra problem

$$Mx = 0, \tag{4.17}$$

where $M$ is the matrix superoperator representing the semigroup generator $L$ and $x$ is a vector representation of $\rho$. For example, one way to represent $\rho$ as a vector $x$ is simply to 'flatten' $\rho$ into a length $d = N^2$ vector, $x = \{\rho_{11}, \rho_{12}, \ldots, \rho_{1N}, \rho_{21}, \rho_{22}, \ldots, \rho_{NN}\}$,* in which case (4.17) is a $d \simeq 1300$ dimensional linear algebra problem, which is not very large in absolute terms, but it is large enough to make it worthwhile to think a little about how to solve it efficiently, especially since we shall be solving it very many times. There are several methods which can solve such problems directly, but these are generally less efficient than solving a

* This is a little wasteful: since $\rho$ is Hermitian we need not solve for all $N^2$ components. The more frugal approach is to flatten only the upper-triangular part of $\rho$, approximately halving the problem size.

problem of the form

$$M'x = y. \tag{4.18}$$

In order to convert (4.17) to the form (4.18) it is sufficient to add the condition $\operatorname{tr}\rho_s = 1$, replacing the first row of $M$ (the row which gives the equation for $\dot{\rho}_{11}$) to give $M'$ and taking $y = \{1, 0, 0, 0, \dots\}$. Equation (4.18) can then be solved using standard packages. The matrix $M'$ is very sparse, unsymmetric, and not so large that iterative methods are needed, consequently a good choice was the multifrontal method as implemented in Mathematica based on UMFPACK [73], for generating an $LU$ decomposition of $M'$.

### 4.6.2 Fitting the experimental data

To reach agreement with the experimentally measured signal for the strongest drive powers, it is necessary to account for a small amount (~ −58 dB) of leakage of the drive past the cavity. In addition, there is a small bias introduced by measuring the intensity as the square of the $I$ and $Q$ quadratures, each of which is subject to noise.* Accordingly, the quantity that corresponds to the experimental signal is

$$A^2 = |2V_0 \operatorname{tr}(a\rho_s) + b\xi|^2 + 2\sigma_n^2, \tag{4.19}$$

where $b \in \mathbb{C}$ describes the amplitude and phase of the leakage of the drive bypassing the cavity, and $\sigma_n$ is the measurement noise in each of the $I$ and $Q$ channels.

Fits are obtained by minimizing the mean squared deviation between experiment and calculation over the full power range and over the full frequency range, with unconstrained fit parameters being $b$ and the two scaling factors describing the attenuation and amplification for input and output signals. To obtain optimal agreement, we also make adjustments to the system parameters $\gamma_\pm$, $\kappa_\pm$, $\gamma_\varphi^{\mathrm{C}}$, $\gamma_\varphi^{\Phi}$, $g$, $\omega_{\mathrm{r}}$, $E_{\mathrm{J}}$, and $E_{\mathrm{C}}$. These parameters can be measured to some degree in separate experiments, and the values from the fits are consistent within the experimental uncertainties. Once obtained, the same set of parameters is used for generating figures 4.12 and 4.14.

The master equation (3.52) can be rewritten with the parameter dependence indicated

---

* In the first datasets, this 'small bias' was actually a very large bias, a problem which was traced to the fact that the experimental data acquisition chain was recording the heterodyne intensity as the average of the sum of the squares of the quadratures, as opposed to the sum of the squares of the average of the quadratures.

explicitly, in order to show exactly what was fitted:

$$A = \sqrt{|2V_0 \operatorname{tr}(a\rho_s) + b\xi|^2 + 2\sigma_n^2}, \quad \text{where } \rho_s \text{ satisfies} \tag{4.20a}$$

$$\begin{aligned} 0 = &-\mathrm{i}\left[H, \rho_s\right] \\ &+ C_\kappa\left[N(\omega_r, T) + 1\right]\mathcal{D}[a]\rho_s + C_\kappa N(\omega_r, T)\mathcal{D}[P_W \cdot a^\dagger \cdot P_W]\rho_s \\ &+ C_\gamma\left[N\big(\omega_{01}(E_J(\tilde\Phi), E_C), T\big) + 1\right]\mathcal{D}\Big[\sum_j n_{j,j+1}(E_J(\tilde\Phi), E_C)\,|j\rangle\langle j+1|\Big]\rho_s \\ &+ C_\gamma N\big(\omega_{01}(E_J(\tilde\Phi), E_C), T\big)\mathcal{D}\Big[\sum_j n_{j+1,j}(E_J(\tilde\Phi), E_C)P_W\,|j+1\rangle\langle j|\,P_W\Big]\rho_s \\ &+ C_\varphi^{C}\mathcal{D}\Big[\sum_j \epsilon_j(E_J(\tilde\Phi), E_C)P_W T\,|j\rangle\langle j|\,P_W\Big]\rho_s \\ &+ C_\varphi^{\Phi}\sin\left(\frac{\pi\tilde\Phi}{\Phi_0}\right)\mathcal{D}\Big[\sum_j 2j\,P_W\,|j\rangle\langle j|\,P_W\Big]\rho_s, \end{aligned} \tag{4.20b}$$

$$\begin{aligned} H = &\big(\omega_r - \omega_d\big)a^\dagger a + \big(a\xi^* + a^\dagger\xi\big) \\ &+ \sum_{j=0}\Big\{\big[\omega_j(E_J(\tilde\Phi), E_C) - j\omega_d\big]|j\rangle\langle j| + \beta\big(n_{j+1,j}(E_J(\tilde\Phi), E_C)a\,|j+1\rangle\langle j| + \text{h.c.}\big)\Big\}, \end{aligned} \tag{4.20c}$$

$$N(\omega, T) = \frac{1}{e^{\omega/T} - 1}, \tag{4.20d}$$

$$E_J(\tilde\Phi) = E_J^{\max}\cos(\pi\tilde\Phi/\Phi_0). \tag{4.20e}$$

It is (hopefully) obvious how to relate the 'derived' parameters we have been using until now, in terms of the raw parameters $V_0$, $b$, $\sigma_n$, $E_C$, $E_J^{\max}$, $C_\kappa$, $C_\gamma$, $C_\varphi^{\Phi}$, $C_\varphi^{C}$, $\omega_r$, $T$, $\tilde\Phi$, $\omega_d$, $\xi$. For example $g = \beta n_{01}\big(E_J(\tilde\Phi), E_C\big)$, evaluated at the value of $\tilde\Phi$ such that $\omega_{01}\big(E_J(\tilde\Phi), E_C\big) = \omega_r$.

The Levenberg–Marquardt method is ideal for performing these fits. Such Gauss–Newton methods are much more efficient when there is direct access to the Jacobian, as opposed to using numerical differentiation, which means we would like to be able to calculate expressions of the form $\partial x/\partial\theta$, where $\theta$ represents a parameter of the master equation, such as $\omega_r$ or $E_C$. These can be found as solutions of

$$M' \cdot \frac{\partial x}{\partial\theta} = -\frac{\partial M'}{\partial\theta} \cdot x, \tag{4.21}$$

which is also of the form (4.18), and in fact it is possible to reuse the same $LU$ decomposition as was constructed for solving for $x$, so this is quite efficient. The Mathematica code which was used for doing these numerical calculations is given in appendix A. The Levenberg–Marquardt algorithm works very well once it gets 'close' to a local minimum, but given the highly nonlinear nature of the present problem, it is certainly necessary to do quite a bit of fitting by hand and to spend some time feeding the data to the algorithm in pieces, first fitting the low-power section of the data using a reduced set of parameters, and then using these fitted parameters as a starting value for a fit using a larger subset of the data with more of the parameters unlocked, and so on.

Now that we know how to perform numerical fits, the next section attempts to interpret the values of the parameters thus found.

### 4.6.3 Parameter analysis

**Dephasing.** The fitted values of $\gamma_\varphi^{\mathrm{C}}$ and $\gamma_\varphi^{\Phi}$ are consistent with zero, so these terms were dropped for figures 4.12 to 4.14. This shows that there is very little pure dephasing of the lower two levels of the transmon.

**Temperature.** The fitted values of $\kappa_+$ and $\gamma_+$ are consistent with zero, indicating that the effective temperature is very low, and these terms were dropped for figures 4.12 to 4.14. In fact we can be a little more quantitative than this and say that the largest value for the ratio $r$ from (3.54) that is still consistent with the data is approximately 0.003, corresponding to a upper bound on the reservoir temperature of ~ 55 mK. Although this is somewhat higher than the ~ 15 mK base temperature of the refrigerator, it is still the most stringent bound to date that has been placed on the temperature of a transmon sample (the reason that no stronger bound has been placed is that 0.003 thermal photons are hard to see on top of the ~ 20 photons of amplifier noise).

We have some justification for taking this temperature somewhat seriously, despite all the caveats given in chapter 3 regarding the derivation of the master equation: there was a secondary experiment for which the circulator connected to the output port of the cavity was placed on the '100 mK stage' of the refrigerator, the actual temperature of which is typically around 110–120 mK. A representative cavity response is shown in figure 4.15, where a new feature is a pair of broad peaks in between the multiphoton peaks. There is good agreement with the theoretical calculation using an effective temperature of 130 mK, although it is hard to fit these thermal peaks over the full power range and this is a 'by eye' fit.

The problem with making detailed fits at these higher temperatures is that the thermal peaks are due to multiple overlapping transitions between highly excited states—once the temperature is high enough and the driving strong enough there is a sort of bistability effect, where once the system fluctuates into a sufficiently excited state then anharmonicity is reduced and the system can be driven to very highly excited states comparable to the coherent states that would exist in the absence of the transmon (as shown in figure 4.9). Obviously, our numerical technique, with a truncation to $W = 7$ excitations is ill-equipped to simulate this situation. Rau *et al.* [74] considered the *linear-response* regime of the finite-temperature vacuum Rabi splitting, for which the numerics are much more tractable, and showed theoretically that there are three regimes, shown also in figure 4.16: at low temperatures the vacuum Rabi peaks dominate, with additional discrete $|n,\pm\rangle \leftrightarrow |n+1,\pm\rangle$ peaks being visible between them; at intermediate temperatures the $|n,+\rangle \leftrightarrow |n+1,+\rangle$ peaks overlap and similarly the $|n,-\rangle \leftrightarrow |n+1,-\rangle$ peaks overlap, thus are a pair of broad peaks separated by $\sim g\bar{n}^{-1/2}$, where $\bar{n} \simeq T/\omega_r$ is the mean number of thermal excitations; and at very high temperatures the qubit disappears and the response is a single peak at the cavity frequency. We interpret the thermal peaks visible in figure 4.15 as being of the same type as the intermediate-temperature peaks in [74].

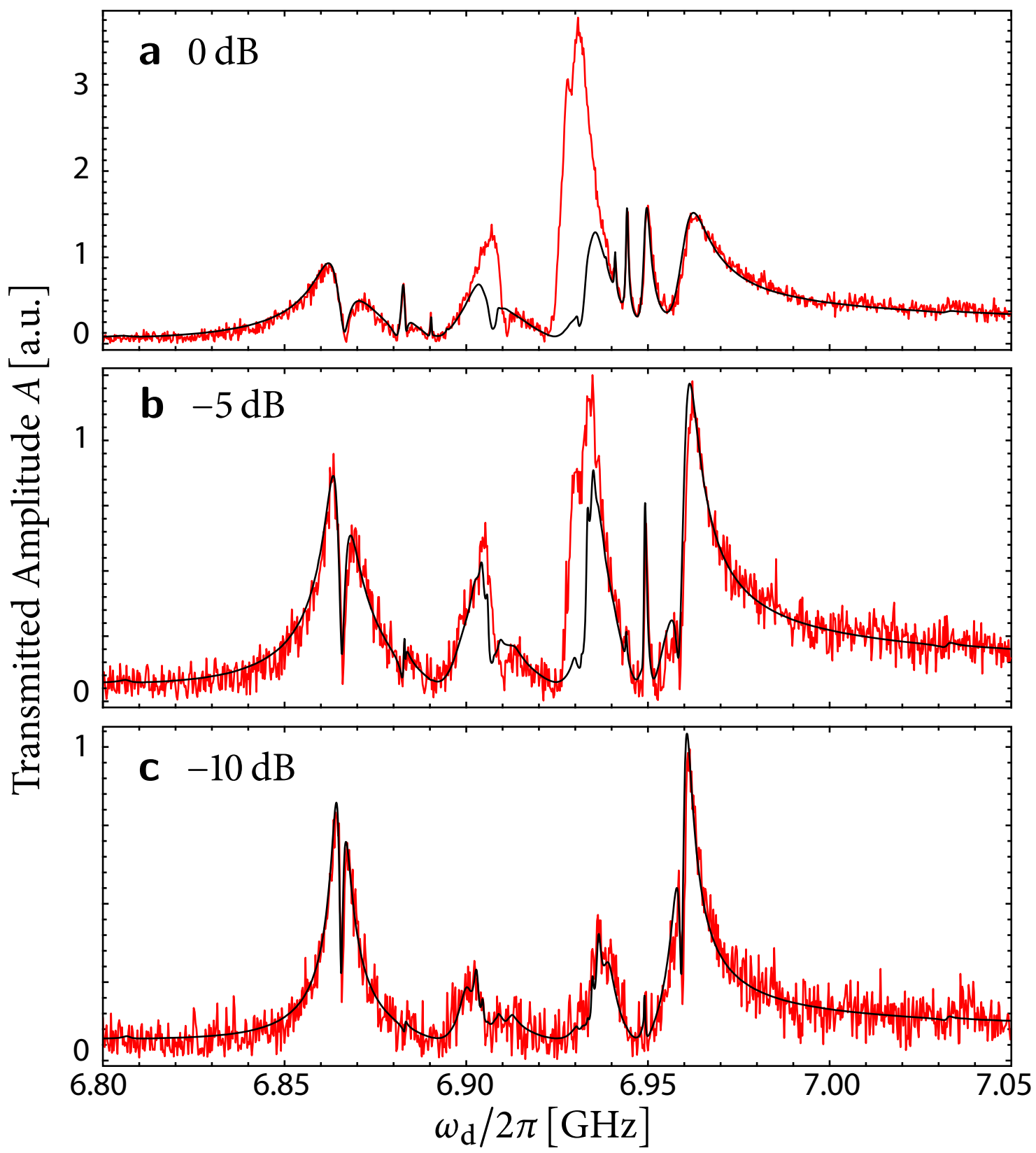

**Figure 4.15: Strongly-driven vacuum Rabi response at elevated temperature.** For this run of the experiment, the 50 Ω termination on the circulator at the output port of the sample was kept at a temperature of ~ 110 mK. The theoretical response (black) was calculated for an effective temperature of 130 mK, showing good agreement with moderate driving, **c**. For the stronger driving of **a** and **b**, the theory and experiment disagree due to the truncated Hilbert space used in the simulations.

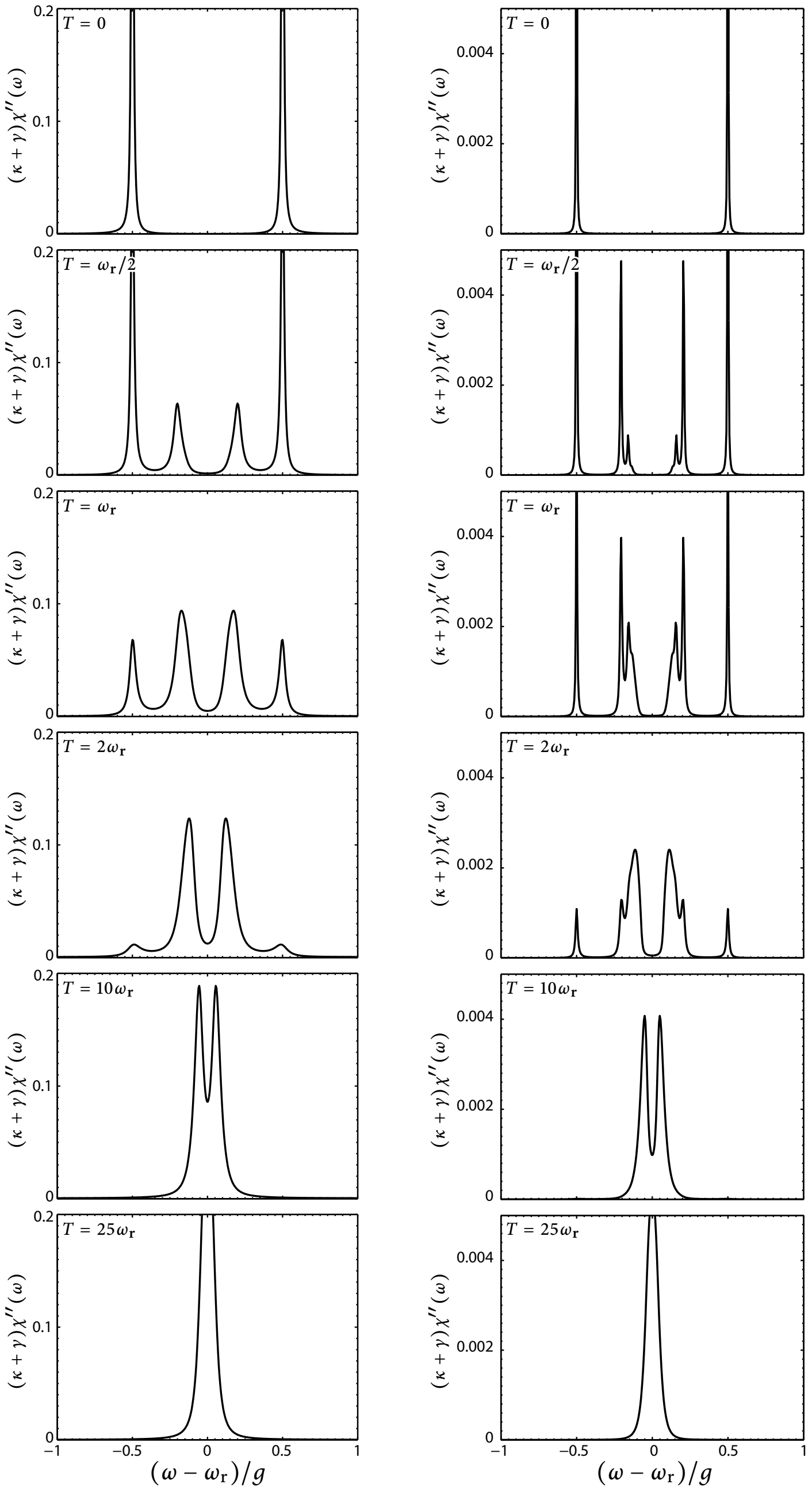

**Figure 4.16: Vacuum Rabi splitting at elevated temperature.** This is a linear-response calculation, showing the imaginary part of the cavity susceptibility, $\chi = \chi' + i\chi''$ ,versus frequency, for different temperatures. The parameters are $g = 0.01\omega_r$, cavity quality factor $Q = 10^4$ ($Q = 10^5$) for the left (right) set of graphs, and qubit dissipation $\gamma_- = 0.08\kappa_-$. (Figure used with permission from [74]. See Copyright Permissions.)

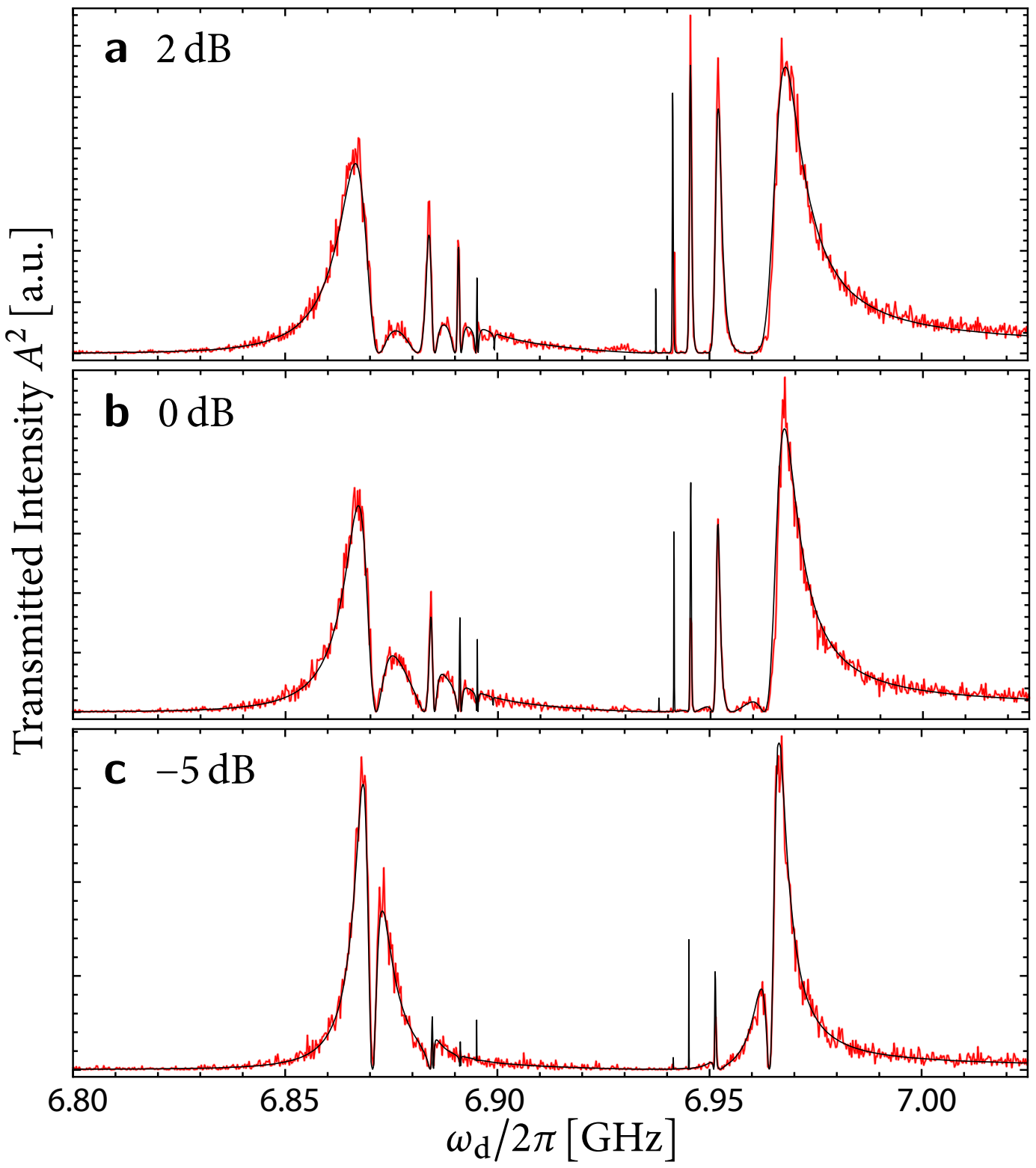

**Figure 4.17: The shape of the $\sqrt{n}$ peaks in the limit of low dissipation.** Compared to figure 4.12c–e, we see that many of the same features are present, indicating that these are purely caused by the strong driving, in the same way as figure 4.11 shows that the supersplitting is caused by strong driving rather than by dissipation. The lack of dissipation means that the unsaturated transitions are much sharper in this plot than in figure 4.12.

**Decoherence parameters.** Looking at figures 4.12 and 4.14 and seeing the excellent agreement between theory and experiment, over a huge range of powers, it is natural to think that this must mean we are able to know the parameters of the system to an extremely high level of precision, and perhaps we can use this to draw some conclusions about the poorly-understood relaxation processes affecting the higher levels of the transmon. Unfortunately, this is not the case, for three reasons. The first reason is that although we keep higher levels of the transmon in the calculations, these higher levels are not significantly occupied for the parameter range of the experiment—the effect of the higher transmon levels is mostly just to cause frequency shifts.* The second reason is that much of the detailed structure in figures 4.12 and 4.14 is unrelated to dissipation. As we saw in section 4.5.1 the supersplitting can be explained entirely as a strong driving effect on the two-level system, without including any dissipation (except as an averaging over the rotation angle). We can perform a similar calculation using the full driven generalized Jaynes–Cummings Hamiltonian (2.62). Figure 4.17 shows the results of such numerical calculations, with the same parameters $g$, $\omega_r$, $E_J$, $E_C$ as in figure 4.12. As could be expected, the lineshapes are correct for those transitions which are fully saturated (and thus have their width set by power-broadening) but quite wrong for the unsaturated transitions, which show up in the theoretical curves as extremely sharp features. The third reason why the fitted parameters do not tell us anything significant about the relaxation of higher transmon levels is that the experiment is performed in a regime where the multimode Purcell effect dominates over any intrinsic transmon decay mechanisms.

**Hamiltonian parameters.** By contrast to the dissipation parameters, the Hamiltonian parameters $g$, $\omega_r$, $E_J$ and $E_C$ *are* quite strongly constrained by the experimental data. However, even these cannot be directly interpreted as the bare parameters of the uncoupled resonator and transmon, as was discussed in section 3.6.

### 4.6.4 Final thoughts

In conclusion, we have shown that the dephasing and intrinsic relaxation of the (lowest 2 levels of the) transmon is very low. We have placed a stringent upper bound on the effective temperature of the system. We have also shown that we do not need to invoke any new effects beyond the generalized Jaynes–Cummings model with the usual photon leakage and multimode Purcell effect, in order to explain the detailed response of the system over a range of five orders of magnitude in drive power.

Before finishing this chapter about the strongly-driven vacuum Rabi resonance, it is worth noting that there has been an alternative analysis due to Peano and Thorwart [75] in terms of a semiclassical quasienergy surface. This is an interesting viewpoint because it allows analogies between the driven Jaynes–Cummings model and the quantum Duffing oscillator.

* See also the footnote on page 87.

# CHAPTER 5

## Generating and detecting Greenberger–Horne–Zeilinger states

Previous chapters have discussed the quantum optics of superconducting circuits, but much of the interest in these systems relates to their potential for quantum information processing tasks. In this chapter, we move in such a direction and consider a circuit containing several artificial atoms, focussing on their potential to behave as qubits. We discuss a proposal for generating and detecting a particular class of maximally entangled states, first pioneered by Greenberger, Horne, and Zeilinger (GHZ). Instead of creating the GHZ state by multi-qubit gates, we employ the idea of entanglement generation by measurement—an elegant way to create entangled states of two or more qubits without the need for high-fidelity two-qubit gates [76–80]. The basic idea is: first prepare an initial state, then perform a measurement. Repeat the protocol many times, keeping only the subensemble of states which produced a particular measurement outcome. In order to produce entangled states in this way, either the preparation step has to involve two-qubit interactions or the measurement needs to be a *joint readout* of several qubits. In cQED, high-fidelity single-qubit rotations are available for the preparation step [27], and the dispersive readout constitutes a very natural joint multi-qubit measurement [52, 81]. It is a joint readout because the qubits are coherently coupled to the voltage inside the resonator, giving rise to a state-dependent shift of the resonator frequency. Heterodyne or homodyne detection of the phase of microwave radiation transmitted through the resonator can thus be used to gain information about the state of the qubits. If the linewidth $\kappa$ of the cavity is much larger than the dispersive shift due to the qubits, $\kappa \gg \chi$, (and if we ignore relaxation processes), then the corresponding measurement operator can be arranged to be a weighted sum of the qubit Pauli operators $\sigma^z$, where the weights are conveniently adjusted by the detunings of the respective qubits from the resonator frequency.

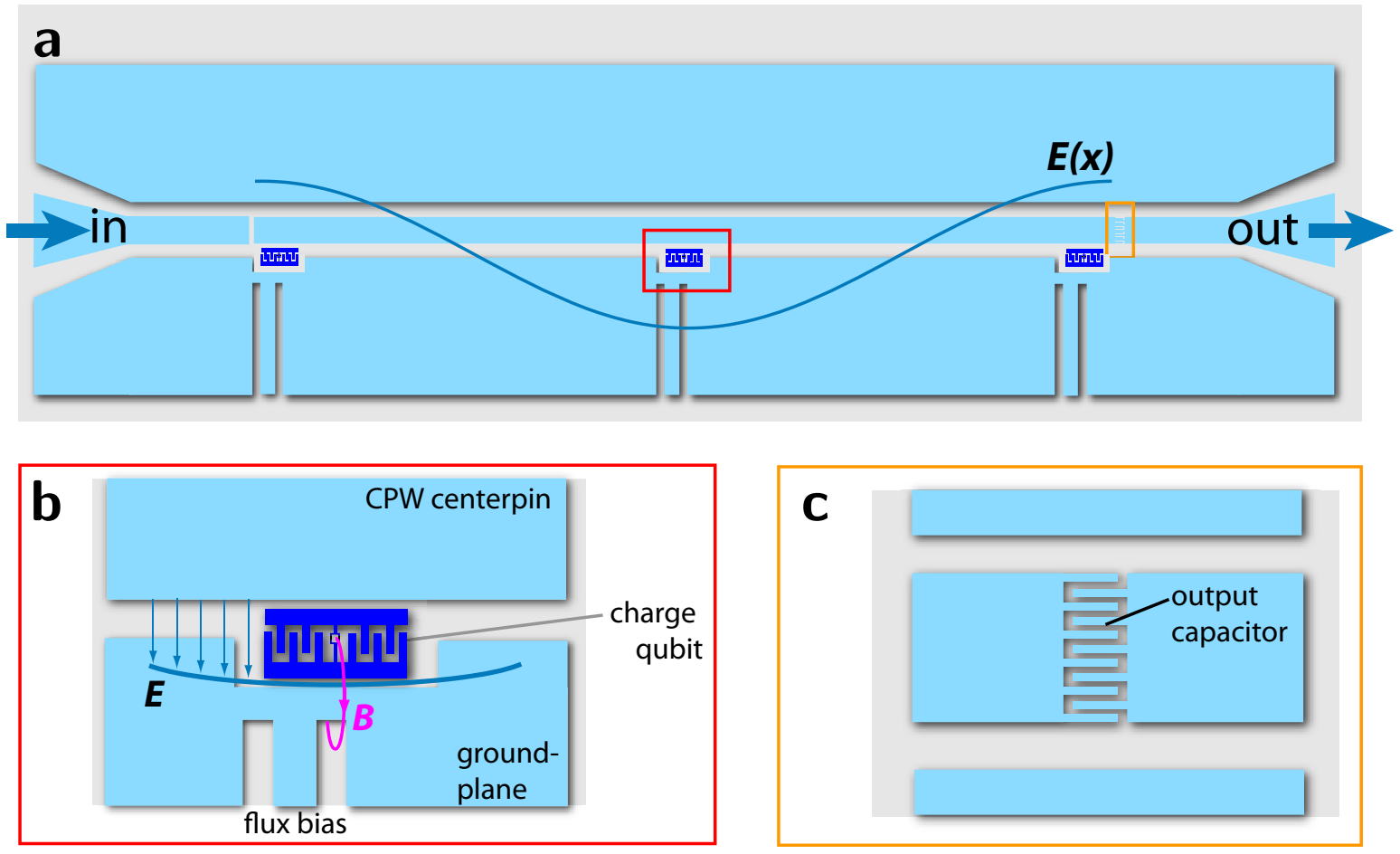

**Figure 5.1: Sketch of the circuit QED architecture.** **a**, The system consists of a superconducting coplanar waveguide resonator, with charge qubits, shown magnified in **b**, placed within the resonator at positions corresponding to antinodes of the microwave electric field $\mathbf{E}(x)$. Each qubit can be addressed by an individual flux bias line, which is used to tune the local magnetic field **B**. The coplanar waveguide is coupled capacitively to an input and output port, such that microwave signals can be coupled into the resonator, and microwave photons leaking out of the resonator can be detected. The input and output capacitors, **c**, are chosen asymmetric so that photon leakage occurs preferentially through the output side.

This idea of *probabilistic state-preparation by measurement* has recently been applied to a 2-qubit cQED system by Hutchison *et al.* [82]. Their theoretical study included the adverse effects of qubit relaxation and dephasing, and showed the practical applicability of the method, even for realistic decay and decoherence rates as currently realized in cQED experiments. Here, we extend this method to the generation of multi-qubit GHZ states [83] (a similar extension has now been performed by Helmer and Marquardt [84]). For superconducting qubit systems there have been successful demonstrations of Bell-state preparation [85–88], and various proposals for creating GHZ states, mainly focusing on the generation via two-qubit gates and qubit-qubit interactions, see e.g. [89–91]. Instead of employing such entangling gates for generating a GHZ state, we propose a scheme tailored to cQED, consisting of one-qubit rotations and a dispersive measurement only. Based on quantum trajectory simulations, we show that currently attainable values for qubit decoherence and decay allow for the creation of three-qubit GHZ states in cQED with high fidelity and high degree of entanglement. The degree of entanglement can be increased at the cost of lowered production rate.

In order to verify the production of the desired GHZ state, we propose to use a second dispersive readout. Because the GHZ state is maximally entangled, this verification is related to proving the violation of a Bell-type inequality [92, 93]. However, proving such a violation in a *loophole-free* fashion turns out to be a much more challenging task in cQED. Given the required measurement time of hundreds of nanoseconds, space-like distances (in the sense of special relativity) are of the order of tens of meters and thus difficult to achieve in a cQED setup, and the dispersive readout is in fact inherently nonlocal. Accepting that the communication loophole therefore cannot strictly be closed, we discuss the potential of the dispersive readout for observing quantum correlations in a 3-qubit GHZ state, as well as the potential for devising a factorizing measurement that is local in the no-signalling sense [94]. Using quantum trajectory simulations including the measurement imperfections caused by qubit decay, we show that a convincing violation of the Bell inequality would require a signal-to-noise ratio which is currently out of experimental reach, but may be approached once efficient methods for protecting qubits from decay have been devised, or with improvements in the noise performance of microwave amplifiers [95].

This chapter is organized as follows: The following section discusses what is meant by a Bell test in general and in the context of cQED. Section 5.3 presents the central idea of generating and detecting multi-qubit GHZ states by dispersive measurements, starting with the idealized situation of no qubit decoherence and decay. The remainder of the chapter is devoted to the consequences of imperfections introduced by decay during the measurement process. In section 5.4, we specify the treatment of qubit decay and continuous homodyne detection using an effective stochastic master equation previously introduced in [82]. Quantitative results from solving this master equation for the situation of GHZ-state generation are presented in section 5.5. We describe different protocols for accepting or rejecting a generated state as a GHZ state, and show in particular that nonlinear filtering offers a significant advantage over simple boxcar filters. Section 5.6 discusses the detection of GHZ states within the dispersive measurement scheme and comments on the potential to violate a Bell-type inequality. Finally, conclusions are presented in section 5.7.

## 5.1 Bell tests

A Bell test is an experiment that attempts to prove Bell's theorem: *quantum mechanics is incompatible with local realism*. Generally these tests take the form of attempting to violate a *Bell inequality*, which is an upper bound on the correlations of results of distant measurements. These inequalities are obeyed by any Local Hidden Variable (LHV) theory, namely a theory that uses local variables with objective values. There are many such inequalities, including those of Bell [96]; Clauser, Horne, Shimony and Holt (CHSH) [97]; Greenberger, Horne and Zeilinger (GHZ) [83]. At the present time there are rather few physicists who seriously doubt the validity of quantum mechanics. However, proving non-classicality of a given experimental platform, via a Bell inequality violation, is widely viewed as excellent benchmark for such a platform for quantum information processing. Despite rapid progress in the field of cQED, a clear-cut demonstration that the system violates classicality was outstanding until the very recent work described in Refs. [98, 99].

### 5.1.1 Idealized Bell test

An idealized Bell test is conceptually simple, if rather abstract, involving a number of participants, measuring devices, settings and outcomes. We have at least two participants. If there are exactly two then it is traditional to name them Alice and Bob. For generality however there may be any number $\ell \geq 2$ of participants. Each participant is in possession of a device with at least two settings. To remain general we define the device belonging to the $i$th participant as having $m_i \geq 2$ settings. Each device in each setting can produce at least two outcomes. Again, for generality let us say that the $i$th device in its $j$th setting can produce one out of a set of $n_{ij}$ outcomes. There is no requirement that each participant's device have the same number of settings and outcomes, nor that the number of outcomes be fixed independent of the setting, but in the special case that all the participants have identical devices, for which the number of allowed results is setting-independent, we can describe the experiment in shorthand by the triple $\ell \times m \times n$. The simplest and most-studied situation is the $2 \times 2 \times 2$ case, with two participants ('bipartite'), two possible settings ('binary settings'), two outcomes ('binary outcomes'). The simplest version of the Mermin inequality, described in section 5.3.2, corresponds to the multipartite, binary-setting, binary outcome situation, $\ell \times 2 \times 2$ for $\ell \geq 3$.

This framework is very abstract. A concrete example of the type that typically people have in mind is an optical experiment where the participants are each measuring the polarizations of photons belonging to entangled pairs, where the device might consist of a polarizing filter in front of a photomultiplier tube and where the $m_i$ settings are a set of rotation angles for the filter, and binary outcomes are that a photon is or is not detected. However, nothing in the test protocol requires that the devices actually *measure* anything. It would not be cheating if, for example, for one setting of one of the machines it always gave a particular result with unit probability. Similarly, one of the settings of the machine might cause it to roll a dice or toss a coin in order to produce the measurement outcome. These devices are all legal, although they are rather crippled from the point of view of violating a Bell inequality. As far as the Bell test protocol is concerned, the physical implementation of the devices, settings and outcomes are irrelevant and only the correlations between the outcomes are important.

**Table 5.1: A demonstration 2×2×2 Bell test correlation table.** The numbers in the *occurrences* column are random, for illustrative purposes.

| Alice | | Bob | | |
|---|---|---|---|---|
| setting | outcome | setting | outcome | occurrences |
| 0 | 0 | 0 | 0 | 123 |
| 0 | 0 | 0 | 1 | 90 |
| 0 | 0 | 1 | 0 | 21 |
| 0 | 0 | 1 | 1 | 34 |
| 0 | 1 | 0 | 0 | 232 |
| 0 | 1 | 0 | 1 | 3 |
| 0 | 1 | 1 | 0 | 77 |
| 0 | 1 | 1 | 1 | 99 |
| 1 | 0 | 0 | 0 | 42 |
| 1 | 0 | 0 | 1 | 42 |
| 1 | 0 | 1 | 0 | 523 |
| 1 | 0 | 1 | 1 | 121 |
| 1 | 1 | 0 | 0 | 0 |
| 1 | 1 | 0 | 1 | 17 |
| 1 | 1 | 1 | 0 | 45 |
| 1 | 1 | 1 | 1 | 11 |

Given our set of participants and their devices, the protocol is that the participants spread themselves far enough apart from each other that they cannot communicate. To make sure that they cannot communicate even in principle, they must have a space-like separation in the sense of special relativity. Now, each participant chooses one of the $m_i$ settings for their device, and records the setting and the associated outcome in some immutable form. This must be done *simultaneously* in the sense that all of the participants must have finished recording their outcomes before they come into causal contact with each other. The protocol is repeated many times, after which the participants bring their lists of recorded settings and outcomes to a convenient location and compile a summary, of the form of table 5.1, showing all the possible correlations. For an $\ell \times m \times n$ experiment, this table has $m^\ell n^\ell$ rows.

The final step in the Bell test protocol is to examine the table and decide if there exists *any* conceivable LHV theory that is able to explain the results. If not, then the test has successfully disproved the ability of LHV theories to explain the universe.

### 5.1.2 How many repetitions of the protocol are needed?

The proof of a Bell inequality violation will necessarily be statistical in nature. Even a very highly correlated set of results could in principle result by lucky chance from an entirely random classical process. However, if the protocol is repeated sufficiently many times it can convince even skeptics. A fair coin might by fluke turn up heads 10 times in a row, but who is willing to believe that it would do so 1000 times?

The statistical nature of Bell tests notwithstanding, a certain mystique has arisen surrounding a particular class of tests, of which the GHZ was the first, that have been termed Bell tests *without inequalities*. This consists of a set of settings and outcomes for which the quantum probabilities are all either 0 or 1, while for LHV this is impossible. This should be contrasted with, for example, the CHSH [97] and CH [100] schemes, where the quantum probabilities exceed the classical ones but do not reach the extremal 0 and 1. The fact that all the quantum probabilities are 0 or 1 has led some people to claim that the experiment need only be repeated one time in order to rule out local realism. This is clearly untrue, as was stated quite strongly by Peres [101]:

> This is sheer nonsense: a single experiment can only verify one occurrence of one of the terms. . . .

Nevertheless, this is a popular misconception, again quoting Peres [101]:

> The list of authors is too long to be given explicitly and it would be unfair to give only a partial list.

To be explicit: the reason why a single experiment is insufficient is that we can imagine a LHV theory that simply adopts the uniform random distribution over all outcomes. Although quantum mechanics is correct, the resulting sequence of observations does not have zero probability under the classical theory, simply because *no* sequence of results has zero probability under the classical theory. It is only by repeating the experiment many times that we can show that the results are vastly *more likely* under quantum mechanics than LHV.

Still, there is a gut feeling that the 'without inequality' class of Bell tests should be somehow *stronger* than the ordinary kind. In fact van Dam *et al.* [102] were able to prove exactly this, using a game-theoretical analysis and an argument based on the Kullback–Leibler divergence (also known as the *relative entropy*). They showed that the 3-qubit GHZ test is around 4.5 times stronger than the CHSH test, which in turn is the strongest of the 2-qubit tests they examined. The sense of this statement is that if, say, 200 repetitions of the GHZ test would be sufficient to persuade someone to discard a classical viewpoint, then $4.5 \times 200 = 900$ repetitions of the CHSH test would be needed in order to have an equivalently persuasive effect. The outline of their argument is that they imagine a game played between two players: an experimentalist who believes in quantum mechanics and a theorist who believes in local realism. In order to convince the local realist to change his mind, the experimentalist should choose to perform the set of experiments such that the *best* local realistic model explains the data *worst*, when compared to the quantum mechanical description. The relative entropy quantifies the divergence between the probability distributions that the two models predict.

### 5.1.3 Loopholes

As should be clear from the idealized description in section 5.1.1, a Bell test is a remarkably difficult experiment to perform. There are some obvious tricky questions such as 'at what stage can we consider the result to be recorded immutably?' There are also some more prosaic problems such as finite detector efficiencies. Because of these problems, real implementations invariably have *loopholes*. This section examines the effect of the three main loopholes: *superdeterminism, detection* and *communication.* Generally only detection and communication are given serious consideration, and proposals for so-called *loophole-free* experiments really mean that they simultaneously avoid both the detection and communication loopholes. So far such a loophole-free experiment does not yet exist, although both loopholes have been closed in independent experiments [103, 104].

**Superdeterminism.** One fundamental problem with performing a Bell test is that the participants and their devices will never truly be causally disconnected (for the reason that the entire visible universe is causally connected). This has been termed the 'superdeterminism' loophole by Bell and, although it is recognized as a conceptual problem, it is usually not considered to be a problem in practice, on the basis that any theory that would be able to take advantage of superdeterminism would likely be much less plausible than quantum mechanics.*

**Detection.** The detection loophole is caused by the problem of finite detector efficiency. For most Bell tests, the detectors need to be very efficient in order to have a chance to violate the inequality, for example for the CHSH test the threshold is $2(\sqrt{2}-1) \simeq 82\%$. For less efficient detectors, one obtains results which are consistent with quantum mechanics, but unfortunately are also consistent with certain classical theories. These classical theories are somewhat strange, however, because in order to reproduce the quantum mechanical result, the probability of the detector to fail to detect a given event has to be correlated with what the result would have been if it were detected. If we make an additional *fair sampling assumption*, namely that the detector efficiency is independent of what the result would have been, then we can calibrate away the detector efficiency.

**Communication.** The communication loophole is caused by the fact that maintaining delicate quantum correlations across large distances is quite difficult, especially for experiments that do not use photons. Generally it is necessary to *assume* that even though the parts of the experiment are close enough that they could in principle interact, that they are still sufficiently separated to make such an interaction quite unlikely.

### 5.1.4 So what are we trying to do?

In cQED we have no chance of avoiding the communication loophole, given the space constraints of a dilution refrigerator and that the dispersive readout is *inherently* nonlocal, since a photon must 'bounce around a few times in the cavity' in order for us to talk about the cavity even having a resonance frequency. Since the loophole is unavoidable, our purpose cannot be to convince skeptics of the superiority of quantum mechanics over LHV, nor to

* LHV theories that take advantage of the superdeterminism loophole have sometimes been presented as having mysterious philosophical consequences, implying the non-existence of free will [105]. I feel compelled to point out, however, that allowing for the much more mundane possibility of malicious tampering with the experiment, the superdeterminism loophole is quite easily exploited: for example two devices fitted with 'cryptographic' pseudorandom number generators can produce results which appear to be entirely random, to someone not in possession of the key, but there can be arbitrarily strong correlations between the devices. This requires zero communication between the devices beyond initializing them with the same cryptographic keys.

perform the long-sought loophole-free Bell test. Rather, as stated in the introduction, our purpose is to perform the Bell test as a benchmark—to be blunt we wish to demonstrate a Bell test merely *because this is known to be difficult*. In order to make the achievement as impressive as possible, then, we should avoid using the fair sampling assumption. We should also ensure that the communication loophole is only opened *in principle*, due to the non-space-like separations. In other words we should avoid using a measurement scheme that behaves in an explicitly non-local fashion. A necessary condition [106] for this is that the measurement scheme should be *non-signalling* which simply means the measurement scheme should forbid the participants to send each other messages purely by choosing a particular sequence of measurement settings.* Therefore, we adopt as our goal the rather ambitious task of performing a Bell test in cQED, without making the fair-sampling assumption, and using a measurement that explicitly obeys a non-signalling property.

## 5.2 Quantum trajectories

Stochastic wavefunctions [108–110] and quantum trajectories [111], are *stochastic unravelings* of the master equation. This means that they provide a stochastic description of the system dynamics, which reduces to the master equation description once an averaging is performed over the ensemble of stochastic realizations. One reason to prefer such a description is for numerical efficiency—for a $d$-dimensional Hilbert space, storing a wavefunction requires $\mathcal{O}(d)$ space, compared to $\mathcal{O}(d^2)$ for the full density matrix. It might seem that this is a time-for-space tradeoff, given the need to ensemble average over stochastic realizations, but it can be shown [112] that for large $d$, the time requirements of the stochastic solution scale no worse than the master equation solution. Depending on the precise nature of the problem, the stochastic solution can be vastly faster, for example when there is a majority of the state space where large time-steps can safely be taken and a small part of the space where very small time-steps are required. In such a case, the integrator of the stochastic wavefunction can generally take large timesteps, only adaptively shrinking them when encountering the region of rapid dynamics, whereas the integrator of the master equation describes the dynamics of the whole ensemble and must therefore always adapt to the *smallest* timescale of the problem.

* The non-signalling requirement is also a *sufficient* condition for the measurement scheme to be local if we additionally assume the validity of quantum mechanics [94]. A recent candidate for a necessary *and* sufficient criterion for separating quantum mechanics from *post-quantum* theories is the criterion of *information causality*, namely the requirement that communication of $m$ classical bits should result in an information gain of at most $m$ bits [107].

Even in the absence of numerical efficiency considerations, quantum trajectories are useful for describing the dissipative dynamics of the master equation as being due a continuous measurement of the system. This can be related to the input-output theory of section 4.3, being more specific about the exact nature of the measurement of $b_{\text{out}}$. For example, if there is a term $\kappa\mathcal{D}[a]\rho$ in the master equation, leading to an outgoing field $b_{\text{out}}(t) = \sqrt{\kappa}a(t)$, then if we monitor the $b_{\text{out}}$ channel with a photomultiplier, each time the photomultiplier registers a 'click', the state of the system *conditioned on our observing the click* updates as

$$\rho \mapsto \frac{a\rho a^\dagger}{\operatorname{tr}[a^\dagger a\rho]}, \tag{5.1}$$

which is called a *jump unraveling*. Similarly there is a *quantum state diffusion* unraveling, corresponding to a homodyne or heterodyne monitoring of the output channel. The technical details of the quantum trajectories approach are covered very nicely in standard texts [28, 30] so we refrain from further discussion here.

In the calculations of this chapter, we use a somewhat unusual *stochastic master equation* approach. This corresponds to monitoring only one of the relaxation channels of the system, namely the photon leakage through the output port of the cavity. This does not provide the numerical speed-up compared to simply integrating the master equation. However it provides a much more convenient way to describe time-domain processing of the output signal, compared to the alternative of using the ordinary master equation and calculating high-order multi-time correlation functions via the quantum regression theorem.

## 5.3 Idealized preparation and detection of GHZ states

The $N$-qubit GHZ state [83] is the maximally entangled multi-qubit state of the form

$$|\text{GHZ}\rangle = \left(\bigotimes_{j=1}^{N}|\uparrow\rangle_j + \bigotimes_{j=1}^{N}|\downarrow\rangle_j\right)\Big/\sqrt{2} \tag{5.2a}$$

$$= \big(|\uparrow\uparrow\cdots\uparrow\rangle + |\downarrow\downarrow\cdots\downarrow\rangle\big)\big/\sqrt{2}, \tag{5.2b}$$

where $|\cdot\rangle_j$ denotes the state of the $j$th qubit. GHZ states have received much attention in the context of violation of Bell-type inequalities, see e.g. [113–117], ruling out classical local hidden variable (LHV) theories as a valid description of nature. They are also of interest as optimal resource states for measurement-based computation [118].

In this section, we lay out the essential ideas behind the preparation and detection of GHZ states using the joint dispersive readout typical for cQED. To keep the discussion as clear as possible, the exposition in this section ignores the adverse effect of qubit decay and decoherence, and other possible sources of measurement imperfections. We shall return to the full discussion of the realistic situation including these effects in the following sections.

### 5.3.1 Preparation scheme

With the dispersive readout of cQED, the measurement outcomes are inferred from the detection of the homodyne signal for the microwaves transmitted through the resonator. Recall from (4.6) that $b_{\mathrm{out}}^{\dagger}$ and $b_{\mathrm{out}}$ denote the creation and annihilation operator for a photon in the output line. Homodyne measurement is much like the heterodyne measurement described in section 4.4, except that $\omega_{\mathrm{IF}} = 0$. Thus, for a particular setting of the phase between the RF drive and the LO, the homodyne signal is proportional to $\langle b_{\mathrm{out}}(t) + b_{\mathrm{out}}^{\dagger}(t)\rangle$.

For each measurement, this time-dependent signal can then be reduced to a single number, the time-integrated signal $s \propto \int_0^t \mathrm{d}t' \, \langle b + b^{\dagger}\rangle$. In the absence of qubit decay and decoherence, the probability distribution $p(s)$ for this integrated signal $s$ takes the form of Gaussian peaks, which initially overlap strongly, indicating that we begin with little information about the state of the system, and subsequently separate with increasing measurement time $t$ [81, 87, 119–121], as the signal-to-noise ratio gradually increases with longer time-averaging. In the limit of negligible overlap between peaks, the dispersive readout corresponds to a projective measurement of the operator

$$X = \sum_j \delta_j \sigma_j^z, \qquad (5.3)$$

where the weights $\delta_j = \chi_j/\bar{\chi}$ are the fractional contributions to the mean dispersive shift, $\bar{\chi} = \sum_j \chi_j/N$. In detail, the preparation scheme can be described as follows:

(i) Arrange all qubit detunings such that the system is dispersive, and mutual qubit detunings are large compared to the qubit-qubit interaction strengths. The initial state is the state with each qubit in its ground state, $\otimes_{j=1}^{N} |{\downarrow}\rangle_j$, which is easy to reach because, as we showed in the section 4.6.3, the effective temperature of cQED circuits can be arranged to correspond to *at most* 0.003 thermal photons on average;

(ii) Perform $\pi/2$ rotations (see section 2.10.1) on each of the $N$ qubits, preparing the state $2^{-N/2} \otimes_{j=1}^{N} (|{\downarrow}\rangle_j + |{\uparrow}\rangle_j)$;

(iii) While keeping the system dispersive and the mutual qubit detunings sufficiently large, adjust the qubit detunings such that their dispersive shifts assume the ratio,

$$\chi_1 : \chi_2 : \ldots : \chi_{N-1} : \chi_N = 1 : 1 : \ldots : 1 : N-1, \tag{5.4}$$

and perform a dispersive measurement. Ideally, this corresponds to a projective measurement of the observable $X = \sum_j \delta_j \sigma_j^z$. Conditioned on the measurement result being '0', see figure 5.2a, we thus obtain the pre-GHZ state

$$|\mathrm{pGHZ}\rangle = \left(|\downarrow\downarrow\cdots\downarrow\uparrow\rangle + |\uparrow\uparrow\cdots\uparrow\downarrow\rangle\right)/\sqrt{2}; \tag{5.5}$$

(iv) In the final step, a $\pi$ rotation is applied to qubit $N$, yielding the GHZ state, (5.2). Alternatively, one may choose a different computational basis by interchanging the '↑' and '↓' labels for qubit $N$.

The necessary adjustment of the $\chi_j$ ratios is possible in cQED samples employing local flux-bias lines [88], which allow for the fine tuning of individual qubit frequencies. We note that even though there are $2^N$ different states, the scheme requires the resolution of only $\sim 2N$ different peaks, which should be compared to the need for application of $(N-1)$ two-qubit gates for the preparation of the same GHZ state via gates, see e.g. [122]. The *probability of success* of the above idealized procedure is $2^{-N+1}$.

### 5.3.2 Detection scheme

Ideally, the confirmation of the GHZ state production and the verification of its quantum correlations proceed by a measurement of the Bell–Mermin operator [113],

$$M = 2^{N-1}\mathrm{i}\left(\prod_{j=1}^{N}\sigma_j^- - \prod_{j=1}^{N}\sigma_j^+\right). \tag{5.6}$$

For the $N$-qubit GHZ state, this operator takes on the value $M = 2^{N-1}$, while the corresponding combination of correlations for LHV theories predicts an outcome $M \leq 2^{N/2}$ if $N$ is even, and $M \leq 2^{(N-1)/2}$ if $N$ is odd [113]—thus leading to a violation that grows exponentially in the qubit number.

In the general case, the Bell–Mermin operator is not amenable to a direct measurement. However, for $N$ qubits, it can be decomposed into $2^{N-1}$ $N$-qubit parity operators, which are more easily accessible by experiment, and the GHZ state is a simultaneous eigenstate of all the

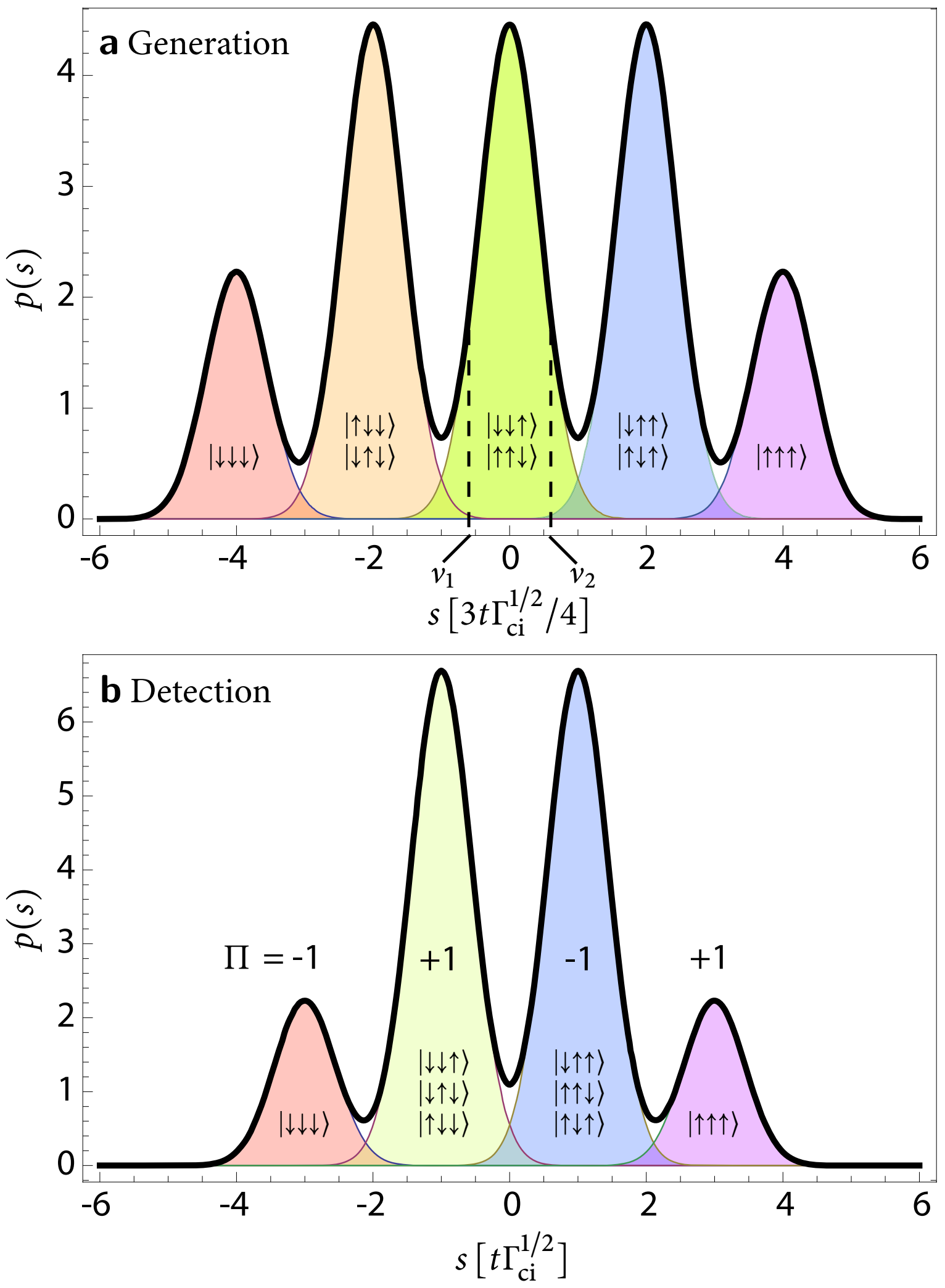

**Figure 5.2: Dispersive measurements for generating and detecting GHZ states.** Dispersive measurements employed for **a** generating a GHZ state, and **b** detecting the parity $\Pi = \prod_j \sigma_j^z$. Both panels show the probability density $p(s)$ for the integrated homodyne signal $s$ for the concrete example of a 3-qubit system. **a**, For the generation of a 3-qubit pre-GHZ state, the dispersive shifts are fixed at ratios $\chi_1 : \chi_2 : \chi_3 = 1 : 1 : 2$. Ideally, the Gaussian peaks belonging to the 5 measurement results $\{\pm 4, \pm 2, 0\}$ separate with increasing measurement time $t$ (here: $t = 5/\Gamma_{ci}$), allowing for a reliable projective measurement when using appropriate thresholds, e.g. $v_1$, $v_2$ for the selection of the measurement outcome '0'. **b**, The dispersive parity measurement requires identical dispersive shifts, $\chi_1 : \chi_2 : \chi_3 = 1 : 1 : 1$. The four measurement outcomes $x_i \in \{\pm 3, \pm 1\}$ then allow the inference of the parity value by $\Pi_i = -\sin(x_i \pi/2)$.

relevant parity operators. The specific form of the Bell–Mermin operator in the three-qubit case is given by

$$M = \sigma_1^x \sigma_2^x \sigma_3^x - \sigma_1^x \sigma_2^y \sigma_3^y - \sigma_1^y \sigma_2^x \sigma_3^y - \sigma_1^y \sigma_2^y \sigma_3^x, \tag{5.7}$$

obtained from (5.6) by setting $N = 3$ and using $\sigma_j^\pm = (\sigma_j^x \pm i\sigma_j^y)/2$. In the ideal case, one would perform the $2^{N-1}$ parity measurements, using a quantum non-demolition method on one and the same state.

Since the dispersive readout does not realize exact parity measurements, we accept the necessity to repeat measurements. Instead of the parity, the dispersive readout can easily access the operator $X = \sum_j \sigma_j^z$ (obtained from the general expression (5.3) for $X$ by setting all the dispersive shifts equal, $\delta_j = 1$). Once $X$ is known, the value of the parity $\prod_j \sigma_j^z$ can be uniquely inferred, see figure 5.2b. Specifically, the measurement results of the operator $X$, given by $x_i \in \{\pm 3, \pm 1\}$ also reveal the parity $\prod_j \sigma_j^z$ of the states: for $x_i = -3, 1$ there is an odd number of 'spin-downs' (↓) and the parity is negative, whereas for $x_i = -1, 3$ the number of 'spin-downs' is even and the parity is positive. Using single-qubit rotations mapping the appropriate $x$ and $y$ axes to $z$ [123], all the required parities can be measured dispersively.

The crucial step thus consists in tuning all dispersive shifts to be identical. As with the preparation step, this can be achieved by adjusting qubit detunings using local flux-bias lines. Compared to the setting employed for the GHZ state generation, it is in fact only the detuning of the $N$th qubit that needs to be changed. Ideally, the measurement of $X$ then leads to the measurement outcomes $x_i \in \{\pm N, \pm(N-2), \ldots, \pm\ell\}$, terminating with $\ell = 1$ if $N$ is odd and with $\ell = 0$ if $N$ is even. The inferred parity outcomes simply alternate in sign according to $\Pi_i = -\sin(x_i\pi/2)$ for odd $N$ and by $\Pi_i = \cos(x_i\pi/2)$ for even $N$. It is important to note that, while the number of required different measurements grows exponentially with $N$, the number of measurement outcomes that need to be resolved is given by $N+1$, only growing linearly with the qubit number. This should be compared to the situation of a full state readout, which would require resolution of $2^N$ different peaks and dispersive shifts to be spread over an exponentially large frequency range, $\chi_j = 2^j \chi_0$.

Both the generation and detection scheme will obviously suffer from qubit decoherence and decay. The subsequent sections take into account these effects and study quantitatively how the idealized proposal performs under more realistic conditions.

## 5.4 Model

For the generation and subsequent detection of a multi-qubit GHZ state we consider a cQED system comprising three superconducting charge qubits coupled to the fundamental mode of a microwave resonator. The model of the system and notation follow those in reference [82]. Neglecting the possible influence of levels beyond the two-level approximation for the superconducting qubits, the system is described by a driven Tavis–Cummings Hamiltonian [124], which is simply the Jaynes–Cummings Hamiltonian (2.50), extended to more than one qubit

$$H = \omega_r a^\dagger a + \sum_j \frac{\omega_{q,j}}{2}\sigma_j^z + \sum_j g_j(a\sigma_j^+ + a^\dagger\sigma_j^-) + (a\xi^* e^{i\omega_d t} + a^\dagger \xi e^{-i\omega_d t}), \tag{5.8}$$

where, as before, $\omega_r/2\pi$ denotes the resonator frequency and $\xi$ the strength of the measurement drive. The qubit frequencies $\omega_{q,j}/2\pi$ are considered to be tunable individually, as realized by local flux-bias lines in recent cQED experiments [88]. The qubit-resonator couplings are given by $g_j$, whose signs are determined by the location of the respective qubit within the resonator. For concreteness, we will focus on the case of a half-wave coplanar waveguide resonator, with two qubits placed close to one end, and the third qubit on the opposite end, leading to relative signs $\mathrm{sgn}(g_1) = \mathrm{sgn}(g_2) = -\mathrm{sgn}(g_3)$.

The system is to be operated in the dispersive regime described in section 2.10, where the detuning is large compared to the coupling, $|\lambda_j| = |g_j|/|\omega_{q,j} - \omega_r| = |g_j/\Delta_{q,j}| \ll 1$, and the photon occupation remains small compared to the critical photon number [125], $\langle a^\dagger a\rangle \ll n_{\mathrm{crit}} = \Delta^2/4g^2$. Under these conditions the interaction term in (5.8) can be adiabatically eliminated [81], such that the effective Hamiltonian that generalizes (2.72), in the frame rotating with the measurement drive frequency $\omega_d/2\pi$ reads

$$H_{\mathrm{eff}} = \Delta_r a^\dagger a + \sum_j \frac{\Delta_{q,j} + \chi_j}{2}\sigma_j^z + \sum_j \chi_j a^\dagger a \sigma_j^z + (a\xi^* + a^\dagger\xi), \tag{5.9}$$

where, as before, $\Delta_r = \omega_r - \omega_d$ is the detuning between measurement drive and resonator, and as in (2.74) $\chi_j = g_j^2/\Delta_{q,j}$ gives the dispersive shift* due to qubit $j$. Here, the qubit-qubit coupling $\sim J$ via virtual photons has been neglected, as is appropriate for sufficient detuning between qubits, $J \ll \left|\Delta_{q,j} - \Delta_{q,j'}\right|$. The effects of qubit decay and cavity photon leakage are taken into account within a master equation description. Specifically, we include intrinsic

* A structurally identical Hamiltonian is also obtained in the case of transmons; merely the dispersive shifts $\chi_j$ are modified, as given in (2.73).

qubit relaxation with rates $\gamma_j$, (single-mode) Purcell-induced relaxation with rates $\gamma_{pj}$ [41, 42], and photon decay from the cavity with rate $\kappa$. As we saw in the last chapter, pure dephasing can be ignored for transmons [43], and will be neglected here. (We have checked that inclusion of pure dephasing at small rates, comparable to those achieved in [43], does not significantly alter the results.)

As demonstrated in [82], one can dramatically simplify the resonator-qubit master equation and reach an effective master equation for the qubits only, given that photon decay is fast. Specifically, we assume that the drive is on resonance to the cavity, $\Delta_r = 0$, and require

$$\kappa \gg \max\left\{\xi, \textstyle\sum_j |\chi_j|\right\}. \tag{5.10}$$

Under these conditions, an analogous separation of qubit and resonator degrees of freedom can also be reached on the level of the stochastic master equation (SME), appropriate for the situation of continuous homodyne detection of the emitted microwave radiation [82]. The effective SME for the qubit density matrix $\rho_J$ conditioned on the measurement record

$$J(t) = \sqrt{\Gamma_{ci}} \sum_j \langle \delta_j \sigma_j^z \rangle + \zeta(t) \tag{5.11}$$

is given by

$$\dot{\rho}_J = L\rho_J + \sqrt{\Gamma_{ci}}\zeta(t)\, \mathcal{M}\Big[\sum_j \delta_j \sigma_j^z\Big]\rho_J, \tag{5.12}$$

where we are using similar notation to [82]: $\mathcal{M}[c]$ is the measurement operator given by $\mathcal{M}[c]\rho_J = (c - \langle c \rangle)\rho_J/2 + \rho_J(c - \langle c \rangle)/2$, $\zeta(t)$ represents Gaussian white noise with zero mean and $\langle \zeta(t)\zeta(t')\rangle = \delta(t - t')$, and $\Gamma_{ci} = \eta\Gamma_m$ denotes the effective measurement rate, reduced by an efficiency factor* with respect to the maximum rate $\Gamma_m = 64\bar{\chi}^2|\xi|^2\kappa^{-3}$. The generator $L$ is defined as

$$\begin{aligned} L\rho = -\mathrm{i}\Bigg[\sum_j \frac{\omega_{q,j} + \chi_j}{2}\sigma_j^z + \frac{4\bar{\chi}|\xi|^2}{\kappa^2}\sum_j \delta_j \sigma_j^z, \rho\Bigg] \\ + \sum_j (\gamma_j + \gamma_{pj})\mathcal{D}[\sigma_j^-]\rho + \frac{\Gamma_d}{2}\mathcal{D}\Big[\sum_j \delta_j \sigma_j^z\Big]\rho \end{aligned} \tag{5.13}$$

---

* As in [82], we have checked that additional measurement-induced dephasing does not alter our results, and that the relevant parameter is the ratio of coherent information rate $\Gamma_{ci}$ and decay rates. As a result, we may set $\Gamma_d = \Gamma_{ci}/2$.

with the measurement-induced dephasing rate $\Gamma_d = \Gamma_m/2$. Assuming mutually distinct qubit frequencies, we have treated the Purcell effect in the secular approximation. This distinguishes the current discussion from the work of Hutchison *et al.*—whereas they were able to make use of a *dark state* that was protected from Purcell decay due to symmetry, we are not so fortunate. Specifically, we neglect the cross-terms in $\mathcal{D}[\sigma_1^- + \sigma_2^- - \sigma_3^-]$ which are interference effects for radiation from different qubits, and which only become important if the qubit frequencies are sufficiently close, *i.e.*, $|\Delta_{q,i} - \Delta_{q,j}| \ll \gamma_{pi} + \gamma_{pj}$ [82, 120]. With this approximation, Purcell-induced decay and intrinsic decay (which might itself be due to multimode Purcell effect) can be treated on the same footing, and in the following we will assume similar decay rates for all qubits and subsume them under the shorthand $\gamma = \gamma_j + \gamma_{pj}$. Finally, the integrated signal $s$ is simply given as the time integral of the measurement record for the full measurement time $t$,

$$s = \int_0^t \mathrm{d}t' \, J(t'). \tag{5.14}$$

## 5.5 Preparation of the GHZ state under realistic conditions

We now turn to the situation of GHZ state preparation in the presence of qubit decay, which we study using quantum trajectory simulations based on the stochastic master equation (5.12). Following the steps (i)–(iii) described in section 5.3, the system is initialized and dispersive shifts are adjusted for the measurement step. The interplay of measurement-induced dephasing, gradual state projection, and the simultaneous qubit decay are captured by the conditional density matrix $\rho_J$, where each simulation run generates a particular measurement record $J(t)$ up to a final measurement time $t$, corresponding to the experimentally accessible homodyne signal.

Since preparation of the correct pre-GHZ state is probabilistic (ideally, state generation succeeds with probability $P = 1/4$ in the present case), one has to define a criterion ('filter') for success of preparation, and postselect the corresponding subensemble [119]. In principle, the information available to the filter is the full measurement record. In the following, we will discuss two different filters, the linear boxcar filter and the full nonlinear Bayesian filter and compare their performance in selecting high-fidelity GHZ states under realistic conditions.

The simple filter already outlined in section 5.3 is the linear boxcar filter. It compresses each measurement record into a single number, the integrated signal $s = \int_0^t \mathrm{d}t' \, J(t')$, and declares successful pre-GHZ state preparation whenever $s$ falls within the limits of appropriately chosen thresholds, $\nu_1 \leq s \leq \nu_2$. Otherwise, the state is rejected.

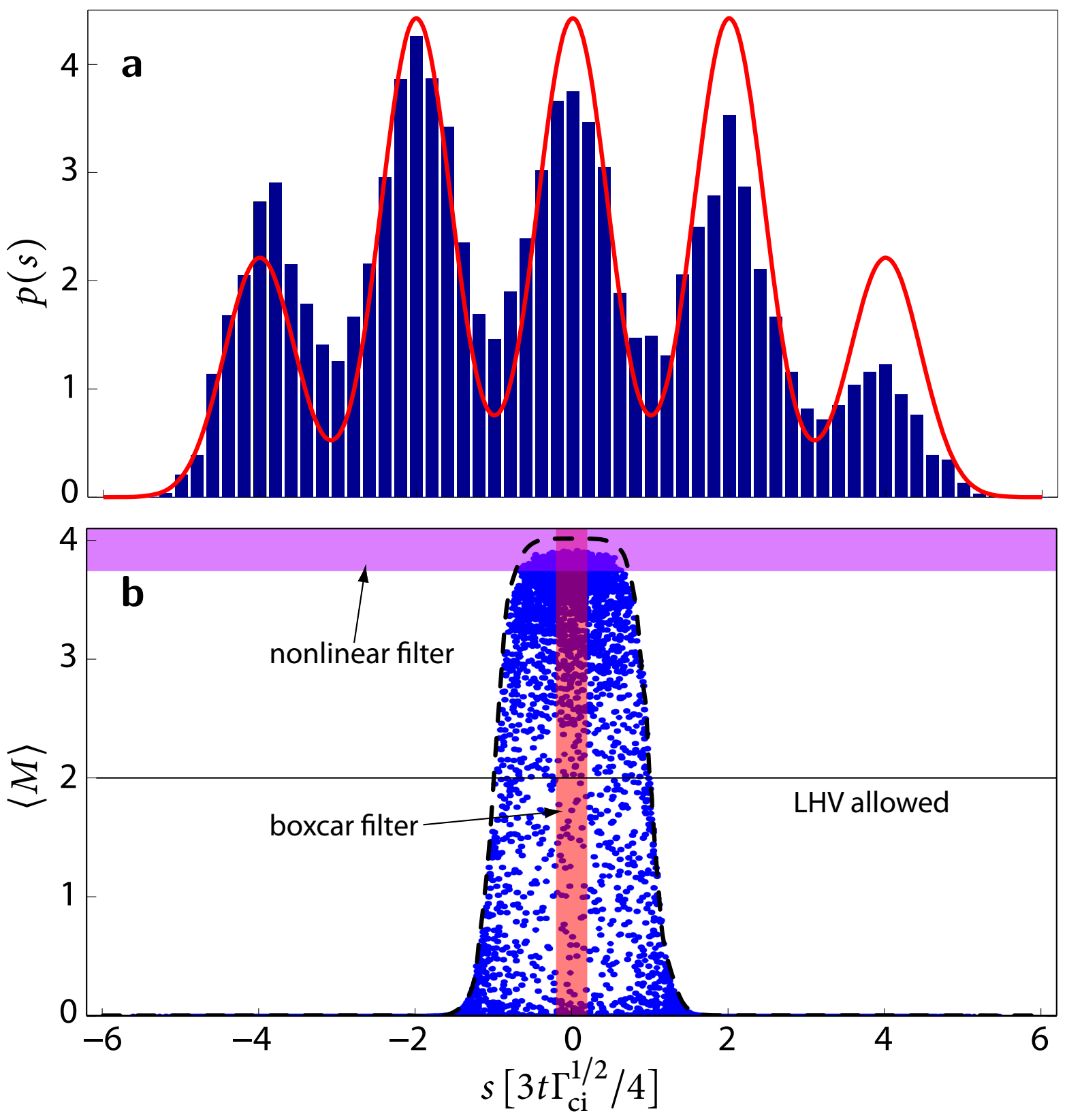

**Figure 5.3: GHZ state preparation in presence of decay. a**, Histogram of the integrated signal after a measurement time $t = 5/\Gamma_{ci}$, and probability distribution $p(s)$ in the absence of any decay (red curve). **b** Scatterplot (blue dots) showing the correlation between the expectation value of the Mermin operator $\langle M \rangle$ and the integrated signal $s$ for $t = 5/\Gamma_{ci}$. Each point corresponds to one of 10 000 trajectories. For comparison, the correlation in the ideal case of no decay is shown as the dashed black curve. The boxes indicate the action of the boxcar and the nonlinear filtering scheme, where the nonlinear filter selects all points lying in the purple box and the boxcar filter the ones in the red box. The Mermin bound is $\langle M \rangle = 2$, and local hidden variable (LHV) theories only permit values $\langle M \rangle \leq 2$. Parameters are chosen as $\Gamma_d = \Gamma_{ci}/2$, $\gamma/\Gamma_{ci} = 1/35$ and $\delta_1 = \delta_2 = 3/4$ and $\delta_3 = 3/2$.

The results for the integrated signal of many such measurements are conveniently plotted in form of a histogram, see figure 5.3a. When compared to the probability distribution expected in the ideal case of no decay, one observes that qubit decay leads to a distortion of the probability density with an overall shift of probability density towards the left-most peak,

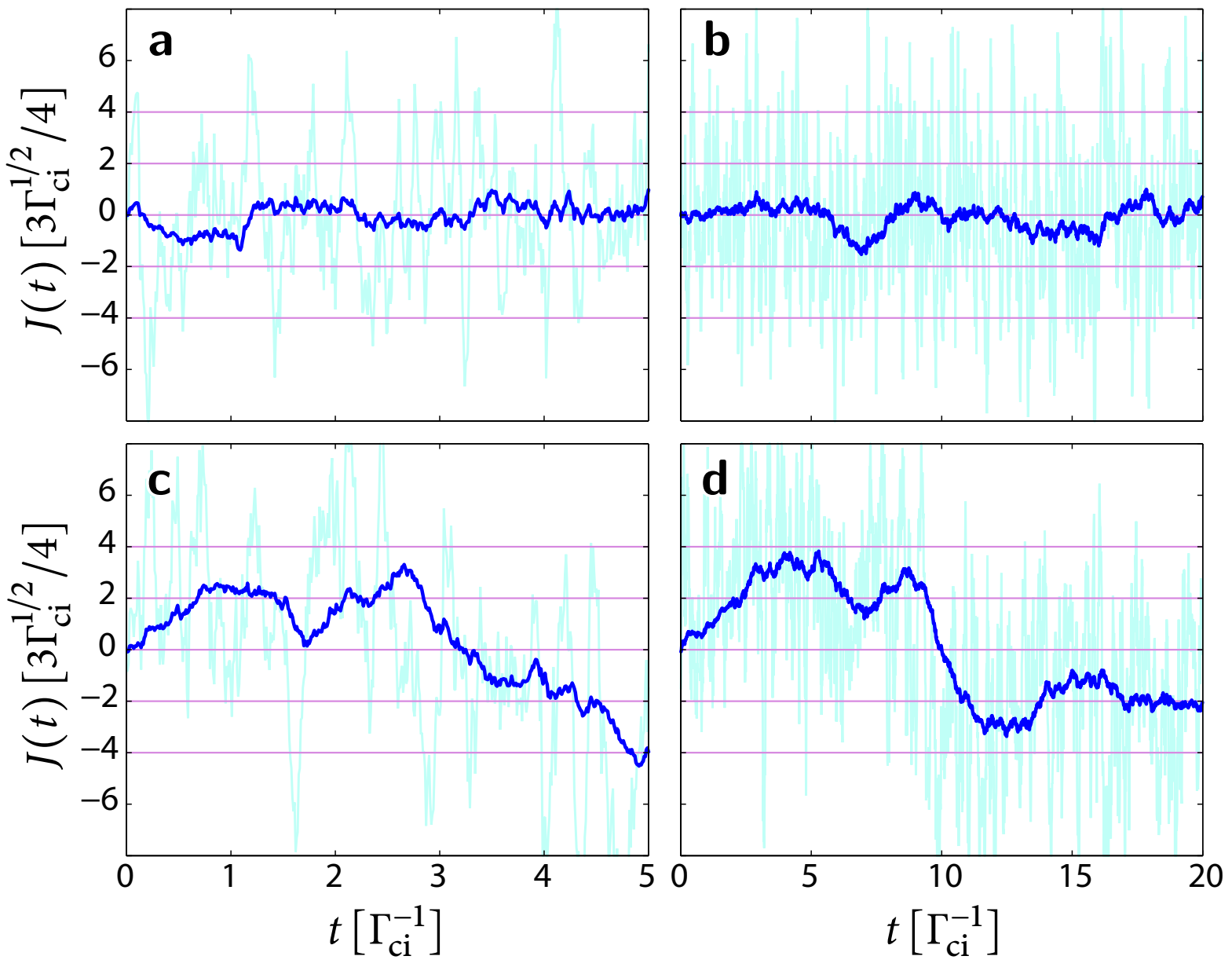

**Figure 5.4: Time traces of the signal $J(t)$ for individual quantum trajectories.** The traces are smoothed over time $0.1\Gamma_{ci}^{-1}$ (cyan) and $\Gamma_{ci}^{-1}$ (blue). For **a** and **b** the expectation of the Mermin operator is large, $\langle M \rangle > 3.9$, whereas for **c** and **d** it is small, $\langle M \rangle < 0.1$. The horizontal lines indicate the values $J(t)$ would take on average for the integrated signal $s$ to be at the peaks of figure 5.3a. All 4 traces are selected by boxcar filter on the integrated signal, such that they all lie close to the middle of the center peak, $s \simeq 0$. For **b**, **d** the relaxation is low, $\Gamma_{ci}/\gamma = 142$, and trajectories with extremal values of $\langle M \rangle$ can be distinguished by eye. For **a**, **c** the measurement time is shorter and relaxation is faster $\Gamma_{ci}/\gamma = 35$, nevertheless the nonlinear filter is still able to reliably estimate $\langle M \rangle$, as is demonstrated in figure 5.5.

*i.e.*, towards the signal associated with the ground state. The shift is thus easily understood as a consequence of decay processes acting during the finite measurement time.

As a benchmark for the quality of the generated states and its correlation with the integrated signal, figure 5.3b shows a scatterplot of the expectation value of the Bell–Mermin operator $\langle M \rangle$ versus the integrated signal for 10 000 individual measurement trajectories. For comparison, the corresponding scatterplot in the ideal case of no decay is shown to collapse to a single curve. The scatter in the nonideal case results in trajectories of the same integrated signal, but very different values of $\langle M \rangle$, and thus in a significant number of falsely accepted states within the simple boxcar filtering.

The essential mechanism for false acceptance of states is illustrated in figure 5.4, showing the measurement record $J(t)$ as a function of time for four individual trajectories. Speaking loosely, the trajectories with integrated signal $s$ close to 0 can be divided into two categories:

trajectories with measurement records $J(t)$ fluctuating around $J(t) = 0$, see figure 5.4a,b and measurement records showing larger variations of $J(t)$ which accidentally average to $s = 0$ upon integration. Trajectories of the first category correspond to the correct pre-GHZ state with high probability. On the other hand, an example from the second category consists of trajectories which, with high probability, initially assume the state $|\downarrow\uparrow\uparrow\rangle$ with $\langle X\rangle = 2$, and then suffer a decay process in qubit 3 at some intermediate time, thus transitioning to the state $|\downarrow\uparrow\downarrow\rangle$ with $\langle X\rangle = -2$, see figure 5.4c,d.

This insight also points to a remedy for the boxcar filter. The full measurement record can, when spaced densely enough, be used to reconstruct the actual underlying quantum trajectory $\rho_J(t)$ in the following way: Given that the state before the onset of the measurement [see step (ii) in section 5.3] as well as the parameters entering the stochastic master equation are known with sufficient accuracy, one can successively determine the Wiener increments $\mathrm{d}W(t) = \zeta(t)\mathrm{d}t$ from the measurement record. These, in turn, can then be used to propagate $\rho_J$ from the initial time to the measurement time $t$, and the resulting $\rho_J(t)$ encodes the expected value of the Bell–Mermin operator via $\langle M\rangle = \mathrm{tr}[\rho_J(t)M]$. This procedure corresponds to a nonlinear filter [119], with an acceptance criterion based on the value of $\langle M\rangle$ itself, see figure 5.3b.

The advantage of using the nonlinear filter is highlighted by figure 5.5, which compares the performances of boxcar and nonlinear filter. For acceptance probabilities smaller than the ideally attainable $P = 1/4$, we find that the nonlinear filter constitutes a significant improvement over the boxcar filter. Specifically, for ratios $\Gamma_{\mathrm{ci}}/\gamma \lesssim 4$ currently supported by experiments, the nonlinear filter will be crucial in order to reliably exceed the value $\langle M\rangle = 2$, which is the relevant Mermin bound for violation of local-hidden variable theories in this case. Figure 5.5 demonstrates that, when exploiting the trade-off between large expectation values of $\langle M\rangle$ and high acceptance probabilities, high-fidelity GHZ states can be prepared under realistic conditions.

## 5.6 GHZ state detection under realistic conditions

The measurement of the Bell–Mermin operator via parity detection, presented in section 5.3, requires the resolution of $\sim N$ peaks in the probability density $p(s)$ of the integrated signal. While clearly advantageous relative to the resolution of $\sim 2^N$ peaks needed for a full readout, the parity detection remains difficult with current experimental parameters due to the qubit relaxation within the measurement time. In the following, we discuss a scheme that avoids this problem.

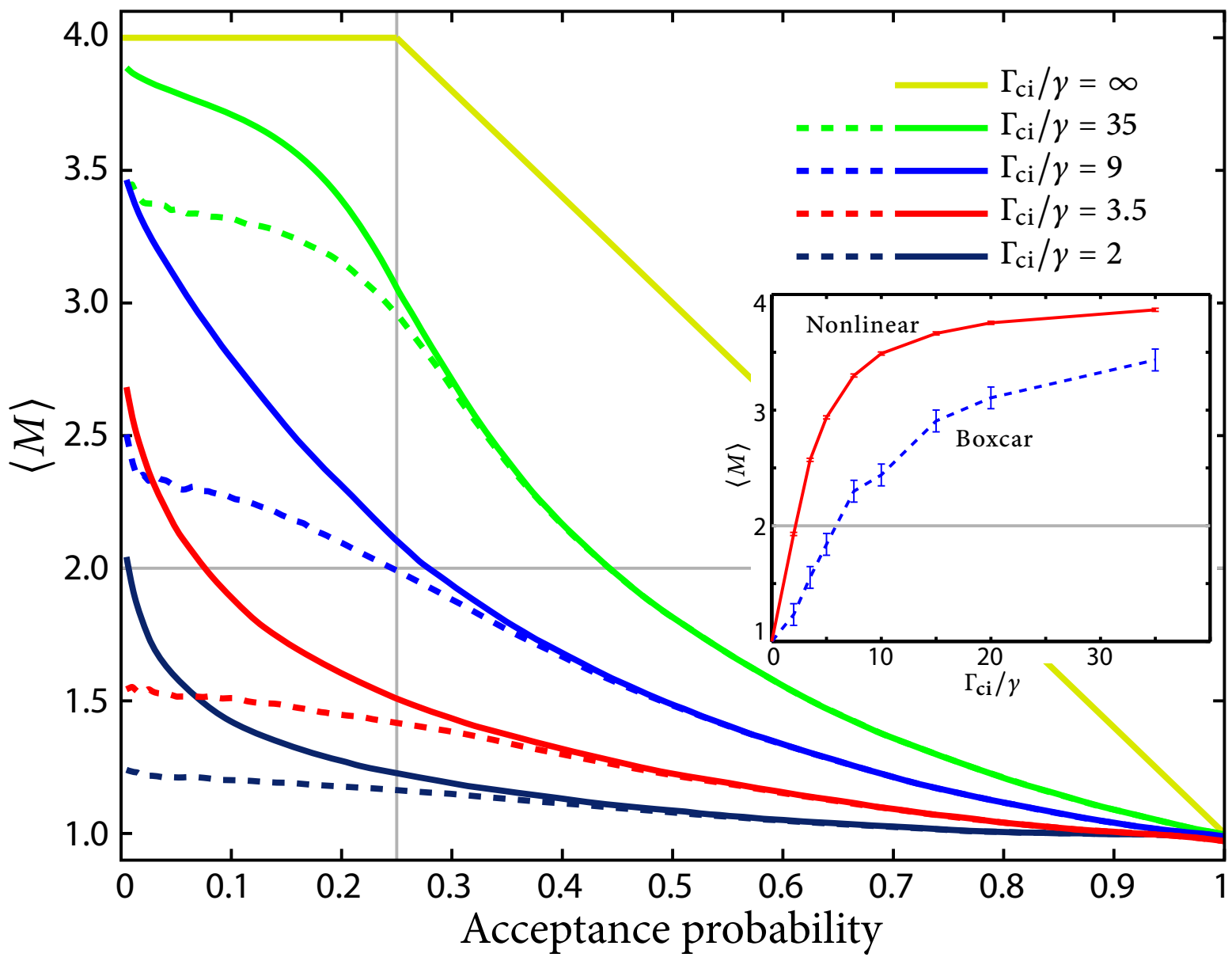

**Figure 5.5: Expectation value of the Mermin operator $\langle M \rangle$ as a function of acceptance probability, for several ratios $\Gamma_{ci}/\gamma$.** Parameters are chosen as in figure 5.3. Solid (dashed) lines show the results using the nonlinear (boxcar) filter. (See text for details.) Using nonlinear filtering, the fraction of accepted trajectories with high $\langle M \rangle$-value can be substantially increased. For an acceptance probability $\lesssim 1/4$ the advantage of the nonlinear scheme becomes apparent. For each point, $\langle M \rangle$ is obtained by averaging over 20 000 trajectories and optimizing with respect to measurement time $t$ and boxcar thresholds. The inset shows the expectation value $\langle M \rangle$ as a function of the ratio $\Gamma_{ci}/\gamma$ for an acceptance probability of 1%.

The key of this scheme lies in the fact that at low temperatures, decay into the state $|\Uparrow\rangle = |\uparrow\uparrow \cdots \uparrow\rangle$ is negligible. This is similar to Kofman and Korotkov's use of the 'negative result outcomes' to avoid the effects of measurement crosstalk in Bell tests using superconducting phase qubits [123]. False positive events in the detection of the state $|\Uparrow\rangle$ can thus be suppressed by setting the acceptance threshold $\nu$ for the integrated homodyne signal sufficiently high. Using this insight, we construct a measurement $B$ by assigning the measurement outcomes '0', '1' to the cases where the signal is respectively smaller or larger than a preset threshold. We can describe $B$ in the language of generalized observables as a positive operator valued

mapping (POVM) [37], by specifying its effects

$$E_1 = \alpha \left|\Uparrow\right\rangle \left\langle\Uparrow\right|, \tag{5.15a}$$

$$E_0 = \mathbb{1} - E_1. \tag{5.15b}$$

Here, $\alpha = P_{|\Uparrow\rangle}(s > v)$ is the probability that the signal exceeds the threshold $v$ given the system was prepared in $|\Uparrow\rangle$. This probability is set by the decay of the $|\Uparrow\rangle$ state during the measurement time, and is analogous to the detector efficiency in quantum optics. As a result, $1 - \alpha$ can be described as a 'false negative' probability that the measurement fails to detect a valid $|\Uparrow\rangle$ state. Experimentally, $\alpha$ can be determined by repeatedly preparing the system in $|\Uparrow\rangle$ (using single-qubit $\pi$ rotations), and subsequently performing the measurement. This procedure yields the expectation value $\langle\Uparrow|B|\Uparrow\rangle$, which is identical to the fraction of the cases where $s > v$, and hence to $\alpha$. In general, complete characterization of a POVM via detector tomography [126–129] requires many measurements and a numerical optimization procedure to ensure the resulting POVM remains physical. Due to the simple structure of the measurement $B$, it may be conveniently characterized by determining only a single parameter $\alpha$.

The measurement $B$ can now be combined with single-qubit rotations to determine the parity. We perform all combinations of $n$-qubit bit flips, $0 \le n \le N$, and sum the measured $\langle B\rangle$ with relative sign $(-1)^n$. For clarity we specialize to the 3-qubit case, and define

$$\begin{aligned} f_{zzz} =& \langle B\rangle - \langle\sigma_1^x B \sigma_1^x\rangle - \langle\sigma_2^x B \sigma_2^x\rangle - \langle\sigma_3^x B \sigma_3^x\rangle \\ &+ \langle\sigma_2^x\sigma_3^x B \sigma_2^x\sigma_3^x\rangle + \langle\sigma_1^x\sigma_3^x B \sigma_1^x\sigma_3^x\rangle + \langle\sigma_1^x\sigma_2^x B \sigma_1^x\sigma_2^x\rangle - \langle\sigma_1^x\sigma_2^x\sigma_3^x B \sigma_1^x\sigma_2^x\sigma_3^x\rangle. \end{aligned} \tag{5.16}$$

The value of $f_{zzz}$ is proportional to the parity measured in the $z$-basis, $f_{zzz} = \alpha\langle\sigma_1^z\sigma_2^z\sigma_3^z\rangle$, with the proportionality constant being $\alpha$ as defined above. It is straightforward to extend this scheme to the actual parities required for determining the value of the Bell–Mermin operator by prepending additional single-qubit rotations.

The expectation of the Bell–Mermin operator can now be related to the actual measurements via $F = \alpha\langle M\rangle$, where

$$F = f_{xxx} - f_{xyy} - f_{yxy} - f_{yyx}. \tag{5.17}$$

Thus, the measurement of the 32 expectation values entering into $F$ and determination of $\alpha$ allow for the extraction of $\langle M \rangle = F/\alpha$, with no restrictions on the qubits' decay rates.*

As explained in section 5.1.4 the nature of the dispersive measurement prevents us in principle from a strict violation of a Bell-type inequality. However, in the limit where the measurement effects factorize into tensor products over the single-qubit Hilbert spaces, *i.e.*,

$$E_{ijk} = E_i^{(1)} \otimes E_j^{(2)} \otimes E_k^{(3)}, \tag{5.18}$$

the measurement can be considered local in the sense of the no-signalling property [94, 130]. The effect defined in (5.15) obeys such a factorization

$$E_1 = \alpha \big( |\uparrow\rangle_1 \langle \uparrow|_1 \big) \otimes \big( |\uparrow\rangle_2 \langle \uparrow|_2 \big) \otimes \big( |\uparrow\rangle_3 \langle \uparrow|_3 \big). \tag{5.19}$$

Similarly, the rotated measurements entering into $F$ factorize in this sense, provided the rotations themselves also factorize. This additional requirement holds not only for perfect single-qubit rotations [123], but also for imperfect rotations, as long as there is no coupling or crosstalk between qubits during the rotation pulse. For example, independent single-qubit relaxation processes during a finite-duration rotation pulse do not spoil the factorization property. By contrast, a rotation of qubit $b$ caused by a rotation pulse on qubit $a$ no longer factorizes. In the following, we will assume that such crosstalk is negligible. In that case, the argument of Mermin applies, which states that a local hidden variable theory has bounds on the allowed $F$, $-2 \le F \le 2$ [113]. Meanwhile quantum mechanics allows for $\langle M \rangle = 4$ and hence if $\alpha > 1/2$ there is the possibility to violate Mermin's version of the Bell inequality.

Figure 5.6 shows the variation of the false negative probability $(1 - \alpha)$ with threshold $\nu$, so that the required threshold for $\alpha > 1/2$ can be read off. Since the derivation of the Bell inequality required factorization of measurement effects, we estimate the corrections to (5.15). In our case, the largest correction will be due to misidentification of states from the subspace $\{|\downarrow\uparrow\uparrow\rangle, |\uparrow\uparrow\downarrow\rangle, |\uparrow\downarrow\uparrow\rangle\}$, for which $X = \sum_j \sigma_j^z = 1$. We put an upper bound on this misidentification probability $\beta$ by assuming that there is no decay out of this subspace and thus assume that $P_{X=1}(s)$, the distribution of the homodyne signal arising from this subspace, is Gaussian. Under these conditions, one obtains a worst-case estimate of the 'false positive' probability $\beta$ as a function of $\nu$.

Figure 5.6 shows that with a low rate of qubit decay, $\Gamma_{\text{ci}}/\gamma = 20$, we find $\alpha > 1/2$ and a low

* If decay is fast, however, $\alpha$ may become so small that the time required for gathering sufficient statistics may become impractically long.

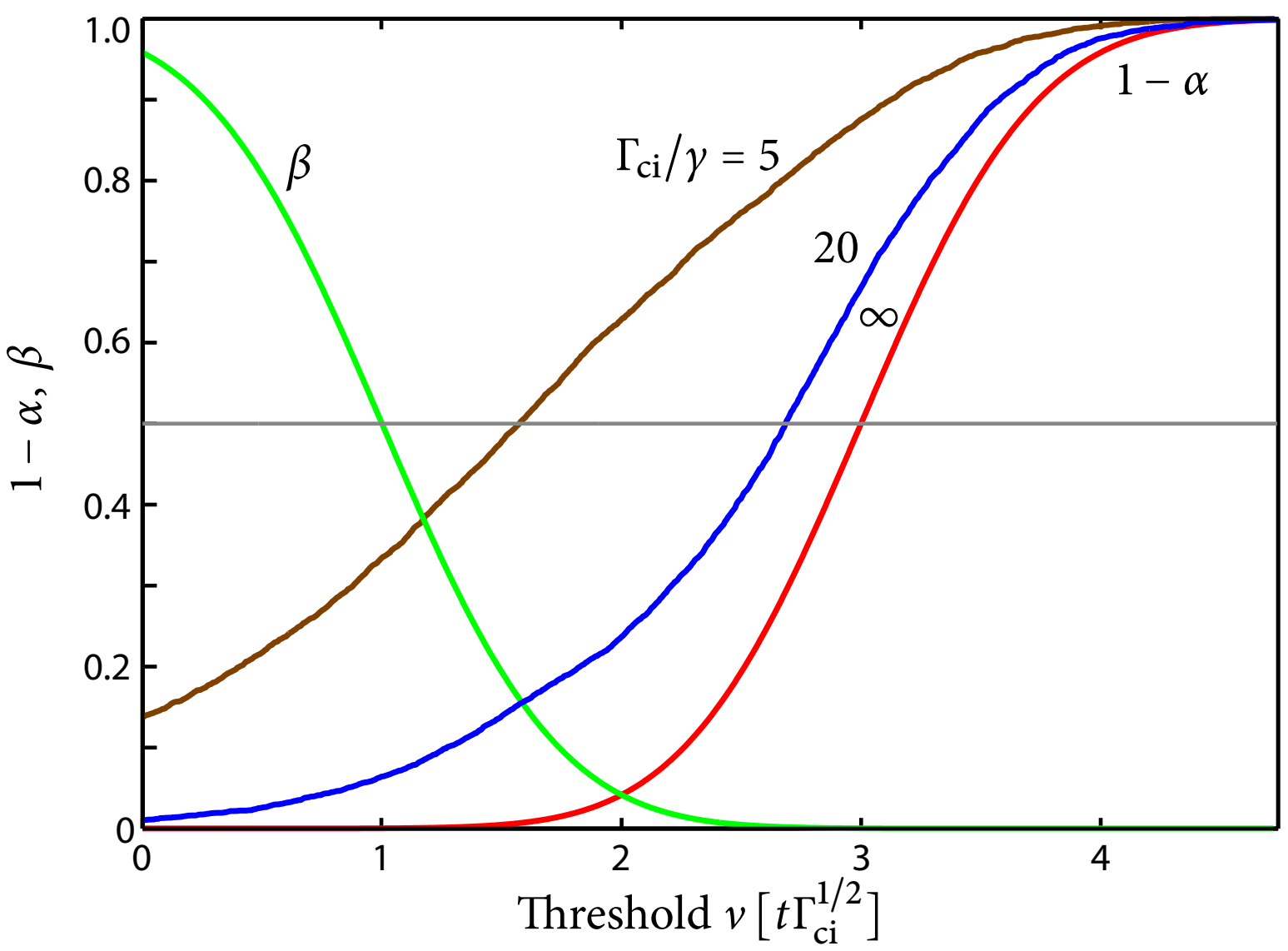

**Figure 5.6: False negative probability $1-\alpha$ and worst-case value for the false positive probability $\beta$ (definition see text) versus threshold $\nu$.** The horizontal line indicates the necessary constraint on $\alpha$ to violate the Mermin inequality. The measurement time is chosen as $t = 3/\Gamma_{ci}$.

probability of false positives, $\beta \simeq 0.002$, meaning that a meaningful violation of a Bell-type inequality should be possible. Conversely, for a more realistic rate of qubit decay $\Gamma_{ci}/\gamma = 5$, the requirement $\alpha > 1/2$ leads to significant false positive rates $\beta \simeq 0.16$, and factorization of $E_1$ breaks down. We note that the required $\Gamma_{ci}/\gamma \simeq 20$ for the violation of the Bell inequality is much more stringent than the experimentally realistic $\Gamma_{ci}/\gamma \simeq 4$ that was shown in the previous section to be sufficient for producing states with $\langle M \rangle > 2$.

## 5.7 Conclusions

In conclusion, we have presented a concrete proposal for efficient statistical production of multi-qubit GHZ states by dispersive measurement in a cQED setup, taking into account the realistic conditions of decoherence and decay. Our proposal is based on the possibility of adjusting the dispersive shifts of individual qubits, which effectively modifies the measurement operator and allows for the generation of entanglement starting from separable input states. Our simulations show that even with experimentally achievable values of $2 < \Gamma_{ci}/\gamma < 4$ it is possible to achieve a 1% efficiency in preparing states with values of the Bell–Mermin operator exceeding its classical bound, $\langle M \rangle > 2$.

By using the global dispersive measurement in the same setup, we have also proposed a scheme for implementing parity measurements on the prepared state. Using these measurements, we have studied the sufficient conditions for verifying that such states indeed violate the Bell–Mermin inequality. We find that a signal-to-noise ratio of $\Gamma_{ci}/\gamma = 20$ will be sufficient to observe a violation of the Mermin bound. While this ratio is larger than currently demonstrated, we hope that the present limits on detector efficiencies in semiconductor amplifiers ($1/20$ of the quantum limit) will soon be improved by using superconducting pre-amplifiers [95].

## CHAPTER 6

# Conclusions and outlook

SOME reflections on the possibilities for extending this work, followed by some general predictions for the future of the field, including an idea for a novel qubit design. Finally, a somewhat tenuous metaphor.

## 6.1 Vacuum Rabi splitting

This thesis has explained how to formulate a quantum description of electrical circuits, and given an introduction to a particular circuit that behaves as an artificial atom. It has explained the general framework for describing such a quantum circuit in contact with its environment, and in chapter 4 these ideas were shown to provide a superb description of an experimental scenario which is borrowed from atomic quantum optics, but in a regime which is extraordinarily difficult to access using real atoms. It is interesting to consider what possibilities we have for making use of this very accurate description of our physical system. Because this work shows that the excited states of the Jaynes–Cummings Hamiltonian appear to be just as well-behaved as the states of the transmon itself, there is the intriguing possibility of using these excited states as computational basis states for quantum computing. Quantum computing is usually formulated in terms of qubits (2 levels), but qutrits (3 levels) and more generally qudits ($n$ levels) have some attractive features [131, 132]. Neeley *et al.* [133] have demonstrated 5-level qudits using the normally-ignored higher levels of the phase qubit.

The higher levels of the transmon could be used in the same way, but there are possibly advantages in using the Jaynes–Cummings states. For example the higher transmon levels will have much increased charge dispersion, whereas the Jaynes–Cummings states do not suffer from this problem. This idea is closely related to the *cavity-stabilized* qubits of Koch and coworkers [134, 135].

The results presented in section 4.6.3 regarding the temperature dependence of the strongly-driven vacuum Rabi splitting suggest that there is the possibility to use the system as an extremely sensitive thermometer. Preliminary experimental work at ETH has already shown that this is a promising idea, although the problems with simulating a large Hilbert space mean that more sophisticated theoretical techniques are needed than just solving the master equation. For example the quantum trajectory approach outlined in section 5.2 could be useful here, and an extension of the formalism of Rau *et al.* [74] to include the influence of higher levels of the transmon has already shown some promise.

The 'switching' behavior hinted by figures 4.15 and 4.16, where beyond a certain number of excitations, the transmon effectively disappears and the cavity becomes very highly excited, suggests the possibility to make high-fidelity single-shot readouts. Similar *latching* readout schemes, involving the bifurcation of a driven nonlinear oscillator, have already proved very effective [136–138], with single-shot fidelities as high as 70%. By using the qubit for the readout, we may have the advantage of a circuit with a smaller number of 'moving parts' and we can leverage all the effort that has gone into reducing the dissipation of the qubit.

## 6.2 Future trends

Chapter 5 considered a circuit with 3 qubits, and it is clear that in the near future we shall soon see experiments involving 4 or more qubits. Finding efficient ways to deal with such circuits will be a challenging problem. As the Hilbert space grows exponentially with each added qubit, even describing the state of the system becomes a non-trivial task, and fully characterizing the dynamics becomes truly daunting. Fortunately, it is rarely the case that we are interested in knowing every detail, rather we wish to know more 'macroscopic' quantities such as the entanglement of a state, or the fidelity of a gate. These quantities can be found in principle without reconstructing the full density matrix or the full set of Kraus operators. Techniques to support this *less-is-more* approach [139] will become increasingly important as the number of qubits increases. I see the readout scheme presented in section 5.6, for reducing the detector characterization to measuring a single parameter, as a (tiny) step in this direction.

With devices containing only two qubits, the primary difficulty has been in arranging a sufficiently strong interaction between the qubits to be able to perform gate operations. With multiple qubits the task will be to maintain the strong interaction when it is needed to perform a gate between a particular pair of qubits, but to be able to switch all the other interactions off. The 100:1 on-off ratio for 2-qubit interaction strength already demonstrated in [88] is a promising start, but more is needed. It remains to be seen whether this ratio can be maintained or, ideally, exceeded for circuits involving more than 2 qubits.

Another direction of increased circuit complexity results from adding, not qubits, but additional resonators to the circuit. An example of such a proposal suggests that the qubit can act as a quantum switch [140], but this is surely not the only interesting use for such a circuit, and experiments involving a two-cavity device are currently underway. There are even proposals to build a *lattice* of Jaynes–Cummings Hamiltonians, which resembles the Bose–Hubbard model [141] and can display a superfluid–Mott insulator transition [142].

## 6.3 New qubit designs

In order to keep up with Schoelkopf's law,* continuous development of qubit technology is necessary. The general development of the charge qubit shows a very consistent trend where each advance has involved two simultaneous aspects: (1) removing a control channel from the qubit in order to avoid relaxation via that channel; and (2) finding a (necessarily more indirect) scheme for exerting influence over the qubit and measuring it, that manages to work without the control channel. The first charge qubits [18] used a charge-based readout, for example via an RF single-electron transistor [144, 145]. Unfortunately, in the regime where the charge-based readout works, the qubit frequency is first-order sensitive to stray electric fields, leading to rapid dephasing. The first great improvement in coherence times came with the *quantronium* qubit [22], which operates at the optimal point $n_g = 1/2$, at which point the transition frequency becomes only second-order sensitive to electric fields. However, since the environment can now no longer measure the qubit this way, neither can we. Instead, a second-order readout scheme is used, based on the state-dependent susceptibility. This

* The prediction made in 2004 that solid-state qubit coherence times will continue to increase 5 dB/year. See also [143].

leads to readouts using the quantum inductance and quantum capacitance. The second great improvement came with the invention of the transmon qubit [20, 43], described in detail in chapter 2, for which the charge dispersion is exponentially suppressed, and the environment can no longer interrogate the qubit via its susceptibility but again, neither can we. However, the anharmonicity remains finite, allowing the dispersive readout described in section 2.10. A new design of qubit, termed *fluxonium* [146] (not strictly a charge qubit) has a similar behavior to the transmon, being insensitive to charge noise, but has the advantage that the anharmonicity can remain large.

With the transmon, the primary source of decoherence is the Purcell effect. Therefore the next obvious step is to eliminate the matrix element for a transition between the ground and the first excited state caused by the coupling to the resonator. In other words we need to construct a circuit whose dipole moment is zero, rather than the strong coupling of section 4.1. As we saw in figure 4.2, the dipole moment can be reduced by increasing the geometric symmetry of the circuit. Taking this to a logical conclusion we could make a circuit where the capacitor plates are concentric circles. This would indeed have close to zero dipole moment, but it would also have no interaction with a (uniform) electric field with which to measure or control it, so a more subtle approach is needed.

Figure 6.1 shows one possible circuit which has the desired protected state, but remains measurable. (In fact, it is the same circuit as in [10, figure 3], except that the inductor $L_3$ is absent and the inductors $L_1$, $L_2$ are now Josephson junctions.) The intention is that the circuit should be fabricated as symmetrically as possible, so that $C_1 \simeq C_2$, $E_{J1} \simeq E_{J2}$. To get a feel for this circuit, note that if the coupling capacitor $C_t$ is absent then we just have two uncoupled Cooper pair boxes, and in the limit of large coupling $C_t \to \infty$ the two qubits are effectively shorted by the capacitor and we have a single qubit with effective capacitor $C_{\text{eff}} = C_1 + C_2$ and inductor $L_{\text{eff}}^{-1} = L_1^{-1} + L_2^{-1}$, and in the perfectly symmetric case this looks like double the capacitance and half the inductance of the single qubit, such that the frequency $\omega_{\text{eff}} = \sqrt{8E_{C\text{eff}}\,E_{J\text{eff}}}$ is unchanged. The advantage of this circuit is that it has both symmetric and antisymmetric excitations, and selection rules that forbid certain transitions. This makes it very reminiscent of real atoms, where certain transitions are *dipole-forbidden* and hence very long-lived. By choosing appropriate values for the capacitors and Josephson energies, we can arrange for the lowest symmetric excitation to have the same highly-suppressed charge dispersion as the transmon, and be tunable over the same frequency as a transmon, but unlike the transmon the dipole matrix element between this state and the ground state will be zero. So far this is nothing special: so-called *dark states* are generally present when two qubits are symmetrically coupled to a resonator, as was seen for example in [120] with the qubits at opposite ends of the cavity. These can be understood as resulting from the fact that an excitation of the dark state has *two* possible pathways via which to decay, and these interfere destructively. However, this type of interference will in general only hold for decay via one specific channel. On the other hand, the geometrically symmetric design has the great advantage that it can be dark to not only the single-mode Purcell effect, but to the full multi-mode Purcell effect, as well as to *any* decay channel that effectively acts via a linear electric field, for example spurious modes of the box containing the sample. This means that the $T_1$ lifetime of such a qubit should be extremely long, presumably limited by dielectric losses in the substrate or the junction oxide.

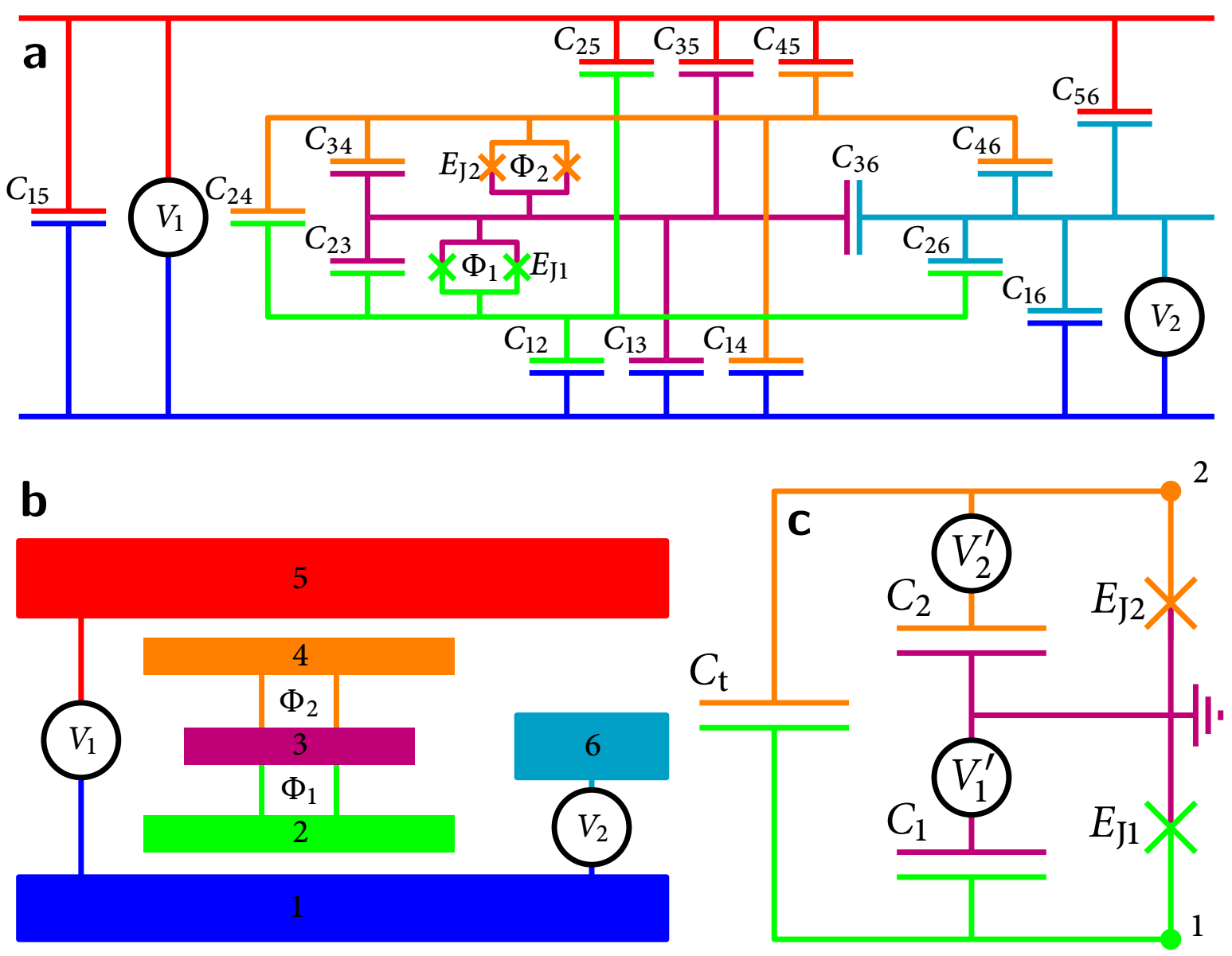

**Figure 6.1: An artificial atom with a quadrupole transition.** The capacitance network in **a** for the geometric layout of the atom indicated in **b**, which can be interpreted as comprising a pair of back-to-back transmon-style qubits. The voltage $V_1$ couples to the dipole moment of the circuit and the voltage $V_2$ couples to the quadrupole moment. **c** shows a simplified equivalent circuit.

Despite the fact that the lowest symmetric state cannot decay via the cavity, we will still be able to measure the state of the qubit, via the same dispersive readout as discussed in section 2.10. This is because although there is no allowed transition to any *lower-energy* state, thus preventing decay, there are still allowed transitions to *higher-energy* states, causing a state-dependent dispersive shift. Control over the state of the qubit will be more challenging, however, because the same suppression of the interaction with the environment also affects our attempts at control. One possible solution is to allow the symmetry to be broken dynamically, with a fast flux bias line for example. Then, the qubit can be kept in a long-lifetime *memory* configuration until we need it for a computation, at which time we break the symmetry and quickly perform any needed rotations. An alternative approach is to overcome the suppression of the matrix element via brute force: we saw in chapter 4 that there are no particular adverse results from driving these superconducting circuits extremely strongly. Yet another option, more closely related to atomic quantum optics, is to perform rotations between our computational states *indirectly*, going via an intermediate state having dipole-allowed transition to both computational states. Work towards choosing an optimal scheme is currently under way.

## 6.4 A metaphor

Superconducting charge qubits are only 10 years old, but they are maturing quickly. As they have grown up, they have responded well to being allowed gradually more and more independence. They have learned to interact well with their peers on a one-on-one basis and they are beginning to form larger circles of friends. In other words, they seem to be like any normal 10-year-olds. We are apprehensive that they are reaching an age where in the near future we should not expect to know all the details of their lives, and we worry that upcoming physical changes might make them hard to control, but we hope that not too many years from now, we shall proudly be reading their doctoral thesis, perhaps on the topic of factoring the largest numbers.

❧

# APPENDIX A

## Mathematica code for strongly-driven vacuum Rabi

Numerical code for solving the transmon–cavity master equation (4.20) follows. After some initialization tasks, the first part of the code concerns solving the transmon Hamiltonian (2.25) in a truncated charge basis (2.30) by exact diagonalization. The entry point for this diagonalization is the function `egtrans[]`. Because the diagonalization is a somewhat expensive procedure, and since the resulting energies and matrix elements are smooth functions of $E_J$ and $E_C$ (and almost independent of $n_g$), the next part of the code, entered via `makeinterp[]`, constructs an opaque interpolation object that can evaluate these energies and matrix elements for a range of $E_J/E_C$ ratios. Next there is some code which checks the previous calculations for convergence.

The purpose of the next block of code is to construct driven Jaynes–Cummings Hamiltonian (4.20c), storing it in `H0s`. Several utility functions for finding the eigenvalues of this Hamiltonian are also created. These functions are used for determining $E_C$ from two-tone pump-probe experiments, and for correlating features in the full nonlinear spectrum with the associated multi-photon transitions.

The generator $L$ of the semigroup is constructed next, firstly as the functional operator `lindblad[]` and then in the matrix form $M'$ of (4.18). The final step is to solve (4.18) to obtain $\rho_s$, and reuse the factorization for solving (4.21). These last steps are performed by the function `steadystatevalue[]` which returns a function that is optimized to factorize $M'$.

For bookkeeping convenience, various parts of the code make use of the qmatrix package by T. Felbinger [147] for setting up the problem, although native Mathematica matrices are significantly faster to manipulate. The former are stripped and packed to produce the latter, for all the numerically intensive algorithms, losing convenience but gaining efficiency.

# Initialization

```
$HistoryLength = 0;
```

## Load packages

```
<< qmatrix.m
```

```
Needs["Notation`"]
```

## Define symbols

```
Symbolize[γ1];
Symbolize[γϕ];
Symbolize[HJ-C];
Symbolize[H2];
Symbolize[σx];
Symbolize[σy];
Symbolize[σz];
Symbolize[σ+];
Symbolize[σ-];
Symbolize[â];
Symbolize[â†];
```

```
Symbolize[ωa];
InfixNotation[·, NonCommutativeMultiply];
Symbolize[T1];
Symbolize[T2];
Symbolize[t1];
Symbolize[t2];
Symbolize[H0];
Symbolize[ωr];
Symbolize[ωd];
Symbolize[Hd];
Symbolize[Δd];
```

```
Symbolize[n̂];
Symbolize[q̂];
Symbolize[αr];
Symbolize[EJ];
Symbolize[EC];
Symbolize[ng];
Symbolize[HQ];
Symbolize[Hg];
Symbolize[ℒ_];
Symbolize[ρ̂];
Symbolize[ĝ];
```

## System modes

```
qubitletter = Characters["GEFH"] ~ Join ~ CharacterRange["J", "Z"];
```

```
levels::usage =
  "levels represents the number of levels kept in the truncation of the
    qubit and cavity Hilbert spaces. Change it only using setlevels[]";
```

```
setlevels::toofew = "Too few levels `1`; at least 2 needed";
setlevels::usage = "setlevels[n] sets things
    up to keep n transmon levels and n cavity levels";
setlevels[n_Integer ? (# > 1 || Message[setlevels::toofew, #] &)] := (
  Unprotect[levels];
  levels = n;
  Protect[levels];
  setSystem[qubit, cavity];
  setModeType[qubit, {bosonic, levels}];
  setModeType[cavity, {bosonic, levels}];
  "System set to dimension: " <> ToString@dimension[system])
```

## Notations

### Superoperators

```
𝒟 /: 𝒟[A_matrix ? properMatrixQ][ρ_matrix ? properMatrixQ] :=
 A · ρ · hc[A] - hc[A] · A · ρ / 2 - ρ · hc[A] · A / 2
```

### Operators

```
σ⁺ := matrix[op[ad, qubit]];
σ⁻ := matrix[op[a, qubit]];
â† := matrix[op[ad, cavity]];
â := matrix[op[a, cavity]];
n̂ := â† · â;
q̂ := σ⁺ · σ⁻;
```

```
AddInputAlias["sp" → σ⁺];
AddInputAlias["sm" → σ⁻];
AddInputAlias["ad" → â†]
AddInputAlias["nh" → n̂];
AddInputAlias["qh" → q̂];
```

### Options

```
SetOptions[Manipulator, Appearance → "Labeled"];
```

# Transmon Calculations

### Do the matrix solve

This function egtrans[] gives the eigenenergies $e_j$ and the coupling terms $g_{ij}$ and then also calculates the derivative of these wrt $E_J / E_C$.

Because it calculates the derivative by 1st-order perturbation theory, it has problems with degeneracies when $E_J / E_C$ is low enough compared to cutoff that there are levels with (almost) degeneracies at $n_g \in \{0, 1/2\}$. Consider using $n_g = 0.5 + \epsilon$ instead.

We have to manually correct the signs of the $g_{ij}$ because `Eigensystem[]` doesn't guarantee a consistent phase for the eigenvectors.

I haven't checked whether it's better to use a sparse solver or the dense one, but either way we need to get all of the eigenstates for the perturbation theory, so we should not use Krylov methods.

We also normalize things so that $e_0 \equiv 0$, $e_1 \equiv 1$, $g_{12} = g_{21} \equiv 1$.

```
egtrans::usage =
  "egtrans[ng, EjEc, cutoff] gives {e, g, ∂e/∂(Ej / Ec), ∂g/∂(Ej / Ec)}";
egtrans::toofew = "Cutoff `1` is too low; must be at least 2";
Block[{fx, gx, hx, part, x},
```

```
Hold[egtrans[ng_?NumericQ, EjEc_?NumericQ,
    cutoff_Integer?(# > 1 || Message[egtrans::toofew, #] &)] := Module[
    {h = SparseArray[{Band[{1, 1}] → 4 (Range[-cutoff, cutoff] - ng)^2}],
     hv = SparseArray[
       {Band[{1, 2}], Band[{2, 1}]} → -1., {2 cutoff + 1, 2 cutoff + 1}],
     n = SparseArray[{Band[{1, 1}] → Table[m - ng, {m, -cutoff,
          cutoff}]}], e, v, e2, v2, o, g, de, dv2, dg, sgn},
    {e, v} = Eigensystem[h + EjEc/2 hv];
    o = Ordering@e;
    e2 = e[[o]];
    v2 = v[[o]];
    g = v2.n.v2^T;
    sgn = Sign@g;
    g = sgn g;
    de = #.hv.# & /@ v2 / 2;
    dv2 = Table[Sum[If[i == j, 0, v2[[j]] (v2[[j]].hv.v2[[i]]) / (e2[[i]] - e2[[j]])],
       {j, 2 cutoff + 1}], {i, 2 cutoff + 1}];
    dg = sgn (dv2.n.v2^T + v2.n.dv2^T) / 2;
    {(e2 - e2[[1]]) / (e2[[2]] - e2[[1]]),
     (D[(fx[x] - gx[x]) / (hx[x] - gx[x]), x] /. {fx'[x] → de, fx[x] → e2,
          gx'[x] → part[de, 1], gx[x] → part[e2, 1],
          hx'[x] → part[de, 2], hx[x] → part[e2, 2]}) //
       FullSimplify // Experimental`OptimizeExpression,
     g / g[[1, 2]],
     (D[fx[x] / gx[x], x] /. {fx[x] → g, fx'[x] → dg,
          gx[x] → part[g, 1, 2], gx'[x] → part[dg, 1, 2]}) //
       FullSimplify // Experimental`OptimizeExpression}
   ];
  ] /. x_Experimental`OptimizeExpression :> RuleCondition[x] /.
   Experimental`OptimizedExpression[x_] :> x /.
  HoldPattern[part] → Part // ReleaseHold;
]
```

Now we need to interpolate the results of the numerical calculation of $e_i$ and $g_{ij}$.

The indices i,j are zero-based...

## Interpolation of the solutions

```
energyinterp::usage =
  "energyinterp[{f2, f3, ...}, ng, cutoff, {min, max, step}] represents
    a function that interpolates the transmon energies.";
couplinginterp::usage = "energyinterp[{f2, f3, ...}, ng,
     cutoff, {min, max, step}] represents a
    function that interpolates the transmon couplings.";

interpf::level =
  "Tried to calculate for transmon level:`1`, but interpolating
    function was only defined for levels 0..`2`";
interpf::dom = "Tried to calculate for EJ/EC of `1`, but
    interpolating function was only defined for `2`≤ EJ/EC ≤`3`";
interpf::invalidform = "Invalid form for a transmon interpolation";
```

```
Unprotect[energyinterp, couplinginterp];
```

```
idx::usage = "idx[] has the attribute NHoldAll";
SetAttributes[idx, NHoldAll];
```

```
transmoninfo[ng_, c_, {min_, max_, step_}] :=
  Column[{"Ei[EJ/EC]", "i:0.." <> ToString[c], HoldForm[min ≤ "EJ/EC" ≤ max],
    "interp step: " <> ToString@step, HoldForm["ng" == ng]}];
```

## ■ energyinterp[]

```
energyinterp[a__][i : Except[_idx]] := energyinterp[a][idx@i];

energyinterp[___][idx@0] = 0. &;
energyinterp[___][idx@1] = 1. &;

energyinterp[_, _, c_, _][idx@i_] /;
   (If[NumericQ[i] && ! TrueQ[0 ≤ i ≤ c && i ∈ Integers],
     Message[interpf::level, i, c]; Abort[]];
    False) := None;

energyinterp[l_, _, c_, {min_, max_, _}][idx@i_][x_] /;
   (If[NumericQ[x] && ! TrueQ[min < x < max],
     Message[interpf::dom, x, min, max]; Abort[]];
    NumericQ[x] && NumericQ[i] && min ≤ x ≤ max && 2 ≤ i ≤ c) := l⟦i - 1⟧[x];

Derivative[d_Integer /; d ≥ 1][
     energyinterp[l_, _, c_, {min_, max_, _}][idx@i_]][x_] /;
   (If[NumericQ[x] && ! TrueQ[min < x < max],
     Message[interpf::dom, x, min, max]; Abort[]];
    NumericQ[x] && NumericQ[i] && min ≤ x ≤ max && 2 ≤ i ≤ c) :=
  Derivative[d][l⟦i - 1⟧][x];

Format[energyinterp[l : {__InterpolatingFunction},
      ng_?NumericQ, c_Integer?(2 ≤ # &), mm : {min_, max_, step_} /;
        0 < min < max && 0 < 10 step < max - min][idx@i_] /; Length[l] + 1 == c] :=
  Tooltip[HoldForm["E"i], transmoninfo[ng, c, mm]];

Format[energyinterp[l : {__InterpolatingFunction},
    ng_?NumericQ, c_Integer?(2 ≤ # &), mm : {min_, max_, step_} /;
     0 < min < max && 0 < 10 step < max - min]] := energyinterp["<>", ng, c, mm];
```

## ■ couplinginterp[]

```
couplinginterp[___][idx@1, idx@0] =
  couplinginterp[___][idx@0, idx@1] = 1. &;

couplinginterp[a_, b_, c_, d_][i : Except[_idx], j : Except[_idx]] :=
 couplinginterp[a, b, c, d][idx@i, idx@j]

couplinginterp[_, _, c_, _][idx@i_, idx@j_] /;
   (If[NumericQ[i] && ! TrueQ[0 ≤ i ≤ c && i ∈ Integers],
     Message[interpf::level, {i, j}, c]; Abort[]];
    If[NumericQ[j] && ! TrueQ[0 ≤ j ≤ c && j ∈ Integers],
     Message[interpf::level, {i, j}, c]; Abort[]];
    False) := None;

couplinginterp[l_, _, c_, {min_, max_, _}][idx@i_, idx@j_][x_] /;
   (If[NumericQ[x] && ! TrueQ[min < x < max],
     Message[interpf::dom, x, min, max]; Abort[]];
    NumericQ[x] && NumericQ[i] && NumericQ[j] && min ≤ x ≤ max &&
     0 ≤ i ≤ c && 0 ≤ j ≤ c) := l⟦i + 1, j + 1⟧[x];
Derivative[d_][couplinginterp[l_, _, c_, {min_, max_, _}][idx@i_, idx@j_]][
    x_] /;
   (If[NumericQ[x] && ! TrueQ[min < x < max],
     Message[interpf::dom, x, min, max]; Abort[]];
    NumericQ[x] && NumericQ[i] && NumericQ[j] && min ≤ x ≤ max &&
     0 ≤ i ≤ c && 0 ≤ j ≤ c) := Derivative[d][l⟦i + 1, j + 1⟧][x];

Format[couplinginterp[l_, ng_?NumericQ, c_Integer?(2 ≤ # &),
      mm : {min_, max_, step_} /; 0 < min < max && 0 < 10 step < max - min]
     [idx@i_, idx@j_] /; Dimensions[l] == {c, c} + 1] :=
  Tooltip[HoldForm[g_ij], transmoninfo[ng, c, mm]];

Format[couplinginterp[l_?MatrixQ, ng_?NumericQ, c_Integer?(2 ≤ # &),
    mm : {min_, max_, step_} /; 0 < min < max && 0 < 10 step < max - min]] :=
  couplinginterp["<>", ng, c, mm];
```

## ■ Finish up defining tags

```
(e : energyinterp[___][_][_])* ^:= e;
```

```
(c : couplinginterp[___][__][_])* ^:= c;
SetAttributes[{energyinterp, couplinginterp}, {NHoldAll}];
Protect[energyinterp, couplinginterp];
```

```
SetAttributes[evalinterp, HoldAll];
evalinterp[x_] := x /. {idx[i_] :> i,
   energyinterp[{l__}, __] :> ({0, 1, l}[[# + 1]] &),
   couplinginterp[l_, __] :> (l[[#1 + 1, #2 + 1]] &)}
```

### Construct interpolations

```
makeinterp::usage =
  "makeinterp[ng, cutoff, levels, {min, max, step}] gives e[i, Ej/Ec],
    g[i, j, Ej/Ec] for min ≤ Ej/Ec ≤ max, (i,j = 0,...,levels-1),
    using 2cutoff+1 charge-basis transmon levels in the calculation";
makeinterp::levcut = "Require 2≤levels≤cutoff , but levels=`1`, cutoff=`2`";
makeinterp::step = "Require 0 < 10*step < max-min";
makeinterp::minmax = "Require 0<min<max but min=`1`, max=`2`";

Options[makeinterp] = {InterpolationOrder → 7};
```

```
makeinterp[ng_?NumericQ, cutoff_Integer, levels_Integer, mms :
    {min_?NumericQ, max_?NumericQ, step_?NumericQ}, OptionsPattern[]] /;
  (2 ≤ levels ≤ cutoff || Message[makeinterp::levcut, levels, cutoff]) &&
   (0 < min < max || Message[makeinterp::minmax, min, max]) &&
   (0 < 10 step < max - min || Message[makeinterp::step]) :=
 Module[{egtab, x},
  egtab = Table[{N@x, egtrans[ng, N@x, cutoff]}, {x, min, max, step}];

  {energyinterp[
    Table[
     Interpolation[Cases[egtab, {x_, {e_, de_, _, _}} :> {{x}, e[[i]], de[[i]]}],
      InterpolationOrder → OptionValue[InterpolationOrder]],
     {i, 3, levels}], ng, levels - 1, mms],
   couplinginterp[Table[Interpolation[
      Cases[egtab, {x_, {_, _, g_, dg_}} :> {{x}, g[[i, j]], dg[[i, j]]}],
      InterpolationOrder → OptionValue[InterpolationOrder]],
     {i, levels}, {j, levels}], ng, levels - 1, mms]}
 ]
```

## Check the transmon calculations

### What does it look like?

Spectrum vs $E_J / E_C$

```
Manipulate[Module[
  {x = Transpose[Table[egtrans[ng, ejec, cut][[1, ;; ls]], {ejec, 10, 100, 4},
      {ng, {0.0001, 0.5001}}], {2, 3, 1}]},
  Show[
   Table[
    ListLinePlot[Flatten[x[[δ + 1 ;;]] - x[[ ;; -(δ + 1)]], {1, 3}], PlotRange → All,
     AxesLabel → {"EJ/EC", "Ei-Ej"}, Filling → Table[2 n - 1 → {2 n}, {n, ls - δ}],
     DataRange → {20, 100}], {δ, 1, ls - 1}]]],
 {{ls, 4, "# levels to show"}, 3, cut, 1},
 {{cut, 15, "Charge basis cutoff"}, 10, 30, 1}]
```

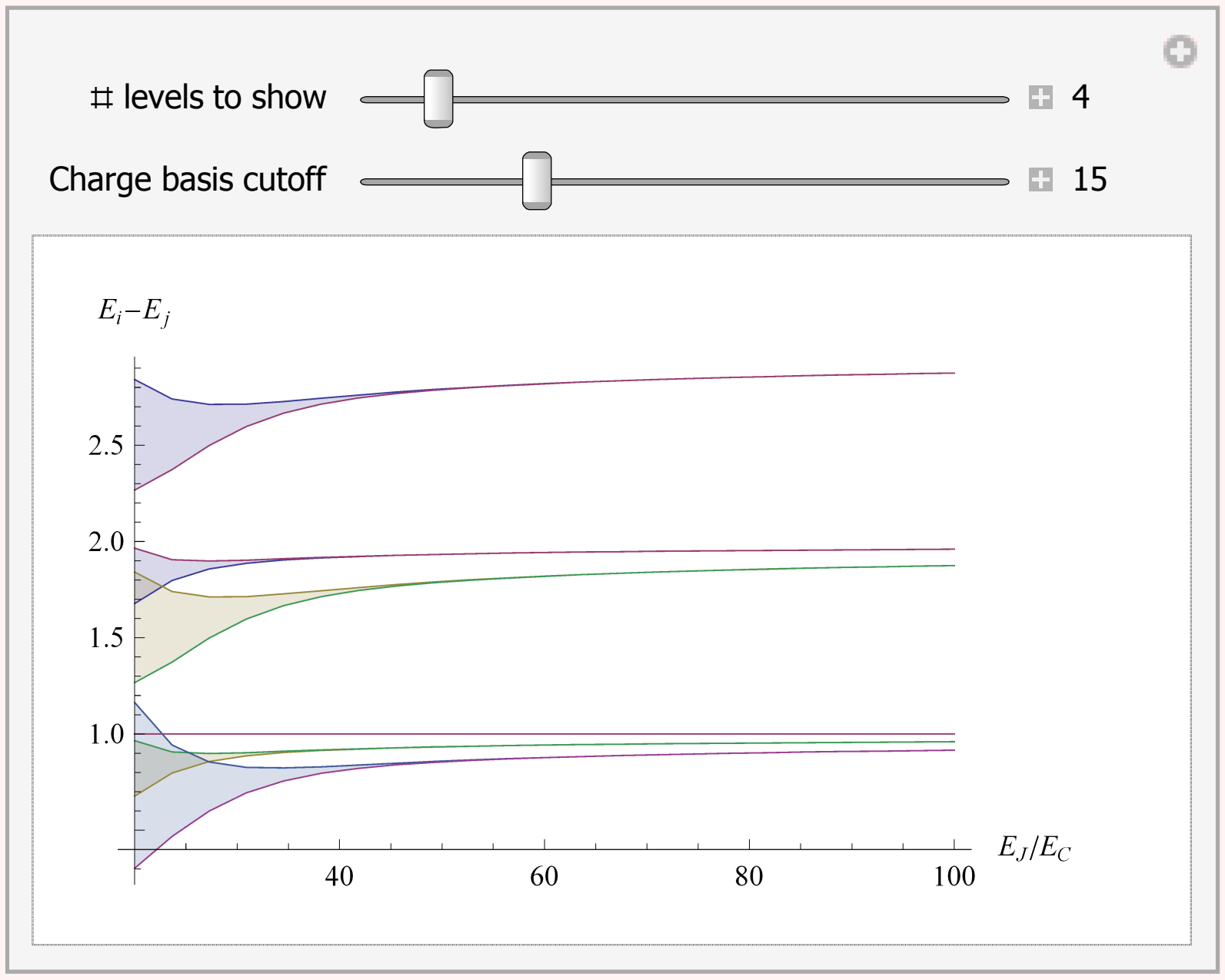

Energy levels and spectra vs $n_g$

```
Manipulate[Module[
  {x = Table[egtrans[ng, egec, cut][[1, ;; ls]], {ng, -.4999, .5, .05}], xxx},
  xxx = Flatten[Table[x, {3}], 1]ᵀ;
  GraphicsRow[{
    ListLinePlot[xxx, AxesLabel → {"n_g", "E_i"}],
    ListLinePlot[xxx[[δ + 1 ;;]] - xxx[[ ;; -(δ + 1)]], PlotRange → All,
     AxesLabel → {"n_g", With[{δ = δ}, HoldForm["E"_i - "E"_i-δ]]}]}]],
 {{egec, 50., "E_J/E_C"}, 10., 100.},
 {{ls, 4, "Levels to show"}, 3, cut, 1},
 {{cut, 15, "Charge basis cutoff"}, 10, 30, 1},
 {{δ, 1, "# of quanta of energy to exchange"}, 1, ls - 1, 1}]
```

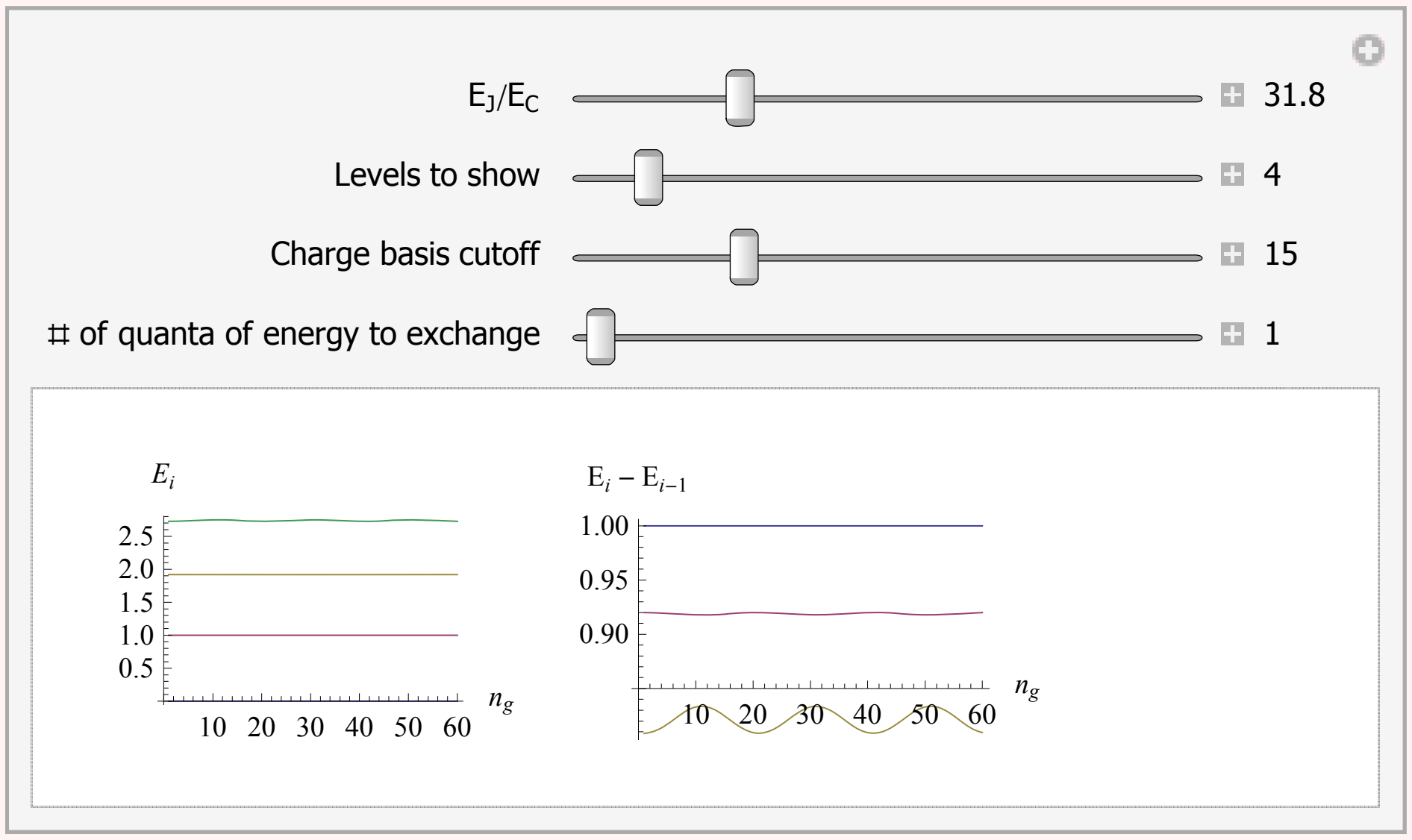

- **Choose a cutoff**

```
et[ng_?NumericQ, EjEc_?NumericQ, cutoff_?IntegerQ] := Module[{e, v, v2},
   {e, v} = Eigensystem[SparseArray[
      {{i_, i_} :→ 4 (i - Floor[cutoff / 2] - ng - 1)^2,
       {i_, j_} /; Abs[i - j] == 1 → -EjEc / 2}, {cutoff, cutoff}]];
   v2 = v[[Ordering[e]]];
   v2.DiagonalMatrix[Table[m - Floor[cutoff / 2], {m, 0, cutoff - 1}]].v2ᵀ];
```

```
Manipulate[Module[{ccm = 60, etm, ll = 4},
  etm = Abs[et[.5, ejec, ccm][[ ;; ll,  ;; ll]]];
  GraphicsGrid@Map[ListLinePlot[#, PlotRange → {0, 10^-12},
      Ticks → Dynamic@{{{marker, "", {.5, 0}}}, None}, ClippingStyle → Red] &,
    Transpose[Table[Abs[Abs[et[.5, ejec, cc][[ ;; ll,  ;; ll]]] - etm],
      {cc, 2 ll + 1, ccm}], {3, 1, 2}], {2}]],
 {{ejec, 72}, 30, 130}, {{marker, 30}, 10, 60, 1}]
```

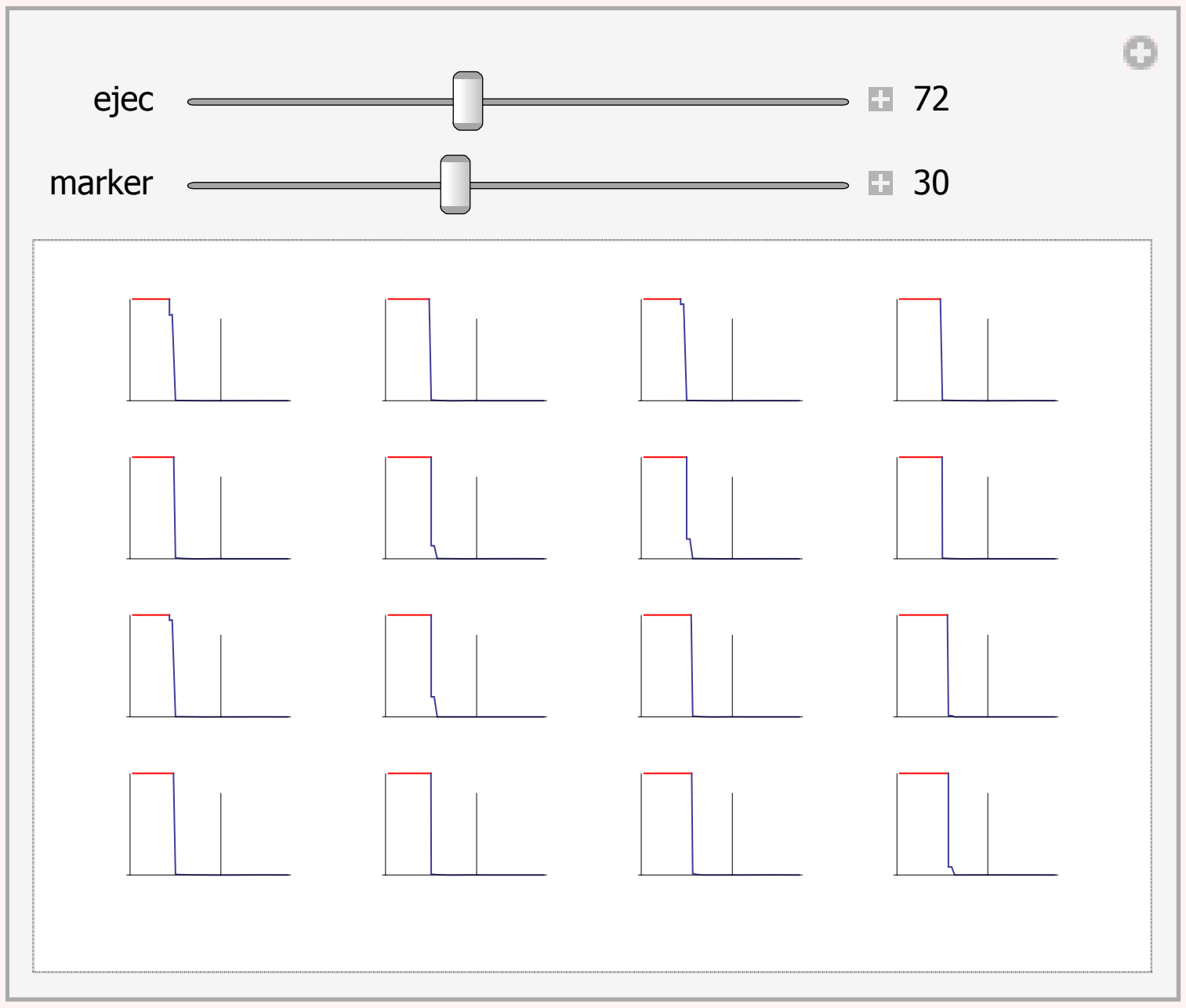

- **Mathieu function calculation, for comparison**

```
k[m_, ng : _] := Sum[(Round[2 ng + l / 2] ~Mod~ 2)
    (Round[ng] - l (-1)^m ((m + 1) ~Quotient~ 2)), {l, {-1, 1}}];
aν_[x_] := MathieuCharacteristicA[ν, x];
Em_[ng : _, EJ : _, EC : _] := EC a-2 (ng-k[m,ng])[-EJ/(2 EC)];
```

- **Quantities derived from the transmon solutions**

$$\epsilon_{m_}[E_J:_, E_C:_] := (-1)^m E_C \frac{2^{4m+5}}{m!} \sqrt{\frac{2}{\pi}} \left(\frac{E_J}{2 E_C}\right)^{\frac{m}{2}+\frac{3}{4}} e^{-\sqrt{8 E_J/E_C}}$$

$$\tilde{\epsilon}_{m_}[E_J:_, E_C:_] := \mathrm{Abs}[E_m[0.0001, E_J, E_C] - E_m[0.4999, E_J, E_C]]$$

```
ϵ[_] := Σ_{m}^{levels} ϵ_m[72, 1] matrix@op[basis, qubit, m]
```

```
E_{m_,n_} = E_m[.0001, EjEc, 1] - E_n[.0001, EjEc, 1];
En[EjEc_?NumericQ]_{m_,n_} := Module[{q = etrans[.5, EjEc]}, q[[m + 1]] - q[[n + 1]]]
```

```
H_Q[EjEc_] := Σ_{m=0}^{levels-1} (En[EjEc]_{m0} / En[EjEc]_{10}) matrix@op[basis, qubit, m + 1];
```

- **Asymptotic expression compared with exact**

```
Show[LogPlot[Evaluate@Table[Tooltip[Abs[ϵ_m[E_J, 1]], m], {m, 0, 5}],
   {E_J, 10, 100}, PlotRange → All],
  LogPlot[Evaluate@Table[Tooltip[Abs[ϵ̃_m[E_J, 1]], m], {m, 0, 5}],
   {E_J, 10, 100}, PlotRange → All, PlotStyle → Dashed],
  Plot[x, {x, 0, 100}]]
```

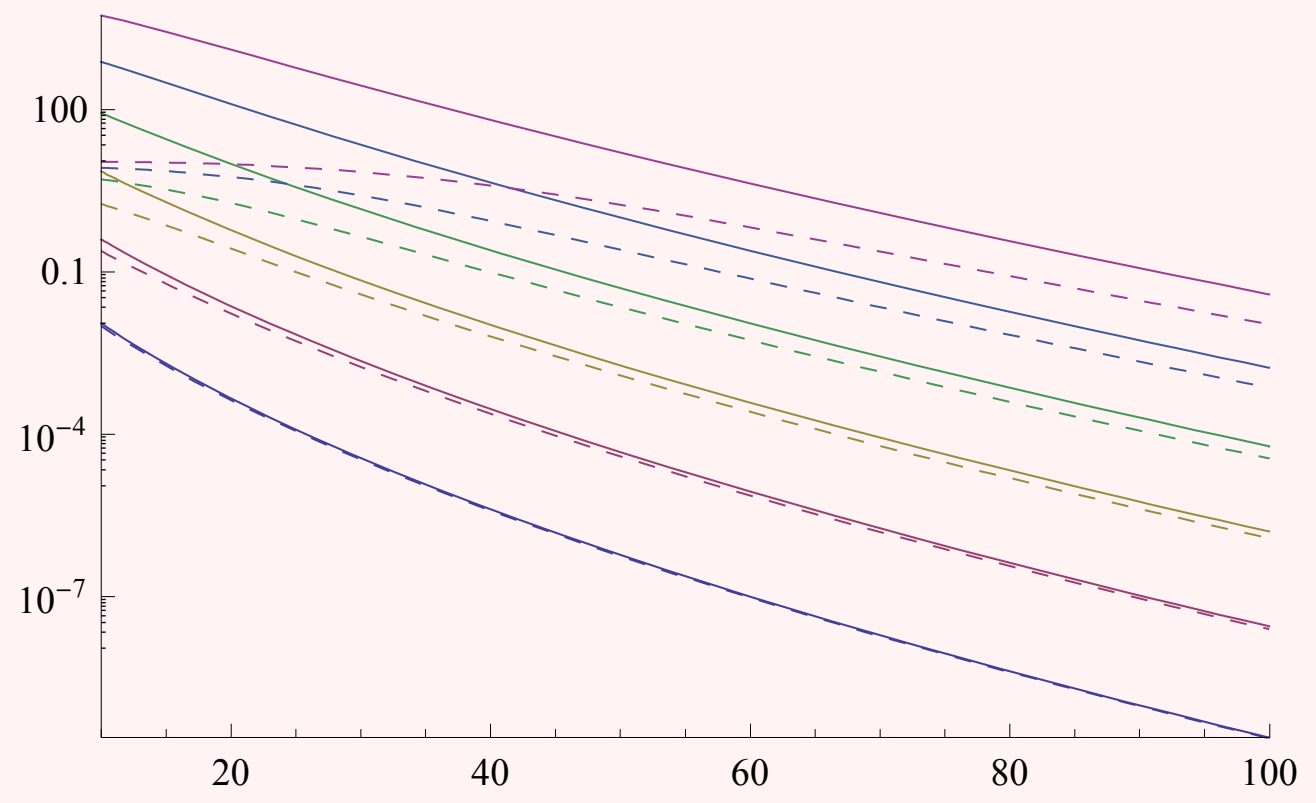

### Transmon dispersion

```
Plot[Evaluate@Table[{
    Tooltip[Em[0.00001, EJ, 1] - E0[0.00001, EJ, 1], m],
    Tooltip[Em[0.4999, EJ, 1] - E0[0.00001, EJ, 1], m]}, {m, 1, 7}],
 {EJ, 0, 100}, PlotRange → All, Filling → Table[2 n - 1 → {2 n}, {n, 7}]]
```

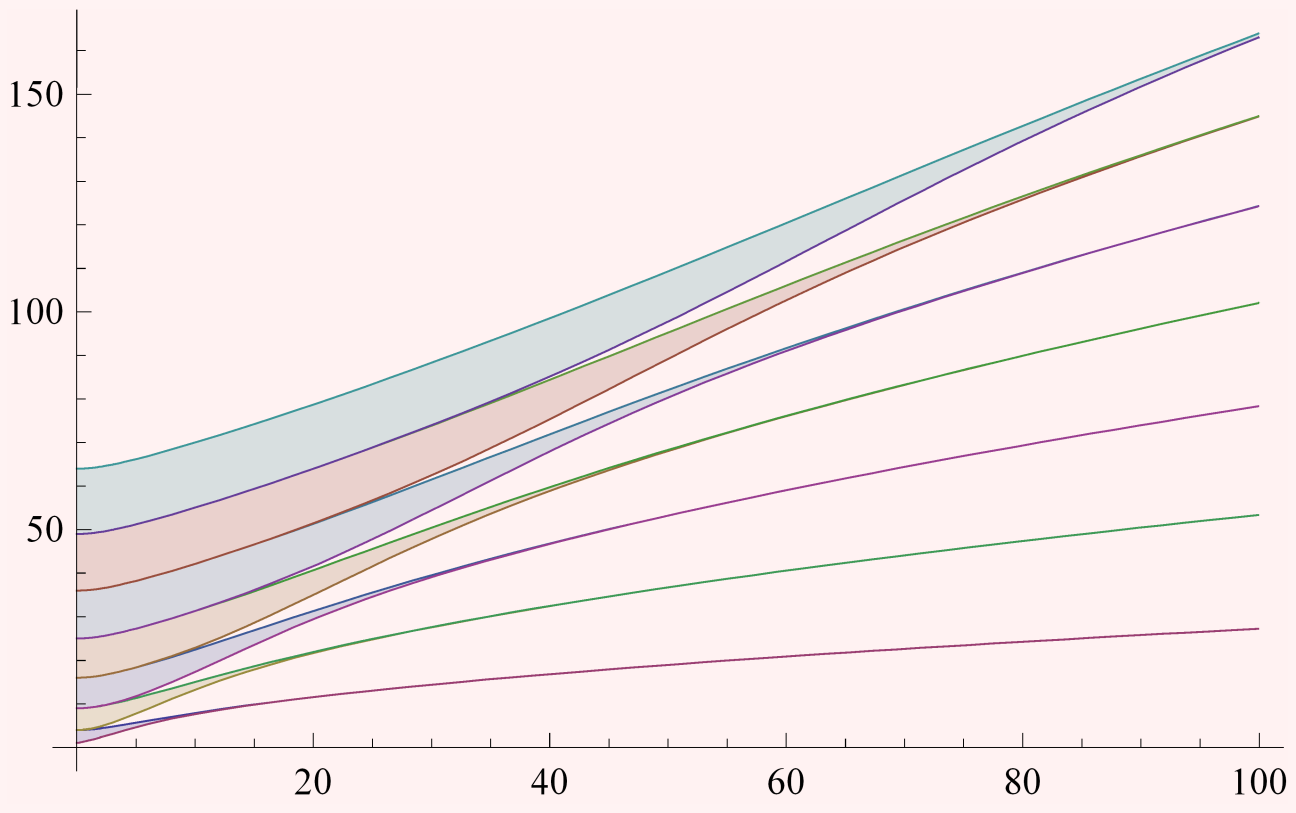

## Solve the system

### Parameters

NB: These quantites are protected because everything here depends on them being symbols.
They should only have values assigned to them in a Block[] or similar structure.

```
params =
  {ωr, ωd, δ, g, ξ, ejec, γϕ, (*γϕ2,*)γ, pm, κ(*,pf1,pf2,pf3,pf4,pf5*)};
ωr::usage = "ωr is cavity frequency";
ωd::usage = "ωd is the frequency of the drive";
δ::usage = "δ is given by ωr-ωqubit == δ";
g::usage =
  "g is the coupling strength g01 (between the 0↔1 transition of the
    transmon and the cavity annihilation operator)";
ξ::usage = "ξ is the drive strength";
ejec::usage = "ejec is the EJ/EC ratio for the transmon";
γϕ::usage = "γϕ is the transmon dephasing strength";
γ::usage = "γ is the transmon relaxation rate";
κ::usage = "κ is the cavity relaxation rate";
pf1* ^= pf1;
pf2* ^= pf2;
pf3* ^= pf3;
pf4* ^= pf4;
pf5* ^= pf5;
Protect[Evaluate@params];
$Assumptions = params ∈ Reals && ℏ > 0;
```

## Hamiltonian

### Do the normal transmon interpolations

This is the standard interpolation:

```
$maxlevels::usage =
  "$maxlevels is the number of transmon levels calculated so
    far. We need to recalculate the interpolations
    and some other stuff if we want to go higher...";
Unprotect[$maxlevels];
$maxlevels = 8;
Protect[$maxlevels];

{ef1, gf1} = makeinterp[0.4999, 15, $maxlevels, {10, 200, 1}];
{ef2, gf2} = makeinterp[0.0001, 15, $maxlevels, {10, 100, 1}];
{ef1, gf1}
ef1[3][72]
```

```
{energyinterp[<>, 0.4999, 7, {10, 200, 1}],
 couplinginterp[<>, 0.4999, 7, {10, 200, 1}]}
```

```
2.84936
```

## ▪ Subspace

Set up the basis states (sstates), the projectors onto the degenerate subspaces (psstates) and the size of the Hilbert space for subsequent calculations (nn):

```
ClearAll["bket*"];
sstates :=
 Table[Symbol["bket" <> qubitletter⟦j⟧ <> ToString[i - j]], {i, levels}, {j, i}]
states := Flatten@sstates;
Array[
  (Evaluate@Symbol["bket" <> qubitletter⟦#1⟧ <> ToString[#2 - 1]] := basisKet[
       qubit, #1 ] · basisKet[cavity, #2]) &, {$maxlevels, $maxlevels}];
psstates := projector /@ sstates;
nn := Length@states;
```

## ▪ Set up the Hamiltonian

$H_Q$ is in units of $\omega_{01}$

```
setlevels[3]
{ef, gf} = {ef1, gf1};
```

```
System set to dimension: 9
```

```
H_Q := ℏ Sum[ef[m][ejec] matrix@op[basis, qubit, m + 1], {m, 0, levels-1}];
(*like (â·σ⁺+â†·σ⁻) *)
ĝ := Sum[gf[i - 1, i][ejec] matrix@op[basis, qubit, i, i + 1], {i, levels-1}];
H_g := ℏ g (# + hc[#] &@ (ĝ · â†));
```

We are in the rotating frame and make the RWA:

```
H_d := ℏ ξ (â + â†);
(* H₀=(ω₀₁H_Q-ω_d q̂)+ℏ (ω_c- ω_d)n̂+g H_g *)
H₀ := ((ω_r - δ) H_Q - ℏ ω_d q̂) + ℏ (ω_r - ω_d) n̂ + H_g;
```

Here's the matrix version of the Hamiltonian (a list of the matrices in each n-excitation subspace, n=1...levels) :

```
H0s := Table[Simplify@Table[trace[hc[sstates⟦n, i⟧] · H₀ · sstates⟦n, j⟧],
     {i, n}, {j, n}], {n, levels}];
```

### Diagonalizing the Hamiltonian

```
diagfns::usage =
  "diagfns[] returns {energies[...],vectors[...]} functions.";
diagfns[] := Block[{ωd = 0, ℏ = 1, ef = ef1, gf = gf1,
   Eigenvalues, Eigenvectors, PadRight, map},
  With[{H0s = H0s, levels = levels, nn = nn},
   With[{sp = {ωr, δ, g, ejec}},
    {Function[Evaluate@sp, Evaluate[Eigenvalues /@ evalinterp[H0s]]],
     Function[Evaluate@sp,
       Evaluate[Table[With[{ic = i (i - 1) / 2}, PadRight[#, nn, 0., ic] &~
           map~Eigenvectors[H0s[[i]]]], {i, levels}]]] /. map → Map}]]]
```

```
diagfns2::usage =
  "diagfns2[] returns {energies[...],vectors[...]} functions.";
diagfns2[] := Block[{ℏ = 1, ef = ef1, gf = gf1,
   Eigenvalues, Eigenvectors, PadRight, map},
  With[{H0s = H0s, levels = levels, nn = nn},
   With[{sp = {ωr, ωd, δ, g, ejec}},
    {Function[Evaluate@sp, Evaluate[Eigenvalues /@ evalinterp[H0s]]],
     Function[Evaluate@sp,
       Evaluate[Table[With[{ic = i (i - 1) / 2}, PadRight[#, nn, 0., ic] &~
           map~Eigenvectors[H0s[[i]]]], {i, levels}]]] /. map → Map}]]]
```

Show the energy levels and transitions:

```
transAnn::usage =
  "transAnn[i1,j1,i2,j2] is a tag representing the transition
    between the j1th level of the i1-excitation subspace
    and the j2th level of the i2-excitation subspace";
levelAnn::usage = "levelAnn[i,j] is a tag representing
    the jth level of the i-excitation subspace";
Protect[transAnn, levelAnn];

$hilited::usage =
  "$hilited contains the tag of the currently selected item";

flash::usage = "flash[list,t] flashes
    between styles in the list l, over a total time t";
flash[l_List, t_] := l[[Clock[{1, Length@l, 1}, t]]];
flashing[s_] :=
  flash[{Directive[s, Dashed], Directive[s, Dashing[{}]]}, 1];
maybeflashing[a_, s_] := Dynamic@If[a === $hilited, flashing@s, s];
handlemouse[g_] :=
  EventHandler[g, "MouseClicked" :> ($hilited = MouseAnnotation[]),
   PassEventsDown -> Automatic];

With[{x1 = 1, x2 = 2, x4 = 0.2`, x5 = 0.15`, x6 = 0.1`, x7 = 0.2`},
  leveldiagram[e0_List, e1_List, ls_Integer] :=
   DynamicModule[{q1, q2},
    {q1, q2} = (5 (ls - 1) # / #[[-1, -1, -1]]) &@{e0, e1};
    handlemouse@
     Graphics[Dynamic@Flatten[{Antialiasing -> False,
         Table[{Line[{{0, q1[[i, j]]}, {x1, q1[[i, j]]}}]}, {i, ls}, {j, i}],
         Table[{Gray,
           Line[{{x1, q1[[i, j]]}, {x2, q2[[i, j]]}}]}, {i, ls}, {j, i}],
         Module[{xx = x2 - x4 - x5 - x6 - x7},
          Flatten[{Table[
              xx += KroneckerDelta[i, j1, j2, 1] x4 +
                KroneckerDelta[j1, j2, 1] x5 + KroneckerDelta[j2, 1] x6 + x7;
              With[{s = transstyle[i, j1, i + k, j2], a =
                  transAnn[i, j1, i + k, j2]},
               {maybeflashing[a, s],
                Annotation[Line[
                   {{xx, q2[[i, j1]]}, {xx, q2[[i + k, j2]]}}], a, "Mouse"]}],
              {k, ls - 1}, {i, ls - k}, {j1, i}, {j2, i + k}],
             Table[
              With[{s = levelstyle[i, j], a = levelAnn[i, j]}, {{{maybeflashing[
                     a, s], Annotation[Line[{{x2, q2[[i, j]]}, {xx, q2[[i,
                          j]]}}], a, "Mouse"]}}}], {i, ls}, {j, i}]}, 4]]}, 1]]]];
```

```
transstyle[i1_, j1_, i2_, j2_] := Directive[
   Flatten@{ColorData[1][j1], If[i1 == 1 && i2 == 2, {Thick, Black}, {}],
     If[MatchQ[$hilited, levelAnn[i1, j1] | levelAnn[i2, j2]], Red, {}]}];
levelstyle[i_, j_] := ColorData[1][j];
```

```
setlevels[4]
{energiestt, vectorstt} = diagfns[];
```

```
System set to dimension: 16
```

```
Manipulate[
 DynamicModule[{evtab},
  evtab = Table[{g, energiestt[wr0, d0, g / 2, ej0]}, {g, 0, g0 + .2, g0 / 10}];
  Deploy@GraphicsRow[
    {handlemouse@Graphics[
       Dynamic@Flatten[Table[
          With[{s = transstyle[i, j1, i + k, j2], a = transAnn[i, j1, i + k, j2]},
           {maybeflashing[a, s],
            Annotation[Line[{evtab[[All, 1]], (evtab[[All, 2, i + k, j2]] -
                   evtab[[All, 2, i, j1]]) / k}^T], a, "Mouse"]}],
          {k, ls - 1}, {i, ls - k}, {j1, i}, {j2, i + k}], 3],
       Frame → True, AspectRatio → 1, PlotRangeClipping → True, PlotRange →
        Dynamic[If[zoom, {{g0 - .1, g0 + .1}, {6.9, 7.4}}, All]], GridLines →
        {{g0}, (*{7.365,7.11,7.175,7.31}*){7.355, 7.103, 7.168, 7.093}}],
     leveldiagram[energiestt[wr0, d0, 0, ej0],
      energiestt[wr0, d0, g0 / 2, ej0], ls]}]],
 {{ls, 3}, 2, levels, 1},
 {{wr0, 6.9077}, 6.89, 6.92},
 {{d0, 0.}, -.5, .1},
 {{g0, .346}, 0, .5},
 {{ej0, 72.}, 20, 100},
 {zoom, {True, False}},
 TrackedSymbols → Full,
 Bookmarks → {
   "get Ec" :→
    {ls = 3, wr0 = 6.917458, d0 = -.44265, g0 = 93.88 / 1000, ej0 = 52.12},
   "expt" :→ {ls = 3, wr0 = 6.915, d0 = -.006, g0 = 93.88 / 1000, ej0 = 50}}]
```

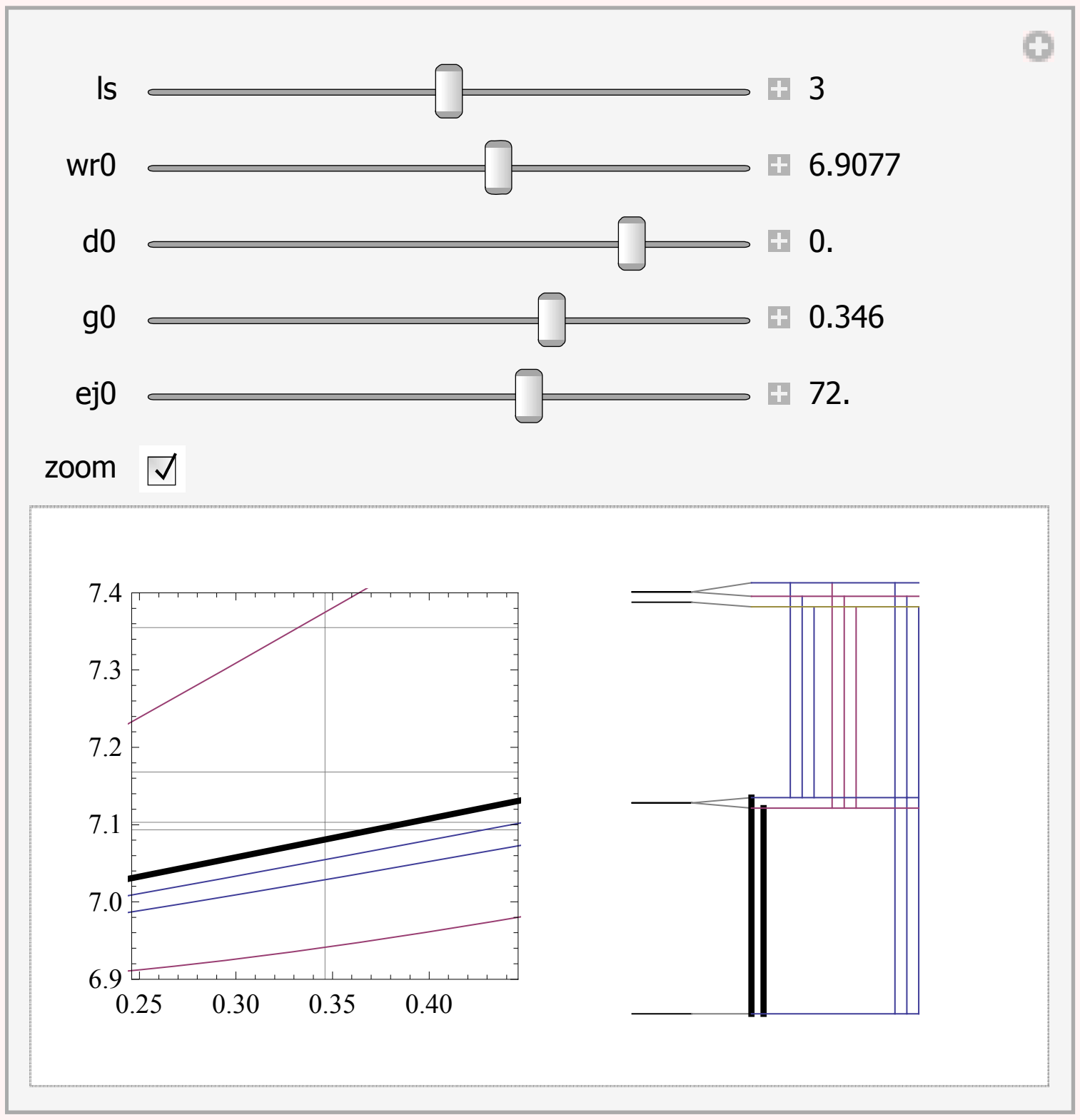

```
Manipulate[
 DynamicModule[{evtab},
  evtab = Table[{d0, energiestt[wr0, d0, g0 / 2, ej0]}, {d0, -1, 1, .01}];
  Deploy@GraphicsRow[{
     handlemouse@Graphics[
       Dynamic@Flatten[Table[
          With[{s = transstyle[i, j1, i + k, j2], a = transAnn[i, j1, i + k, j2]},
           {maybeflashing[a, s],
            Annotation[Line[{evtab[[All, 1]], (evtab[[All, 2, i + k, j2]] -
                   evtab[[All, 2, i, j1]]) / k}ᵀ], a, "Mouse"]}],
          {k, ls - 1}, {i, ls - k}, {j1, i}, {j2, i + k}], 3],
       Frame → True, AspectRatio → 1, PlotRangeClipping → True, PlotRange →
        If[zoom, {All, {6.5, 7.2}}, All], GridLines → {{d0}, None}],
     leveldiagram[energiestt[wr0, d0, 0, ej0],
      energiestt[wr0, d0, g0 / 2, ej0], ls]}]],
 {{ls, 3}, 2, levels, 1},
 {{wr0, 6.9077}, 6, 7},
 {{d0, 0.}, -1, 1},
 {{g0, .346}, 0, .5},
 {{ej0, 72.}, 20, 100},
 {zoom, {True, False}},
 TrackedSymbols → Full]
```

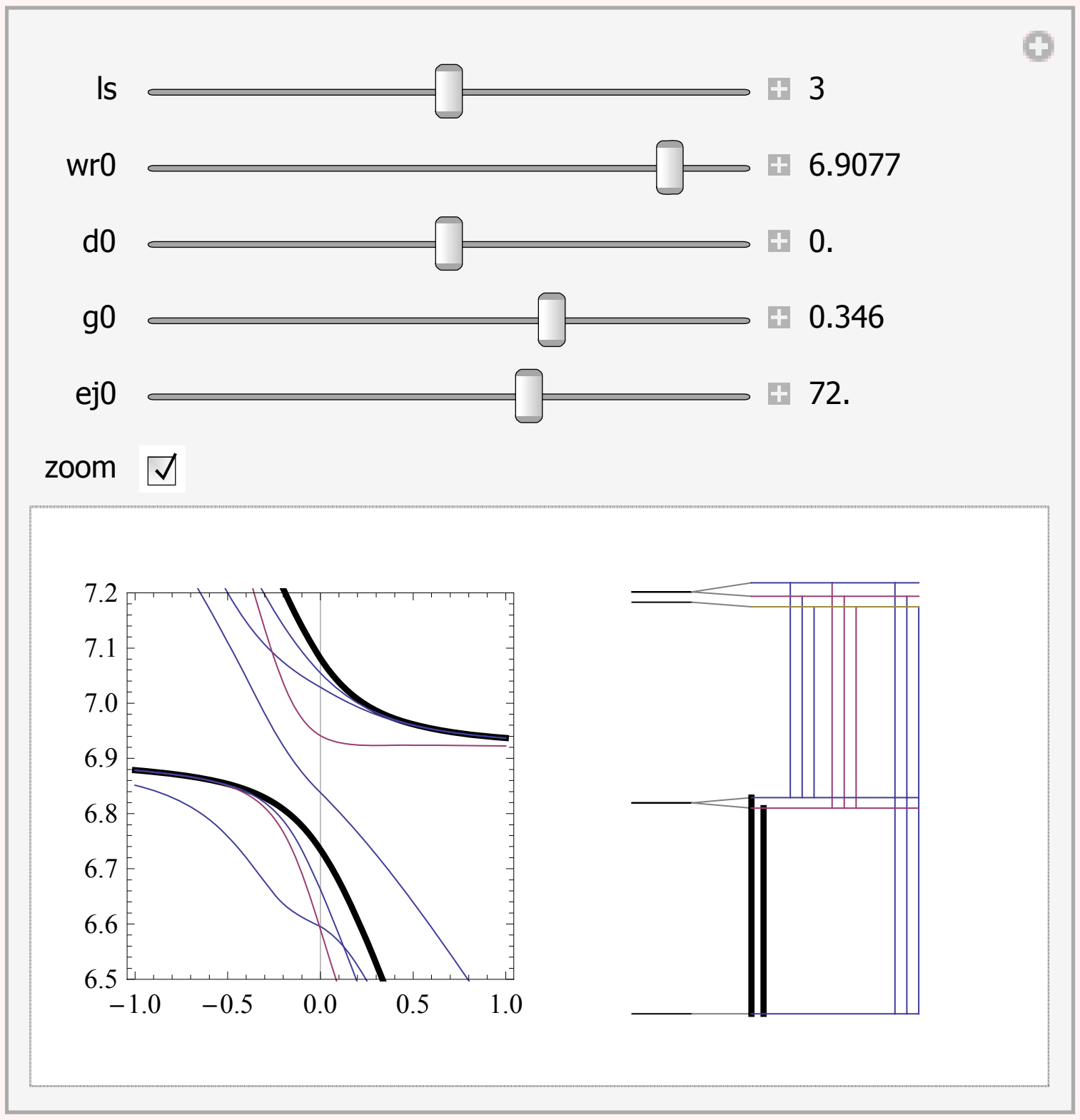

## Density matrices

### Lindblad operators

Here's the Lindblad form of the RHS of the master equation for $\dot{\rho}$:

```
pr := projector[states]
```

```
ℒ1[ρ_?operatorMatrixQ] := -(i/ℏ) commutator[H0 + Hd, ρ] +
  κ 𝒟[â][ρ] + γ 𝒟[σ⁻][ρ] + γ pm 𝒟[pr · σ⁺ · pr][ρ] + γϕ 𝒟[q̂][ρ] / 2
```

```
ℒ2[ρ_?operatorMatrixQ] :=
 -(i/ℏ) commutator[H0 + Hd, ρ] + κ 𝒟[â][ρ] + γ 𝒟[ĝ][ρ] + γ pm 𝒟[pr · hc[ĝ] · pr][ρ] + 10^7
  γϕ 𝒟[ Sum_{m=0}^{levels-1} (ef1[m][ejec] - ef2[m][ejec]) matrix@op[basis, qubit, m + 1]][ρ]
```

```
ℒ3[ρ_?operatorMatrixQ] :=
 -(ⅈ/ℏ) commutator[H0 + Hd, ρ] + κ 𝒟[â][ρ] + γ 𝒟[ĝ][ρ] + κ pm 𝒟[pr · â† · pr][ρ] + 10^7 γϕ
   𝒟[Sum[(ef1[m][ejec] - ef2[m][ejec]) matrix@op[basis, qubit, m + 1], {m, 0, levels-1}]][ρ]
```

```
ℒ4[ρ_?operatorMatrixQ] := -(ⅈ/ℏ) commutator[H0 + Hd, ρ] + κ 𝒟[â][ρ] +
  γ 𝒟[ĝ][ρ] + γ pm 𝒟[pr · hc[ĝ] · pr][ρ] + κ pm 𝒟[pr · â† · pr][ρ] + 10^7 γϕ
   𝒟[Sum[(ef1[m][ejec] - ef2[m][ejec]) matrix@op[basis, qubit, m + 1], {m, 0, levels-1}]][ρ]
```

```
ℒ5[ρ_?operatorMatrixQ] :=
 -(ⅈ/ℏ) commutator[H0 + Hd, ρ] + κ 𝒟[â][ρ] + γ 𝒟[ĝ][ρ] + γ pm 𝒟[pr · hc[ĝ] · pr][ρ] +
  κ pm 𝒟[pr · â† · pr][ρ] + 10^7 𝒟[Sum[pϕ[[m]] matrix@op[basis, qubit, m + 1], {m, 1, levels-1}]][ρ]
pϕ = {pf1, pf2, pf3, pf4, pf5};
```

```
ℒ6[ρ_?operatorMatrixQ] := -(ⅈ/ℏ) commutator[H0 + Hd, ρ] + κ 𝒟[â][ρ] +
  γ 𝒟[ĝ][ρ] + γ pm 𝒟[pr · hc[ĝ] · pr][ρ] + κ pm 𝒟[pr · â† · pr][ρ] +
  10^7 γϕ 𝒟[Sum[(ef1[m][ejec] - ef2[m][ejec]) matrix@op[basis, qubit, m + 1], {m, 0, levels-1}]][
    ρ] + γϕ2 𝒟[q̂][ρ] / 2
```

Now put it in matrix form and project onto our reduced Hilbert space:

```
lindblad::trnz = "The trace of ρ̇ was not zero!";
lindblad::usage =
  "lindblad[ℒ] returns {ρ̂, ρij, ρ̇ij} for a given Lindblad operator ℒ[ρ̂]";

lindblad[ℒ_] := With[{nn = nn, states = states},
   Module[{ρs, ρ, Π, ℒρ, Πℒρ, δρ},
    ρs =
     Table[Symbol["ρ" <> ToString[i] <> "x" <> ToString[j]], {i, nn}, {j, nn}];
    ρ = Simplify[Sum[ρs[[i, j]] states[[i]] · hc[states[[j]]], {i, nn}, {j, nn}]];
    Π = Simplify@projector[states];
    ℒρ = ℒ[ρ];
    Πℒρ = Π · ℒρ · Π;
    δρ = Table[trace[hc[states[[i]]] · Πℒρ · states[[j]]], {i, nn}, {j, nn}];
    If[! TrueQ[Chop@FullSimplify@Tr@δρ == 0], Message[lindblad::trnz]];
    {ρ, ρs, δρ}]];
```

## Steady state solver

```
steadystatevalue[op_?operatorMatrixQ,
  pt : {(_? (MemberQ[params, #] &) → _?NumericQ) ...}] :=

 Block[Evaluate[Join[{sol, vparms}, params]],
  Evaluate[params] = params /. pt;
  vparms = Select[params, ! NumericQ[#] &];

  lusolve := 0^0;
  oldvec := 0^0;

  With[{sparms = Sequence @@ vparms, nn = nn},
   Module[{crys = CoefficientArrays[
       {Tr@ρs - 1} ~ Join ~ Rest[Flatten[ddd = δρ]], Flatten@ρs],

      M1, M2, c1, c2, cf1, cf2, cff1, cff2, M1c, M2c, cfm1,
      cfm2, cffm1, cffm2, ope, opc1, opc2, opm, ss, rparms, nparms,
      repparms, ρte, dρte, nrm, bb, cb, cfb, cffb, bbc, occ},

    nparms := Sequence @@ (Pattern[#, _?NumericQ] & /@ vparms);
    rparms = {#, _Real} & /@ vparms;

    M1 = FullSimplify[crys[[1]]];
    M2 = FullSimplify[-crys[[2]]^TT];
    bb = FullSimplify[Flatten[D[M2, {vparms}], {{3, 1}, {2}}]];

    c1 = M1 /. HoldPattern@SparseArray[__, {__, a_}] :→ a;
    c2 = M2 /. HoldPattern@SparseArray[__, {__, a_}] :→ a;
    cb = bb /. HoldPattern@SparseArray[__, {__, a_}] :→ a;

    repparms = Thread[vparms → Unique[vparms]];
    cf1 = Compile[Evaluate@rparms, Evaluate@Developer`ToPackedArray@
         evalinterp@c1, CompileOptimizations → All] /. repparms;
    cf2 = Compile[Evaluate@rparms, Evaluate@Developer`ToPackedArray@
         evalinterp@c2, CompileOptimizations → All] /. repparms;
    cfb = Compile[Evaluate@rparms, Evaluate@Developer`ToPackedArray@
         evalinterp@cb, CompileOptimizations → All] /. repparms;

    M1c = (M1 /. HoldPattern@SparseArray[a__, {b__, c_}] :→
        SparseArray[a, {b, cff1[sparms]}]);
    M2c = (M2 /. HoldPattern@SparseArray[a__, {b__, c_}] :→
        SparseArray[a, {b, cff2[sparms]}]);
    bbc = (bb /. HoldPattern@SparseArray[a__, {b__, c_}] :→
        SparseArray[a, {b, cffb[sparms]}]);

    cfm1 = Compile[Evaluate@rparms,
      Evaluate@Developer`ToPackedArray@evalinterp@Normal@M1];
    cfm2 = Compile[Evaluate@rparms, Evaluate@
```

```
    Developer`ToPackedArray@evalinterp@Normal@M2];

 ope = trace[op · ρ];
 occ = {sparms,
   ρs /. Thread[Flatten@ρs → Table[ss[sol, i], {i, Length@Flatten@ρs}]]};
 opm = ope /. Thread[Flatten@ρs → Table[ss[sol, i],
      {i, Length@Flatten@ρs}]];
 nrm = Total@Diagonal@ρs /. Thread[Flatten@ρs →
     Table[ss[sol, i], {i, Length@Flatten@ρs}]];
 {opc1, opc2} = CoefficientArrays[ope, Flatten@ρs];

 ReleaseHold[
  Hold[
    ρte[nparms] := Module[{sol, m1, o1},
      mmm = mat; (*m1=mat;
      o1=off;
      Quiet@Check[
        oldvec=sol=LinearSolve[m1,Normal@o1,Method→
            {"Krylov","Preconditioner"→(lusolve[#]&),MaxIterations→10,
             "StartingVector"→oldvec,Tolerance→10^-4}],

        numlu++;
        lusolve=LinearSolve[m1,Method→"Multifrontal"];
        oldvec=sol=lusolve[o1]];*)
      sol = LinearSolve[mat, off];
      Sow[occ1];
      result];

    dρte[nparms] := Module[{y, c, sol, ls},
      ls = LinearSolve[mat, Method → "Multifrontal"];
      c = off;
      y = ls[c];
      sol = -ls[Partition[B.y, nn^2]^T];
      {c2.y + c1, c2.sol}
     ];

   ] /. {HoldPattern[ρ] → ρ,
    HoldPattern@off → M1c,
    HoldPattern@mat → M2c,
    HoldPattern@B → bbc,
    HoldPattern@offm → cffm1[sparms],
    HoldPattern@matm → cffm2[sparms],
    HoldPattern@result → opm,
    HoldPattern@occ1 → occ,
    HoldPattern@normalize → nrm,
    HoldPattern@c1 → opc1,
    HoldPattern@c2 → opc2,
    HoldPattern@nn → nn
   } /.
```

```
      {ss → Part,
       cff1 → cf1,
       cff2 → cf2,
       cffb → cfb,

       cffm1 :→ cfm1,
       cffm2 :→ cfm2}];
   {vparms, ρte, dρte}]]]
```

# Copyright Permissions

- Figures 2.7 and 2.8 reproduced with permission from:
  Jens Koch, Terri M. Yu, Jay Gambetta, A. A. Houck, D. I. Schuster, J. Majer, Alexandre Blais, M. H. Devoret, S. M. Girvin, and R. J. Schoelkopf, *Physical Review A* **76**, 042319 (2007).
  Copyright (2007) by the American Physical Society.

- Figure 3.1 reproduced (with altered formatting) with permission from:
  A. A. Houck, J. A. Schreier, B. R. Johnson, J. M. Chow, Jens Koch, J. M. Gambetta, D. I. Schuster, L. Frunzio, M. H. Devoret, S. M. Girvin, and R. J. Schoelkopf, *Physical Review Letters* **101**, 080502 (2008).
  Copyright (2008) by the American Physical Society.

- Figures 4.2 to 4.8 and 4.10 to 4.14 reproduced (with altered formatting) with permission from:
  Lev S. Bishop, J. M. Chow, Jens Koch, A. A. Houck, M. H. Devoret, E. Thuneberg, S. M. Girvin, and R. J. Schoelkopf, *Nature Physics* **5**, 105–109 (2009).
  Copyright © 2009, Nature Publishing Group.

- Figures 5.1 to 5.6 reproduced with permission from:
  Lev S. Bishop, Lars Tornberg, David Price, Eran Ginossar, Andreas Nunnenkamp, Andrew A. Houck, Jay M. Gambetta, Jens Koch, Göran Johansson, Steven M. Girvin, and Robert J. Schoelkopf, *New Journal of Physics* **11**, 073040 (2009).
  © IOP Publishing Ltd and Deutsche Physikalische Gesellschaft.

- Figure 4.16 reproduced (with some curves removed) with permission from:
  Ileana Rau, Göran Johansson, and Alexander Shnirman, *Physical Review B* **70**, 054521 (2004).
  Copyright (2004) by the American Physical Society.